\documentclass[11pt,a4paper]{article}
\pdfoutput=1

\usepackage{jheppub-nosort,amsthm}

\usepackage{amsmath,amssymb,amsthm,amscd}
\usepackage{physics}
\usepackage{tikz}
\usetikzlibrary{decorations.pathmorphing}
\tikzset{node distance=2cm, auto}
\tikzset{snake it/.style={decorate, decoration=snake}}
\usepackage{epsfig}
\usepackage{psfrag,comment}
\usepackage{graphicx}
\usepackage{caption}
\usepackage{subcaption}
\usepackage{mathrsfs}
\usepackage{xcolor}
\usepackage[normalem]{ulem}
\usepackage{bm}
\usepackage{lscape}
\usepackage{diagbox}
\usepackage{upgreek}

\usepackage{url}
\newcommand\doilink[1]{\href{http://dx.doi.org/#1}{#1}}
\newcommand\arxivlink[1]{\href{http://arxiv.org/abs/#1}{#1}}

\newcommand{\CB}{{\mathcal B}}

\newcommand{\CS}{{\mathcal S}}

\def\BC{{\mathbb C}}
\def\BP{{\mathbb P}}

\newcommand{\be}{\begin{equation}}
	\newcommand{\ee}{\end{equation}}
\newcommand{\ba}{\begin{aligned}}
	\newcommand{\ea}{\end{aligned}}
\newcommand{\bea}{\begin{eqnarray}}
	\newcommand{\eea}{\end{eqnarray}}
\newcommand{\bean}{\begin{eqnarray*}}
	\newcommand{\eean}{\end{eqnarray*}}

\def\r{\right\rangle}

\def\1{\mathbf{1}}
\def\0{|\1\r}

\newcommand{\rme}{{\mathrm{e}}}
\newcommand{\rmi}{{\mathrm{i}}}
\newcommand{\rmd}{{\mathrm{d}}}

\def\XXint#1#2#3{{\setbox0=\hbox{$#1{#2#3}{\int}$}
		\vcenter{\hbox{$#2#3$}}\kern-.5\wd0}}

\makeatletter
\newsavebox\myboxA
\newsavebox\myboxB
\newlength\mylenA

\newcommand*\widebar[2][0.75]{%
	\sbox{\myboxA}{$\m@th#2$}%
	\setbox\myboxB\null% Phantom box
	\ht\myboxB=\ht\myboxA%
	\dp\myboxB=\dp\myboxA%
	\wd\myboxB=#1\wd\myboxA% Scale phantom
	\sbox\myboxB{$\m@th\overline{\copy\myboxB}$}%  Overlined phantom
	\setlength\mylenA{\the\wd\myboxA}%   calc width diff
	\addtolength\mylenA{-\the\wd\myboxB}%
	\ifdim\wd\myboxB<\wd\myboxA%
	\rlap{\hskip 0.8\mylenA\usebox\myboxB}{\usebox\myboxA}%
	\else
	\hskip -0.5\mylenA\rlap{\usebox\myboxA}{\hskip 0.5\mylenA\usebox\myboxB}%
	\fi}
\makeatother

\newdimen\tableauside\tableauside=1.0ex
\newdimen\tableaurule\tableaurule=0.4pt
\newdimen\tableaustep
\def\phantomhrule#1{\hbox{\vbox to0pt{\hrule height\tableaurule width#1\vss}}}
\def\phantomvrule#1{\vbox{\hbox to0pt{\vrule width\tableaurule height#1\hss}}}
\def\sqr{\vbox{%
		\phantomhrule\tableaustep
		\hbox{\phantomvrule\tableaustep\kern\tableaustep\phantomvrule\tableaustep}%
		\hbox{\vbox{\phantomhrule\tableauside}\kern-\tableaurule}}}
\def\squares#1{\hbox{\count0=#1\noindent\loop\sqr
		\advance\count0 by-1 \ifnum\count0>0\repeat}}
\def\tableau#1{\vcenter{\offinterlineskip
		\tableaustep=\tableauside\advance\tableaustep by-\tableaurule
		\kern\normallineskip\hbox
		{\kern\normallineskip\vbox
			{\gettableau#1 0 }%
			\kern\normallineskip\kern\tableaurule}%
		\kern\normallineskip\kern\tableaurule}}
\def\gettableau#1{\ifnum#1=0\let\next=\null\else
	\squares{#1}\let\next=\gettableau\fi\next}
\title{Parametric Resurgences of the Second Painlev\'e Equation and Minimal Superstrings}

\author[a]{Roberto~Vega\,}
\affiliation[a]{CAMGSD, Departamento de Matem\'atica, Instituto Superior T\'ecnico,\\ Universidade de Lisboa, 1049-001 Lisboa, Portugal\\}
\emailAdd{roberto.vega@tecnico.ulisboa.pt}

\abstract{The aim of this paper is to study the resurgent transseries structure of the inhomogeneous and $q$-deformed Painlev\'e II equations. Appearing in a variety of physical systems we here focus on their description of $(2,4)$-super minimal string theory with either D-branes or RR-flux backgrounds. In this context they appear as double scaled string equations of matrix models, and we relate the resurgent transseries structures appearing in this way with explicit matrix model computations. The main body of the paper is focused on studying the transseries structure of these equations as well as the corresponding resurgence analyses. Concretely, the aim will be to give a recursion relation for the transseries sectors and obtain the non-perturbative transmonomials---{\it i.e.}, the instanton actions of the systems. From the resurgence point of view, the goal is to obtain Stokes data. These encode how the transseries parameters jump at the Stokes lines when turning around the complex plane in order to produce a global transseries solution. The main result will be a conjectured form for the transition functions of these transseries parameters. We explore how these equations are related to each other via the Miura map. In particular, we focus on how their resurgent properties can be translated into each other. We study the special solutions of the inhomogeneous Painlev\'e II equation and how these might be encoded in the transseries parameters. Specifically, we have a discussion on the Hastings--McLeod solution and some results on special function solutions. Finally, we discuss our results in the context of the matrix model and the (2,4)-minimal superstring theory.}


\arxivnumber{}

\begin{document}

%%%%%%%%%%%%%%%%%%%%%%%%%%%%%%%%%%%%%%%%%%%%%%%%%%%%%%%%%%%%%%%%%
%%%%%%%%%%%%%%%%%%%%%%%%%%%%%%%%%%%%%%%%%%%%%%%%%%%%%%%%%%%%%%%%%
\maketitle
%%%%%%%%%%%%%%%%%%%%%%%%%%%%%%%%%%%%%%%%%%%%%%%%%%%%%%%%%%%%%%%%%
%%%%%%%%%%%%%%%%%%%%%%%%%%%%%%%%%%%%%%%%%%%%%%%%%%%%%%%%%%%%%%%%%

\vfill

\eject

\allowdisplaybreaks

%\tableofcontents

%%%%%%%%%%%%%%%%%%%%%%%%%%%%%%%%%%%%%%%%%%%%%%%%%%%%%%%%%%%%%%%%%
%%%%%%%%%%%%%%%%%%%%%%%%%%%%%%%%%%%%%%%%%%%%%%%%%%%%%%%%%%%%%%%%%
\section{Introduction}
%%%%%%%%%%%%%%%%%%%%%%%%%%%%%%%%%%%%%%%%%%%%%%%%%%%%%%%%%%%%%%%%%
%%%%%%%%%%%%%%%%%%%%%%%%%%%%%%%%%%%%%%%%%%%%%%%%%%%%%%%%%%%%%%%%%

In recent years, the study of matrix models and their dual theories has been a very fruitful field of research \cite{asv11,bk90,ciny16,djmw92,djm92,d91,d92,d03,dss90,ds90,em08,ez93,gz91,gs21,gm90,gm90b,kms03,lm93,m06,m08,msw07,m90,ps09,sv13}. In particular, one of their main features is that they are dual to some non-critical and minimal (super) string theories. These low-dimensional models, corresponding to $c\leq 1$ conformal field theories (CFTs) coupled to worldsheet quantum gravity, are very interesting systems since they constitute toy models for quantum gravity.   Furthermore, the dual matrix models are completely solvable, so one can hope to study non-perturbative effects in string theory.  In particular, we will consider the worldsheet quantum gravity to be given by super-Liouville theory.

It is well established that at the level of non-perturbative contributions, D-branes play a fundamental role. Both in minimal string theory and from the point of view of the matrix model, there are two types of branes. These are ZZ-branes, constructed by Zamolodchikov and Zamolodchikov in \cite{zz01}; and FZZT branes, discovered by Fateev, Zamolodchikov, Zamolodchikov and Teschner \cite{fzz00,t00}. ZZ-branes correspond to a boundary state obtained by quantisation of a classical solution for which the Liouville field $\phi$ tends to infinity on the boundary of the worldsheet. At the level of the matrix model, these branes correspond to configurations in which one or more eigenvalues have tunnelled to a non-perturbative saddle, leading to subleading contributions in the matrix integral. Thus, we should interpret the sum over multi-instantons in the transseries as a sum over ZZ-brane backgrounds. FZZT-branes, on the other hand, appear in minimal string theories labelled by a continuous parameter $\mu_B=x$, the boundary cosmological constant, which is interpreted as a target space coordinate. The FZZT-brane wavefunction suggests that it is a D-brane in $\phi$ space stretching from $\phi=-\infty$ (weak coupling region) and dissolving at $\phi\sim-\frac 1b \log x$, where  $b$ is a parameter of the corresponding quantum theory $\hbar=b^2$. In the matrix model picture, these FZZT-branes are described by insertions of the exponential loop operator
\begin{equation}
	\psi(x)=\rme^{\text{Tr}\log(x-M)}=\det(x-M)
\end{equation} 
into the matrix integral. We refer the reader to \cite{n04,t01} for very nice reviews on Liouville theory, while \cite{ciny16,ss04} offer all the necessary ingredients to understand the brane content in the (super)minimal string theories that we will discuss herein.

In the present paper, we shall focus on two-dimensional (2d) quantum supergravity, or pure supergravity. This is, (2,4)-superminimal $\hat c=0$ CFT coupled to $\hat c=10$ super-Liouville. This theory corresponds to 0A and 0B fermionic string theories with $\hat c=0$ and it is known as $(2,4)$-minimal superstring theory. Klebanov, Maldacena and Seiberg \cite{kms03} analysed this system in detail. Concretely, they found that the smooth transition between weakly coupled string backgrounds with D-branes, and backgrounds with Ramond--Ramond fluxes, were due to non-perturbative effects. This transition relates the $\mu>0$ region with the $\mu<0$, where $-\mu$ is the Fermi level of the string theory. This goes as follows:
\begin{align*}
	\text{0A model: }&\mu>0\Longrightarrow\text{we can have RR flux but no charged D-branes,}\\
	&\mu<0\Longrightarrow\text{we can have D-branes but no RR flux;}\\
	&\\
	\text{0B model: }&\mu>0\Longrightarrow\text{we can have D-branes but no RR flux,}\\
	&\mu<0\Longrightarrow\text{we can have RR flux but no charged D-branes.}
\end{align*}
Their interpretation of the perturbative expansion is that, for negative $\mu$, we have D-branes as boundaries in the spherical worldsheet. For positive $\mu$ these boundaries are interchanged by insertions of even number of RR fields. Therefore, our objective is to understand a bit more about the non-perturbative effects entering these smooth transitions. We will also try to give more insight on these regimes. Specifically, we shall study (or, at least, mention) all the possible solutions to the $q$-deformed\footnote{We stress to the reader that ``$q$-deformed'' is just a name which has absolutely nothing to deal with quantum groups. Instead, as we show bellow, it is simply that ``$q$'' is the convention for the parameter deforming the Painlev\'e II equation.} Painlev\'e II equation and how these can be associated with either RR fluxes or D-branes.

From the matrix model point of view, we will focus in the case of a rectangular complex matrix model. The theory is described in the double-scaling limit by the $q$-deformed Painlev\'e II equation \cite{kms03}
\begin{equation}
	\label{eq:q-PII}
	u''(z)-\frac 12 u^3(z)+\frac 12 z\,u(z)+\frac{q^2}{u^3(z)}=0.
\end{equation}
Note that the above equation corresponds to a deformation of the Painlev\'e II equation in which one introduces the last $q$-dependent term, where $q$ is just a mere constant. Its solution can be related to its free energy via the relation 
\begin{equation}
	F''(z)=\frac{u(z)^2}4.
\end{equation}
One of the objectives of this article is to explore the resurgent properties of the solutions to this equation. First, we will give a brief description of the matrix model and how the double-scaling limit of certain quantities in this model can be directly translated into transseries terms. Then, we will study the large-order growth of the perturbative solution. From there, one should be able to determine quantities such as the absolute value of the instanton action associated with the non-perturbative terms. As a check, we will also compute it from a two-parameter transseries ansatz. Motivated by \cite{asv11,sv13}, we will perform some changes of variable on our equation. These will make the resurgent analysis cleaner and easier to describe, since the map to the Borel plane becomes straightforward. In general, obtaining non-perturbative information from string theory requires very technical knowledge of the different components of the theory plus carrying out hard CFT computations. One way to circumvent this issue in these models is by applying a transseries ansatz to the would-be solutions of the equation. On the one hand, one gets these non-perturbative contributions almost trivially by means of a recursion relation. On the other hand, what these correspond to, at the level of the physical theory, is very much still an open problem that is sometimes called semi-classical decoding. Nonetheless, our main goal is to give a form for this non-perturbative information and how it evolves when going around the $\mu$ plane. These relations, which are commonly denoted {\it connection formulae} or {\it transition functions}, are encoded in the resurgent information of the transseries. These take the form of vectors, which are known as {\it Stokes data} ---alternatively, Stokes vectors. Therefore, our main objective will be the computation of Stokes data, since this will allow us to construct global, continuous solutions on the whole complex plane. With these, we will also be able to write the aforementioned transition functions \cite{bssv21}. We refer the reader to the nice and complete review on resurgent analysis \cite{abs18}, while for the explicit methods and discussions \cite{bssv21} is the closest example, with \cite{asv11,sv13} as precedents.

On the other hand, it is well-known that this equation \eqref{eq:q-PII} is related to the inhomogeneous Painlev\'e II equation via the Miura map \cite{djmw92,kms03}. For us, the relevancy of this connection lies in the fact that the amount of results and properties that are known for the inhomogeneous Painlev\'e II equation far exceeds those of the $q$-deformed one. In this line of thought, we are also interested in how the resurgent properties of these equations are related. Thus, the second aim of this work is to do a complete resurgent analysis on the inhomogeneous Painlev\'e II equation
\begin{equation}
	\label{eq:first_PII_with_alpha}
	u''(z)-2 u^3(z)- z\,u(z)-\alpha=0,
\end{equation}
where the novelty with respect to previous works \cite{sv13}, lies in the fact that we now allow for the constant term $\alpha$ to be non-vanishing. 
This equation itself is a very interesting object of study. Not only it appears in diverse Physical scenarios, but it has also been largely studied in the Mathematics literature for its very interesting properties---see, {\it e.g.}, \cite{djmw92,djm92,kms03,sv13,ss04,bssv21,admn97,a99,blmst17,ck07,ckv05,c03,c06,c19,d18,fw14,fw15,fw1b5,gg18,i96,i03,ik03,jk92a,k04a,k04b,t02, emm23}.
We will start by analising the Hermitian matrix model whose double-scaling limit is described by this equation. This is the two-cut case of a quartic Hermititan matrix model with a deformation factor $\det(M)^\alpha$ \cite{admn97,ck07,ckv05}. Matrix models are full of techniques to compute different quantities in various limits. In particular, we will be interested in the one-loop free energy, which will yield the simplest Stokes vector. This is not a new feature, and one can find several examples of these computations, {\it e.g.},  \cite{asv11,d92,msw07,sv13,msw08}. Using the results in \cite{sv13} we have combined two methods to obtain the simplest Stokes coefficient: on the one hand, we will use the method of orthogonal polynomials in combination with a transseries ansatz; on the other hand, we will apply a saddle-point analysis. These two procedures analytically compute the first orders of the one-loop amplitude and the comparison will allow us to determine the simplest Stokes coefficient \cite{msw07,msw08,sv13}. We will then proceed with presenting the derivation of the inhomogeneous Painlev\'e II equation as a double-scaling limit of the aforementioned model, very much in the same spirit of \cite{sv13}. Due to the additional factor of the determinant in the matrix model, one cannot use the well-known loop equations for general potentials. This is because $\alpha$ does not scale, which means that this determinant term cannot be absorbed as a leading part of the potential in $g_{\text{s}}$. Therefore, one needs to go back to the original references \cite{a95,ackm93} and rederive these loop equations for this particular situation. Notice that this is not necessary for orthogonal polynomials, but it is for the case of saddle-point analysis. We have reserved these technical computations to some appendices.

The second part of the analysis will be focused on the resurgent properties of the inhomogeneous Painlev\'e II equation. We will repeat everything we did for the $q$-deformed version, but this time to a higher depth, since obtaining coefficients for the inhomogeneous equation is computationally much faster. After both equations are studied, we will proceed with the description of their relation via the Miura map. In particular, we will make the map between the asymptotic expansions explicitly and we will describe the caveats behind this procedure. Regarding the global structure of this inhomogeneous Painlev\'e II equation, most of the works in this area---see references in the latter paragraph---are based on finding particular asymptotic expansions of this equation and some Stokes phenomena are known for some specific scenarios. For example, these methods include Riemann--Hilbert problems, isomonodromy, asymptotic analyses, etc. Our aim, as for the case of the $q$-deformed equation, will therefore be to find general transition functions that would give a complete and straightforward framework that would reproduce all of these scenarios. In this sense, the last part of the paper is thus reserved to the study of both the Hastings--McLeod solution \cite{hm80}, the special function solutions, and the rational solutions of the inhomogeneous Painlev\'e II equation---see, {\it e.g.}, \cite{d18,c06}. In particular, we will be interested in how these are encoded in our transseries ansatz and how they are therefore constructed globally. We will see that our resurgence prescription is a very powerful machinery up to an important caveat that we will point out that requires a deeper understanding of the Riemann sheet that governs the perturbative sector, see \cite{a99}.

This paper is structured as follows. Section \ref{chap:qPII} is reserved to the study of the $q$-deformed Painlev\'e II equation and the rectangular complex matrix model that it describes. In subsection \ref{sec:qorigPII} we study some preliminary properties of the equation, such as the asymptotic growth of the coefficients of the perturbative sector or the associated instanton action. Given these preliminary results, we extend our analysis with a transseries ansatz and study its properties in section \ref{sec:qPII}. Section \ref{chap:alphaPII} is dedicated to the inhomogeneous Painlev\'e II equation and the Hermitian matrix model that double-scales into it. Subsections \ref{subsec:HMMOP}, \ref{subsec:HMMSP} and \ref{subsec:HMMDSL} study the one-loop amplitude of this matrix model, as well as its connection to the inhomogeneous Painlev\'e II equation. Subsection \ref{sec:origPII} again examines the properties of the perturbative solution to the equation in its original normalisation, while section \ref{sec:PII} delves into the larger resurgent transseries analysis. Section \ref{chap:Miura&SpecSol} studies the Miura map, while subsection \ref{sec:compspeccases} is dedicated to examine the special solutions of the inhomogeneous Painlev\'e II equation. Finally, conclusions are presented in the last section \ref{sec:concl}. With the purpose of keeping the main discussions clean, we have some appendices dedicated to long computations. Appendix \ref{app:freeEnergy} discusses the translation of our transseries solutions into the free energies of the matrix models. Appendix \ref{app:alphaMM} contains technical computations that assist on the calculation of the one-loop amplitude of the Hermitian matrix model that double-scales into the inhomogeneous Painlev\'e II equation. Appendix \ref{app:Hratio} is dedicated to analyses on our main asymptotic quantities as well as some preliminary results on transasymptotics. Finally, we perform appendix \ref{app:iso} some illustrative toy computation based on isomonodromy theory that indicates a connection with our approach.

%\subheader{\hfill
%{\small{\textsf{PRE-PRINT-YY-NNN}}}
%}
%%%%%%%%%%%%%%%%%%%%%%%%%%%%%%%%%%%%%%%%%%%%%%%%%%%%%%%%%%%%%%%%%
%%%%%%%%%%%%%%%%%%%%%%%%%%%%%%%%%%%%%%%%%%%%%%%%%%%%%%%%%%%%%%%%%
\section{q-Deformed Painlev\'e II Equation and Matrix Models}\label{chap:qPII}
%%%%%%%%%%%%%%%%%%%%%%%%%%%%%%%%%%%%%%%%%%%%%%%%%%%%%%%%%%%%%%%%%
%%%%%%%%%%%%%%%%%%%%%%%%%%%%%%%%%%%%%%%%%%%%%%%%%%%%%%%%%%%%%%%%%

It is well known that the $q$-deformed Painlev\'e II equation appears in the double-scaling limit (dsl) of different complex, Hermitian and unitary matrix models \cite{kms03}. Here, we will focus on the case of the rectangular complex matrix model
\begin{equation}
	Z=\int \text{d}M \text{d}M^\dagger \rme^{-\frac{N}{\gamma}V(MM^\dagger)},
\end{equation}
with a $U(N + q) \times U(N)$ gauge symmetry, where $M$ is a rectangular $(N + q) \times N$ matrix with complex entries. In our case, we consider $\gamma$ to be a small parameter that controls, in combination with $N$, our double-scaling limit. Recall that this matrix model describes the dynamics of open 0A strings on  $(N + q)$ D0-branes and $N$ anti D0-branes. In the eigenvalue picture, one can bring the matrix integral to a ``diagonal'' form
\begin{align}
	Z&=\prod_{i=1}^N\int_0^{+\infty} \text{d}\lambda_i\,\lambda_i^{1+2q}\,\rme^{-\frac N\gamma V(\lambda_i^2)}\,\Delta^2(\lambda^2)=\\
	&=\prod_{i=1}^N\int_0^{+\infty} \text{d}y_i\, y_i^{q}\,\rme^{-\frac N\gamma V(\lambda_i)}\,\Delta^2(y),
\end{align}
where $\Delta(y)=\prod_{i<j}(y_i-y_j)$ stands for the Vandermonde determinant. The $q$-deformed Painlev\'e II equation appears for a quadratic potential, {\it e.g.}, we will use
\begin{equation}
	\label{eq:MMpotential}
	V(y)=-y+\frac{y^2}2.
\end{equation}

The method of orthogonal polynomials deals with the fact that the Vandermonde determinant is hard to integrate analytically. The main idea is based on finding polynomials which are orthogonal with respect to the measure $\text{d}\mu=y^q\rme^{-\frac N\gamma V(y)}$
\begin{equation}
	\int \text{d}\mu(z)\, p_n(z)\, p_m(z)=h_n\, \delta_{n,m}.
\end{equation} 
Thus, they would constitute a basis for expanding the Vandermonde determinant. An additional feature is that they can be used to derive the string equation. Originally done in \cite{biz80}, the reader can find extra details about this specific case in \cite{kms03}. Let us summarise some findings herein. Defining\footnote{The notation $\langle n|f|m\rangle$ is a short-cut for writing 
\begin{equation}
	\int \text{d}\mu(z)\, p_n(z)\, f(z)\, p_m(z).
\end{equation}}
\begin{align}
	\Omega_n&\equiv \frac 1{h_n}\langle n|V'|n\rangle,\\
	\tilde{\Omega}_n&\equiv \frac 1{h_{n-1}}\langle n-1|V'|n\rangle-x_n,\qquad\qquad x_n=\frac{n\gamma}{N},
\end{align}
Klebanov, Maldacena and Seiberg arrived to the following equations
\begin{align}
	&\tilde{\Omega}_n+\tilde{\Omega}_{n+1}=-s_n\Omega_n+\frac{\gamma q}N,\\
	&r_n\Omega_n\Omega_{n-1}=\tilde{\Omega}_n^2-\frac{q\gamma}N\tilde{\Omega}_n,\\
	&r_n-x_n=\tilde{\Omega}_n,\\
	&-1+s_n=\Omega_n.
\end{align}
Since the $q$-deformed Painlev\'e II equation is obtained in the double-scaling limit from the $\tilde{\Omega}_n$ term above, the aim is to solve this system in such a way that we can obtain an expression exclusively involving the functions $\tilde{\Omega}_n$'s. This procedure yields
\begin{align}
	0=q\,\frac{\gamma}N \tilde{\Omega}_n - \tilde\Omega_n^2 + \frac 14 &\left(1 - \sqrt{1 + 4 q \,\frac\gamma N - 4 \tilde \Omega_{n-1} - 4\tilde\Omega_n}\right)\times\nonumber\\
	\times&\left(1 - \sqrt{1 + 4 q\,\frac\gamma N - 4 \tilde\Omega_n - 4 \tilde\Omega_{n+1}}\right) \left(\tilde\Omega_n + x_n\right).
\end{align}
We are interested in the large $N$ limit of the above equation. Note that we also want to take into account that the parameter $\frac\gamma N$ will play a role in the double-scaling limit. The first step is just a mere substitution, $x_n\to x$, $\tilde\Omega_n\to\tilde\Omega(x)$, obtaining the following equation for $\tilde\Omega(x)$
\begin{align}
	0=q\,\frac\gamma N \tilde\Omega(x) - \tilde\Omega(x)^2 + \frac 14 \left(x + \tilde\Omega(x)\right) &\left(1 - \sqrt{ 1 + 4q\,\frac\gamma N - 4 \tilde\Omega(x) - 4 \tilde\Omega\left(x-\frac{\gamma}N\right)}\right)\times\nonumber\\
	\times&\left(1 - \sqrt{ 1 + 4 q\,\frac\gamma N - 4 \tilde\Omega(x) - 
		4 \tilde\Omega\left(x+\frac\gamma N\right)}\right).\label{eq:tildeOmegaRecRel}
\end{align}
Now, note that the usual method of orthogonal polynomials is not quite enough to grasp the non-perturbative features of the model in the large $N$ limit. Nonetheless, it can be upgraded via the addition of a {\it transseries ansatz}. This would constitute the second ingredient towards analysing this model at the non-perturbative level
\begin{equation}
	\tilde{\Omega}(x)=\sum_{n,m=0}^{+\infty}\sigma_1^n\sigma_2^m\rme^{-(n-m)A\frac{N}\gamma}\tilde\Omega^{(n,m)}(N), \qquad\qquad\tilde\Omega^{(n,m)}(N)\simeq\sum_{g=0}^{+\infty}\tilde\Omega_g^{(n,m)}(x)\left(\frac\gamma N\right)^g.
\end{equation}
Plugging these expressions into \eqref{eq:tildeOmegaRecRel} we can recursively compute the coefficients $\tilde{\Omega}^{(n,m)}_g(x)$. Notice that the expansion of the exponential terms is given by
\begin{equation}
	\exp\left(-\frac{A\left(x\pm \frac\gamma N\right)}{\frac\gamma N}\right)=\exp\left(-\frac{A(x)}{\frac\gamma N}\right)\rme^{\mp A'(x)}\sum_{l'=0}^{+\infty}\frac{1}{l'!}\left(-\sum_{l=0}^{+\infty}(\pm 1)^l \left(\frac\gamma N\right)^{l-1}\frac{A^{(l)}(x)}{l!}\right)^{l'}.
\end{equation}
For the perturbative sector, the first two coefficients are
\begin{align}
	\tilde\Omega^{(0,0)}_0(x)=&\frac1{18} \left(1 - 12 x + \sqrt{1 + 12 x}\right),\\
	\tilde\Omega^{(0,0)}_1(x)=&\frac q{12-48x} \left(5 + \frac 2{\sqrt{1 + 12 x}} - \sqrt{5 + 48 x - 4 \sqrt{1 + 12 x}} \right.-\nonumber\\
	&\left.-\frac{2 (\sqrt{5 + 48 x - 4 \sqrt{1 + 12 x}} + 4 x (-2 + \sqrt{1 + 12 x})}{\sqrt{1 + 12 x}}\right).
\end{align}
Note that the first coefficient agrees with the one in \cite{kms03}, where we have taken the determination that guarantees to have a proper double-scaling limit. Now, for the sake of readability, we define $p^2=1+12x$. The higher orders read
\begin{align}
	\tilde\Omega^{(0,0)}_0(p)&=-\frac1{18} (1 + p)(p - 2),\\
	\tilde\Omega^{(0,0)}_1(p)&=\frac q6\left(\frac{1 + p}{p}\right),\\
	\tilde\Omega^{(0,0)}_2(p)&=\frac1{(p-2)^2}\left(\frac{(1+p)^2}{12 p^4}(4-7p+4p^2)-q^2\frac{(1+p)}{p^3}(2-4p+3p^2)\right),\\
	\tilde\Omega^{(0,0)}_3(p)&=\frac{(1+p)}{(p-2)^3}\left(-q\frac{(1 - p)(32 -18 p + 3 p^2 + 8 p^3)}{4 p^6}+q^3\frac{(-6 + 9 p)(1-p+p^2)}{p^5}\right),\\
	\tilde\Omega^{(0,0)}_4(p)&=\frac{(1+p)}{(p-2)^5}\left\{\frac{(1+p)(1568-4656p+5946 p^2 - 4445 p^3 + 2364 p^4 - 1065 p^5 + 368 p^6)}{16 p^9}-\right.\nonumber\\
	&\hspace{2.4cm}-q^2\frac{3 (384 - 1044 p + 1142 p^2 - 664 p^3 + 273 p^4 - 157 p^5 + 116 p^6)}{4  p^8}+\nonumber\\
	&\hspace{2.4cm}\left.+q^4\frac{9  (-20 + 70 p - 102 p^2 + 82 p^3 - 38 p^4 + 3 p^5)}{2p^7}\right\};
\end{align}
where now the double-scaling limit $x\mapsto \frac14+\varepsilon z$, $\varepsilon\to0$ gets translated into $p\mapsto 2+3\varepsilon z$, where $\varepsilon$ is given in terms of $\gamma$ and $N$ via the relation $N = \gamma\varepsilon^{-3/2}$. An interesting exercise is to recover the coefficients of the $q$-deformed Painlev\'e II transseries from those of the $\tilde{\Omega}$ ones. Using the latter limit, the result for the first coefficients read
\begin{align}
	\tilde\Omega^{(0,0)}_0(p)&\to z\varepsilon\left(\frac 12\right),\\
	\tilde\Omega^{(0,0)}_1(p)&\to \frac q4,\\
	\tilde\Omega^{(0,0)}_2(p)&\to \frac 1{z^2\varepsilon^2}\left(\frac{1 - 4 q^2}{32}\right),\\
	\tilde\Omega^{(0,0)}_3(p)&\to\frac 1{z^3\varepsilon^3}\left(\frac{q - 4 q^3}{32}\right),\\
	\tilde\Omega^{(0,0)}_4(x)&\to\frac1{z^5\varepsilon^5}\left(-\frac9{256} + \frac{5}{32}\;q^2 - \frac1{16}\;q^4\right).
\end{align}
Notice that we still need to add some power of $\varepsilon$ to make the complete connection. Comparing these results with the expansions in \cite{kms03}\footnote{Notice that we have changed notation with respect to this article. For them, our $u$ is a rescaled version of their $f$ (the solution to the $q$-deformed Painlev\'e II equation), while our $v$ is written as $u$.}, we see that the double-scaling limit for the {\it even} perturbative coefficients correspond with the expansion of
\begin{equation}
	v(z)=u(z)^2+z,
\end{equation} 
whose first orders read
\begin{equation}
	v(z)\simeq\frac z4+\frac1{z^2}\left(- \frac18 + \frac{q^2}2\right) +\frac1{z^5}\left(-\frac 98  + 5 q^2 - 2 q^4\right)+\mathcal{O}(z^{-8}).
\end{equation}
This phenomenon has an easy explanation. We still need to include the factor of $\varepsilon$ that will complete the connection between the two sets of coefficients. The desired quantity reads\footnote{Note that when writing this double-scaling limit, we have considered all powers of $g$ in such a way that we can explicitly see how even-$g$ matrix-model-coefficients yield the nonvanishing coefficients $v_{2g'}^{(0,0)}$ for $v(z)$, while the odd-$g$ coefficients are mapped to $v_{2g'-1}^{(0,0)}=0$. Therefore, we have ``artificially'' added these zeros for completeness.}
\begin{equation}
	-(2\varepsilon)^{-1 + \frac32 g}\,\,\tilde\Omega^{(0,0)}_g(p)\xrightarrow{\qquad\text{dsl}\qquad} v^{(0,0)}_g z^{1-\frac32 g}.
\end{equation}
Notice that this prefactor cancels the $\varepsilon$ contributions of the even-$g$ coefficients, while the odd-$g$ coefficients are left with a contribution $\varepsilon^{\frac12}$. Hence, in the double-scaling limit $\varepsilon\to0$, only the even-$g$ contributions survive, thus finding $v^{(0,0)}_g=0$ for odd $g$---this explains why only the even coefficients contribute to $v(z)$.
\subsection{Perturbative Solution and Large-Order Analysis}\label{sec:qorigPII}
\begin{figure}[hbt!]
	\centering
	\begin{subfigure}[t]{0.5\textwidth}
		\centering
		\includegraphics[height=1.6in]{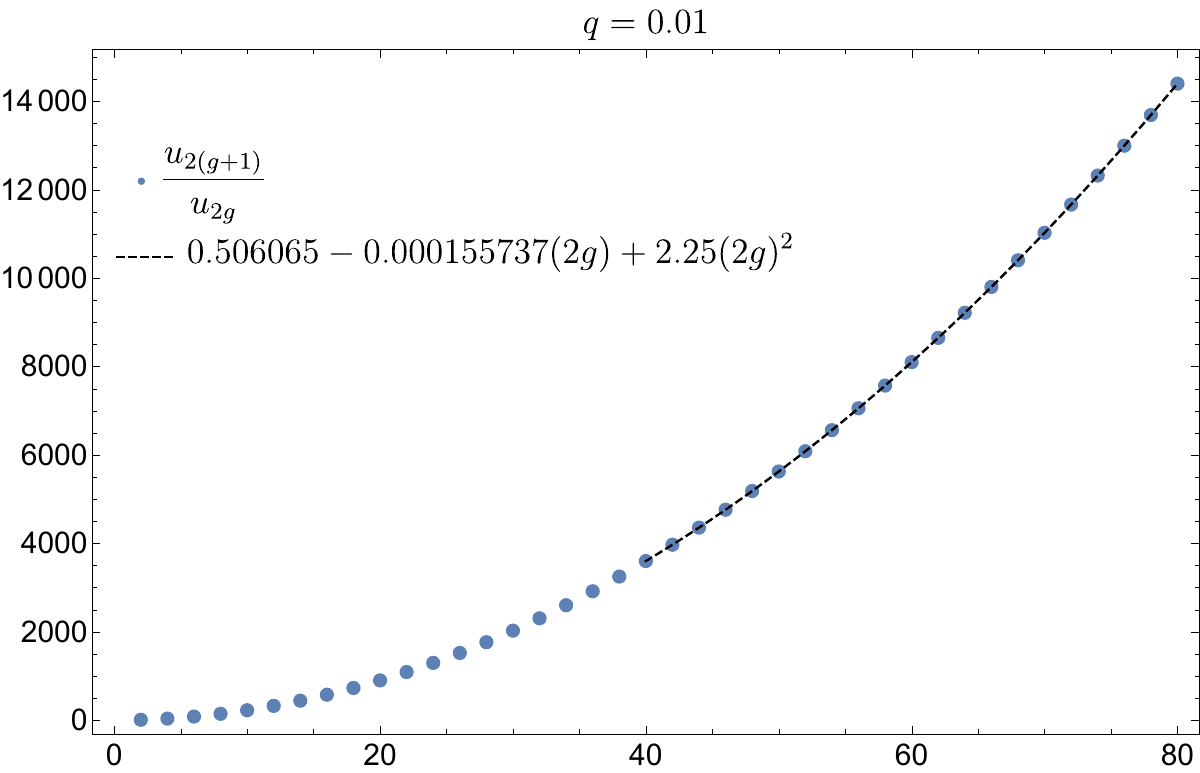}
		%\caption{}
	\end{subfigure}%
	~ 
	\begin{subfigure}[t]{0.5\textwidth}
		\centering
		\includegraphics[height=1.6in]{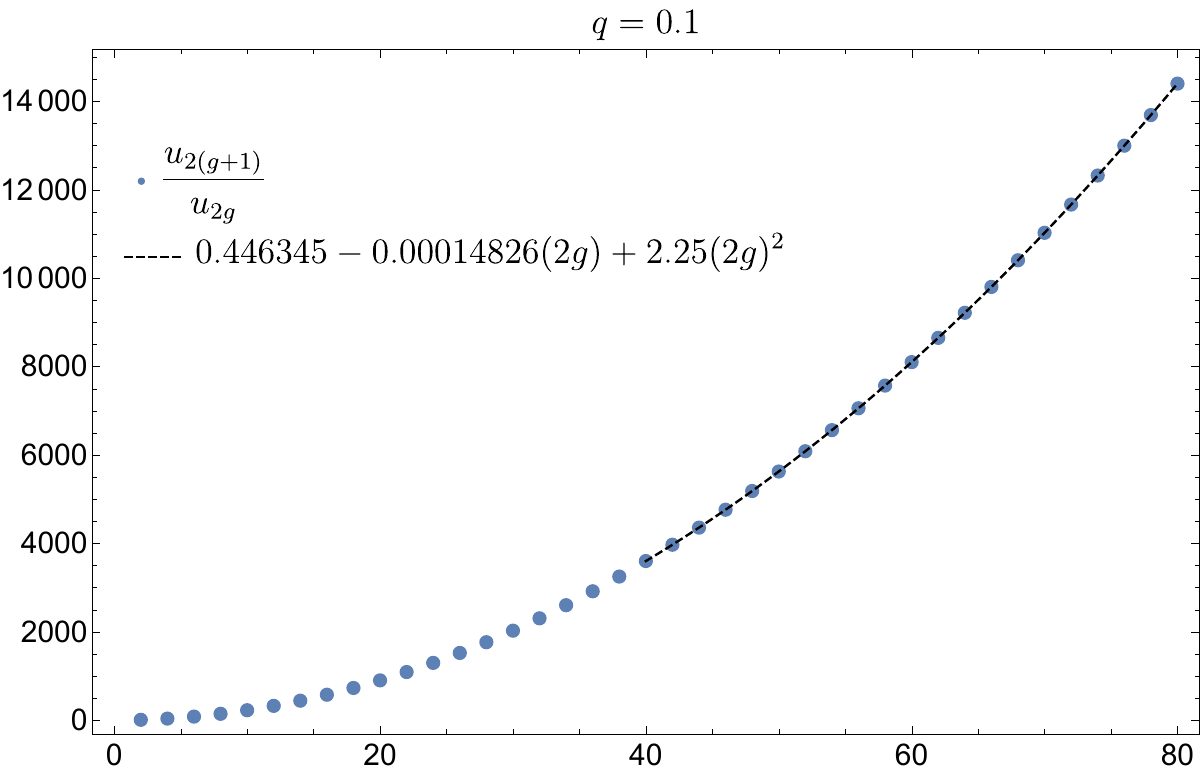}
		%\caption{Conformal transformations}
	\end{subfigure}\\
	\begin{subfigure}[t]{0.5\textwidth}
		\centering
		\includegraphics[height=1.6in]{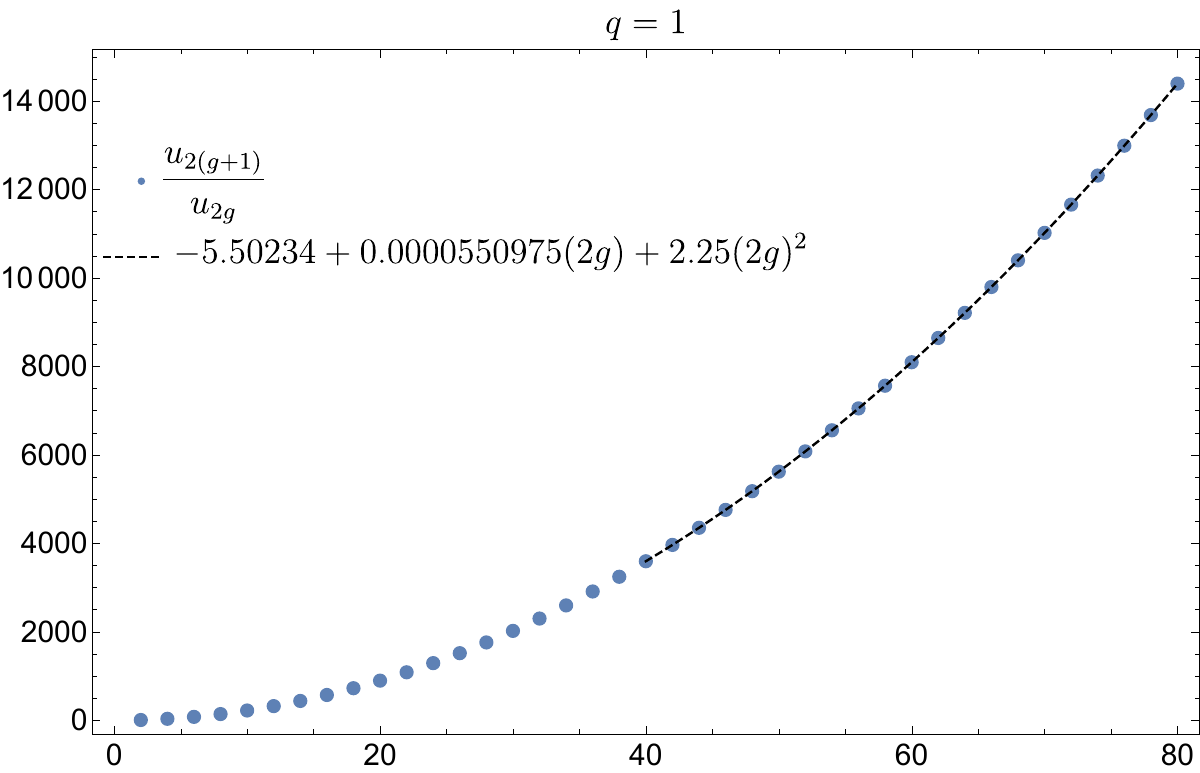}
		%\caption{\cite{asv11} computation }
	\end{subfigure}%
	~ 
	\begin{subfigure}[t]{0.5\textwidth}
		\centering
		\includegraphics[height=1.6in]{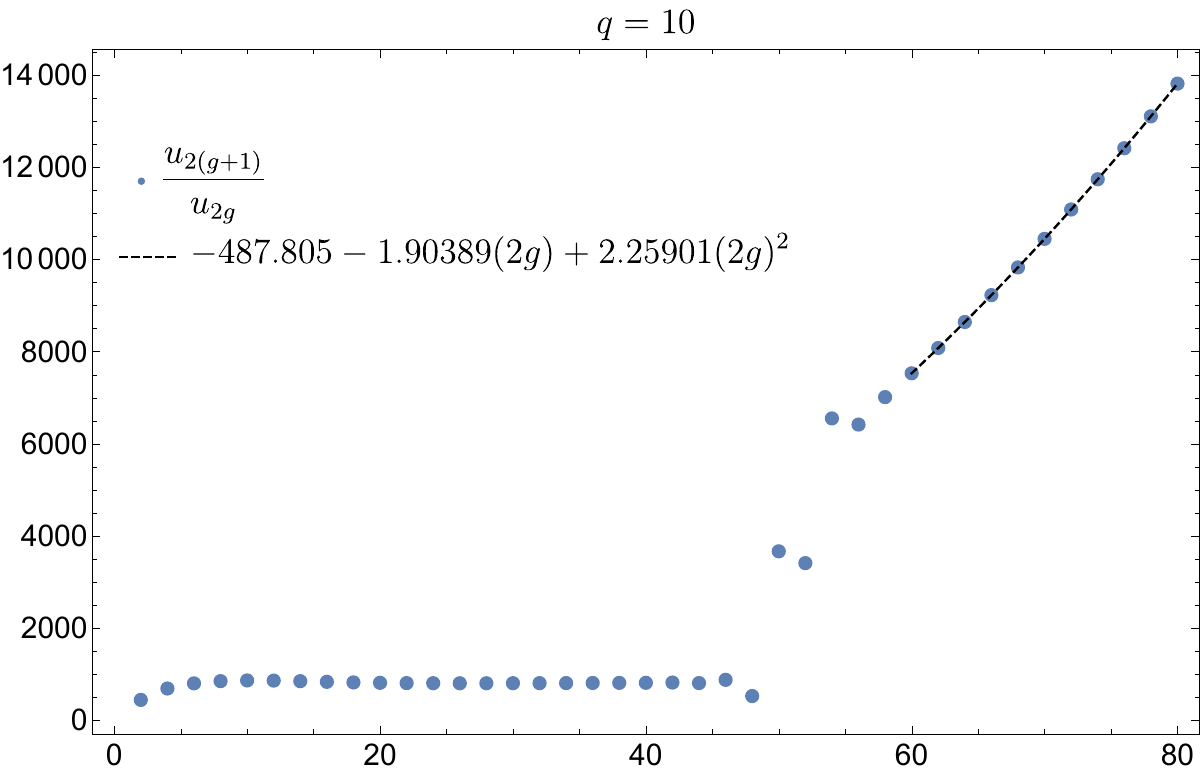}
		%\caption{Conformal transformations}
	\end{subfigure}
	\caption{Checks on the instanton action for different values of $q$. As explained, the Gevrey-1 property kicks in later as we increase the value of $q$. Also, notice that each of the fits gives a quadratic term $\sim \left(\frac23\right)^{-2}=2.25$.}
	\label{fig:qchecksaction}
\end{figure}

%%%%%%%%%%%%%%%%%%%%%%%%%%%%%%%%%%%%%%%%%%%%%%%%%%%%%%%%%%%%%%%%%
%%%%%%%%%%%%%%%%%%%%%%%%%%%%%%%%%%%%%%%%%%%%%%%%%%%%%%%%%%%%%%%%%
Now that we have done the most straightforward analyses of the matrix model, it is time to delve into the equation itself. We start by writing it in its original form---as given in \cite{kms03}
\begin{equation}
	\label{eq:original_PII_q2}
	u''(z)-\frac 12 u^3(z)+\frac 12 z\,u(z)+\frac{q^2}{u^3(z)}=0.
\end{equation}
Since $u(z)=0$ is not a solution, it will become very convenient to multiply the whole equation by $u^3(z)$. Trivially, the result is given by
\begin{equation}
	\label{eq:PII_with_q2}
	u''(z)\,u^3(z)-\frac 12 u^6(z)+\frac 12 z\,u^4(z)+q^2=0.
\end{equation}
The first step towards understanding its structure is to study the perturbative sector. For that, we will use the following ansatz
\begin{equation} 
	\label{eq:pert_ansatz_with_q2}
	u(z)\simeq\sum_{g=0}^{+\infty} u_g \; z^{-g\gamma+\beta}.
\end{equation}
Plugging \eqref{eq:pert_ansatz_with_q2} into equation \eqref{eq:PII_with_q2} we get\footnote{\label{fnote:otherqsol}Note that there other four solutions for the perturbative sector that behave as $\sim\frac {(-2q^2)^{\frac14}}{z^{\frac14}}$.}
\begin{equation}
	\begin{array}{lll}
		\beta=\displaystyle\frac{1}{2}, &\hspace{1cm} & u_0=\pm 1,\\
		\gamma=3, & & u_1=\pm\left(q^2-\displaystyle\frac{1}{4}\right),
	\end{array}
\end{equation}
and a recursion relation for the rest of the coefficients. By choosing the positive sign\footnote{There is no special reason to choose the positive determination of the square root. In fact, there is not even a reason for us to have chosen the double-cover solution instead of the quadruple-cover and it it clear that the analyses of all solutions will require understanding the solutions associated to the quadruple-cover. Nonetheless, our choice can be argued to be the simplest one computationally, since more cumbersome powers of $z$ would imply closer examination. We will also cover some ground towards the complete solution when discussing the Miura map. Recall that our choice also impacts the nature of the exponential terms that we will consider when dealing with the transseries.} one gets
\begin{align}
	\label{eq:PIIexp}
	u(z)\simeq&z^{\frac12}+\left(q^2-\frac{1}{4}\right)z^{-\frac52}+\left(-\frac{9\,q^4}{2}+\frac{41\,q^2}{4}-\frac{73}{32}\right)z^{-\frac{11}2}\nonumber\\
	&+\left(\frac{65\,q^6}{2}-\frac{1827\,q^4}{8}+\frac{12419\,q^4}{32}-\frac{10657}{128}\right)z^{-17/2}+\cdots\simeq\\
	\simeq&\sqrt{z}\left\{1+\left(q^2-\frac{1}{4}\right)z^{-3}+\left(-\frac{9\,q^4}{2}+\frac{41\,q^2}{4}-\frac{73}{32}\right)z^{-6}+\right.\nonumber\\
	&\left.+\left(\frac{65\,q^6}{2}-\frac{1827\,q^4}{8}+\frac{12419\,q^4}{32}-\frac{10657}{128}\right)z^{-9}+\cdots\right\}.
\end{align}
Notice that this obviously coincides with the original analysis in \cite{kms03}.
The instanton action for this problem can be obtained from two approaches:
\begin{itemize}
	\item Fitting the ratio
	\begin{equation}
		\label{eq:ratio}
		\frac{u_{g+1}}{u_g}.
	\end{equation}
	We have checked that the perturbative series is a Gevrey-1 series and thus resurgent\footnote{See \cite{abs18,s14} for further information on resurgence, Gevrey series and their relations.}. Being Gevrey-1 means that the coefficients have the following large-order behaviour
	\begin{equation}
		\label{eq:qGevrey-1_prop}
		|u_g|\sim \frac{C}{A^g}g!.
	\end{equation}
	From there, one can compute the instanton action $A$ from the ratio \eqref{eq:ratio}\footnote{Noticing that one has half of the coefficients being 0 (check section \ref{subsec:qtransans}) one would have to fit the ratio 
		\begin{equation}
			\frac{u_{2(g+1)}}{u_{2g}},
		\end{equation}
		to a polynomial of order two with coefficients approaching the behaviour of Gevrey-1 property
		\begin{equation}
			\frac{u_{2(g+1)}}{u_{2g}}\sim\frac{(2g+2)(2g+1)}{A^2}.
		\end{equation}
		Notice, also, that we have plotted these quantity with respect to the variable $2g$ in figure \ref{fig:qchecksaction}.}.
	\item Computing it from a one-parameter or two-parameter transseries approach, which yields a second-order algebraic equation for $A$. The result reads
	\begin{equation}
		\label{eq:PII_with_q2_action}
		A=\pm \frac{2}{3},
	\end{equation}  
	which, as expected, coincides with the previous approach\footnote{Notice that upon choosing the perturbative sector as explained in footnote \ref{fnote:otherqsol}, this action gets modified into
	\begin{equation}
		A=\pm\frac{2\sqrt2}3\rmi.
	\end{equation}	
	We will explore the connection explicitly in the example of the inhomogeneous Painlev\'e II equation---see footnotes \ref{fnote:otheru0s}, \ref{fnote:otherA}.}.
\end{itemize}
It is worth mentioning that if one keeps track of all the changes of variables, this result coincides with previous works on the $q=0$ case \cite{m08,sv13}. The fact that the instanton action does not depend on $q$ should not be a surprise. Recall that these instanton actions appear as singularities of the appropriate Borel Transform of the solution. Borel transforms appear to deal with factorially divergent power series and their purpose is to remove this behaviour, thus obtaining a power series with a nonvanishing radius of convergence on the Borel plane, which we will parametrise with the variable $s$. Applying the Borel transform to the whole equation, the $q$ factor will be just represented, on the Borel plane, by an additional factor $q\, s^p$, for some power $p$. The whole solution can be thus seen as this $q$-factor in convolution with the Borel transform of the $q=0$ case. Given that convolutions do not affect the singularity structure, provided that one of the terms is entire, we do not expect a change in the radius of convergence independently of $q$. This argument also explains why our computation seems consistent, and even why this agrees with the $q=0$ case\footnote{See \cite{s14} for further information on convolutions, Borel plane, its singularities, and other details.}. 

Let us give further details on the first approach. The reader may have realised that computing the ratio \eqref{eq:ratio} is somewhat cumbersome since the coefficients $u_g$ are polynomials in $q^2$. The first thing one realises is that the order of the polynomial is $g$, so we can write 
\begin{equation}
	\label{eq:struc_of_pert_coeffs_q2}
	u_g=\sum_{i=1}^gu_{g,i}\,\left(q^2\right)^{i}.
\end{equation}
In order to check the Gevrey-1 property \eqref{eq:qGevrey-1_prop} of the coefficients $u_g$, we first computed the ratio for fixed $q=0.01,\,0.1,\,1,\,10$ up to $2g=80$---see figure \ref{fig:qchecksaction}. We see that the higher the $q$, the later the Gevrey-1 behaviour starts dominating. 

On the other hand, we have computed the coefficients for arbitrary $q$ up to order $g=100$. The result that comes out of this is that, for fixed $i$, the coefficients $u_{g,i}$ have the Gevrey-1 property. This motivates the fact that the action is not $q$-dependent. In the next section, we will discuss in much more depth all the details of the resurgent structure of this equation. For that, we will define the {\it resurgent variable} $x$, which is nothing more than a variable in which Borel and Laplace transforms ``look nice''. However, when doing computations, we will use the square-root of this quantity $x=w^2$ simply because the transseries becomes a power-series expansion in integer powers.

%%%%%%%%%%%%%%%%%%%%%%%%%%%%%%%%%%%%%%%%%%%%%%%%%%%%%%%%%%%%%%%%%
%%%%%%%%%%%%%%%%%%%%%%%%%%%%%%%%%%%%%%%%%%%%%%%%%%%%%%%%%%%%%%%%%
\subsection{Resurgence Transseries Analysis}\label{sec:qPII}
%%%%%%%%%%%%%%%%%%%%%%%%%%%%%%%%%%%%%%%%%%%%%%%%%%%%%%%%%%%%%%%%%
%%%%%%%%%%%%%%%%%%%%%%%%%%%%%%%%%%%%%%%%%%%%%%%%%%%%%%%%%%%%%%%%%
%%%%%%%%%%%%%%%%%%%%%%%%%%%%%%%%%%%%%%%%%%%%%%%%%%%%%%%%%%%%%%%%%

Starting from  equation (6.9) in \cite{kms03}
\begin{equation}
	f''(z)-f(z)^3-z\,f(z)+\frac{q^2}{f(z)^3},
\end{equation}
we want to obtain an equation with the normalisation of \cite{sv13}. The reason being that we want to have coefficients and instanton actions that do not contain square roots of 2, as they are computationally heavy to carry through. For that, we can change variables as
\begin{equation}
	f=2^{\frac16}u,\qquad z=-2^{\frac13}t,
\end{equation}
which yields, after multiplying by $\sqrt{2}\,u^3(t)$, the equation
\begin{equation}
	\label{eq:PII_with_q2_final}
	u''(t)\,u^3(t)-2 u^6(t)+2 t\,u^4(t)+q^2=0.
\end{equation}
The perturbative sector for $u(t)$ is given by a power-series expansion in $t^{-3}$, with an overall factor of $t^{\frac12}$. Since the perturbative sector is expected to be a {\it closed} string expansion, we expect the natural parameter to be the {\it open} string coupling $x=t^{-3/2}$. However, this will yield non-perturbative sectors with initial fractional powers. It is therefore more convenient to define the square-root of this natural parameter as the quantity to work with, $x=w^2$. As such, we will transform equation \eqref{eq:PII_with_q2_final} by means of \cite{asv11,sv13}
\begin{equation}
	\label{eq:change_of_variables_q}
	u(w)\equiv \left.\frac{u(t)}{\sqrt{t}}\right|_{t=w^{-4/3}}.
\end{equation}
The final form of the equation, in this variable, reads:
\begin{equation}
	\label{eq:PIIfinalform}
	\frac 9{16} w^6 u(w)^3 u''(w)+\frac 9{16} w^5 u(w)^3 u'(w)-2u(w)^6+\left(2-\frac {w^4}4\right) u(w)^4+q^2 w^4=0.
\end{equation}
The next step in understanding the structure of the solutions is to attack the above equation with a two-parameter\footnote{Note that we have a second-order differential equation, which will give a second-order algebraic equation for the instanton action, meaning two possible values which one has to take into account.} transseries ansatz.
%%%%%%%%%%%%%%%%%%%%%%%%%%%%%%%%%%%%%%%%%%%%%%%%%%%%%%%%%%%%%%%%%
\subsubsection{Transseries Ansatz: Recursion Relation and Coefficient Structure}\label{subsec:qtransans}
%%%%%%%%%%%%%%%%%%%%%%%%%%%%%%%%%%%%%%%%%%%%%%%%%%%%%%%%%%%%%%%%%
%%%%%%%%%%%%%%%%%%%%%%%%%%%%%%%%%%%%%%%%%%%%%%%%%%%%%%%%%%%%%%%%%
As stated in the end of the last subsection, given that \eqref{eq:PIIfinalform} is a second-order differential equation, the appropriate ansatz to tackle it is a two-parameter transseries one. This is given, in our case, by
\begin{equation}
	\label{eq:q2ptransansatz}
	u(w,\sigma_1,\sigma_2)=\sum_{n,m=0}^{+\infty}\sigma_1^n\sigma_2^m\rme^{-\frac {(n-m)A}{w^2}}\Phi_{(n,m)}(w),
\end{equation}
where the actions $A_1=-A_2\equiv A=\frac43$\footnote{Note that this coincides with \eqref{eq:PII_with_q2_action} upon taking the appropriate changes of variables in $z$.} can be computed directly from plugging the transseries ansatz into the equation\footnote{Notice that the action can depend, in general, in the determination of the value $u^{(0,0)[0]}_0$, which we have fixed to be $u^{(0,0)[0]}_0=+1$. There are, however, three possible choices, $u^{(0,0)[0]}_0=-1,0,+1$. The $\pm1$ choices give the same action, while the 0 one produces a different one. We will discuss a bit more about this in footnotes \ref{fnote:otheru0s}, \ref{fnote:otherA}.}. 
Futhermore, notice that the system is resonant \cite{gikm10,asv11,sv13,gs21,mss22,bssv21,sst23,eggls23}. This is, there is a linear combination of these instanton actions with coefficients in $\mathbb{Z}$ which can vanish. This means that {\it several different} non-perturbative sectors have the {\it same exponential weight} associated with them. This also means that the recursion relation for the coefficients cannot be solved unless one introduces logarithmic sectors to our transseries ansatz \cite{asv11,bssv21,gikm10,sv13}\footnote{Note that these logarithmic sectors are a useful trick to solve the equation. They can be easily resummed as shown in the aforementioned references. Alternatively, one could have also started with the resummed version as an ansatz.}. Thus, the sectors in \eqref{eq:q2ptransansatz} have the form
\begin{equation}
	\label{eq:qasympsector}
	\Phi_{(n,m)}(w)\simeq\sum_{k=0}^{k_{n,m}}\left(\log(w)\right)^k\sum_{g=0}^{+\infty}u_{2g}^{(n,m)[k]}w^{2g+2\beta^{[k]}_{(n,m)}},
\end{equation}
where $\beta^{[k]}_{(n,m)}$ is a factor that encodes the leading exponent of each the sector, while $k_{n,m}$ is dictated by the maximum power of logarithms that we can have. These quantities are given by\footnote{These were originally found from numerical explorations \cite{asv11,sv13}. In the case of the $k_{n,m}$, there were also some physical motivations for the existence of these logarithms, albeit in the context of the Painlev\'e I equation and 2d quantum gravity---see \cite{asv11,bssv21}. It is also worth mentioning that these beta-factors only hold for the determination in which $u^{(0,0)[0]}_0=\pm1$, while for $u^{(0,0)[0]}_0=0$, the beta-factor would be different and $q$-dependent. This will be evident when we make the connection between the $q$-deformed and the $\alpha$-inhomogeneous Painlev\'e II equations. }
\begin{equation}
	\label{eq:knm}
	k_{n,m}=\min(n,m)-m\delta_{n,m},\qquad\qquad 2\beta^{[k]}_{(n,m)}=n+m-2\left[\frac{k_{n,m}+k}2\right]_I;
\end{equation}
where $[\cdot]_I$ stands for the integer part. Thus, by plugging \eqref{eq:qasympsector} and \eqref{eq:q2ptransansatz} into equation \eqref{eq:PIIfinalform}, we obtain a recursion relation \eqref{eq:recrelation} for the $u_g^{(n,m)[k]}$,

\begin{landscape}
	\tiny
	\begin{align}
		\label{eq:recrelation}
		0=&\frac 94 A^2\sum_{n_1=0}^{n}\sum_{n_2=0}^{n-n_1}\sum_{n_3=0}^{n-n_1-n_2}\sum_{m_1=0}^{m}\sum_{m_2=0}^{m-m_1}\sum_{m_3=0}^{m-m_1-m_2}\sum_{k_1=0}^{k}\sum_{k_2=0}^{k-k_1}\sum_{k_3=0}^{k-k_1-k_2}\sum_{g_1=0}^{g}\sum_{g_2=0}^{g-g_1}\sum_{g_3=0}^{g-g_1-g_2}(n_1-m_1)^2u^{(n_1,m_1)[k_1]}_{g_1}u^{(n_2,m_2)[k_2]}_{g_2}u^{(n_3,m_3)[k_3]}_{g_3}u^{(n-n_1-n_2-n_3,m-m_1-m_2-m_3)[k-k_1-k_2-k_3]}_{g-g_1-g_2-g_3}\nonumber\\
		&+\frac 94 A\sum_{n_1=0}^{n}\sum_{n_2=0}^{n-n_1}\sum_{n_3=0}^{n-n_1-n_2}\sum_{m_1=0}^{m}\sum_{m_2=0}^{m-m_1}\sum_{m_3=0}^{m-m_1-m_2}\sum_{k_1=0}^{k}\sum_{k_2=0}^{k+1-k_1}\sum_{k_3=0}^{k+1-k_1-k_2}\sum_{g_1=0}^{g-2}\sum_{g_2=0}^{g-2-g_1}\sum_{g_3=0}^{g-2-g_1-g_2}(n_1-m_1)k_1u^{(n_1,m_1)[k_1+1]}_{g_1}u^{(n_2,m_2)[k_2]}_{g_2}u^{(n_3,m_3)[k_3]}_{g_3}u^{(n-n_1-n_2-n_3,m-m_1-m_2-m_3)[k+1-k_1-k_2-k_3]}_{g-2-g_1-g_2-g_3}\nonumber\\
		&+\frac 98 A \sum_{n_1=0}^{n}\sum_{n_2=0}^{n-n_1}\sum_{n_3=0}^{n-n_1-n_2}\sum_{m_1=0}^{m}\sum_{m_2=0}^{m-m_1}\sum_{m_3=0}^{m-m_1-m_2}\sum_{k_1=0}^{k}\sum_{k_2=0}^{k-k_1}\sum_{k_3=0}^{k-k_1-k_2}\sum_{g_1=0}^{g-2}\sum_{g_2=0}^{g-2-g_1}\sum_{g_3=0}^{g-2-g_1-g_2}(n_1-m_1)(2g_1-3)u^{(n_1,m_1)[k_1]}_{g_1}u^{(n_2,m_2)[k_2]}_{g_2}u^{(n_3,m_3)[k_3]}_{g_3}u^{(n-n_1-n_2-n_3,m-m_1-m_2-m_3)[k-k_1-k_2-k_3]}_{g-2-g_1-g_2-g_3}\nonumber\\
		&+\frac 9{16}  \sum_{n_1=0}^{n}\sum_{n_2=0}^{n-n_1}\sum_{n_3=0}^{n-n_1-n_2}\sum_{m_1=0}^{m}\sum_{m_2=0}^{m-m_1}\sum_{m_3=0}^{m-m_1-m_2}\sum_{k_1=0}^{k+2}\sum_{k_2=0}^{k+2-k_1}\sum_{k_3=0}^{k+2-k_1-k_2}\sum_{g_1=0}^{g-4}\sum_{g_2=0}^{g-4-g_1}\sum_{g_3=0}^{g-4-g_1-g_2}k_1(k_1-1)u^{(n_1,m_1)[k_1]}_{g_1}u^{(n_2,m_2)[k_2]}_{g_2}u^{(n_3,m_3)[k_3]}_{g_3}u^{(n-n_1-n_2-n_3,m-m_1-m_2-m_3)[k+2-k_1-k_2-k_3]}_{g-4-g_1-g_2-g_3}\nonumber\\
		&+\frac 9{16} \sum_{n_1=0}^{n}\sum_{n_2=0}^{n-n_1}\sum_{n_3=0}^{n-n_1-n_2}\sum_{m_1=0}^{m}\sum_{m_2=0}^{m-m_1}\sum_{m_3=0}^{m-m_1-m_2}\sum_{k_1=0}^{k+1}\sum_{k_2=0}^{k+1-k_1}\sum_{k_3=0}^{k+1-k_1-k_2}\sum_{g_1=0}^{g-4}\sum_{g_2=0}^{g-4-g_1}\sum_{g_3=0}^{g-4-g_1-g_2}k_1(2g_1-1)u^{(n_1,m_1)[k_1]}_{g_1}u^{(n_2,m_2)[k_2]}_{g_2}u^{(n_3,m_3)[k_3]}_{g_3}u^{(n-n_1-n_2-n_3,m-m_1-m_2-m_3)[k+1-k_1-k_2-k_3]}_{g-4-g_1-g_2-g_3}\nonumber\\
		&+\frac 9{16} \sum_{n_1=0}^{n}\sum_{n_2=0}^{n-n_1}\sum_{n_3=0}^{n-n_1-n_2}\sum_{m_1=0}^{m}\sum_{m_2=0}^{m-m_1}\sum_{m_3=0}^{m-m_1-m_2}\sum_{k_1=0}^{k}\sum_{k_2=0}^{k-k_1}\sum_{k_3=0}^{k-k_1-k_2}\sum_{g_1=0}^{g-4}\sum_{g_2=0}^{g-4-g_1}\sum_{g_3=0}^{g-4-g_1-g_2}g_1(g_1-1)u^{(n_1,m_1)[k_1]}_{g_1}u^{(n_2,m_2)[k_2]}_{g_2}u^{(n_3,m_3)[k_3]}_{g_3}u^{(n-n_1-n_2-n_3,m-m_1-m_2-m_3)[k-k_1-k_2-k_3]}_{g-4-g_1-g_2-g_3}\nonumber\\
		&+\frac 9{8} A \sum_{n_1=0}^{n}\sum_{n_2=0}^{n-n_1}\sum_{n_3=0}^{n-n_1-n_2}\sum_{m_1=0}^{m}\sum_{m_2=0}^{m-m_1}\sum_{m_3=0}^{m-m_1-m_2}\sum_{k_1=0}^{k}\sum_{k_2=0}^{k-k_1}\sum_{k_3=0}^{k-k_1-k_2}\sum_{g_1=0}^{g-2}\sum_{g_2=0}^{g-2-g_1}\sum_{g_3=0}^{g-2-g_1-g_2}(n_1-m_1)u^{(n_1,m_1)[k_1]}_{g_1}u^{(n_2,m_2)[k_2]}_{g_2}u^{(n_3,m_3)[k_3]}_{g_3}u^{(n-n_1-n_2-n_3,m-m_1-m_2-m_3)[k+1-k_1-k_2-k_3]}_{g-2-g_1-g_2-g_3}\nonumber\\
		&+\frac 9{16} \sum_{n_1=0}^{n}\sum_{n_2=0}^{n-n_1}\sum_{n_3=0}^{n-n_1-n_2}\sum_{m_1=0}^{m}\sum_{m_2=0}^{m-m_1}\sum_{m_3=0}^{m-m_1-m_2}\sum_{k_1=0}^{k+1}\sum_{k_2=0}^{k+1-k_1}\sum_{k_3=0}^{k+1-k_1-k_2}\sum_{g_1=0}^{g-4}\sum_{g_2=0}^{g-4-g_1}\sum_{g_3=0}^{g-4-g_1-g_2}k_1u^{(n_1,m_1)[k_1]}_{g_1}u^{(n_2,m_2)[k_2]}_{g_2}u^{(n_3,m_3)[k_3]}_{g_3}u^{(n-n_1-n_2-n_3,m-m_1-m_2-m_3)[k+1-k_1-k_2-k_3]}_{g-4-g_1-g_2-g_3}\nonumber\\
		&+\frac 9{16} \sum_{n_1=0}^{n}\sum_{n_2=0}^{n-n_1}\sum_{n_3=0}^{n-n_1-n_2}\sum_{m_1=0}^{m}\sum_{m_2=0}^{m-m_1}\sum_{m_3=0}^{m-m_1-m_2}\sum_{k_1=0}^{k}\sum_{k_2=0}^{k-k_1}\sum_{k_3=0}^{k-k_1-k_2}\sum_{g_1=0}^{g-4}\sum_{g_2=0}^{g-4-g_1}\sum_{g_3=0}^{g-4-g_1-g_2}g_1u^{(n_1,m_1)[k_1]}_{g_1}u^{(n_2,m_2)[k_2]}_{g_2}u^{(n_3,m_3)[k_3]}_{g_3}u^{(n-n_1-n_2-n_3,m-m_1-m_2-m_3)[k-k_1-k_2-k_3]}_{g-4-g_1-g_2-g_3}\nonumber\\
		&-2 \sum_{n_1=0}^{n}\sum_{n_2=0}^{n-n_1}\sum_{n_3=0}^{n-n_1-n_2}\sum_{n_4=0}^{n-n_1-n_2-n_3}\sum_{n_5=0}^{n-n_1-n_2-n_3-n_4}\sum_{m_1=0}^{m}\sum_{m_2=0}^{m-m_1}\sum_{m_3=0}^{m-m_1-m_2}\sum_{m_4=0}^{m-m_1-m_2-m_3}\sum_{m_5=0}^{m-m_1-m_2-m_3-m_4}\sum_{k_1=0}^{k}\sum_{k_2=0}^{k-k_1}\sum_{k_3=0}^{k-k_1-k_2}\sum_{k_4=0}^{k-k_1-k_2-k_3}\sum_{k_5=0}^{k-k_1-k_2-k_3-k_4}\nonumber\\
		&\sum_{g_1=0}^{g}\sum_{g_2=0}^{g-g_1}\sum_{g_3=0}^{g-g_1-g_2}\sum_{g_4=0}^{g-g_1-g_2-g_3}\sum_{g_5=0}^{g-g_1-g_2-g_3-g_4}u^{(n_1,m_1)[k_1]}_{g_1}u^{(n_2,m_2)[k_2]}_{g_2}u^{(n_3,m_3)[k_3]}_{g_3}u^{(n_4,m_4)[k_4]}_{g_4}u^{(n_5,m_5)[k_5]}_{g_5}u^{(n-n_1-n_2-n_3-n_4-n_5,m-m_1-m_2-m_3-m_4-m_5)[k-k_1-k_2-k_3-k_4-k_5]}_{g-g_1-g_2-g_3-g_4-g_5}\nonumber\\
		&+2 \sum_{n_1=0}^{n}\sum_{n_2=0}^{n-n_1}\sum_{n_3=0}^{n-n_1-n_2}\sum_{m_1=0}^{m}\sum_{m_2=0}^{m-m_1}\sum_{m_3=0}^{m-m_1-m_2}\sum_{k_1=0}^{k}\sum_{k_2=0}^{k-k_1}\sum_{k_3=0}^{k-k_1-k_2}\sum_{g_1=0}^{g}\sum_{g_2=0}^{g-g_1}\sum_{g_3=0}^{g-g_1-g_2}u^{(n_1,m_1)[k_1]}_{g_1}u^{(n_2,m_2)[k_2]}_{g_2}u^{(n_3,m_3)[k_3]}_{g_3}u^{(n-n_1-n_2-n_3,m-m_1-m_2-m_3)[k-k_1-k_2-k_3]}_{g-4-g_1-g_2-g_3}\nonumber\\
		&-\frac 1{4} \sum_{n_1=0}^{n}\sum_{n_2=0}^{n-n_1}\sum_{n_3=0}^{n-n_1-n_2}\sum_{m_1=0}^{m}\sum_{m_2=0}^{m-m_1}\sum_{m_3=0}^{m-m_1-m_2}\sum_{k_1=0}^{k+1}\sum_{k_2=0}^{k+1-k_1}\sum_{k_3=0}^{k+1-k_1-k_2}\sum_{g_1=0}^{g-4}\sum_{g_2=0}^{g-4-g_1}\sum_{g_3=0}^{g-4-g_1-g_2}u^{(n_1,m_1)[k_1]}_{g_1}u^{(n_2,m_2)[k_2]}_{g_2}u^{(n_3,m_3)[k_3]}_{g_3}u^{(n-n_1-n_2-n_3,m-m_1-m_2-m_3)[k+1-k_1-k_2-k_3]}_{g-4-g_1-g_2-g_3}\nonumber\\
		&+q^2\delta_{n,0}\delta_{m,0}\delta_{k,0}\delta_{g,4}.
	\end{align}
\end{landscape}
%\restoregeometry

The coefficients of this recursion relation share a lot of similarities and properties with the original homogeneous Painlev\'e II equation. For that reason, we refer the reader to \cite{bssv21,sv13} for further details. Here, we will limit ourselves to mention the properties and similarities. 

First, notice that the leading term in equation \eqref{eq:recrelation} gets cancelled whenever $|n-m|=1$. This is not an uncommon phenomenon. In fact, this was very much expected from previous cases \cite{asv11,bssv21,gikm10,sv13} and it is due to resonance. Therefore, one has a freedom in choosing the starting coefficients of these sectors. Here we take the same conventions as in \cite{bssv21,sv13}, which basically means that we choose the term proportional to $w^1$ in the $(n+1,n)[0]$, $(n,n+1)[0]$ sectors to be given by $\delta_{n,0}$\footnote{Notice that this assumption was already implemented in the definition of $\beta_{(n,m)}^{[k]}$. Upon changing this condition, one would need to find a new suitable expression.}. Furthermore, the $k$-th sector is related to the 0-th one, with the same proportionality constant,
\begin{equation}
	u^{(n,m)[k]}_{g}=\frac{1}{k!}\left(8(m-n)\right)^ku^{(n-k,m-k)[0]}_g.
\end{equation}
As stated before, this means that the logarithmic sectors can be considered as a useful artefact to solve the problem of resonance, but they can always be resummed so that, in the end, one will not see any logarithmic contributions in $w$.

Secondly, the diagonal $(n,n)$ sectors have a power expansion in powers of $w^4$, which means that the coefficients $u^{(n,n)[0]}_{2(2g+1)}=0$. This will allow us to obtain relations between ``forward'' and ``backward'' Borel residues, as discussed in the next section \cite{bssv21}. Finally, there is also a symmetry of this recursion equation relating the $(n,m)\leftrightarrow(m,n)$ sectors. This symmetry is the same one as has been extracted for the homogeneous case \cite{sv13}
\begin{equation}
	(-1)^{g+\beta^{[k]}_{(n,m)}}u^{(n,m)[k]}_{2g}=(-1)^{\frac{n+m	}{2}}u^{(m,n)[k]}_{2g}.
\end{equation}
Taking all of this into account one can use the recursion relation to generate data for the sectors, {\it e.g.},
\begin{align}
	\Phi_{(0,0)}(w)=&1 + \left(-\frac{1}{16} + \frac{q^2}{4}\right) w^4 + \left(-\frac{73}{512} + \frac{41 q^2}{64} - \frac{9 q^4}{32}\right) w^8+\nonumber\\
	&+ \left(-\frac{10657}{8192} + \frac{12419 q^2}{2048} - \frac{1827 q^4}{512} + \frac{
		65 q^6}{128}\right) w^{12}+\cdots,\\
	\Phi_{(1,0)}(w)=&w + \left(-\frac{17}{96} + q^2\right) w^3 + \left(\frac{1513}{18432} - \frac{53 q^2}{96} + \frac{q^4}{
		2}\right) w^5 +\nonumber \\
	& + \left(-\frac{850193}{5308416} + \frac{17185 q^2}{18432} - \frac{59 q^4}{64} + 
	\frac{q^6}6\right) w^7+\cdots,\\
	\Phi_{(2,0)}(w)=&\frac 12 w^2+ \left(-\frac{41}{96} + q^2\right) w^4 + \left(\frac{5461}{9216} - \frac{113 q^2}{48} + 
	q^4\right) w^6 +\nonumber\\
	&+ \left(-\frac{1734407}{1327104} + \frac{27937 q^2}{4608} - \frac{69 q^4}{16} + \frac{2 q^6}{3}\right) w^8+\cdots,\\
	\Phi_{(1,1)}(w)=&-3w^2+\left(-\frac{291}{128} + 9 q^2\right) w^6 + \left(-\frac{447441}{32768} + \frac{15705 q^2}{256} - \frac{219 q^4}{8}\right) w^{10}+\nonumber \\&+ \left(-\frac{886660431}{4194304} +\frac{2020725 q^2}{2048}  -\frac{601383 q^4}{1024} +\frac{687 q^6}{8} \right) w^{14}+\cdots,\\
	\Phi_{(2,1)}(w)=&w^3 + \left(-\frac{115}{48} + 14 q^2 \right) w^5 + \left(\frac{30931}{18432} - \frac{1279 q^2}{96} + \frac{27 q^4}2\right) w^7 +\nonumber\\
	&+\left(-\frac{4879063}{663552} + \frac{98923 q^2}{2304} - \frac{989 q^4}{24} + \frac{
		20 q^6}3\right) w^9+\cdots+\nonumber\\
	&+\log(w)\left\{-8w + \left(\frac{17}{12} - 8 q^2\right) w^3 + \left(-\frac{1513}{2304} + \frac{53 q^2}{12} - 
	4 q^4\right) w^5\right.+\nonumber\\&\left.+ \left(\frac{850193}{663552} - \frac{17185 q^2}{2304} + \frac{59 q^4}{8} - \frac{4 q^6}{3}\right) w^7+\cdots\right\}.
\end{align}
Note that all of these coincide, as expected, with \cite{sv13} in the limit $q\to 0$. Now that we have established what tools we will be working with, the key part of this paper is to give a conjectural form for Stokes data, while performing a numerical exploration of the {\it ``first''} ones. 

%%%%%%%%%%%%%%%%%%%%%%%%%%%%%%%%%%%%%%%%%%%%%%%%%%%%%%%%%%%%%%%%%
%%%%%%%%%%%%%%%%%%%%%%%%%%%%%%%%%%%%%%%%%%%%%%%%%%%%%%%%%%%%%%%%%
\subsubsection{The Structure of Stokes Data and Borel Residues}\label{subsec:stok-bor}
%%%%%%%%%%%%%%%%%%%%%%%%%%%%%%%%%%%%%%%%%%%%%%%%%%%%%%%%%%%%%%%%%
%%%%%%%%%%%%%%%%%%%%%%%%%%%%%%%%%%%%%%%%%%%%%%%%%%%%%%%%%%%%%%%%%

\begin{figure}[hbt!]
	\centering
	\includegraphics[trim={8cm 4.5cm 0 0},clip,scale=0.4]{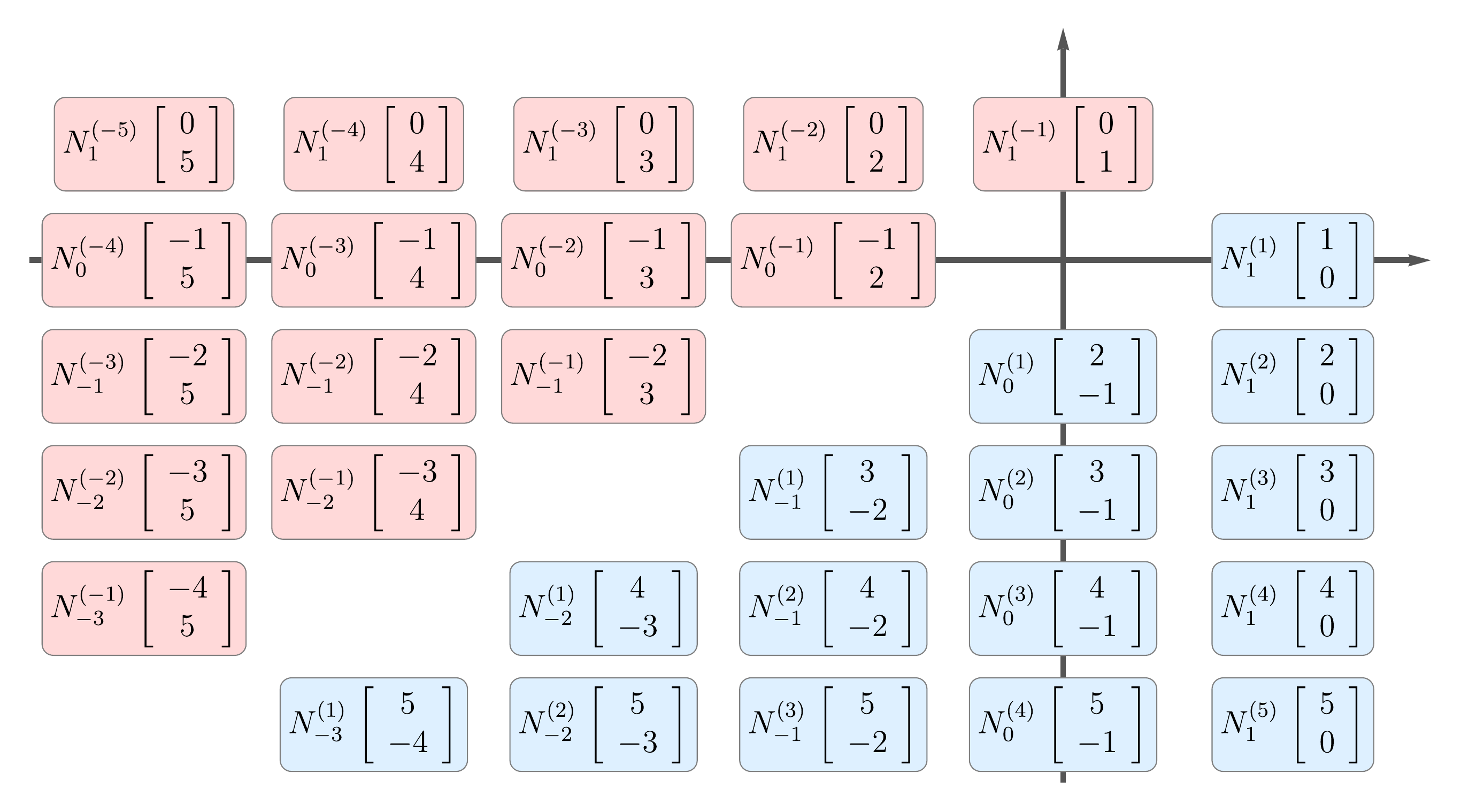}
	\caption{Vectorial structure of Stokes data taken from \cite{bssv21}. In the proportionality factors $N_{\ell-p}^{(\ell)}$ the superscript indicates which diagonal we are on; whereas the subscript indicates the ``depth'' along the selected diagonal. For forward vectors, $\ell-p$ corresponds to the $x$-coordinate on the grid, while for backward it corresponds to the $y$-coordinate. }
	\label{fig:vec_struct}
\end{figure}

As a general idea, the role of Stokes data in our non-perturbative transseries framework is to encode transition functions, {\it i.e.}, to explain how the transseries parameters change upon crossing Stokes lines\footnote{For further details on Borel residues, Stokes data, their connection, alien calculus, \textit{etc.} see \cite{abs18,bssv21,s14}.}. They come in the form of Stokes vectors via the connection to alien calculus; or as Borel residues. Both concepts basically encode the singularity structure of the sectors. In this sense, one can write asymptotic formulae for each Borel sector by using Cauchy's theorem\footnote{Notice that we are using vector notation in order to keep the equations short and convey the general idea; whereas later-on we will use explicit expressions for the two-dimensional case we are analysing.}
\begin{equation}
	\label{eq:cauchythm}
	\Phi_{\boldsymbol{n}}(x) \simeq - \frac1{2\pi\rmi}\int_0^{+\infty} \rmd w\, \frac{\operatorname{Disc}_0 \Phi_{\boldsymbol{n}}(w)}{w-x} - \frac1{2\pi\rmi}\int_0^{-\infty} \rmd w\, \frac{\operatorname{Disc}_\pi\Phi_{\boldsymbol{n}}(w)}{w-x}.
\end{equation}
The key part is that the discontinuity  $\mbox{Disc}_\theta$ along the $\theta$ direction can be written either in terms of Borel residues or via the Stokes automorphism $\underline{\mathfrak{S}}_\theta$ as $\mbox{Disc}_\theta= \1 - \underline{\mathfrak{S}}_\theta$. Let us start with the former: by choosing the appropriate Borel transformation, our sectors become simple resurgent functions (see \cite{bssv21}), 
\begin{equation}
	\label{eq:simple_res_func}
	\CB \left[ \Phi_{\boldsymbol{n}} \right] (s) \Big|_{s = \boldsymbol{\ell} \cdot \boldsymbol{A}} \sim \sum_{\bm{p} \in \ker\mathfrak{P}} \mathsf{S}_{\boldsymbol{n} \to \boldsymbol{n} + \boldsymbol{\ell} + \bm{p}} \left( \frac{\phi_{\boldsymbol{n} + \boldsymbol{\ell} + \bm{p}}}{2\pi\mathrm{i} \left( s-\boldsymbol{\ell} \cdot \boldsymbol{A}\right)} + \CB \left[ \Phi_{\boldsymbol{n} + \boldsymbol{\ell} + \bm{p}} \right] (s-\boldsymbol{\ell} \cdot \boldsymbol{A})\, \frac{\log \left( s-\boldsymbol{\ell} \cdot \boldsymbol{A} \right)}{2\pi\rmi}\right),
\end{equation}
where $\CB \left[ \Phi_{\boldsymbol{n}} \right] (s)$ is the Borel transform and we are evaluating its behaviour in a neighbourhood of the point $\boldsymbol{\ell} \cdot \boldsymbol{A} = (\ell_1, \ell_2)\cdot(A,-A)$. The sum runs over the kernel of the projection map that takes combinations of our actions to zero. In our case, this amounts to $\bm{p}=(p,p)$. Finally, the term $\phi_{\boldsymbol{n} + \boldsymbol{\ell} + \bm{p}}$ is the first coefficient in the asymptotic expansion of the $\Phi_{\boldsymbol{n} + \boldsymbol{\ell} + \bm{p}}$ sector.
Hence, at each singularity, the structure is given by a pole and a logarithmic branch-cut which are multiplied by a number---the Borel residue, $\mathsf{S}_{\boldsymbol{n} \to \boldsymbol{m}}$. Note that due to resonance, at each singularity we get a tower of contributions coming from the sum over the kernel of the resonant projection. Upon resummation, we can take the latter result to the original variable, obtaining 
\begin{align}
	\label{eq:0stokesautborelres}
	\underline{\mathfrak{S}}_0 \Phi_{(n,m)} &= \Phi_{(n,m)} - \sum_{\ell=1}^{+\infty} \rme^{-\ell \frac{A}{x}} \sum_{p=0}^{\min(n+\ell,m)} \mathsf{S}_{(n,m)\to(n+\ell-p,m-p)}\, \Phi_{(n+\ell-p,m-p)}, \\
	\label{eq:pistokesautborelres}
	\underline{\mathfrak{S}}_\pi \Phi_{(n,m)} &= \Phi_{(n,m)} - \sum_{\ell=1}^{+\infty} \rme^{+\ell \frac{A}{x}} \sum_{p=0}^{\min(n,m+\ell)} \mathsf{S}_{(n,m)\to(n-p,m+\ell-p)}\, \Phi_{(n-p,m+\ell-p)},
\end{align}
On the other hand, we can also determine the latter Stokes automorphism in terms of alien calculus. This studies the nature of the singularities on the Borel plane, where the key part is that the alien derivatives $\Delta_{\pm\ell A}$ are determined in terms of Stokes vectors, $\bm{S}_{(a,b)}$,
\begin{align}
	\underline{\mathfrak{S}}_0 &= \exp \left( \sum_{\ell=1}^{+\infty} \rme^{-\ell\frac{A}{x}}\, \Delta_{\ell A} \right), \\
	\underline{\mathfrak{S}}_\pi &= \exp \left( \sum_{\ell=1}^{+\infty} \rme^{+\ell\frac{A}{x}}\, \Delta_{-\ell A} \right),
\end{align}
where
\begin{align}
	\Delta_{\ell A} \Phi_{(n,m)} &= \sum_{p=0}^{\min(n+1,m+1-\ell)} \bm{S}_{(1-p,1-p-\ell)} \cdot \left[ \begin{array}{c}n+1-p\\m+1-p-\ell\end{array} \right] \Phi_{(n+1-p,m+1-p-\ell)}, \\
	\Delta_{-\ell A} \Phi_{(n,m)} &= \sum_{p=0}^{\min(n+1-\ell,m+1)} \bm{S}_{(1-p-\ell,1-p)} \cdot \left[ \begin{array}{c}n+1-p-\ell\\m+1-p\end{array} \right] \Phi_{(n+1-p-\ell,m+1-p)}.
\end{align}
Therefore, we can describe connection formulae by knowing either the Stokes vectors, $\bm{S}_{\boldsymbol{n}}$, or the Borel residues, $\mathsf{S}_{\boldsymbol{n} \to \boldsymbol{m}}$. Notice that our vectors have an interesting structure: one can represent Stokes vectors in a grid, like the one in figure \ref{fig:vec_struct}.
The two ways of writing the Stokes automorphism provide maps between the two sets. Both of them can be found in \cite{bssv21}, section 3.3. Here, we just recall that 
\begin{equation}
	\label{eq:qboreltostokes}
	\mathsf{S}_{(n,n)\to (p+\ell,p)}=-(n+1)N^{(\ell)}_{\ell-n+p}+\mathsf{R}^{(n,n)}_{(p+\ell,p)},
\end{equation}
where the remainder $\mathsf{R}^{(n,n)}_{(p+\ell,p)}$ is made out of products of $N^{(\ell_i)}_{x_i}$ such that for each term $\sum_i\ell_i=\ell$ with $\ell_i<\ell$. While Borel residues are more straightforward to compute, Stokes vectors are less in number and have the advantage that they contain all information in a clearer way, since they can be used to generate the Stokes vector fields that will give rise to connection formulae---one can find an explicit example in section \ref{subsec:transfunct}. Thus, our objective would be to obtain {\it all} of them for our problem. In order to understand the lattice in figure \ref{fig:vec_struct}, one can make use of the so-called {\it backward-forward symmetry}---which we will derive later on, see equation \eqref{eq:qback-forw}. This relates ``forward'' Borel residues with ``backward'' ones\footnote{This notation of ``forward'' and ``backward'' comes form the fact that, on the Borel plane, the singularities locate upon the real line. Thus, any residue Borel residue $\mathsf{S}_{(n,n)\to(a,b)}$ with $a>b$ is located on the positive real line---thus, a ``forward'' one; while the ones with $b>a$ are located on the negative real line, hence a ``backward'' one. The same notion gets translated into Stokes vectors.} and thus tells us that we can obtain half of the Stokes grid, say the red ``backward'' Stokes, from the blue ``forward'' ones. Let us then focus on how to obtain the blue ``forward'' Stokes data. It is clear that two ingredients are needed: a pattern going down each diagonal, and another one going down the first column. For the former, there is a conjecture in \cite{bssv21} which has been called {\it closed-form asymptotics}, which we will review in the next section; while the column pattern has to be guessed from numerical observations and can be arbitrarily complicated. Both closed-form asymptotics and the backward-forward symmetry have their origin in equation \eqref{eq:cauchythm}. The idea is to obtain a large-order formula for each sector. In particular, we will be interested in the diagonal sectors, since they are a sufficient set to derive all Stokes data---see \cite{bssv21}. Let us now define the $H,\,\widetilde{H}$ functions \cite{asv11} 
\begin{align}
	\widetilde{H}_k(g,\ell)&=\int_0^{+\infty}\text{d}z\,z^{g-1}\rme^{-sz}\log(z)^k=\frac{\partial^k}{\partial g^k}\frac{\Gamma(g)}{\ell^g},\\
	\label{eq:H}
	H_k(g,\ell)&=\int_0^{+\infty}\text{d}z\,z^{g-1}\rme^{-sz}\log(-z)^k=\sum_{t=0}^k\left(\begin{array}{c}
		k\\
		t
	\end{array}\right)(\rmi\pi)^{k-t	}\widetilde{H}_t(g,\ell),
\end{align}
where we have used the determination $\log(-z)=\log(z)+\rmi\pi$ for $z$ real and positive\footnote{Notice that this means that we are computing Stokes data for the negative $+\pi$ direction of the Borel plane.}. Therefore, we can use the automorphisms \eqref{eq:0stokesautborelres}-\eqref{eq:pistokesautborelres} into equation \eqref{eq:cauchythm} and use our $H,\,\widetilde{H}$ functions to obtain, after equating powers of $x$ in both sides,
\begin{align}
	\label{eq:asymptunnq}
	u_{2g}^{(n,n)[0]}&\simeq \sum_{\ell=1}^{+\infty}\sum_{h=0}^{+\infty}\sum_{p=0}^{n}\sum_{k=0}^p \,-\,\frac{\mathsf{S}_{(n,n)\to(p+\ell,p)}}{2\pi\rmi}\,\frac{u_{2h}^{(p+\ell,p)[k]}}{(-2)^k}\,\widetilde{H}_k\left(g+n-h-\beta_{(p+\ell,p)}^{[k]},\ell A\right)\nonumber\\
	&\hspace{4mm}-\frac{\mathsf{S}_{(n,n)\to(p,p+\ell)}}{2\pi\rmi}\,\frac{u_{2h}^{(p,p+\ell)[k]}}{(-2)^k}\,(-1)^{h-g-n-\beta_{(p,p+\ell)}^{[k]}} \, H_k\left(g+n-h-\beta_{(p,p+\ell)}^{[k]},\ell A\right).
\end{align}
Notice that $\beta_{(p+\ell,p)}^{[k]}=\beta_{(p,p+\ell)}^{[k]}$. Now, the key parts to deriving a connection between the backward and forward Borel residues is the fact that $u_{2g}^{(n,n)[0]}=0$ for odd-$g$, and the relation between $H$ and $\widetilde{H}$ that appears in the second equality of \eqref{eq:H}. Putting everything together, one finds the final form of the {\it backward-forward symmetry} \cite{bssv21},
\begin{equation}
	\label{eq:qback-forw}
	\mathsf{S}_{(n,n)\to(p,p+\ell)}=\rmi^\ell(-1)^{n+p+\ell}\sum_{r=p}^{n}\frac{(-4\pi\rmi \ell)^{r-p}}{(r-p)!}\mathsf{S}_{(n,n)\to(p+\ell,p)}.
\end{equation} 
In order to further simplify our asymptotic formula \eqref{eq:asymptunnq}, we can implement the above symmetry into the it and evaluate it for even-$g$. This gives
\begin{equation}
	\label{eq:fin_as}
	u_{4g}^{(n,n)[0]} \simeq - \frac{1}{\rmi\pi} \sum_{\ell=1}^{+\infty} \sum_{h=0}^{+\infty} \sum_{p=0}^{n} \sum_{k=0}^{p} \mathsf{S}_{(n,n)\to(p+\ell,p)}\, (-2)^{-k}\, u_{2h}^{(p+\ell,p)[k]}\, \widetilde{H}_k \left( 2g+n-h-\beta^{(k)}_{(p+\ell,p)}, \ell A \right).
\end{equation}
%%%%%%%%%%%%%%%%%%%%%%%%%%%%%%%%%%%%%%%%%%%%%%%%%%%%%%%%%%%%%%%%%
%%%%%%%%%%%%%%%%%%%%%%%%%%%%%%%%%%%%%%%%%%%%%%%%%%%%%%%%%%%%%%%%%
\subsubsection{Closed-Form Asymptotics}\label{subsec:q-closed-form-asymp}
%%%%%%%%%%%%%%%%%%%%%%%%%%%%%%%%%%%%%%%%%%%%%%%%%%%%%%%%%%%%%%%%%
%%%%%%%%%%%%%%%%%%%%%%%%%%%%%%%%%%%%%%%%%%%%%%%%%%%%%%%%%%%%%%%%%
The idea behind the method of closed-form asymptotics \cite{bssv21} is to extract equations for our Borel residues out of the asymptotic expansions \eqref{eq:fin_as}. This amounts to understanding the asymptotic behaviour of the $\widetilde H$ functions. Their leading large-order growth is given by
\begin{equation}
	\label{eq:HtildeHasymptoticgrowth}
	H_k(g,\ell A)\simeq \widetilde{H}_k (g,\ell A)\simeq (g-1)! \log(g)^k (\ell A)^{-g}.
\end{equation}
The key part then would be understanding the role that logarithms play and how they turn out to be a tool to obtain {\it equations} from {\it asymptotic formulae}. Roughly speaking, it consists of isolating the different large-order behaviours until one reaches some way of extracting {\it asymptotic relations}, equations that allow to determine Stokes data analytically. In summary, the method was dissected into three parts \cite{bssv21}: 
\begin{enumerate}
	\item hide on the left-hand side of the asymptotic relation anything that does not give new conditions on Stokes factors;
	\item understand the logarithm structure of the resulting computation;
	\item find a suitable asymptotic limit, inspired by properties of gamma functions and numerical evidence. 
\end{enumerate}
Let us start by defining the truncated series 
\begin{align}
	T_{g,\ell}^{(n,n)} 
	\simeq - \frac{1}{\rmi\pi} \sum_{r=1}^{\ell-1} \sum_{h=0}^{+\infty} \sum_{p=0}^{n} \sum_{k=0}^{p} \mathsf{S}_{(n,n)\to(p+r,p)}\, \frac{1}{k!} \left( 4 r \right)^{k} u_{2h}^{(p+r-k,p-k)[0]}\, \widetilde{H}_k \left( 2g+n-h-\beta^{(k)}_{(p+r,p)}, rA  \right).
\end{align}
We can subtract this series to our asymptotic expansion, then divide by the leading growth and ignore subleading contributions (those associated with $h>0$, $k<p$)
\begin{align}
	\frac{u_{4g}^{(n,n)[0]} - T_{g,\ell}^{(n,n)}}{\widetilde{H}_0 \left( 2g+n-\frac{\ell }{2}, \ell A \right)} \simeq - \frac{1}{\rmi\pi} \sum_{p=0}^{n} \mathsf{S}_{(n,n)\to(p+\ell,p)}\, \frac{1}{p!} \left( 4\ell \right)^{p} u_{0}^{(\ell,0)[0]}\, \frac{\widetilde{H}_p \left( 2g+n-\frac{\ell}{2}, \ell A \right)}{\widetilde{H}_0 \left( 2g+n-\frac{\ell}{2}, \ell A \right)} + \cdots.
\end{align}
Given that we are interested in isolating Borel residues, we define the sequence $\widetilde{D}_{g,\ell}^{(n,n)}$ via
\begin{equation}
\label{eq:PreAsymptoticRelation}
\left( n+1 \right) \widetilde{D}_{g,\ell }^{(n,n)}:=\frac{\rmi\pi}{u_{0}^{(\ell,0)[0]}}\, \frac{u_{4g}^{(n,n)[0]} - T_{g,\ell}^{(n,n)}}{\widetilde{H}_0 \left( 2g+n-\frac{\ell}{2}, \ell A \right)} \simeq - \sum_{p=0}^{n} \mathsf{S}_{(n,n)\to(p+\ell,p)}\, \frac{1}{p!} \left( 4\ell \right)^{p} \frac{\widetilde{H}_p \left( 2g+n-\frac{\ell}{2}, \ell A \right)}{\widetilde{H}_0 \left( 2g+n-\frac{\ell}{2}, \ell A \right)}.
\end{equation}
\noindent
Therefore, understanding $\widetilde{D}_{g,\ell}^{(n,n)}$ is crucial to obtaining equations for the Borel residues. The next move is to understand the ratio of $\widetilde{H}$ functions. This is a cumbersome computation and we have dedicated appendix \ref{app:Hratio} to this calculation. Using the final result \eqref{eq:hratios}, the latter equation reads
\begin{align}
\left( n+1 \right) \widetilde{D}_{g,\ell}^{(n,n)} &\simeq \sum_{p=0}^{n} \mathsf{S}_{(n,n)\to(p+\ell,p)}\, \frac{1}{p!} \left(4\ell \right)^{p} \times \\ 
&
\hspace{-40pt}
\times B_{p} \left( \psi^{(0)} \left( 2g+n-\frac{\ell}{2} \right) - \log \left(\ell A \right), \psi^{(1)} \left( 2g+n-\frac{\ell}{2} \right), \ldots, \psi^{(p-1)} \left( 2g+n-\frac{\ell}{2} \right) \right), \nonumber
\end{align}
\noindent
where $B_{p}$ denotes the complete Bell polynomial of order $p$ and $\psi^{(n)}$ are the polygamma functions. Notice that the left-hand side of this equation has been defined with an artificial factor $(n+1)$. This is because we would like to obtain equations for the Stokes vectors, $N^{(\ell)}_{x}$, and our map \eqref{eq:qboreltostokes} contains precisely this factor. Thus, our intention is to hide the reminder $\mathsf{R}^{(n,n)}_{(p+\ell,p)}$ into a redefinition of $\widetilde{D}_{g,\ell}^{(n,n)}$. Thus, calling this new quantity $D_{g,\ell}^{(n,n)}$, we have
\begin{align}
	\label{eq:AsymptoticRelationWithN}
	D_{g,\ell}^{(n,n)} &\simeq \sum_{p=0}^{n-\ell+1} N_{\ell-n+p}^{(\ell)}\, \frac{1}{p!} \left(4\ell \right)^p \times \\
	&
	\hspace{-30pt}
	\times B_{p} \left( \psi^{(0)} \left( 2g+n-\frac{\ell}{2} \right) - \log \left(\ell A \right), \psi^{(1)} \left( 2g+n-\frac{\ell}{2} \right), \ldots, \psi^{(p-1)} \left( 2g+n-\frac{\ell}{2} \right) \right).
\end{align}
The final and more conjectural part of this procedure is understanding the asymptotic limit of the latter quantity. Interestingly enough, the digamma functions have the following structure for integer values
\begin{equation}
	\psi^{(n)}(x) =
	\begin{cases}
		\,-\gamma_{\text{E}} + \displaystyle\sum\limits_{k=1}^{x-1} \frac{1}{k}, & n = 0, \\
		\,(-1)^{n+1}\, n!\, \zeta \left( n+1 \right) - (-1)^{n+1}\, n!\, \displaystyle\sum\limits_{k=1}^{x-1} \frac{1}{k^{n+1}}, & n\geq 1,
	\end{cases}
\end{equation}
where $\gamma_{\text{E}}$ is the Euler--Mascheroni constant. 
The natural assumption in this conjecture is that all the transcendental part of the Stokes data comes uniquely from the digamma functions. In this line of thought, we consider that in the large $g$ limit we can cancel any contribution that is not the transcendental part of the digammas. Furthermore, due to the logarithmic structure, we still need to fix the first entry of the Bell polynomial and the remaining left-hand side contribution. Now, we need to guess what is the contribution from the $D^{(n,n)}_{g,\ell}$'s such that we can turn the latter into an equation for the highest $N^{(\ell)}_{\ell-n+p}$. It turns out, upon numerical observation, that one can adjust the entries of the Bell polynomials such that we can put the left-hand side to zero with one exception. This is, for the case $n=0$ one should determine the left-hand side, {\it i. e.}, one has to obtain $N^{(\ell)}_1$ by alternative methods. Furthermore, the logarithmic content of the first entry has to be also input by other means. These two details are the ones one has to fix via numerical observation. Putting these details into another renaming of $D_{g,\ell}^{(n,n)}$,  $d^{n}_{\ell}(g)$, we get
\begin{equation}
\label{eq:summedasymptoticsnologslimitinserteds}
\lim\limits_{g\to+\infty} d^{n}_{\ell}(g) = \sum_{p=0}^{n-\ell+1} N_{\ell-n+p}^{(\ell)}\, \frac{1}{p!} \left( 4 \ell\right)^p B_{p} \left(c_{\ell}, \frac{1}{\ell}\psi^{(1)}(1), \ldots, \frac{1}{\ell^{p-1}}\psi^{(p-1)}(1) \right),
\end{equation}
\noindent

\begin{align}
\lim\limits_{g\to+\infty}d^{0}_{\ell}(g) &= N^{(\ell)}_{1}, \\
\lim\limits_{g\to+\infty}d^{n}_{\ell}(g) &= 0, \qquad n > 0,
\end{align}
\noindent
with 
\begin{equation}
	c_{\ell} =  \log \widetilde{A}_\ell + \gamma_{\text{E}}.
\end{equation}
Joining the latter results with the observation in the following sections, we arrive at the asymptotic equations
\begin{align}
	\label{eq:closed-form-asymp}
	N^{(\ell)}_1\delta_{n,0}= \sum_{p=0}^{n-\ell+1}\frac{(4\ell)^p}{p!}N^{(\ell)}_{\ell-n+p}B_p&\left(\frac 1{\ell^0}\psi^{(0)}(1)-\log(8),\frac 1{\ell^1}\psi^{(1)}(1),\dots,\frac 1{\ell^{p-1}}\psi^{(p-1)}(1)\right),
\end{align}
where $N^{(\ell)}_1$ is to be numerically guessed. Thus, we just need a pattern for $N^{(\ell)}_1$ to have all the ingredients in this equation. The next section is therefore dedicated to numerical explorations and checks on these formulae.

%%%%%%%%%%%%%%%%%%%%%%%%%%%%%%%%%%%%%%%%%%%%%%%%%%%%%%%%%%%%%%%%%
%%%%%%%%%%%%%%%%%%%%%%%%%%%%%%%%%%%%%%%%%%%%%%%%%%%%%%%%%%%%%%%%%
\subsubsection{Numerical Checks}\label{subsec:qnumcheck}
%%%%%%%%%%%%%%%%%%%%%%%%%%%%%%%%%%%%%%%%%%%%%%%%%%%%%%%%%%%%%%%%%
%%%%%%%%%%%%%%%%%%%%%%%%%%%%%%%%%%%%%%%%%%%%%%%%%%%%%%%%%%%%%%%%%
This section is dedicated to presenting numerical evidence to support the aforementioned ideas; from the first Stokes coefficient, to the transseries structure, passing through a check on closed-form asymptotics. We have used two numerical methods: asymptotics, and Borel-plane residue computations. The first one has an advantage for the first and simplest checks, while the second one is more general and gives, typically, more precision. Note, however, that we have an additional complication with respect to previous implementations of these methods \cite{abs18,asv11,bssv21,sv13}. Our case is parameter-dependent. Therefore, we need to adjust to the fact that now our coefficients contain powers of this $q$ parameter. For an extensive discussion about asymptotics and the basis of resurgence we refer the reader to \cite{abs18}. The method of Borel plane residues has its theoretical basis in the same paper, but was first developed and extensively explored in \cite{bssv21}.\\

\begin{figure}[hbt!]
	\begin{subfigure}{.5\textwidth}
		\centering
		\includegraphics[width=.94\linewidth]{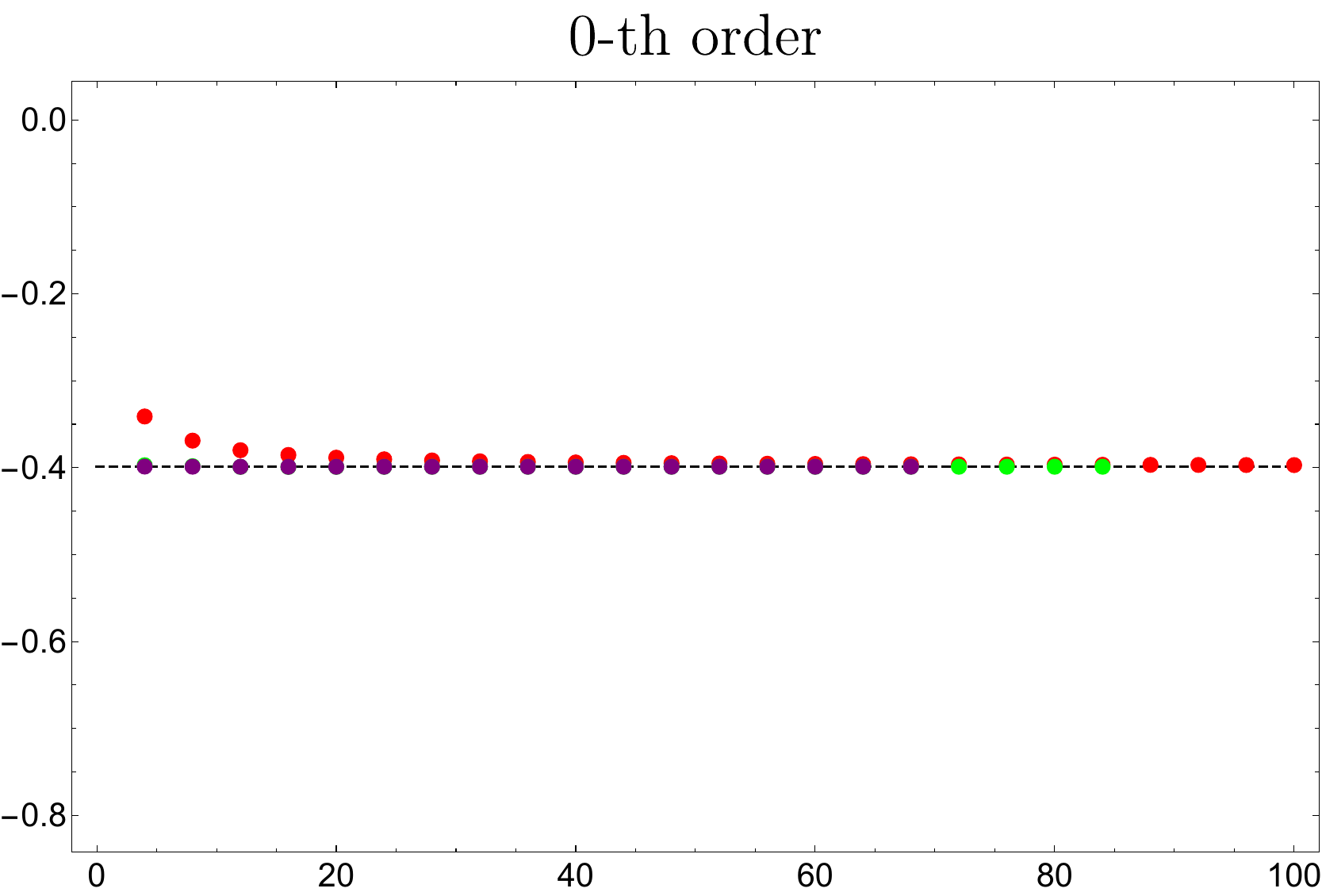}
		%\caption{1a}
		\label{fig:s100thorder}
	\end{subfigure}%
	\begin{subfigure}{.5\textwidth}
		\centering
		\includegraphics[width=.94\linewidth]{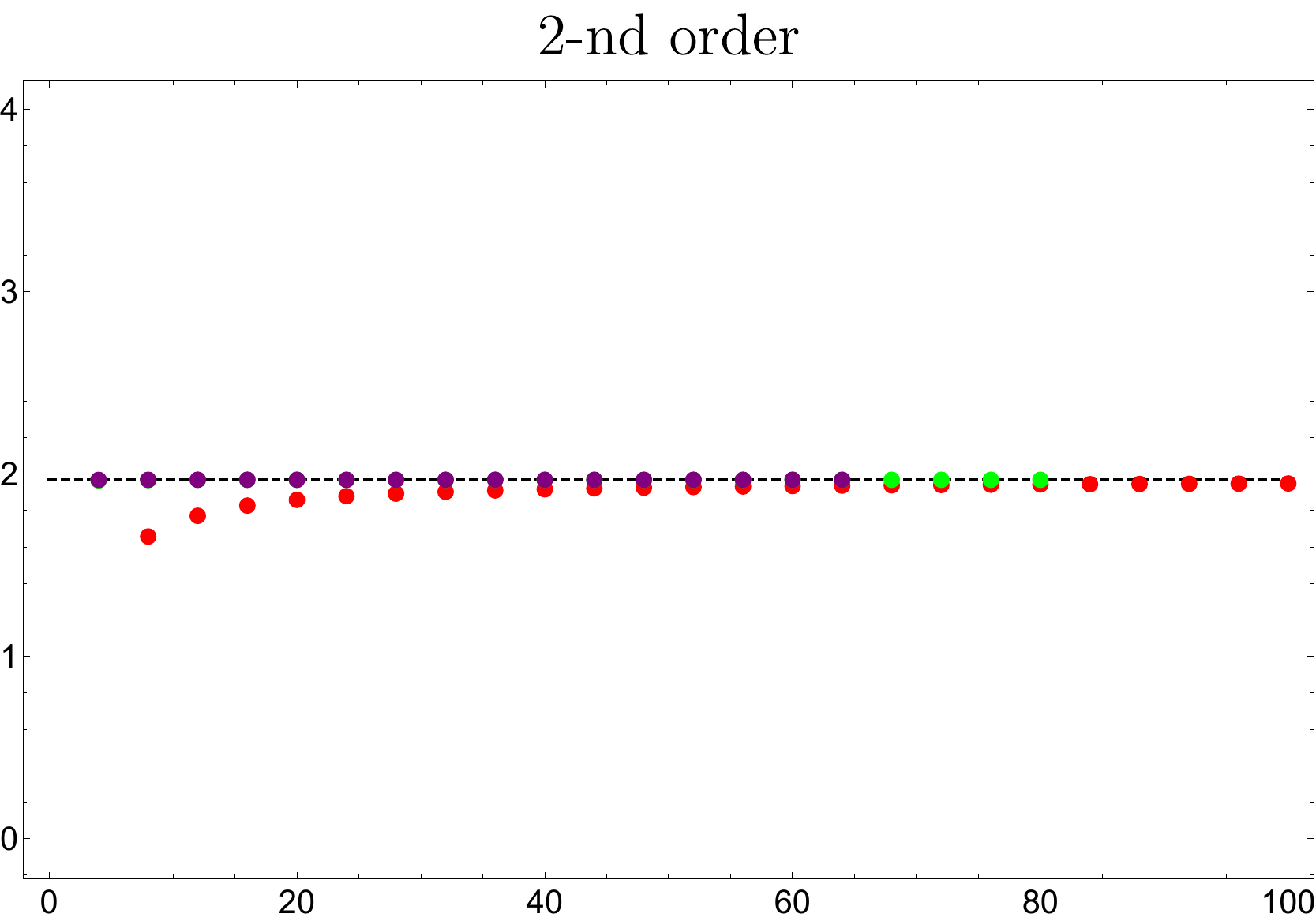}
		%\caption{1b}
		\label{fig:s102ndorder}
	\end{subfigure}\\
	\begin{subfigure}{.5\textwidth}
		\centering
		\includegraphics[width=.94\linewidth]{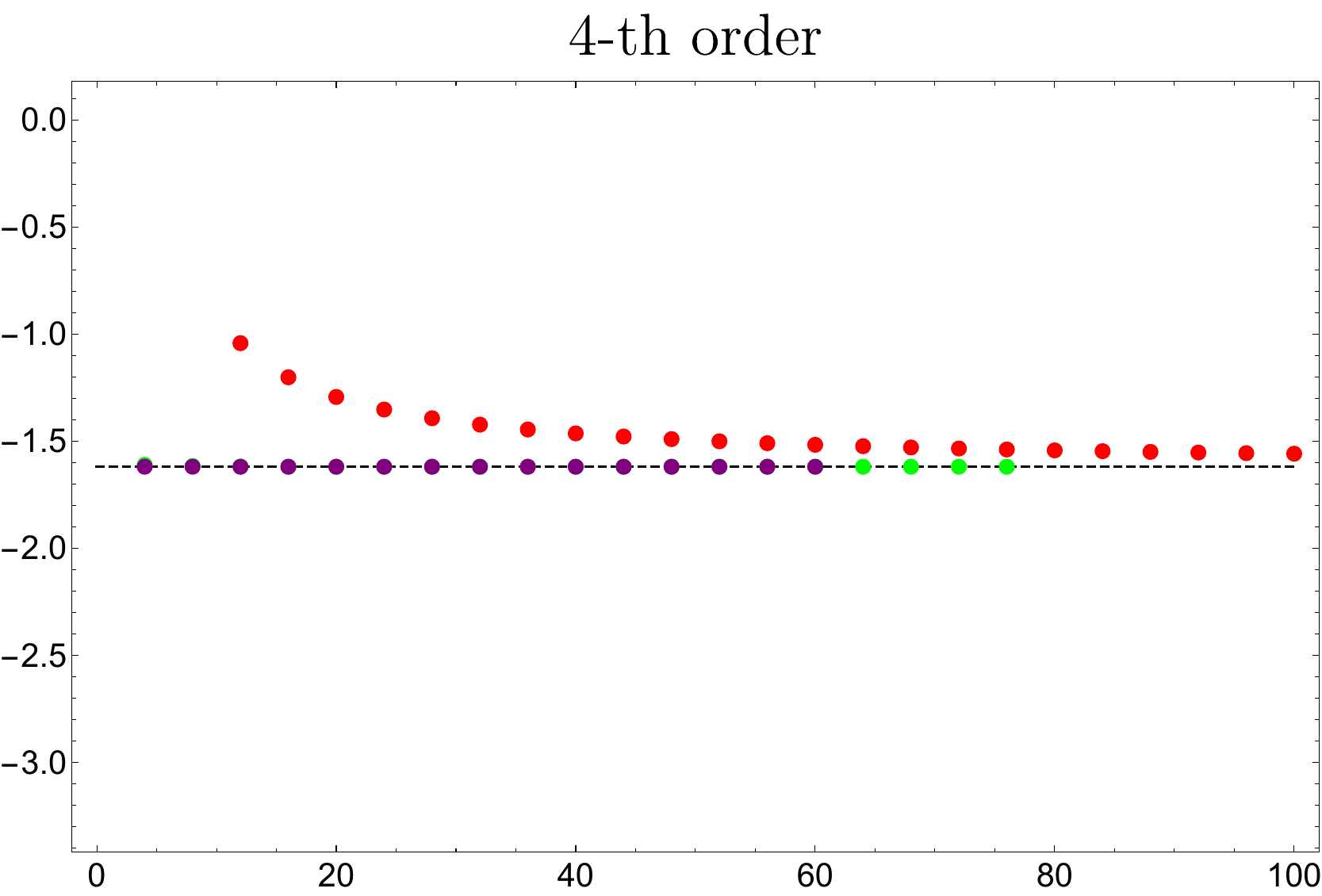}
		%\caption{1a}
		\label{fig:s104thorder}
	\end{subfigure}%
	\begin{subfigure}{.5\textwidth}
		\centering
		\includegraphics[width=.94\linewidth]{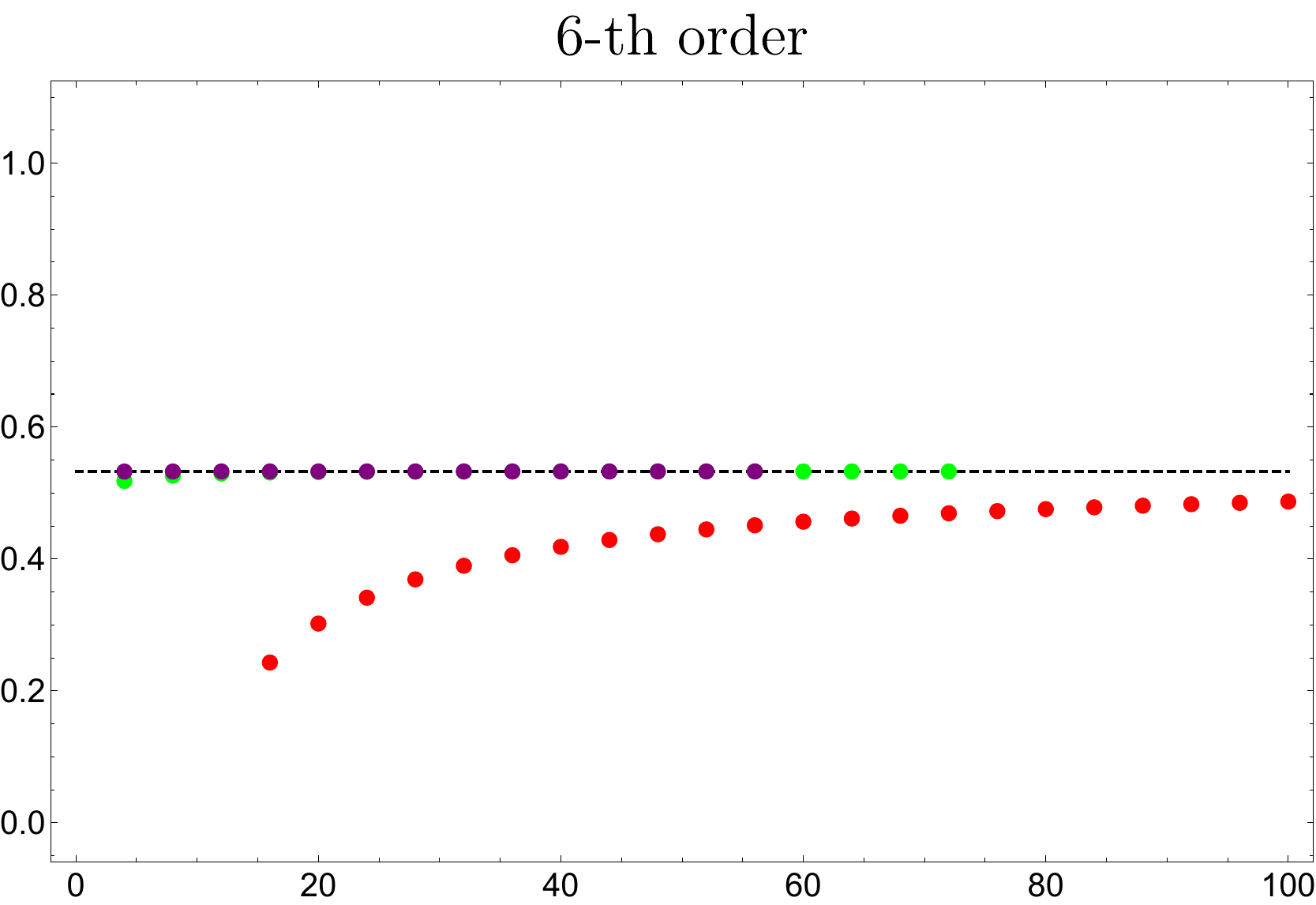}
		%\caption{1b}
		\label{fig:s106thorder}
	\end{subfigure}
	\caption{Asymptotic checks on the different orders of the Stokes coefficient $S^{(0)}_1=-\frac{\rmi}{\sqrt{2\pi}}\cos(\pi q)\simeq-\frac{\rmi}{\sqrt{2\pi}}-\left(-\frac{\rmi}{\sqrt{2\pi}}\frac{\pi^2}{2!}\right)q^2 + \left(-\frac{\rmi}{\sqrt{2\pi}}\frac{\pi^4}{4!}\right)q^4-\left(-\frac{\rmi}{\sqrt{2\pi}}\frac{\pi^6}{6!}\right)q^6+\cdots $ in $q$. The dotted black line is the analytic result; red dots correspond to the pure asymptotic prediction, while green and purple correspond to 4 and 8 Richardson transforms, respectively. These checks have been made using 25 data points, which translates into $g=100$.\label{fig:s10orders}}
\end{figure}

\noindent\textbf{Asymptotics}: Asymptotics has the advantage that it is relatively fast for the easiest case, so it is perfect to do some preliminary tests. It is also an interesting method since it allows us to check the resurgent transseries structure, as we will discuss later on. On the other hand, its precision decreases rapidly for the heavier computations \cite{bssv21}. In this sense, we have used this method to check the easiest Stokes coefficient\footnote{Here, we display the different notations that have been used in the literature \cite{abs18,asv11,bssv21,sv13}, as well as the relation with the Borel residue.}
\begin{equation}
	\label{eq:qS10}
	S_1^{(0)}(q)=N^{(1)}_1(q)=-\mathsf{S}_{(0,0)\to(1,0)}(q)=-\frac{\rmi}{\sqrt{2\pi}}\cos(\pi q),
\end{equation}
as well as to check the transseries structure. Both tests are based on the asymptotic formula\footnote{Again, in order to make the connection with other notations, we specify that 
	\begin{equation}
		\widetilde{S}_{-1}^{(0)}=N^{(-1)}_1=-\mathsf{S}_{(0,0)\to(0,1)}=\frac{1}{\sqrt{2\pi}}\cos(\pi q).
\end{equation}}
\begin{equation}
	\small
	\label{eq:asympt}
	u_{2g}^{(0,0)[0]}\simeq\sum_{k=0}^{+\infty}\frac{\left(S^{(0)}_1\right)^k}{2\pi \rmi}\sum_{h=0}^{+\infty}\frac{\Gamma\left(g-\beta_{(k,0)}^{[0]}-h+1\right)}{(kA)^{g-\beta_{(k,0)}^{[0]}-h+1}}u_{2h}^{(k,0)[0]}+\sum_{k=0}^{+\infty}\frac{\left(\widetilde{S}^{(0)}_{-1}\right)^k}{2\pi \rmi}\sum_{h=0}^{+\infty}\frac{\Gamma\left(g-\beta_{(0,k)}^{[0]}-h+1\right)}{(-kA)^{g-\beta_{(0,k)}^{[0]}-h+1}}u_{2h}^{(0,k)[0]}.
\end{equation}
Following \cite{asv11,bssv21,sv13}, one easily finds that, in the large $g$ limit, the leading contributions will come from the $k=1,\,h=0$ terms. It is also important to get rid of the divergent behaviour if one wants to be able to use Richardson transforms \cite{abs18}. Thus, the combination that yields a convergent quantity plus sub-leading terms is\footnote{Note that the $(0,0)$ sector is diagonal, we have a power expansion in $u^{(0,0)[0]}_{4g}$, while $u^{(0,0)[0]}_{2(2g+1)}=0$. However, this last fact allows us to relate backward and forward Stokes data as in \cite{asv11,bssv21,sv13}.}
\begin{equation}
	\label{eq:preparedasymp}
	\frac{2\pi \rmi\, u^{(0,0)[0]}_{4g}\, A^{2g-\beta_{(1,0)}^{[0]}+1}}{\Gamma\left(2g-\beta_{(1,0)}^{[0]}+1\right)}\simeq S_1^{(0)}\, u_1^{(1,0)[0]}+\rmi \widetilde{S}^{(0)}_{-1}\, u^{(0,1)[0]}_1+\mathcal{O}(g^{-1}),
\end{equation}
where the right-hand side can be set to be equal to $2S_1^{(0)}$ when one takes into account the values of the $u$'s (recall that we set them to be 1) and the {\it backward-forward} symmetry \eqref{eq:qback-forw}\footnote{In particular, for this case, \eqref{eq:qback-forw} gives $S_1^{(0)}=\rmi \widetilde{S}^{(0)}_{-1}=0$.}. When dealing with the $q$-parameter, we have simply taken the Taylor expansion. The results are presented both in figure \ref{fig:s10orders} and in table \ref{tab:qs10precasymp}. The figure illustrates how Richardson transformations improve the precision and accuracy for the same number of coefficients, while the table quantitatively specifies numerical precision. It is worth mentioning that, for this method, the fact that we know coefficients depending on a parameter limits the number of Richardson transforms that we can use before the method becomes time-inefficient. The explanation being that the idea behind Richardson transforms is to approximate the asymptotic behaviour faster by using combinations of coefficients to generate new ones that converge faster. Henceforth, this means that the expressions for the new coefficients get much more cumbersome with every iteration. Rapidly, the computation of highly complicated coefficients and their Taylor expansion becomes very time-consuming, rendering the method computationally intensive.

\begin{figure}[hbt!]
	\centering
	\begin{subfigure}[t]{0.5\textwidth}
		\centering
		\includegraphics[height=1.7in]{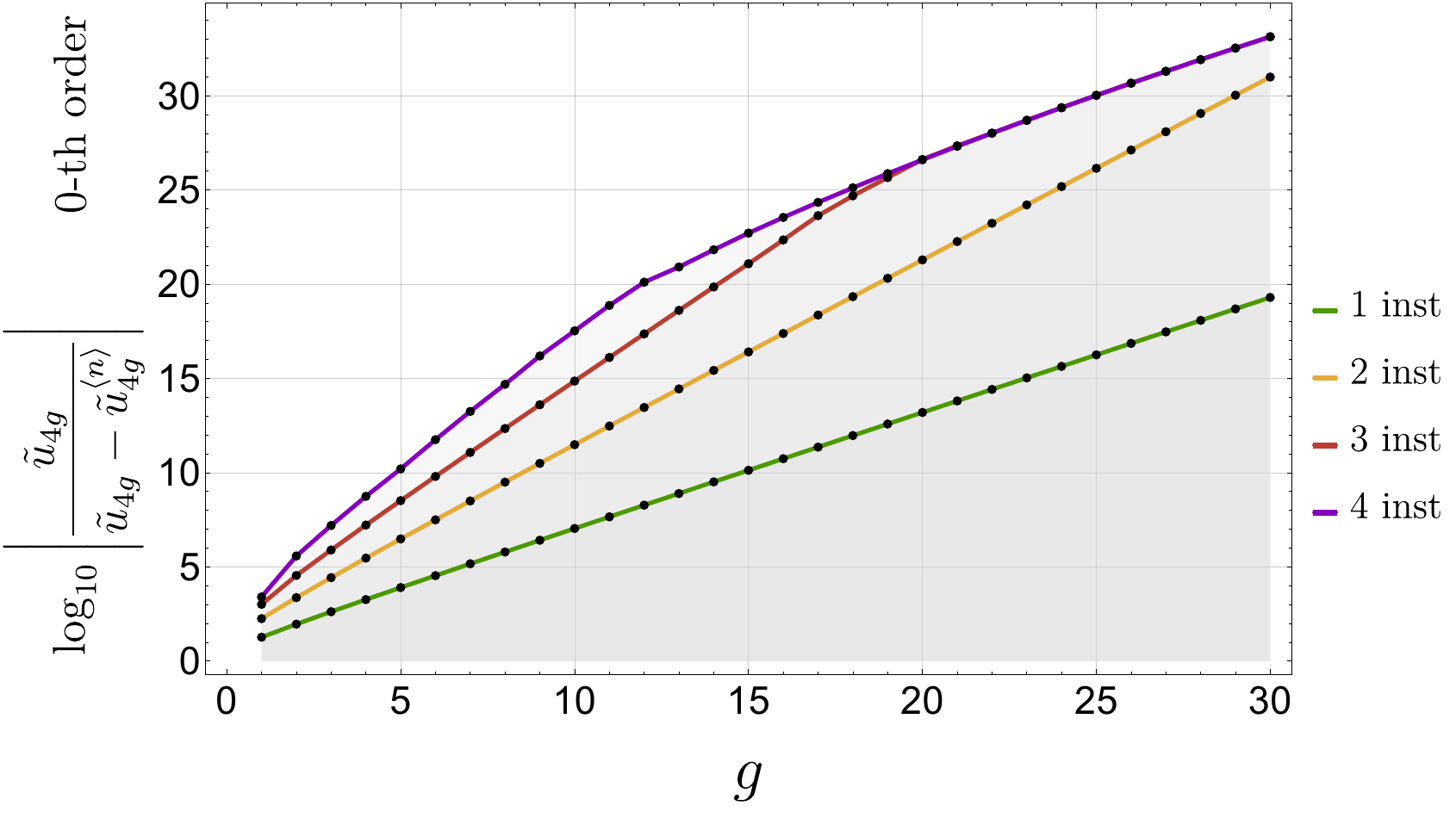}
		%\caption{}
	\end{subfigure}%
	~ 
	\begin{subfigure}[t]{0.5\textwidth}
		\centering
		\includegraphics[height=1.7in]{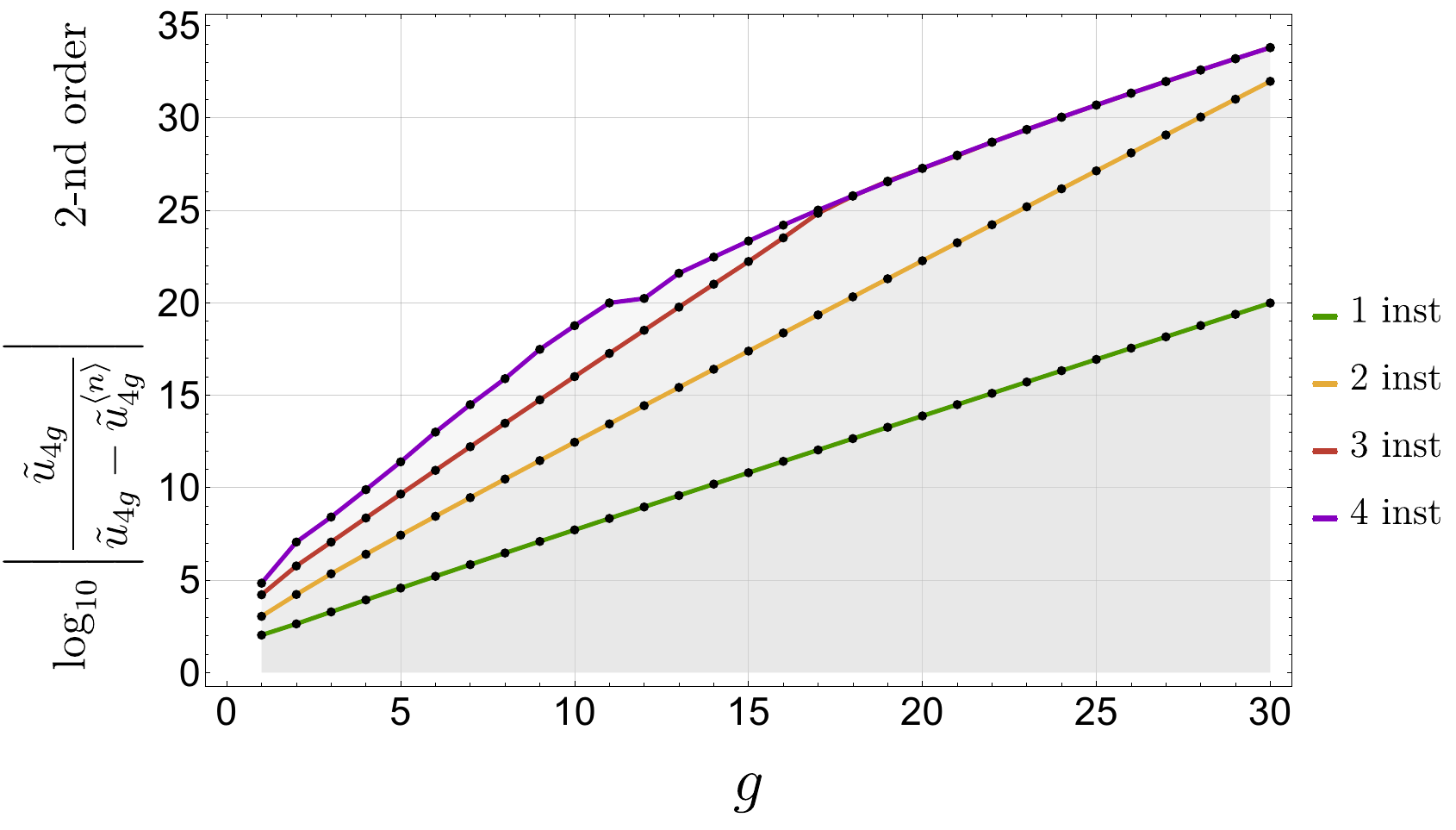}
		%\caption{Conformal transformations}
	\end{subfigure}\\
	\begin{subfigure}[t]{0.5\textwidth}
		\centering
		\includegraphics[height=1.7in]{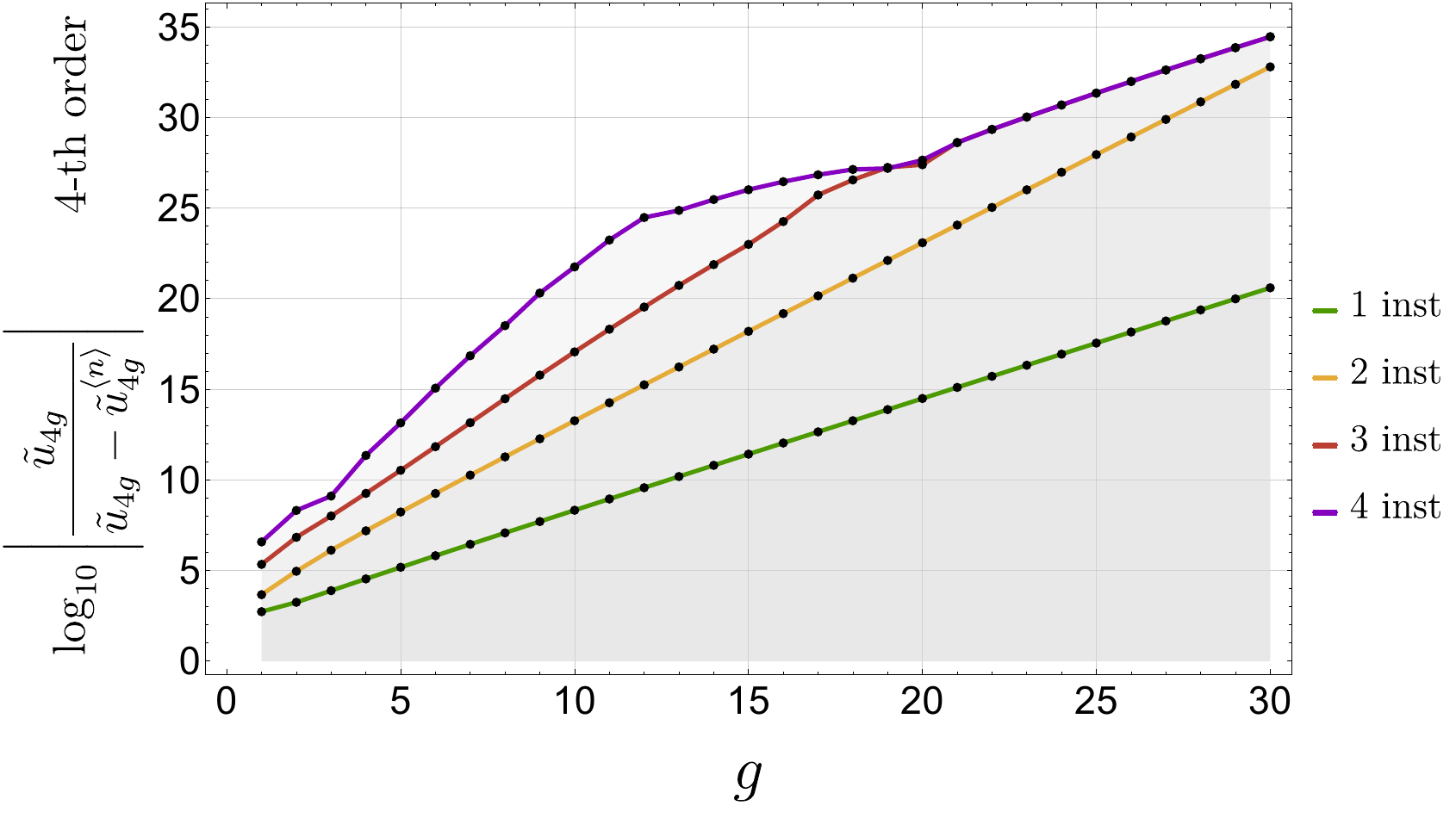}
		%\caption{\cite{asv11} computation }
	\end{subfigure}%
	~ 
	\begin{subfigure}[t]{0.5\textwidth}
		\centering
		\includegraphics[height=1.7in]{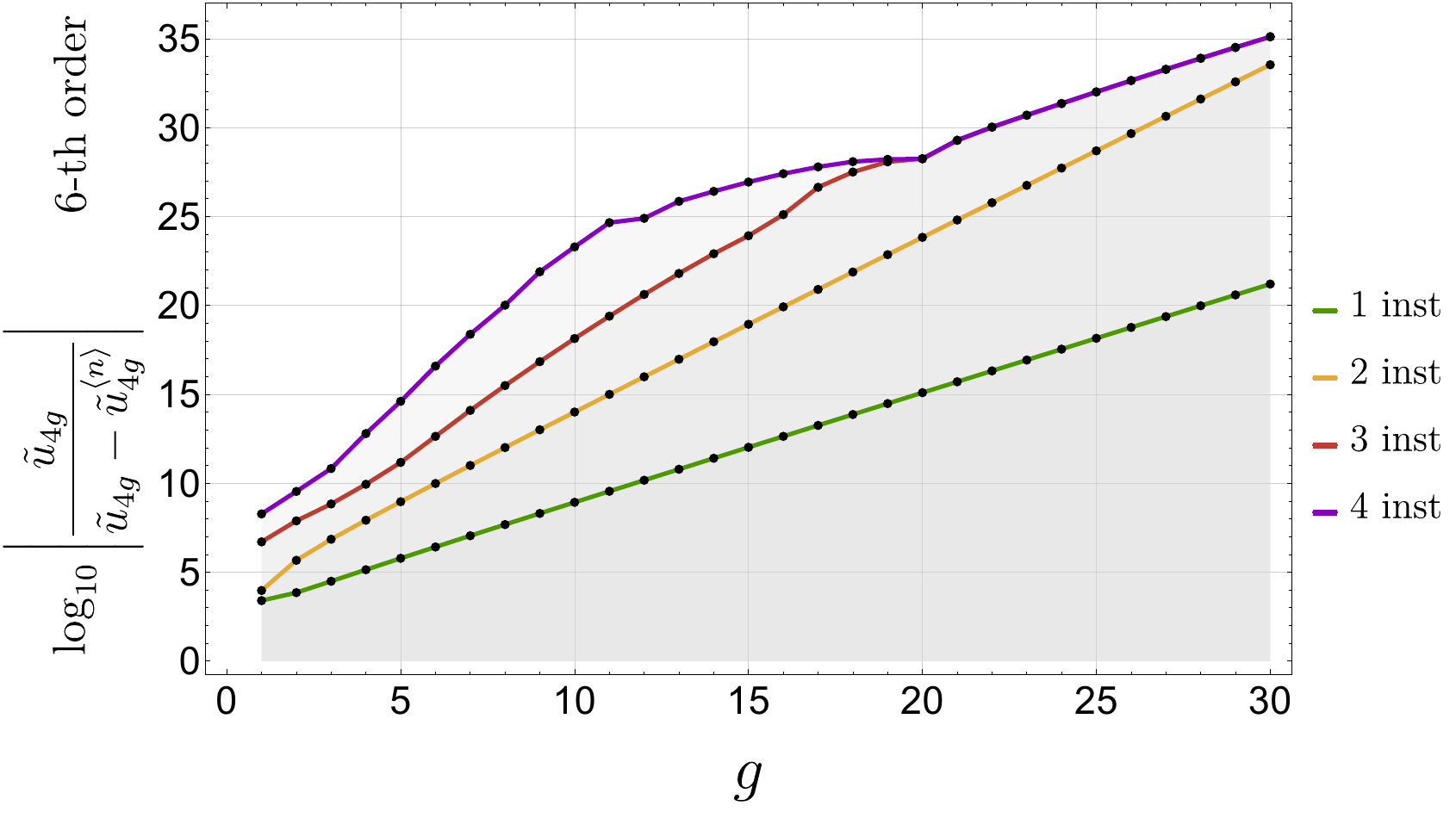}
		%\caption{Conformal transformations}
		\\
	\end{subfigure}
	\begin{subfigure}[t]{0.5\textwidth}
		\centering
		\includegraphics[height=1.7in]{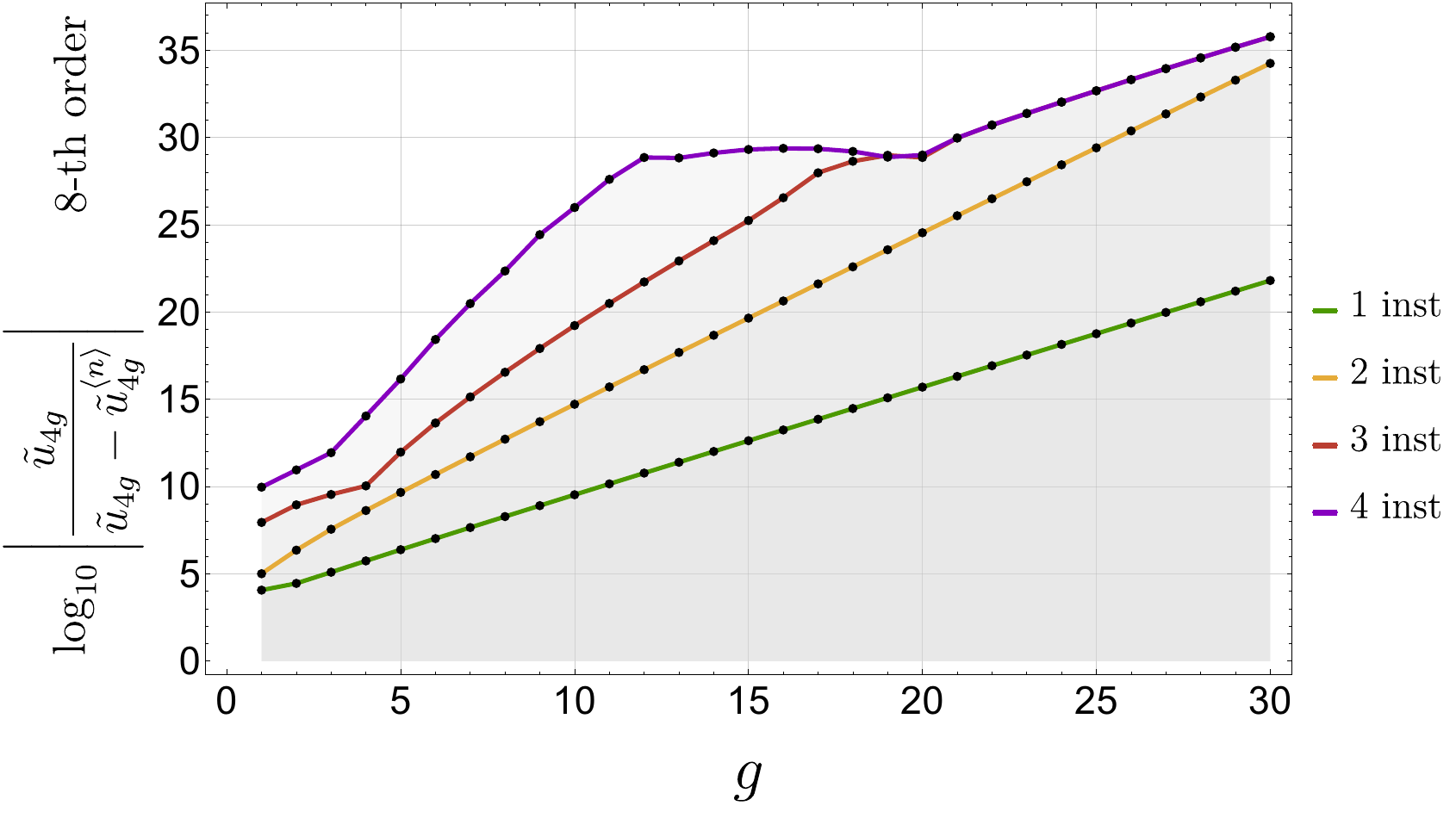}
		%\caption{\cite{asv11} computation }
	\end{subfigure}%
	~ 
	\begin{subfigure}[t]{0.5\textwidth}
		\centering
		\includegraphics[height=1.7in]{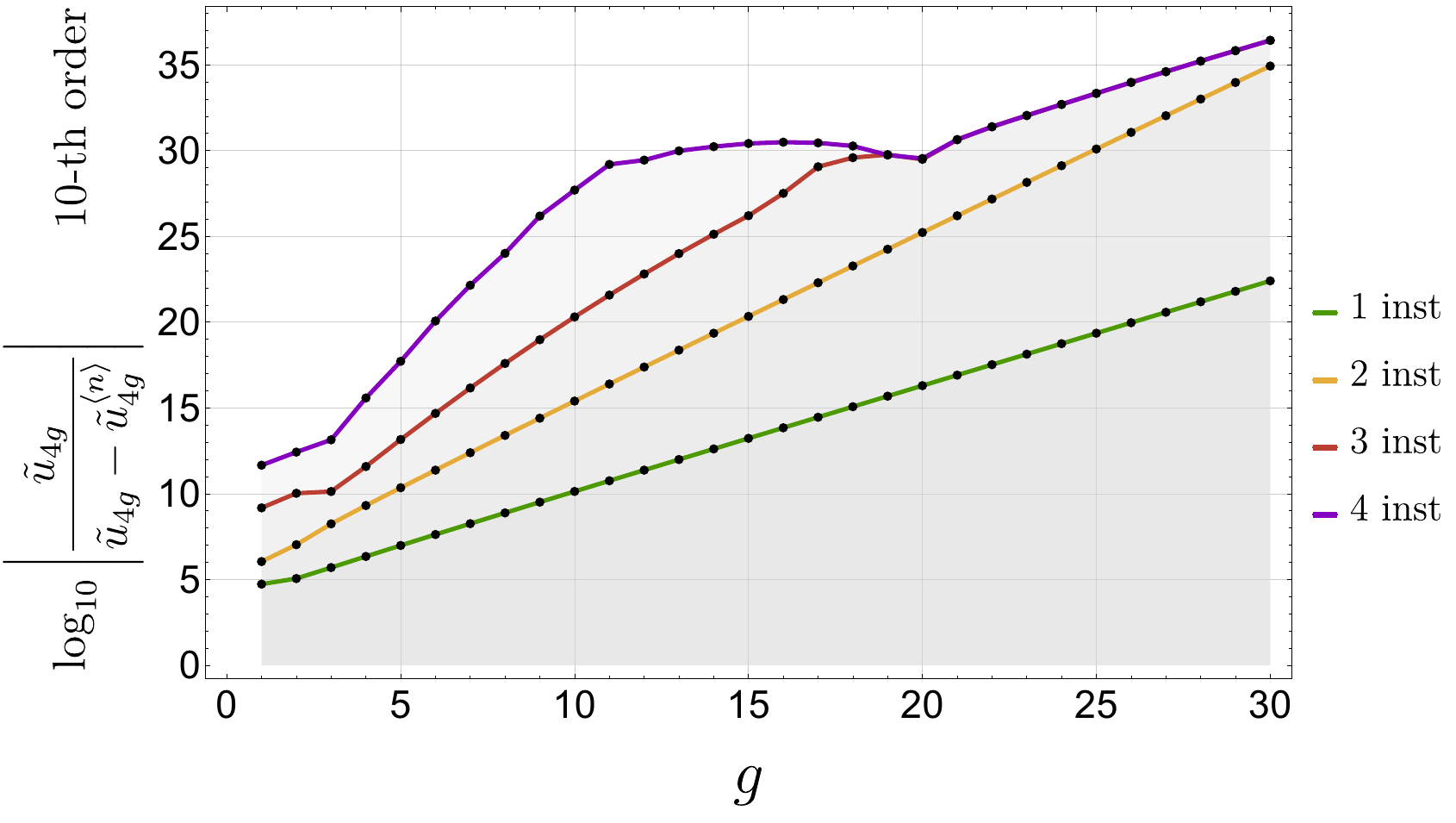}
		%\caption{Conformal transformations}
	\end{subfigure}
	\caption{Checks on the large-order formula. Notice that we have used $h=50,\,40,\,30,\,20$ for $k=1,\,2,\,3,\,4$, respectively. The precision is not astonishing as in other tests \cite{asv11,bssv21} since we have used relatively low numbers for $h$. This is due to the fact that the recursion relation \eqref{eq:recrelation} contains highly recursive sums that make computational time for coefficients much larger than any other previous cases.}
	\label{fig:checksasymp}
\end{figure}

\begin{table}[hbt!]
	\begin{center}
		\scriptsize
		\begin{tabular}{||c||c|c|c|c|c|c|c||} 
			\hline
			Expression & 0th order & 2nd order & 4th order & 6th order & 8th order & 10th order & 12th order\\ [0.5ex] 
			\hline\hline
			& & & & & & & \\
			$\displaystyle \frac{S^{(0)\mbox{\tiny num}}_1-S^{(0)\mbox{\tiny an}}_1}{S^{(0)\mbox{\tiny an}}_1}$ & $-1.19\cdot 10^{-8}$ & $7.01\cdot10^{-9}$ & $4.15\cdot 10^{-8}$ & $7.98\cdot 10^{-8}$ & $4.19\cdot10^{-8} $ & $-2.88\cdot 10^{-7}$ & $-3.41\cdot 10^{-6}$ \\
			& & & & & & & \\
			\hline\hline
		\end{tabular}
		\caption{Precision for the asymptotic checks of the different orders of $S^{(0)}_1$ in $q$. The checks have been made with 25 data points (therefore $g=100$) and 12 Richardson transforms.\label{tab:qs10precasymp}}
	\end{center}
\end{table}
Once this result has been checked, one may take it for granted in order to check the transseries structure of the model. This is done in the following way: take the left-hand side of equation \eqref{eq:preparedasymp}, but now keep all terms as in \eqref{eq:asympt}. This means that we will have two sums in instanton sectors---which can be reduced to one if one takes into account the properties of the $u$'s, $\beta$'s and Stokes coefficients---and other two---again, can be made into one---sums in $h$. The procedure is now to resum the $h$ sums and start adding instanton sector corrections, to see how well the ``resurgent-transseries-like'' right-hand side approaches the analytically known left-hand side. We have labelled this quantity as 
\begin{equation}
	\tilde{u}_{4g}\equiv\frac{\pi \rmi \, u^{(0,0)[0]}_{4g}\,  A^{2g-\beta_{(1,0)}^{[0]}+1}}{\Gamma\left(2g-\beta_{(1,0)}^{[0]}+1\right)}.
\end{equation}
The agreement in number of digits of the asymptotic formula with the analytic version is presented, in orders of $q$, in figure \ref{fig:checksasymp}. Here, we have to adapt the method due to the presence of a parameter in our transseries. Each power of $q$ is multiplied by a polynomial in $w$. The idea is therefore to resum these polynomials individually and sum the result instead of trying to resum the whole thing. The reason for it being that Pad\'e approximants are highly inefficient with parameters, again for the same reason as Richardson transforms. Moreover, the Laplace transform reduces to a numerical integral and these cannot be performed whenever the integrand depend on a parameter. For these reasons, one has to resum these polynomials that go with each power of $q$ and obtain the result as a power expansion in $q$\footnote{Alternatively, one could simply choose several values for $q$, which is much less general but gives better precision and is, indeed, much faster.}. Notice, however, that some of these polynomials will consist in a couple of monomials with a relatively large starting power. In this case, Pad\'e approximants lose most of their value, so one can simply resum these polynomials by omitting this particular step. The reader should keep in mind this solution to the problem of resummations with a parameter, since it will also be applied in the next section. \\  	

\begin{figure}[hbt!]
	\centering
	\begin{tikzpicture}[xscale=1]
		\draw[thick, darkgray] (-0.8, 1.9)--(-0.5, 1.9);
		\draw[thick, darkgray] (-0.5, 1.9)--(-0.5, 2.2);
		\node at (-0.68, 2.1) {$s$};
		\draw[->, ultra thick, darkgray] (-0.5,0) -- (13.5,0);
		\draw[->, ultra thick, darkgray] (0,-1) -- (0,2);
		\foreach \n in {1,...,11}
		{
			\draw[violet, fill=violet] (3+\n^2/15,0) circle (3pt);
		};
		\foreach \n in {1,...,10}
		{
			\draw[blue, fill=blue] (6+\n^2/15,0) circle (3pt);
		};
		\foreach \n in {1,...,7}
		{
			\draw[black!60!green, fill=black!60!green] (9+\n^2/15,0) circle (3pt);
		}
	\end{tikzpicture} 
	\caption{Illustration of poles of the Pad\'e-approximant to a simple resurgent function \cite{bssv21}. Purple points are poles representing the first branch-cut, while blue and green are poles describing the second and third branch-cuts, respectively.}
	\label{fig:ResurgentFunctionPadeApproximation}
\end{figure}
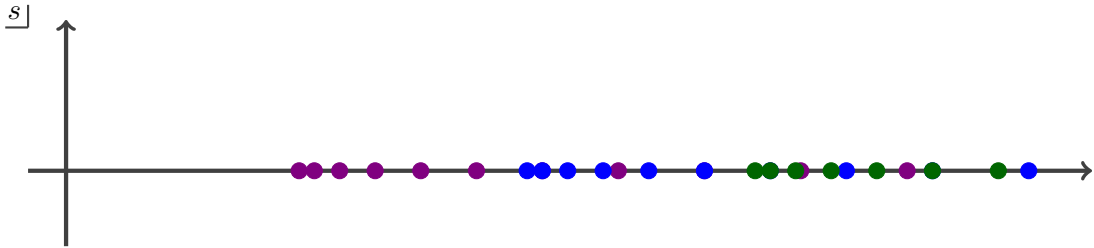

\noindent\textbf{Borel Plane Residues}: One of the main features of the transseries solution is that upon taking an appropriate Borel transform, the sectors $\Phi_{(n,m)}^{[0]}(w)$ turn out to display a simple resurgent function behaviour---check \eqref{eq:simple_res_func}. This allows us to use a method of computing Borel residues which was first developed in \cite{bssv21}. One of its nicest features is that it is based on isolating singularities while imposing continuity on the resummations, so it is indeed a very natural approach. This method starts by taking our sectors to the Borel plane\footnote{Note that there is a consistent way of taking both beta-factors and logarithms to the Borel plane. Furthermore, this method is equivalent to leaving them aside and incorporating them at the end after the resummation---see \cite{bssv21}.}. We can then approximate the Borel transform by a rational function, the Pad\'e approximant of order $(p,q)$, where $p$ (resp., $q$) is the order of the polynomial in the numerator (resp. denominator). In general, if one has a Borel transform with $g$ coefficients, then a good choice would be a $(\frac g2,\frac g2)$-Pad\'e aproximant. Indeed, this will be our choice from here on. In this setup, the logarithmic cuts are represented as poles of the denominator and they tend to accumulate in the closer singularities. We illustrate this scenario over the positive real line in figure \ref{fig:ResurgentFunctionPadeApproximation}.
With this setup, we want to compute the Borel residues by computing discontinuities while zooming-in on the closest singularity. A natural way of doing this is just integrating with an exponential weight. This is nothing else than the Laplace transform of the sector,
\begin{equation}
	\CS_{\theta} \Phi (x) = \int_{0}^{\rme^{\rmi\theta} \infty} \rmd s\, \CB \left[ \Phi \right] (s)\, \rme^{-\frac{s}{x}} \equiv \mathcal{L}_{\theta} \left\{ \CB \left[ \Phi \right] (s)\right\} (x).
\end{equation}
Thus, this method is purely based on imposing continuity on crossing the Stokes line. Computing the discontinuity along the positive real yields, at leading order,
\begin{equation}
	\label{eq:DampingOutOfTheSecondBranchCuts}
	\rme^{\frac{A}{\check{x}}} \left( \mathcal{S}_{+} - \mathcal{S}_{-} \right) \Phi_{\boldsymbol{n}} (\check{x}) = \mathsf{S}_{\boldsymbol{n} \to \boldsymbol{n}+(1,0)}\, \Big( \phi_{\boldsymbol{n}+(1,0)} + \mathcal{S}_{+} \Phi_{\boldsymbol{n}+(1,0)} (\check{x}) \Big) + o \left( \rme^{- \frac{A}{\check{x}}} \right),
\end{equation} 
where $\check{x}$ is a point of the resummation that has been chosen small and real such that the exponential factor is damped with respect to the leading orders. Form here, we would be able to compute $\mathsf{S}_{\boldsymbol{n} \to \boldsymbol{n}+(1,0)}$. The higher-order singularities can be then obtained once we know the first Borel residue by substracting the first quantity and choosing a second suitable point for the resummation $x = \check{\check{x}} \gg \check{x}$ small and real. This procedure gives
\begin{align}
	\label{eq:high_ord_sing}
	\left\{ \left( \mathcal{S}_{+} - \mathcal{S}_{-} \right) \Phi_{\boldsymbol{n}} (x) - \mathsf{S}_{\boldsymbol{n} \to \boldsymbol{n}+(1,0)}\, \Big( \phi_{\boldsymbol{n}+(1,0)} + \mathcal{S}_{+} \Phi_{\boldsymbol{n}+(1,0)} (x) \Big)\, \rme^{-\frac{A}{x}} \right\} \rme^{2 \frac{A}{x}} &=\\
	&
	\hspace{-130pt}
	= \mathsf{S}_{\boldsymbol{n} \to \boldsymbol{n}+(2,0)}\, \Big( \phi_{\boldsymbol{n}+(2,0)} + \mathcal{S}_{+} \Phi_{\boldsymbol{n}+(2,0)} (x) \Big) + o \left( \rme^{- \frac{A}{x}} \right), \nonumber
\end{align}
and can be used to iterate to the next singularities. As we have seen, this computational method relies on performing Pad\'e approximants and Laplace transforms. As discussed on the previous subsection, these parts of the method have to be modified in order to make them effective when having a parameter in the coefficients. Moreover, one still needs to take into account the $\beta$ and logarithm corrections, as well as including resonance. The latter just means that at each singularity we will have a finite number of Borel residues. From those, only two of them at each step will be relevant since the others can be computed from the relation between Borel residues and Stokes vectors. Thus, in order to obtain a second condition, one can just take the derivative of our expressions---see \cite{bssv21} for further details.

The main advantage of this method is that it gives better precision in general and it is also more systematic than the method of asymptotics. In table \ref{tab:s10precborplres} we present the results for the first forward and backward Stokes coefficients, as a comparison with asymptotics.
\begin{table}[hbt!]
	\begin{center}
		\scriptsize
		\begin{tabular}{||c||c|c|c|c|c|c||} 
			\hline
			Expression & 0th order & 2nd order & 4th order & 6th order & 8th order & 10th order\\ [0.5ex] 
			\hline\hline
			& & & & & &  \\
			$\displaystyle \frac{S^{(0)\mbox{\tiny num}}_1-S^{(0)\mbox{\tiny an}}_1}{S^{(0)\mbox{\tiny an}}_1}$ & $1.39\cdot 10^{-14}$ & $4.43\cdot10^{-14}$ & $-3.40\cdot 10^{-9}$ & $-1.05\cdot 10^{-9}$ & $1.10\cdot10^{-5} $ & $5.05\cdot 10^{-4}$ \\
			& & & & & &  \\
			\hline\hline $\displaystyle \frac{\widetilde{S}^{(0)\mbox{\tiny num}}_{-1}-\widetilde{S}^{(0)\mbox{\tiny an}}_{-1}}{\widetilde{S}^{(0)\mbox{\tiny an}}_{-1}}$ & $-1.39\cdot 10^{-14}$ & $-4.43\cdot10^{-14}$ & $3.40\cdot 10^{-9}$ & $1.05\cdot 10^{-9}$ & $-1.10\cdot10^{-5} $ & $-5.05\cdot 10^{-4}$  \\
			& & & & & &  \\
			\hline\hline
		\end{tabular}
		\caption{Precision for the Borel residue checks of the different orders of $S^{(0)}_1$ in $q$.\label{tab:s10precborplres}}
	\end{center}
\end{table}
We have also computed the second Borel residue/Stokes vector\footnote{Notice that for the first diagonal, the relation between the Borel residue and the Stokes vector entries is a linear one, so computing one or the other is a triviality. Check \cite{bssv21} for extra details.} in the main diagonal---see table \ref{tab:sedondstokprecborplres}. In this case, however, we have opted for exploring the expression for different values of $q$. This is because computing the result as a Taylor expansion turned out to be very time-consuming, while offering almost no precision whatsoever. Since we already have an idea of what the quantity should look like from similar examples \cite{bssv21}, these tests are meant to be more of a check.
\begin{table}[hbt!]
	\begin{center}
		\scriptsize
		\begin{tabular}{||c||c|c||} 
			\hline
			Expression & $q=1$ & $q=\displaystyle\frac14$ \\ [0.5ex] 
			\hline\hline
			& &  \\
			$\displaystyle \mathsf{S}_{(1,1)\to(1,0)}-4 S^{(0)}_1\cos(\pi q)(\gamma_{\text{E}}+\log(16))$ & $7.822\cdot 10^{-8}$ & $4.481\cdot10^{-8}$ \\
			& & \\
			\hline\hline 
		\end{tabular}
		\caption{Precision of the Borel residue $\mathsf{S}_{(1,1)\to(1,0)}$. This Borel residue is related to the Stokes coefficient $N^{(1)}_0$ by a mere negative sign.\label{tab:sedondstokprecborplres}}
	\end{center}
\end{table}
Unfortunately, the construction of the transition functions requires having more details about the structure of the Stokes data for higher diagonals. Following the pattern on the double-cover does not seem to give sensible transitions functions. Thus, further exploration is needed. 
\section{Inhomogeneous Painlev\'e II Equation and Matrix Models}\label{chap:alphaPII}
%%%%%%%%%%%%%%%%%%%%%%%%%%%%%%%%%%%%%%%%%%%%%%%%%%%%%%%%%%%%%%%%%
%%%%%%%%%%%%%%%%%%%%%%%%%%%%%%%%%%%%%%%%%%%%%%%%%%%%%%%%%%%%%%%%%

It is well known that the inhomogeneous Painlev\'e II equation appears in the double-scaling limit of different complex, Hermitian and unitary matrix models \cite{admn97,ck07,ckv05}. In fact, we will provide our own derivation in subsection \ref{subsec:HMMDSL} by following the approach in \cite{sv13}. This is, we will focus on the symmetric two-cut quartic Hermitian matrix model with a term $\det(M)^\alpha$. For simplicity, from now on we shall refer to this model as the $\alpha$-matrix model. 

The objective of this section will be the description of the one-loop amplitude of the $\alpha$-matrix model, culminating in the computation of the simplest Stokes coefficient, $\widetilde{S}_{-1}^{(0)}$\footnote{\label{fnote:typosv13}Note that in \cite{sv13} there is a typo in the matrix model computation. Although it is said that they are computing $S_1^{(0)}$, due to a change of variables latter on, the actual quantity that they computed was $\widetilde{S}_{-1}^{(0)}$. For them, this simply reduces to a mere factor of $\rmi$ and was otherwise inconsequential.}.
This is a two-part procedure. First, the method of orthogonal polynomials yields coupled difference equations that can be solved order-by-order using a transseries ansatz. Second, saddle-point analysis will provide the leading order in the one-loop amplitude of the matrix model. The matrix model partition function for an $s$-cut\footnote{We start with an $s$-cut matrix model because the proper procedure to solve the two-cut matrix model with a quartic potential is considering it as the limiting case of a three-cut model in which one of the cuts gets pinched \cite{msw08,sv13}.} configuration reads \cite{sv13}
\begin{equation}
	\label{eq:matix_int}
	Z_N(N_1,\dots,N_s)=\frac{1}{N_1!\dots N_s!}\int_{\lambda_{k_1}^{(1)}\in \mathcal{C}_1}\dots\int_{\lambda_{k_s}^{(s)}\in\mathcal{C}_s}\prod_{i=1}^{N}\left(\frac{\text{d}\lambda_i}{2\pi \rmi}\right)|\lambda_i|^{2\alpha}\Delta(\lambda_i)^2\rme^{-\frac{1}{g_{\text{s}}}V(\lambda_i)},
\end{equation} 
where $\mathcal{C}_i$ is a contour enclosing the $i$-th cut, $k_i=1,\dots,N_i$ labels each of the $N_i$ eigenvalues in the cut, and the term $|\lambda_i|^{2\alpha}$ is the contribution coming from the determinant insertion. Furthermore, $\Delta(\lambda_i)$ stands for the Vandermonde determinant and the potential we have chosen is
\begin{equation}
	V(z)=-\frac{1}{2!}z^2+\frac{\lambda}{4!}z^4.
\end{equation}
We can reabsorb the $|\lambda_i|^{2\alpha}$ terms in the potential, simply by defining
\begin{equation}
	\label{eq:Wpotential}
	W(\lambda_i)\equiv V(\lambda_i)-2\alpha \, g_{\text{s}} \log(|\lambda_i|).
\end{equation}

%%%%%%%%%%%%%%%%%%%%%%%%%%%%%%%%%%%%%%%%%%%%%%%%%%%%%%%%%%%%%%%%%
%%%%%%%%%%%%%%%%%%%%%%%%%%%%%%%%%%%%%%%%%%%%%%%%%%%%%%%%%%%%%%%%%
\subsection{Stokes Data from Orthogonal Polynomials}\label{subsec:HMMOP}
%%%%%%%%%%%%%%%%%%%%%%%%%%%%%%%%%%%%%%%%%%%%%%%%%%%%%%%%%%%%%%%%%
%%%%%%%%%%%%%%%%%%%%%%%%%%%%%%%%%%%%%%%%%%%%%%%%%%%%%%%%%%%%%%%%%

Let us then proceed to solve this problem via the method of orthogonal polynomials. This method alone does not suffice to obtain non-perturbative contributions of the matrix model. In fact, it cannot even account for models with several saddles since this method is inherently a one-cut solution. Fortunately, by enhancing this method with the addition of transseries ans\"atze, one can bypass this issue \cite{sv13}. Let us start with some well-known results and notations. Orthogonal polynomials are defined like
\begin{equation}
	\int \text{d}\mu(z)\, p_n(z)\, p_m(z)=h_n\, \delta_{n,m},
\end{equation} 
where $\text{d}\mu(z)$ stands for the measure---which reads, for our case, $\text{d}\mu(z)=\text{d}z\, \rme^{-\frac{1}{g_{\text{s}}}W(z)}$. Furthermore, they fulfil the recursion relation\footnote{Since our potential is symmetric, we have $s_n=0$, so we have omitted these terms in the recursion relation.}
\begin{equation}
	z\,p_n(z)=p_{n+1}(z)+r_n p_{n-1}(z), \qquad\qquad r_n=\frac{h_n}{h_{n-1}}.
\end{equation}
One can use these results to obtain the partition function in terms of orthogonal polynomial quantities via 
\begin{equation}
	Z_N=h_0^N\prod_{n=0}^{N-1}r_n^{N-n}.
\end{equation}
Since the orthogonal polynomials for our measure are not known, we will be taking a different route. We will compute the string equation of the model. This will allow us an approach via transseries ansatz to obtain non-perturbative quantities. The equation is given by \cite{biz80}
\begin{equation}
	n\, h_n=r_n\int_{-\infty}^{+\infty}\text{d}z\,\rme^{-\frac{1}{g_{\text{s}}}W(z)}\frac{1}{g_{\text{s}}}W'(z)\,p_n(z)\,p_{n-1}(z).
\end{equation}
For simplicity and readability, we will expand the three contributions coming from the derivative of $W(z)$ independently
\begin{align}
	\int \text{d}\mu(z)\, (-z)\,p_n(z)\,p_{n-1}(z)&=-\int \text{d}\mu(z)\, p_n(z)\Big(p_n(z)+r_{n-1}p_{n-2}(z)\Big)=-h_n,\\
	\frac{\lambda}{3!}\int \text{d}\mu(z)\, z^3\,p_n(z)\,p_{n-1}(z)&=\frac{\lambda}{3!}(r_{n-1}+r_n+r_{n+1})h_n,\\
	-2\alpha \,g_{\text{s}}\int \text{d}\mu(z)\, \frac 1z\,p_n(z)\,p_{n-1}(z)&=-2\alpha \,g_{\text{s}}\left(h_{n-1}-r_{n-1}\int \text{d}\mu(z)\, p_{n-1}(z)\frac 1z p_{n-2}(z)\right)=\cdots=\nonumber\\
	&=-2\alpha g_{\text{s}}\Big(h_{n-1}-r_{n-1}\,h_{n-2}+r_{n-1}\,r_{n-2}\,h_{n-3}-\cdots\Big).
\end{align}
The last contribution can be further simplified by making use of the relation $h_{n+1}=r_{n+1}h_n$,
\begin{equation}
	-2\alpha\,g_{\text{s}} h_n r_{n}^{-1}\left(1-1+1-\cdots\right)=-2\alpha\,g_{\text{s}} h_n r_{n}^{-1}\sum_{m=0}^{n}(-1)^m=-2\alpha\,g_{\text{s}} h_n r_{n}^{-1}\frac 12\Big((-1)^{n+1}+1\Big).
\end{equation}
Putting everything together, the string equation reads
\begin{equation}
	\label{eq:stringeq}
	g_{\text{s}}\left(n+\alpha\Big((-1)^{n+1}+1\Big)\right)=r_n\left(-1+\frac{\lambda}{3!}(r_{n-1}+r_n+r_{n+1})\right).
\end{equation}
When considering a two-cut scenario the solution to the above equation approaches, in the large $N$ limit, two distinct functions, depending if $n$ is even or odd--see \cite{sv13} for a numerical exploration as well as a discussion on the one-cut case--
\begin{align}
	n\mbox{ even:}&\qquad r_n\to \mathcal{P}(x),\\
	n\mbox{ odd:}&\qquad r_n\to \mathcal{Q}(x).
\end{align}
This leads to a pair of coupled equations,
\begin{align}
	\label{eq:coupledeqs1}
	\mathcal{P}(x)\left\{-1+\frac{\lambda}{3!}\left(\mathcal{Q}(x-g_{\text{s}})+\mathcal{P}(x)+\mathcal{Q}(x+g_{\text{s}})\right)\right\}&=x,\\
	\label{eq:coupledeqs2}
	\mathcal{Q}(x)\left\{-1+\frac{\lambda}{3!}\left(\mathcal{P}(x-g_{\text{s}})+\mathcal{Q}(x)+\mathcal{P}(x+g_{\text{s}})\right)\right\}&=x+2\alpha\,g_{\text{s}},
\end{align}
with $x=n\,g_{\text{s}}$. These equations are the perfect set-up to solve with a transseries ansatz. Note that we will have to expand as power series in $g_{\text{s}}$ the terms that depend on $x\pm g_{\text{s}}$, which will become cumbersome in some scenarios.\\

\noindent\textbf{Perturbative Ansatz}: We will now proceed to solve the above coupled equations at the perturbative level. This is, we will use an asymptotic series ansatz of the form
\begin{equation}
	\mathcal{P}(x)\simeq\sum_{g=0}^{+\infty}g_{\text{s}}^{g}\,P_{g}(x),\qquad\qquad\mathcal{Q}(x)\simeq\sum_{g=0}^{+\infty}g_{\text{s}}^{g}\,Q_{g}(x).
\end{equation}
The computation is straightforward and we only have to take into account the expansion around $g_{\text{s}}\sim 0$ of the terms $P(x\pm g_{\text{s}})$, $Q(x\pm g_{\text{s}})$. Just for the sake of being explicit, we write the first orders
\begin{equation}
	\label{eq:pmexpansion}
	P_{g}(x\pm g_{\text{s}})=P_{g}(x)\pm g_{\text{s}} P_{g}'(x)+\frac1{2!}g_{\text{s}}^2P_{g}''(x)\pm\cdots,
\end{equation}
where $'$ stands for derivatives with respect to $x$. The first coefficients for both functions are
\begin{align}
	P_0(x)=\frac{3}{\lambda}\left(1-\sqrt{1-\frac{2\lambda x}{3}}\right),&\qquad\qquad P_1(x)=-\frac{6(3-\sqrt{9-6\lambda x})}{9-6\lambda x}\alpha,\\
	Q_0(x)=\frac{3}{\lambda}\left(1+\sqrt{1-\frac{2\lambda x}{3}}\right),&\qquad\qquad Q_1(x)=\frac{18}{9-6\lambda x}\alpha.
\end{align}
At this stage, one realises that these functions are real whenever $0\leq \lambda x\leq \frac 32$. Not surprisingly, these bounds will coincide with the ones obtained in the next subsection when computing the endpoints of the cuts. This is a nice consistency check for now. For this reason, and also for readability, it is convenient to rewrite these results in terms of the variable $p^2=9-6\lambda x$, where the range for $\lambda x$ gets translated into $9\geq p\geq 0$ ,

\begin{align}
	P_0(x)=\frac{3-p}{\lambda},&\qquad\qquad P_1(x)=-\frac{6(3-p)}{p^2}\alpha,\\
	Q_0(x)=\frac{3+p}{\lambda},&\qquad\qquad Q_1(x)=\frac{18}{p^2}\alpha.
\end{align}
As a second consistency check, note that these results coincide with \cite{sv13} in the limit $\alpha\to 0$. However, a detail must be mentioned: from the beginning, we have assumed a power-series expansion in powers of $g_{\text{s}}$ instead of $g_{\text{s}}^2$ as in the original reference. This is due to the $\alpha\,g_{\text{s}}$ term in the coupled equations---but we will see that this will be consistent with the inhomogeneous Painlev\'e II expansion.

Up to now, this procedure is the one that one would naturally follow when trying to solve this model at the perturbative level in a one-cut scenario. In the next subsection, we will upgrade this approach by adding extra monomials---exponential contributions---to our ansatz. This will allow us to interpret these new monomials as one-loop contributions arising from tunnelled eigenvalues. \\

\noindent\textbf{Two-Parameter Transseries}: Now, in order to account for the several saddles and to find non-perturbative information we will apply the usual two-parameter transseries ansatz to our functions $\mathcal{P}$ and $\mathcal{Q}$,
\begin{align}
	&\mathcal{P}(\sigma_1,\sigma_2,x)=\sum_{n,m=0}^{+\infty}\sigma_1^n\sigma_2^m \rme^{-(n-m)\frac{A(x)}{g_{\text{s}}}}\sum_{g=0}^{+\infty}g_{\text{s}}^g\,P_{g}^{(n,m)}(x);\\
	&\mathcal{Q}(\sigma_1,\sigma_2,x)=\sum_{n,m=0}^{+\infty}\sigma_1^n\sigma_2^m \rme^{-(n-m)\frac{A(x)}{g_{\text{s}}}}\sum_{g=0}^{+\infty}g_{\text{s}}^g\,Q_{g}^{(n,m)}(x).
\end{align}
Trivially, we recognise the $(0,0)$ sector as the perturbative ansatz from the previous section. Note that, in applying this to the coupled equations, we will need both \eqref{eq:pmexpansion} and the following expansion for the exponential factors \cite{sv13}
\begin{equation}
	\exp\left(-\frac{A(x\pm g_{\text{s}})}{g_{\text{s}}}\right)=\exp\left(-\frac{A(x)}{g_{\text{s}}}\right)\rme^{\mp A'(x)}\sum_{\ell'=0}^{+\infty}\frac{1}{\ell'!}\left(-\sum_{\ell=0}^{+\infty}(\pm 1)^\ell g_{\text{s}}^{\ell-1}\frac{A^{(\ell)}(x)}{\ell!}\right)^{\ell'}.
\end{equation}
In this case, the same phenomenon of resonance happens when trying to solve these coupled equations. This phenomenon is based on the fact that the order  $\sigma_1 g_{\text{s}}^0$ yields an equation that allows us to determine the action $A(x)$ in the exponential\footnote{It allows us to determine the $\cosh^2(A'(x))$.} and a relation between $P_0^{(1,0)}(x)$ and $Q_0^{(1,0)}(x)$, but still undetermined. This is not a problem though, since at the following order, $\sigma_1 g_{\text{s}}^1$, one can determine a ratio between $P_1^{(1,0)}(x)$ and $Q_1^{(1,0)}(x)$. This ratio, when used in the other coupled equation cancels the highest contribution and allows us to determine either $P_0^{(1,0)}(x)$ or $Q_0^{(1,0)}(x)$. This phenomenon repeats for all the sectors with $|n-m|=1$. We will see this occur again when analysing the case of the inhomogeneous Painlev\'e II equation. Now, for the computation of the simplest Stokes vector, we will only need the first orders in the expansion, which read
\begin{align}
	&\cosh^2(A'(x))=\frac{3}{2\lambda x},\\
	&P^{(1,0)}_0=-\sqrt{\frac{3-p}{p}}\left(\frac{3+p}{p^3}\right)^{-\alpha},\\
	&Q^{(1,0)}_0=\sqrt{\frac{3+p}{p}}\left(\frac{3+p}{p^3}\right)^{-\alpha}.
\end{align} 
There are a couple of things to mention. First, note the fact that the instanton action does not depend on $\alpha$. This is expectable from the point-of-view of the matrix model: the $\alpha$ correction is sub-leading in nature and does not enter, for example, in the computation of the spectral curve that will appear in the next subsection. Moreover, the inhomogeneous Painlev\'e II equation also reproduces this feature, both at the level of asymptotic growth of the coefficients as well as at the level of the transseries ansatz, as we will discuss in the next subsection.

Now, having determined these coefficients, one asks how can one extract information about the coefficients of the partition function or free energy. Traditionally, one would make use of the Euler--McLaurin formula to deal with the large-order equations. There is an alternative trick\footnote{Actually, this trick turns out to be just a rewriting of the Euler--McLaurin formula.} that turns out to be more computationally convenient. For the quantity we want to determine, we have \cite{sv13}
\begin{align}
	\mathcal{F}(t+2g_{\text{s}})-2\mathcal{F}(t)+\mathcal{F}(t-2g_{\text{s}})=\log\left(\mathcal{Q}(t+g_{\text{s}})\mathcal{P}^2(t)\mathcal{Q}(t-g_{\text{s}})\right).
\end{align}
The latter expression yields the following equation for $\mathcal{F}^{(1,0)}_0$
\begin{align}
	4\sinh(A'(x))\mathcal{F}^{(1,0)}_0=2\left(\frac{P^{(1,0)}_0}{P^{(0,0)}_0}+\cosh(A'(x))\frac{Q^{(1,0)}_0}{Q^{(0,0)}_0}\right),
\end{align} 
from which one can compute the result 
\begin{align}
	\label{eq:F100}
	\mathcal{F}^{(1,0)}_0=-\frac{\lambda}{2}\sqrt{\frac{3-p}{p^3}}\left(\frac{3+p}{p^3}\right)^{-\alpha}.
\end{align}
Again, this trivially coincides with the results in \cite{sv13} in the limit $\alpha\to0$. 

We finish this subsection by computing some extra terms from the coupled equations. In particular, we will be interested in the $P$ coefficients. This is because when comparing the matrix model Stokes vector with the one of the inhomogeneous Painlev\'e II equation we will need to find the correspondence between the transseries. This is, we will have to relate the $P$ polynomials in the double-scaling limit and the $u$ coefficients of the inhomogeneous Painlev\'e II equation. These factors are
\begin{align}
	P^{(0,0)}_2&=\frac{9\lambda}{2}\frac{9 + 6 x \lambda - 3 \sqrt{9 - 6 x \lambda} + 
		36 \alpha^2 (3 - \sqrt{9 - 6 x \lambda})}{(9 - 6 x \lambda)^{\frac 52}}=\frac{9\lambda}{2}\frac{(
		36 \alpha^2-p) (3 - p)}{p^{5}},\\
	P^{(2,0)}_0&=\frac \lambda 2 \frac{(9 - 6 x \lambda)^{3\alpha-1}(-3 + \sqrt{9 - 6 x \lambda})}{(3 + \sqrt{9 - 6 x \lambda})^{2 \alpha}}=-\frac \lambda 2\left(\frac{p^3}{3+p}\right)^{2\alpha} \frac{(3-p)}{p^2}.
\end{align} 

Those are all the results needed from orthogonal polynomials. In the next subsection, we will find the one-loop contribution to the partition function via saddle-point analysis.
%%%%%%%%%%%%%%%%%%%%%%%%%%%%%%%%%%%%%%%%%%%%%%%%%%%%%%%%%%%%%%%%
%%%%%%%%%%%%%%%%%%%%%%%%%%%%%%%%%%%%%%%%%%%%%%%%%%%%%%%%%%%%%%%%%
\subsection{Stokes Data from Multi-Instantons}\label{subsec:HMMSP}
%%%%%%%%%%%%%%%%%%%%%%%%%%%%%%%%%%%%%%%%%%%%%%%%%%%%%%%%%%%%%%%%%
%%%%%%%%%%%%%%%%%%%%%%%%%%%%%%%%%%%%%%%%%%%%%%%%%%%%%%%%%%%%%%%%%
In this subsection, we will use the same method as in \cite{msw08,sv13} for computing the multi-instanton sectors of the matrix model. The procedure is very much the same and the results therein are valid up to two modifications. First, we have to normalize with respect to the $\alpha$-Gaussian matrix model instead of the ordinary one---see appendix \ref{app:alphaMM} for details and computations. Secondly, when expanding the  multi-instanton amplitude, one has an extra term at order $\mathcal{O}(g_{\text{s}}^0)$. In reference \cite{sv13} this term does not appear since they have an even-power-series expansion of the free energy. With this extra factor, the multi-instanton amplitude can be written in terms of the partial 't-Hooft couplings $t_i=N_i\,g_{\text{s}}$, that can be rearranged as $s_1=\frac12(t_1-t_2-t_3)$, $s_2=\frac12(t_3-t_2-t_1)$. For the eigenvalue distribution $(N_1-n_1,N_2+n_1+n_2,N_3-n_2)$, this is given by\footnote{Notice that, in the notation of \cite{sv13}, our $F_1$ would be written as $F_{\frac{1}{2}}$.}
\begin{align}
	Z^{(n_1,n_2)}=\exp\left(-\frac{1}{g_{\text{s}}}\sum_{i=1}^{2}n_i\partial_{s_i}F_0\right)\exp\left(\frac{1}{2}\sum_{i,j=1}^{2}n_in_j\partial_{s_i}\partial_{s_j}F_0\right)\exp\left(-\sum_{i=1}^{2}n_i\partial_{s_i}F_1\right)\left\{1+\mathcal{O}(g_{\text{s}})\right\}.
\end{align}
Now, for the $\ell$-instanton contribution, we are interested in all configurations which have $\ell$ eigenvalues on the non-perturbative saddle. That is, we are interested in all configurations with $n_1+n_2=\ell$. Changing variables $\ell=n_1+n_2$, $m=n_1-n_2$, one can obtain the $\ell$-instanton amplitude simply by summing over all $m$'s with  $\ell=n_1+n_2$, {\it i.e.}, summing over all configurations with $\ell$ tunnelled eigenvalues. The computation of the terms that consider only the 0-th order in the free energy are due to \cite{sv13} and the end result is\footnote{Note that we used $\partial_{s_1}F_1=\partial_{s_2}F_1\equiv \partial_s F_1$. We will compute these quantities later-on to check that they are indeed the same. }
\begin{equation}
	Z^{(\ell)}=\left(\frac{b^2-a^2}{4\widetilde{M}a^3b^3}\right)^{\frac{\ell^2}{2}}\left(\frac{b-a}{b+a}\right)^{\frac{1-(-1)^\ell}{4}}\exp\Big(\ell\,\partial_{s}F_1\Big)\Big\{1+\mathcal{O}(g_{\text{s}})\Big\},
\end{equation}
where $\pm a$ and $\pm b$ are the endpoints of the perturbative cuts\footnote{Note that the symmetry of the potential makes our saddles be of the form $[-b,-a]\cup[-c,c]\cup[a,b]$, while the double-scaling limit means taking $c\to 0$.} and $\widetilde{M}$ is given by 
\begin{align}
	\widetilde{M}(z)&=\oint_{(\infty)}\frac{\text{d}w}{2\pi\rmi}\frac{V'(w)}{w-z}\frac{1}{\sqrt{(w^2-a^2)(w^2-b^2)}}=\frac\lambda6.
\end{align}
These endpoints may be obtained from saddle-point analysis via the endpoint equation
\begin{equation}
	\oint_\mathcal{C}\frac{\text{d}w}{2\pi\rmi}\frac{w^nV'(w)}{\sqrt{(w^2-a^2)(w^2-b^2)}}=2t \delta_{n,2};\qquad\qquad n=0,1,2,
\end{equation}
which gets translated into
\begin{align}
	a&=\sqrt{\frac6\lambda\left(1-\sqrt{\frac{2\lambda t}3}\right)}; \qquad\qquad	b=\sqrt{\frac6\lambda\left(1+\sqrt{\frac{2\lambda t}3}\right)}.
\end{align}
In an attempt to be self-contained, we have picked the minimal amount of results from \cite{sv13} that were necessary to reproduce the computations herein. However, in order to understand the whole procedure one should take a look at the aforementioned reference. Moving on, the quantity that needs to be properly addressed is the derivative of $F_1$. As we have computed in appendix \ref{app:alphaMM}, we have
\begin{align}
	\partial_{s_1}F_1&=\alpha\int_{-a}^{-c}\text{d}z\left(\frac{z^2}{\sqrt{(z^2-a^2)(z^2-b^2)(z^2-c^2)}}-\frac 1z\right)=\\
	&=\alpha \int_{-a}^{-c}\text{d}z\frac{z^2}{\sqrt{(z^2-a^2)(z^2-b^2)(z^2-c^2)}}-\alpha \log\left(\frac ca\right),\\
	\partial_{s_2}F_1&=-\alpha\int_{c}^{a}\text{d}z\left(\frac{z^2}{\sqrt{(z^2-a^2)(z^2-b^2)(z^2-c^2)}}-\frac 1z\right)=\\
	&=-\alpha \int_{c}^{a}\text{d}z\frac{z^2}{\sqrt{(z^2-a^2)(z^2-b^2)(z^2-c^2)}}+\alpha \log\left(\frac ac\right).
\end{align} 
In the limit of $c\to 0$ these terms become
\begin{align}
	\partial_{s_1}F_1&=\frac \alpha 2 \log\left(\frac{b + a}{b - a}\right)-\alpha\log\left(\frac ca \right),\\
	\partial_{s_2}F_1&=\frac \alpha 2 \log\left(\frac{b + a}{b - a}\right)-\alpha\log\left(\frac ca \right),
\end{align}
where we have kept the second logarithmic term since it will be necessary to cancel contributions that will appear later on in the discussion. We still need to normalise this quantity by means of the $\alpha$-Gaussian matrix model. We have compiled this computation in appendix \ref{app:alphaMM}, where the needed result is just missing the quantity
\begin{align}
	t_2&=\frac1{4\pi\rmi}\oint_{\mathcal{C}_2}\text{d}z y(z)=\\
	&=\frac{1}{2\pi}\int_{-c}^c\text{d}z \widetilde{M}(z)\sqrt{(z^2-a^2)(z^2-b^2)(z^2-c^2)}=\frac{1}4\widetilde{M}abc^2\left\{1+\mathcal{O}(c^2)\right\}.
\end{align}
Therefore, the normalized contribution is 
\begin{align}
	\partial_s\hat{F}_1&=\partial_s F_1-\partial_{t_2}F^{\alpha\text{-G}}_1=\partial_s F_1-\alpha \log(t_2)=\\
	&=\frac \alpha 2 \log\left(\frac{b + a}{b - a}\right)-\alpha\log\left(\frac ca \right)-\alpha \log\left(\frac{1}{4}\widetilde{M} a b\right)-2\alpha \log\left(c\right).
\end{align}
Furthermore, in order to take the integration of the $\ell$-instantons it is not enough to consider the large $N$ limit, but rather evaluate the Gaussian model associated with the collapsing cycle \cite{msw08}. This amounts to adding an extra factor $Z^{\alpha\text{-G}}(\ell)$ to our $\ell$-instanton partition function, as
\begin{equation}
	Z^{(\ell)}=Z^{\alpha\text{-G}}(\ell)\left(\frac{b^2-a^2}{4\widetilde{M}a^3b^3}\right)^{\frac{\ell^2}{2}}\left(\frac{b-a}{b+a}\right)^{\frac{1-(-1)^\ell}{4}+\frac {\ell\alpha} 2}\left(\frac ca \right)^{\ell\alpha}\left(\frac{1}{4}\widetilde{M} a b\right)^{\ell \alpha}\left(c\right)^{2\ell\alpha} \Big\{1+\mathcal{O}(g_{\text{s}})\Big\},
\end{equation}
where the normalisation factor has been computed in appendix \ref{app:alphaMM}. We are interested in the case $\ell=1$ in the double-scaling limit $\lambda\,t\mapsto \frac32$. This gives the contribution
\begin{equation}
	Z^{(1)}=-\frac{2^{-2 \alpha} 9^{-\alpha}}\pi \Gamma\left(\alpha+\frac 12\right)g_{\text{s}}^{\alpha+\frac{1}{2}}\Big\{1+\mathcal{O}(g_{\text{s}})\Big\}.
\end{equation}
From here, noticing that $Z^{(1)}=\widetilde{S}^{(0)}_{-1}F^{(1,0)}_0g_{\text{s}}^{\alpha+\frac12}+\cdots$, and using \eqref{eq:F100} one has that 
\begin{equation}
	\widetilde{S}_{-1}^{(0)}=-\frac{2^{-2 \alpha} 9^{-\alpha}}\pi \, \Gamma\left(\alpha+\frac 12\right).
\end{equation}
This is the simplest Stokes vector for this matrix model. However, we still need to normalise it in order to translate it into the Stokes vector for the inhomogeneous Painlev\'e II equation. 
%%%%%%%%%%%%%%%%%%%%%%%%%%%%%%%%%%%%%%%%%%%%%%%%%%%%%%%%%%%%%%%%%
%%%%%%%%%%%%%%%%%%%%%%%%%%%%%%%%%%%%%%%%%%%%%%%%%%%%%%%%%%%%%%%%%
\subsection{The Double-Scaling Limit}\label{subsec:HMMDSL}
%%%%%%%%%%%%%%%%%%%%%%%%%%%%%%%%%%%%%%%%%%%%%%%%%%%%%%%%%%%%%%%%%
%%%%%%%%%%%%%%%%%%%%%%%%%%%%%%%%%%%%%%%%%%%%%%%%%%%%%%%%%%%%%%%%%

The double-scaling limit for our matrix model is obtained by taking 
\begin{align}
	&g_{\text{s}}\to 0,\qquad\lambda\to \lambda_\text{c}=\frac 32,\\
	&x\to 1,\qquad z=(1-x)g_{\text{s}}^{2/3}\,\,\mbox{ fixed.}
\end{align}
Starting from the coupled equations \eqref{eq:coupledeqs1}-\eqref{eq:coupledeqs2}, we apply the scaling ans\"atze
\begin{align}
	&\mathcal{P}(x)\to 2\Big(1-g_{\text{s}}^{1/3}u(z)+g_{\text{s}}^{2/3}v(z)\Big),\\
	&\mathcal{Q}(x)\to 2\Big(1+g_{\text{s}}^{1/3}u(z)+g_{\text{s}}^{2/3}v(z)\Big).
\end{align}
In order to find the inhomogeneous Painlev\'e II equation we need to take the double-scaling limit of the sum and of the difference of the coupled equations. For this computation, one notices that 
\begin{align}
	\mathcal{Q}(x+g_{\text{s}})+\mathcal{Q}(x-g_{\text{s}})-2\mathcal{Q}(x)&=2\Big(1+g_{\text{s}}^{1/3}(\overbrace{u(z-g_{\text{s}}^{1/3})+u(z+g_{\text{s}}^{1/3})-2u(z))}^{g_{\text{s}}^{2/3}u''(z)}+1-\nonumber\\
	&\hspace{1cm}-2+g_{\text{s}}^{2/3}(\overbrace{v(z-g_{\text{s}}^{1/3})+v(z+g_{\text{s}}^{1/3})-2v(z)}^{g_{\text{s}}^{2/3}v''(z)})\Big)\\
	&=2(g_{\text{s}}u''(z)+g_{\text{s}}^{4/3}v''(z))+\cdots,
\end{align}
and similarly for $\mathcal{P}$. Thus, the sum and the difference read
\begin{align}
	2x-2&=g_{\text{s}}^{2/3}\Big(8v(z)-2u^2(z)\Big),\\
	-4g_{\text{s}}u(z)v(z)+2g_{\text{s}}u''(z)-2g_{\text{s}}^{5/3}u(z)v''(z)+2g_{\text{s}}^{5/3}u''(z)v(z)&=-g_{\text{s}}2\alpha.
\end{align}
Using $(1-x)g_{\text{s}}^{-2/3}=z$ it is easy to solve for $v(z)$ in the second equation (neglecting the $g_{\text{s}}^{5/3}$ term) and introduce the result in the first. The outcome is the expected inhomogeneous Painlev\'e II equation, in the normalisation 
\begin{equation}
	\label{eq:PIIfromMM}
	2u''(z)-u^3(z)+zu(z)+2\alpha=0.
\end{equation}

Now, in order to translate the matrix model Stokes coefficient into the one associated with our above normalisation of the inhomogeneous Painlev\'e II equation, we still need to relate the coefficients $P^{(n,m)[k]}_g$ with the $u^{(n,m)[k]}_g$ that we will display in section \ref{subsec:transans}. Our result is similar to those appearing in \cite{asv11,sv13}. In fact, we can use the aforementioned references and the results of sections \ref{sec:PII} and \ref{subsec:transans} as a
motivation for the following
\begin{align}
	&\left(-\frac 12\right)^g(c_1 \sqrt {g_{\text{s}}})^{n+m}(c_2 \sqrt{g_{\text{s}}})^{-2(n-m)\alpha} g_{\text{s}}^{g - 1/3} P^{(n,m)[0]}_g \xrightarrow[\;\; \mbox{\footnotesize dsl } \;\;]{} 
	w^{2g+2\beta_{(n,m)}^{[0]}} u^{(n,m)[0]}_{2g + 2\beta_{(n,m)}^{[0]}}.
\end{align}
The factor $\frac 12$ can be easily checked for the first $P$'s; $P_0^{(0,0)}$, $P_1^{(0,0)}$ and $P_2^{(0,0)}$. The $g_{\text{s}}^{-(m-n)\alpha}$ can be easily checked from $P_0^{(1,0)}$. In any case, it is trivial to realise that this factor corresponds to the beta-factor associated with the equation that we will present later in section \ref{subsec:transans}. By comparing the results for the $P$'s and the $u$'s we get
\begin{align}
	c_1=2\sqrt 2;\hspace{1cm}	c_2=\frac{\sqrt 2}3.
\end{align}

With this result, we can translate the Stokes coefficient obtained for the quartic matrix model (QMM) into that of the inhomogeneous Painlev\'e II equation (PII). This is
\begin{align}
	\label{eq:tildeS10fromMM}
	\left.\widetilde{S}^{(0)}_{-1}\right|_{\text{PII}}=\frac{\left.\widetilde{S}^{(0)}_{-1}\right|_{\text{QMM}}}{\frac 12 c_1 c_2^{2\alpha}}=-\frac{2^{-\frac 12-3\alpha}}\pi\, \Gamma\left(\alpha+\frac 12\right).
\end{align}
This concludes the computation of the first Stokes vector from the one of the matrix model. As we will see in section \ref{subsec:numcheck}, the result coincides with the Stokes vector found for the inhomogeneous Painlev\'e II, upon the condition that $\alpha\in\mathbb{Z}$, which is a natural constraint from the matrix model point of view. This is because $\alpha$ appears as a difference between the dimensions of the rectangular matrices.

This concludes our preliminary analysis of the $\alpha$-matrix model. However, recent developments in the field of matrix models allow for analytical computation of all Stokes data \cite{mss22}. It would be very interesting to apply these techniques to our case to confirm the numerical results in the next sections. The next subsections will be dedicated to the resurgent study of the inhomogeneous Painlev\'e II equation. First, a quick view on the original normalisation transseries and properties. Then, we will change normalisations into a more suitable form from the resurgence point of view, as in \cite{asv11,sv13}.
%%%%%%%%%%%%%%%%%%%%%%%%%%%%%%%%%%%%%%%%%%%%%%%%%%%%%%%%%%%%%%%%%
%%%%%%%%%%%%%%%%%%%%%%%%%%%%%%%%%%%%%%%%%%%%%%%%%%%%%%%%%%%%%%%%%
\subsection{Perturbative Solution and Large-Order Analysis}\label{sec:origPII}
%%%%%%%%%%%%%%%%%%%%%%%%%%%%%%%%%%%%%%%%%%%%%%%%%%%%%%%%%%%%%%%%%
%%%%%%%%%%%%%%%%%%%%%%%%%%%%%%%%%%%%%%%%%%%%%%%%%%%%%%%%%%%%%%%%%

This subsection is dedicated to some preliminary analysis of the resurgent properties of the inhomogeneous Painlev\'e II equation in the {\it ``classical''} normalisation\footnote{Here, we have followed the normalisation in \cite{c03,c06,c19} since it is one of the most common normalisations.}. In particular, we will analyse the large-order growth of the perturbative solution, the role of the $\alpha$ parameter, and the value of the instanton action. The latter can also be obtained through direct computation, so it would be a nice check on the asymptotic analysis. We start by writing the inhomogeneous Painlev\'e II equation in the {\it ``classical''} variables
\begin{equation}
	\label{eq:PII_with_alpha}
	u''(z)-2 u^3(z)- z\,u(z)-\alpha=0.
\end{equation}
The first step of our analysis is to study the perturbative sector. For that, we will use the following ansatz
\begin{equation} 
	\label{eq:pert_ansatz_with_alpha}
	u(z)\simeq\sum_{g=0}^{+\infty} u_g\; z^{-\gamma g+\beta}.
\end{equation}
Plugging \eqref{eq:pert_ansatz_with_alpha} into equation \eqref{eq:PII_with_alpha} we get\footnote{\label{fnote:otheru0s}Note that there are other two options for the coefficient $u_0$. Apart from the one that we decided for, one can also choose the other square root determination, giving $u_0=\frac{\rmi}{\sqrt2}$ or the third solution $u_0=0$, which produces a perturbative sector proportional to $\alpha$ given by $\sim-\frac\alpha z+\mathcal{O}(z^{-4})$. We will explore these solutions later on.}
\begin{equation}
	\begin{array}{lll}
		\beta=\displaystyle\frac{1}{2}, & \qquad\qquad& u_0= -\displaystyle\frac{\rmi}{\sqrt{2}},\\
		& &\\
		\gamma=\displaystyle \frac 32, & & u_1=\displaystyle\frac \alpha 2,
	\end{array}
\end{equation}
and a recursion relation for the rest of the coefficients yielding (note that this agrees with \cite{gg18})
\begin{align}
	\label{eq:pert_exp}
	u_{\text{pert}}(z)&=-\frac{\rmi}{\sqrt{2}}\sqrt{z}+\frac \alpha 2 z^{-1}-\left(\frac{\rmi}{8\sqrt 2}+\frac{3\,\alpha^2}{4\sqrt 2}\right)z^{-5/2}-\left(\alpha^3+\frac{11\,\alpha}{16}\right)z^{-4}+\cdots=\nonumber\\
	&=\sqrt{z}\left(-\displaystyle\frac{\rmi}{\sqrt{2}}+\frac \alpha 2z^{-3/2}-\left(\frac{\rmi}{8\sqrt 2}+\frac{3\,\alpha^2}{4\sqrt 2}\right)z^{-3}-\left(\alpha^3+\frac{11\,\alpha}{16}\right)z^{-9/2}+\cdots\right).
\end{align}
As in subsection \ref{sec:qorigPII}, the instanton action for this problem can be obtained from two approaches:
\begin{itemize}
	\item Fitting the ratio
	\begin{equation}
		\label{eq:uratio}
		\frac{u_{g+1}}{u_g}.
	\end{equation}
	Again, we expect the perturbative series to be a Gevrey-1 series and thus resurgent\footnote{See \cite{abs18,s14} for further information on resurgence, Gevrey series, and their relation.}. We have indeed checked this, {\it i.e.}, we have obtained that the large-$g$ limit of the perturbative coefficients is 
	\begin{equation}
		\label{eq:Gevrey-1_prop}
		|u_g|\sim \frac{C}{|A|^g}g!.
	\end{equation}
	From there, one can compute the absolute value of the instanton action $|A|$ from the ratio \eqref{eq:uratio}. This method thus yields
	\begin{equation}
		|A|=\frac{2\sqrt 2}{3}.
	\end{equation}
	\item Computing it from a one-parameter (or two-parameter) transseries ansatz. Using one of these ans\"atze and plugging it into our equation yields a second-order algebraic equation for $A$. Setting this equation to zero gives\footnote{\label{fnote:otherA}This would also be the case if one had chosen the other determination of the square-root---see footnote \ref{fnote:otheru0s}. This is because the equation for the action depends on $u_0^2$
		\begin{equation}
			-1 + \frac94 A^2 - 6 u_0^2=0,
		\end{equation} 
		so the square-root determination does not affect the result. Nonetheless, the action gets modified if one chooses $u_0=0$. In that case, one gets
		\begin{equation}
			A=\pm\frac23.
		\end{equation}
	Therefore, in order to study all the possible solutions to the inhomogeneous Painlev\'e II equation one would also need to apply a transseries ansatz to the $u_0=0$ determination and repeat the analyses in this paper.}
	\begin{equation}
		\label{eq:PII_with_alpha_action}
		A=\pm\frac{2\sqrt 2}{3}\,\rmi,
	\end{equation}  
	which, as expected, coincides with the previous approach.
\end{itemize}
It is worth mentioning that if one keeps track of all the changes of variables, this result coincides with previous works in the $\alpha\neq0$ \cite{m08,sv13} and $\alpha=0$ \cite{gg18} cases. Following the same reasoning as for the $q$-deformed Painlev\'e II equation, the fact that the action does not depend on $\alpha$ is expected from the point of view of resurgence. Recall that if one takes the whole equation to the Borel plane, parametrised by $s$, the $\alpha$ factor will be just represented, on the Borel plane, by an additional factor $\alpha s^p$ (for some power $p$). Therefore, one can consider the whole solution as this factor in convolution with the Borel transform of the $\alpha=0$ case. Since the convolution product does not move singularities, provided that one of the factors is entire, then we do not expect a change in the radius of convergence due to this term, thus being consistent with the aforementioned works, in which $\alpha=0$\footnote{Check \cite{s14} for further information on convolutions, Borel plane, its singularities and other details.}. 

\begin{figure}[hbt!]
	\centering
	\begin{subfigure}[t]{0.5\textwidth}
		\centering
		\includegraphics[height=1.7in]{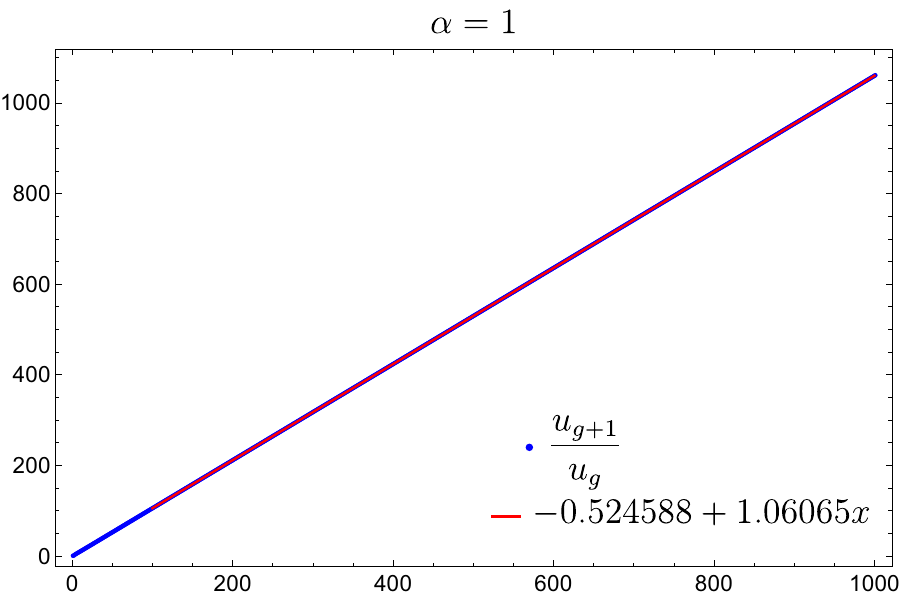}
		%\caption{}
	\end{subfigure}%
	~ 
	\begin{subfigure}[t]{0.5\textwidth}
		\centering
		\includegraphics[height=1.7in]{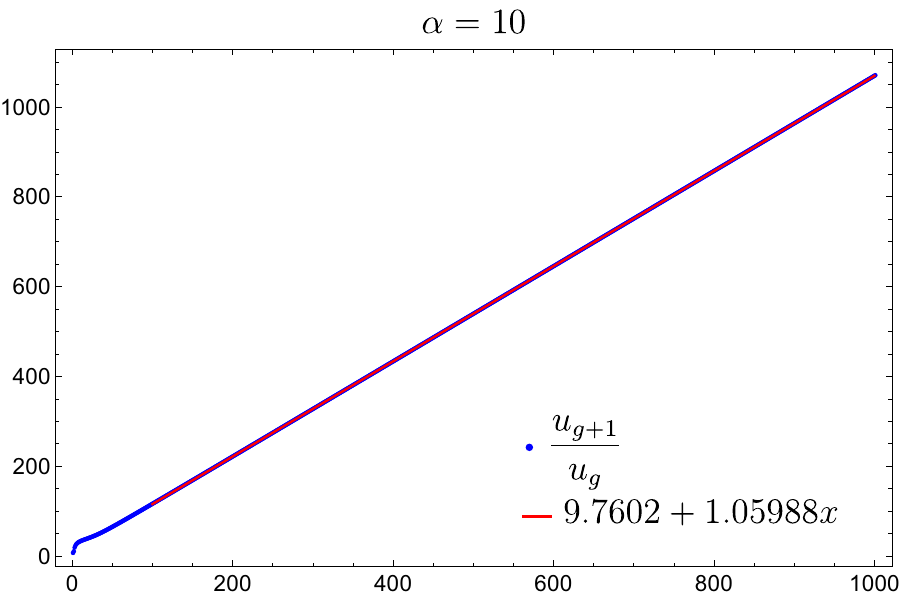}
		%\caption{Conformal transformations}
	\end{subfigure}\\
	\begin{subfigure}[t]{0.5\textwidth}
		\centering
		\includegraphics[height=1.7in]{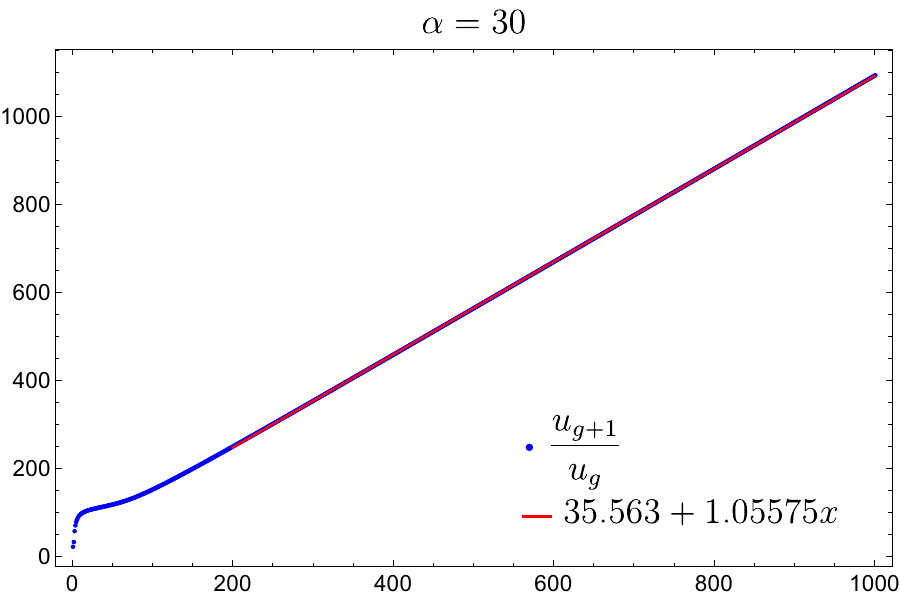}
		%\caption{\cite{asv11} computation }
	\end{subfigure}%
	~ 
	\begin{subfigure}[t]{0.5\textwidth}
		\centering
		\includegraphics[height=1.7in]{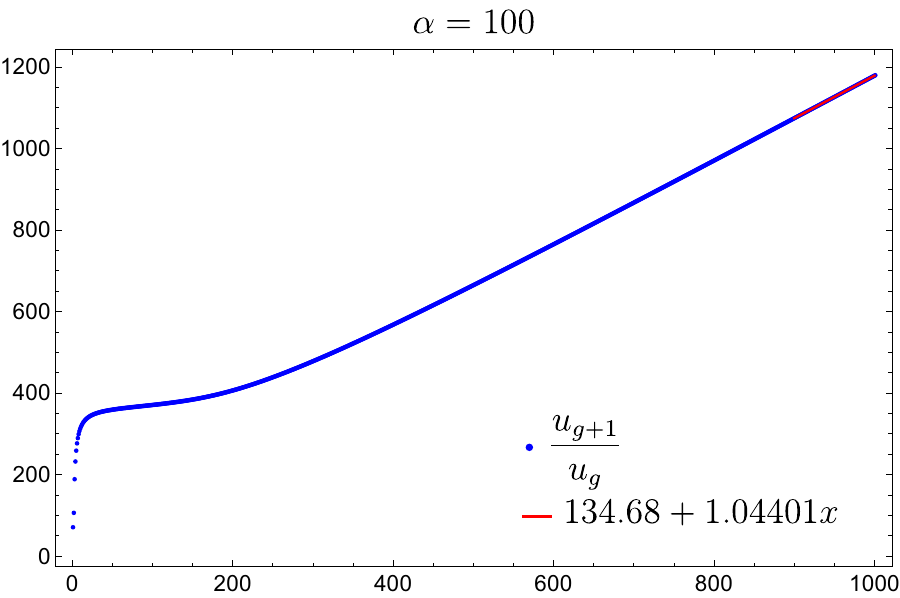}
		%\caption{Conformal transformations}
	\end{subfigure}
	\caption{Checks on the instanton action for different values of $\alpha$. As explained, the Gevray-1 property kicks in later when we increase the value of $\alpha$. Also, notice that each of the fits gives a linear term $\sim \left(\frac{2\sqrt{2}}{3}\right)^{-1}\simeq 1.0606\dots$.}
	\label{fig:checksaction}
\end{figure}

Let us give further details on the first approach. The reader may have realised that computing the ratio \eqref{eq:uratio} is somewhat cumbersome since the coefficients $u_g$ are polynomials in $\alpha$. The first thing one realises is that the order of the polynomial is $g$, so we can write 
\begin{equation}
	\label{eq:struc_of_pert_coeffs}
	u_g=\sum_{i=1}^gu_{g,i}\,\alpha^i.
\end{equation}
In order to check the Gevrey-1 property \eqref{eq:Gevrey-1_prop} of the coefficients $u_g$, we first computed the ratio for fixed $\alpha=1,10,30,100$ up to $g=1000,1000,1000,2000$, respectively. This is depicted in figure \ref{fig:checksaction}. One sees that the higher the $\alpha$, the later the Gevrey-1 behaviour dominates. This hints that, for larger $\alpha$, we will need more data in order to obtain reasonable precision for the resurgent properties\footnote{Indeed, the results presented in tables \ref{tab:s10precasymp}, \ref{tab:sm10precasymp}, \ref{tab:NalphaOneHalf}, \ref{tab:NalphaOne}, \ref{tab:NalphaThreeHalves}, \ref{tab:NalphaTwo}, \ref{tab:NalphaFiveHalves}, \ref{tab:NalphaThree} and \ref{tab:N21prec} support this fact.}. This is a clear difference with respect to the $q$-deformed Painlev\'e II equation, where, recalling \eqref{eq:struc_of_pert_coeffs_q2}, we had an all-powers series expansion for $q^2$ independently of the nature of $g$. 

On the other hand, as a double check, we have computed the coefficients for arbitrary $\alpha$ up to order $g=400$. We first noticed that, when $g$ is even (respectively, odd), then $u_g$ is a polynomial in even (respectively, odd) powers of $\alpha$. Thus, in order to analyse the properties of the $u_{g,i}$'s (recall \eqref{eq:struc_of_pert_coeffs}), we have to compare $u_{g,i}$ with $u_{g+2,i}$. The result that comes out of this comparison is that, for fixed $i$, the coefficients $u_{2g,2i}$ (respectively, $u_{2g+1,2i+1}$) have the Gevrey-1 property.

This concludes our first approach to the resurgent analysis of the inhomogeneous Painlev\'e II equation. Even thought it is possible to apply the whole resurgent analysis to the original form of the equation \cite{abs18,m08}, previous references \cite{asv11,sv13} have shown that with some change of variables, this analysis looks cleaner. We will present these changes in the next subsection as well as a deeper study of the resurgent transseries for our equation.

%%%%%%%%%%%%%%%%%%%%%%%%%%%%%%%%%%%%%%%%%%%%%%%%%%%%%%%%%%%%%%%%%
%%%%%%%%%%%%%%%%%%%%%%%%%%%%%%%%%%%%%%%%%%%%%%%%%%%%%%%%%%%%%%%%%
\subsection{Resurgent Transseries Analysis}\label{sec:PII}
%%%%%%%%%%%%%%%%%%%%%%%%%%%%%%%%%%%%%%%%%%%%%%%%%%%%%%%%%%%%%%%%%
%%%%%%%%%%%%%%%%%%%%%%%%%%%%%%%%%%%%%%%%%%%%%%%%%%%%%%%%%%%%%%%%%

Motivated by the expansion \eqref{eq:pert_exp}, we will perform some changes of variables. First, we will get rid of the global $\sqrt{z}$ term. Second, we would like to go to the ``resurgent variable''. What this means is that for the recursion relation we are interested in having series expansions in integer powers, and we will denote this one by $w$. However, the square-root of this variable will be the one that we will denote as the resurgent variable $x$. One nice feature of the $x$ variable is that our Laplace transform looks nice and easy, but it also encodes how to walk around the complex plane in a more straightforward way, since its connection with the Borel plane is straightforward. But for the moment, let us start with the $w$ variable. For computational reasons, it is more comfortable to have real instanton actions and real coefficients, which can be achieved by sending $z\to -z$\footnote{This change switches $(n,m)\leftrightarrow(m,n)$ and this is the reason why the computation of the matrix model yields $\widetilde{S}_{-1}^{(0)}$ instead of $S_1^{(0)}$---recall footnote \ref{fnote:typosv13}.}. Note that these are the same changes which were performed in \cite{sv13}. The first change amounts to sending the third term of \eqref{eq:PII_with_alpha} to minus itself. Also, we will rescale $u$ and $z$ by factors of $2$ to some adequate fractional power, so that in the end we obtain
\begin{equation}
	\label{eq:PII_with_alpha_prepared_form}
	u''(z)-2 u^3(z)+2z\,u(z)-2\alpha=0.
\end{equation}
\noindent
Then, the more cumbersome change of variables is the same as \eqref{eq:change_of_variables_q}
\begin{equation}
	\label{eq:change_of_var}
	u(w)\equiv \left.\frac{u(z)}{\sqrt{z}}\right|_{z=w^{-4/3}}.
\end{equation}
Applying \eqref{eq:change_of_var} to the equation \eqref{eq:PII_with_alpha_prepared_form} yields the final form of our equation:
\begin{equation}
	\label{eq:PII_with_alpha_Borel_form}
	\frac 9{16} w^6 u''(w) + 
	\frac 9{16} w^5 u'(w)+ \left(2 - \frac{w^4}{4}\right)  u(w) - 2 u(w)^3 -2 w^2 \alpha=0.
\end{equation}
Notice that the above equation coincides, as expected, with the one in \cite{sv13} in the limit $\alpha\to0$. Furthermore, this equation has a lot of similarities with equation \eqref{eq:PII_with_q2_final}, all the terms repeat, but with three powers less of $u(w)$; with the exception of the term $-2w^2\alpha$, that gets substituted by a factor of $q^2w^4$. The following steps constitute the building blocks of the analysis: finding the transseries coefficients for each sector (namely, finding a recursion relation for these coefficients), and computing Stokes data in order to find the transition functions. The reason for using these variables is that, in the next subsection, the transseries sectors will consist of polynomials in integer powers of $w$. Nonetheless, recall that in order to make the connection with the original variable and easily understand how to walk around the complex plane we will be interested in the variable
\begin{equation}
	\label{eq:xtowrelation}
	x=w^2.
\end{equation}

%%%%%%%%%%%%%%%%%%%%%%%%%%%%%%%%%%%%%%%%%%%%%%%%%%%%%%%%%%%%%%%%%
%%%%%%%%%%%%%%%%%%%%%%%%%%%%%%%%%%%%%%%%%%%%%%%%%%%%%%%%%%%%%%%%%
\subsubsection{Transseries Ansatz: Recursion Relation and Coefficient Structure}\label{subsec:transans}
%%%%%%%%%%%%%%%%%%%%%%%%%%%%%%%%%%%%%%%%%%%%%%%%%%%%%%%%%%%%%%%%%
%%%%%%%%%%%%%%%%%%%%%%%%%%%%%%%%%%%%%%%%%%%%%%%%%%%%%%%%%%%%%%%%%
The aim of this subsection is to compute and describe the transseries ansatz for the inhomogeneous Painlev\'e II equation in the resurgent variable, \eqref{eq:PII_with_alpha_Borel_form}. We will present the recursion relation, its properties, and illustrate a couple of examples to get familiar with the structure. Since this analysis is very similar to the one of the $\alpha=0$ case, we will always have in mind the comparison with \cite{sv13}.

Now, since we have a second-order differential equation, we are instructed to use a two-parameter transseries ansatz
\begin{equation}
	\label{eq:2ptransansatz}
	u(w,\sigma_1,\sigma_2)=\sum_{n,m=0}^{+\infty}\sigma_1^n\sigma_2^m\rme^{-\frac {(n-m)A}{w^2}}\Phi_{(n,m)}(w).
\end{equation}
Futhermore, as in \eqref{eq:q2ptransansatz}, the system is resonant \cite{asv11,bssv21,eggls23,gikm10,gs21,mss22,sst23,sv13}; this is, there is a linear combination of intanton actions with coefficients in $\mathbb{Z}$ relating one action with the other. This is relevant for two reasons. First, there will be infinitely many sectors weighted by the same action. This is, the sectors $(n,m)$ and $(n+k,m+k)$ have the same exponential pre-factors. On the other hand, this will require the addition of logarithmic sectors to our transseries ansatz\footnote{Basically, the recursion relation is undetermined unless one includes these logarithmic sectors. Alternatively, one could also bypass this indetermination by modifying the transseries ansatz to include terms like $z^{\gamma(n,m) \sigma_1\sigma_2}$, but this choice seems more unnatural---see \cite{asv11,bssv21,gikm10,sv13} for extra details.}. This yields the form
\begin{equation}
	\label{eq:asympsector}
	\Phi_{(n,m)}(w)\simeq\sum_{k=0}^{k_{nm}}\left(\log(w)\right)^k\sum_{g=0}^{+\infty}u_g^{(n,m)[k]}w^{g+\beta_{(n,m)}^{[k]}},
\end{equation}
where the quantity $k_{n,m}$ is given by\footnote{This limit for the $k$ sum was firstly derived in \cite{asv11} for the Painlev\'e I equation, motivated from a computation in \cite{gikm10}. This example lead to using the same ansatz for the case of the homogeneous Painlev\'e II equation in \cite{sv13}. Alternatively, one could also determine this upper limit by selecting a big artificial $k$ upper bound and seeing from which $k$ on the coefficients start being 0. Listing a table of these coefficients would also lead to the same conclusion about $k_{n,m}$.} \eqref{eq:knm}. Note that these ansatze are identical to \eqref{eq:q2ptransansatz}, \eqref{eq:qasympsector}.

By plugging \eqref{eq:asympsector} and \eqref{eq:2ptransansatz} into equation \eqref{eq:PII_with_alpha_Borel_form}, we obtain the following recursion relation for the $u_g^{(n,m)[k]}$ coefficients
\begin{align}
	\label{eq:PII_rec_rel}
	&
	2 \sum_{n_1=0}^n \sum_{n_2=0}^{n-n_1} \sum_{m_1=0}^m \sum_{m_2=0}^{m-m_1} \sum_{k_1=0}^k \sum_{k_2=0}^{k-k_1} \sum_{g_1=0}^g \sum_{g_2=0}^{g-g_1}\, u_{g_1}^{(n_1,m_1)[k_1]}\, u_{g_2}^{(n_2,m_2)[k_2]}\, u_{g-g_1-g_2}^{(n-n_1-n_2,m-m_1-m_2)[k-k_1-k_2]} = \nonumber \\
	&
	= \left( \frac{9}{4} A^2 (n-m)^2 + 2 \right) u_g^{(n,m)[k]} + \frac{9}{4} A (n-m) (k+1)\, u_{g-2}^{(n,m)[k+1]} \nonumber \\
	&+\frac{9}{4} A (n-m)\left(g+\beta_{(n,m)}^{[k]}-3\right)\, u_{g-2}^{(n,m)[k]}  
	+\frac{9}{16} (k+2)(k+1)\, u_{g-4}^{(n,m)[k+2]}\nonumber \\&+ \frac{9}{8} (k+1)\left(g+\beta_{(n,m)}^{[k+1]}-4\right)\, u_{g-4}^{(n,m)[k+1]} + \frac{140 + 9 \left(g+\beta_{(n,m)}^{[k]}\right) \left(g+\beta_{(n,m)}^{[k]}-8\right)}{16}\, u_{g-4}^{(n,m)[k]}\nonumber\\
	&-2\,\alpha\,\delta_{m,0}\delta_{n,0}\delta_{g,2}\delta_{k,0}
\end{align}
Note that this result coincides with \cite{sv13} up to the last line. Thus, most of the features can be extracted from there and we refer the reader to that paper for further information. With the objective of being pedagogical and self-contained, let us repeat herein the most relevant points, as well as explain what modifications have to be done. First of all, one finds that resonance still plays a role in the recursion relation and one is forced to make a choice for the starting coefficients when $|n-m|=1$. In our case, we have kept the choice in \cite{sv13}, that is, we have set the term proportional to $w^1$ in the $(n+1,n)[0]$, $(n,n+1)[0]$ sectors to be $\delta_{n,0}$. Another feature that is unmodified is the fact that the $k$-th sectors can be obtained in terms of the $k=0$ sector. Of course, the relation is the same
\begin{equation}
	u^{(n,m)[k]}_g=\frac1{k!}\left(8(m-n)\right)^ku^{(n-k,m-k)[0]}_g.
\end{equation}
However, there are three properties which are missing or that one has to modify. First, it is no longer true that the diagonal $(n,n)$ sectors have a genus expansion in powers of $w^4$ (unless $\alpha=0$ of course). This is clear for the $(0,0)$ sector since the first element that has to be 0 for this to happen, $u^{(0,0)}_{2}$, now has a term $-2\alpha$, such that $u^{(0,0)}_{2}=-\frac{\alpha}{2}$. This gets propagated to the rest of diagonal sectors in such a way that this property is unsatisfied. Secondly, the starting powers for each sector, denoted by a power of $\beta_{(n,m)}^{[k]}$ have to be modified to introduce an $\alpha$ dependence. This first happens for the $(1,0)$ sector due to the new $\alpha$ term. If one explores the equation one finds that 
\begin{equation}
	2\beta_{(n,m)}^{[k]}=n+m-2\left[\frac{k_{n,m}+k}2\right]_I-2(n-m)\alpha,
\end{equation} 
where $\left[\cdot\right]_I$ denotes the integer part. Note the extra term with respect to equation \eqref{eq:knm}. One may wonder why this term does not appear in the $q$-deformed Painlev\'e II equation, but it does here. Further explorations show that this $\alpha$, respectively $q$, dependant term appears only in certain perturbative sheets. For the $\alpha$ case, it happens when choosing the double-cover solution, this is, a perturbative solution that behaves as $\sim\pm\sqrt{z}$ ($\sim\pm 1\, w^0$), the one we have chosen in here. On the other hand, for the $q$ case, this term only appears when the perturbative solution behaves as $\sim z^{-\frac14}$, the one we have not chosen in this paper. Finally, the last detail that changes is the relation between the $(n,m)\leftrightarrow(m,n)$ sectors. To this relation one has to add the fact that all the terms proportional to $\alpha$ in the $(n,m)$ sector are now generated in the $(m,n)$ sector by $-\alpha$. Therefore, one only has to take the same relation and send $\alpha\mapsto-\alpha$ to relate the coefficients. This will also get reflected on the relation between forward and backward Stokes coefficients in later subsections.

Taking all of this into account and making use of the recursion relation above, one can write out a couple of transseries sectors
\begin{align}
	\Phi_{(0,0)}(w)=&1- \frac \alpha 2 w^2  - \left(\frac{1}{16}+\frac{3}{8}\alpha^2\right) w^4 - 
	\alpha \left(\frac{11}{32} + \frac{\alpha^2}2\right) w^6  - \left(\frac{73}{512}+\frac{177}{128}\alpha^2+\frac{105}{128}\alpha^4 \right) w^8 +\cdots,\\
	\Phi_{(1,0)}(w)=&w^{1-2\alpha} + \left(-\frac{17}{96} + \frac{3}8\alpha - \frac78 \alpha^2\right) w^{3-2\alpha}+\nonumber\\
	&+ \left(\frac{1513}{18432} - \frac{75}{128}\alpha + \frac{533 }{768}\alpha^2 - 
	\frac{31}{32} \alpha^3 + \frac{49}{128} \alpha^4\right) w^{5-2\alpha} +\cdots,\\
	\Phi_{(2,0)}(w)=&\frac 12 w^{2-4\alpha} + \left(-\frac{41}{96} + \frac 78 \alpha - \frac 78 \alpha^2\right) w^{4-4\alpha}\nonumber+\\
	& + \left(\frac{5461}{9216} - \frac{1645 }{768}\alpha + \frac{1025 }{384}\alpha^2 - \frac{139}{64} \alpha^3 + \frac{49}{64} \alpha^4\right)w^{6-4\alpha}+\cdots,\\
	\Phi_{(1,1)}(w)=&-3 w - \frac{21}{4} \alpha w^3 - \left(\frac{291}{128} + \frac{369}{32} \alpha^2\right) w^5 - 
	\alpha\left(\frac{10371}{512} + \frac{3453 }{128}\alpha^2\right) w^{7}+\cdots,\\
	\Phi_{(2,1)}(w)=&\left(1-\frac{17}{2}\alpha\right) w^{3-2\alpha}+ \left(
	-\frac{115}{48} + \frac{925}{192} \alpha - \frac{231}{16} \alpha^2 + \frac{119}{16} \alpha^3 \right)w^{5-2\alpha}+\cdots+\nonumber\\
	&+\log(w)\left\{-8 w^{1-2\alpha}+  \left(\frac{17}{12} - 3 \alpha + 7 \alpha^2\right) w^{3-2\alpha}\nonumber+\right.\\
	&\left.+ 
	\left(-\frac {1513}{2304} + \frac{75 }{16} \alpha- \frac{533}{96} \alpha^2 + 
	\frac{31}{4} \alpha^3 - \frac{49}{16} \alpha^4\right)w^{5-2\alpha}+\cdots\right\}.
\end{align}

In order understand the structure of {\it any} solution, we also need to compute the transseries sectors for the single-cover initial condition--- the one with $u_0^{(0,0)}[0]=0$, recall footnotes \ref{fnote:otheru0s} and \ref{fnote:otherA}. The recursion relation is the same, while one has to modify both the $\beta_{(n,m)}^{[k]}$ factors to include a term $\delta_{n,m}$ and no $\alpha$ dependence, as well as the initial $(0,0)$ sector. We will denote the sectors associated with the single-cover by $\Phi^{\text{s}}_{(n,m)}(w)$. Hence, the first ones read
\begin{align}
	\Phi^{\text{s}}_{(0,0)}(w)=& w^2 \alpha + w^6 \alpha (-1 + \alpha^2) + 
	w^{10} \alpha (10 - 13 \alpha^2 + 3 \alpha^4)+\nonumber\\
	& +  w^{14} \alpha (-280 + 397 \alpha^2 - 129 \alpha^4 + 12 \alpha^6)+\cdots, \\
	\Phi^{\text{s}}_{(1,0)}(w)=&w - \rmi\left(-\frac5{48\sqrt2} +\frac{2 \alpha^2}{\sqrt2}\right) w^3 + \left(
	 -\frac{385}{9216} + \frac{41 \alpha^2}{48} - \alpha^4\right) w^5+\nonumber\\
	 &+ \rmi\left(-\frac{85085}{1327104 \sqrt2} + \frac{12313 \alpha^2}{4608 \sqrt2} - \frac{55 \alpha^4}{16 \sqrt2} + \frac2{3\sqrt2} \alpha^6\right)w^{7}+\cdots,\\
	 \Phi^{\text{s}}_{(2,0)}(w)=&-\alpha w^4+ \rmi \left(-\frac{77 \alpha}{24\sqrt 2}+ 2\sqrt 2 \alpha^3\right) w^6 + 
	  \left(\frac{17629 \alpha}{2304} - \frac{143 \alpha^3}{12} + 4 \alpha^5\right)w^8+\cdots,\\
	 \Phi^{\text{s}}_{(1,1)}(w)=&6\alpha w^4+  \left(-\frac{1047 \alpha}{32} + 33 \alpha^3\right) w^8 +\nonumber\\& +\left(-\frac{7282152459 \alpha}{262144} + \frac{324506343 \alpha^3}{8192} - \frac{3350529 \alpha^5}{256} + \frac{10053 \alpha^7}8 \right) w^{16}+\cdots.
\end{align}
We will work with the double-cover unless explicitly specified. This concludes the study of the transseries for the inhomogeneous Painlev\'e II equation. Recall that the resummation of the above transseries is not unique in each wedge. In order to understand global solutions we still need to compute Stokes data\footnote{We again refer the reader to \cite{abs18,s14} for nice introductions to resurgence, transseries and related topics.}. These will allow us to compute transition functions at Stokes lines. These transitions make the solutions continuous in the complex plane and are fundamental to understanding global structure. The next subsection presents a method of obtaining Stokes data {\it in closed-form} merely from manipulating asymptotic formulae into equations. This method was first used in \cite{bssv21} and we already have an example in subsection \ref{subsec:q-closed-form-asymp}.

%%%%%%%%%%%%%%%%%%%%%%%%%%%%%%%%%%%%%%%%%%%%%%%%%%%%%%%%%%%%%%%%%
%%%%%%%%%%%%%%%%%%%%%%%%%%%%%%%%%%%%%%%%%%%%%%%%%%%%%%%%%%%%%%%%%
\subsubsection{Closed-Form Asymptotics}\label{subsec:closed-form-asymp}
%%%%%%%%%%%%%%%%%%%%%%%%%%%%%%%%%%%%%%%%%%%%%%%%%%%%%%%%%%%%%%%%%
%%%%%%%%%%%%%%%%%%%%%%%%%%%%%%%%%%%%%%%%%%%%%%%%%%%%%%%%%%%%%%%%%

We start this section by pointing out that, since the recursion relation for $(n,m)\leftrightarrow(m,n)$ fulfils all the properties in \cite{sv13,bssv21} up to sending $\alpha\mapsto-\alpha$ and the fact that our $\beta_{(n,m)}^{[k]}$ factor now has an $\alpha$-dependent term, the ``{\it backward-forward}'' relation is still valid upon sending $\alpha\mapsto-\alpha$ in the final result. Indeed, the derivation is completely analogous to the one in section \ref{subsec:stok-bor} and, as an example, one can compare our results for $S^{(0)}_1$ and $\widetilde{S}_{-1}^{(0)}$ appearing in the next sections. The explicit relation reads
\begin{equation}
	\label{eq:back-forw}
	\mathsf{S}_{(n,n)\to(p,p+\ell)}(\alpha)=(-\rmi)^\ell(-1)^{n+p+\ell-\alpha \ell}\sum_{q=p}^{n}\frac{(-4\pi\rmi \ell)^{q-p}}{(q-p)!}\mathsf{S}_{(n,n)\to(p+\ell,p)}(-\alpha).
\end{equation}
Now, we will try to derive an analytic way of generating all Stokes data by means of closed-form asymptotics---see \cite{bssv21}, and our small recap in subsection \ref{subsec:q-closed-form-asymp}.  
There are some necessary initial ingredients that one needs to take into account. Closed-form asymptotics will, in the end, give us equations or generating functions that will generate the $N_{\ell-p}^{(\ell)}$ along the diagonal direction. Nonetheless, these generating functions will depend on determining the first entry of the diagonal. Thus, one prerequisite is knowing the structure along the first column and what is the residual logarithmic contribution. We assume this to be known, both because we will have some conjecture later coming from the numerical computations as well as because this conjecture agrees with \cite{bssv21} in the case $\alpha=0$. Not only that, but as we will see latter, this pattern will provide with monodromy closure, a more encouraging result. 

We start from the asymptotic relation\footnote{This relation comes from applying Cauchy's theorem to the $(n,n)$ sector and writing the discontinuity in terms of Borel residues \cite{asv11,sv13}. In fact, the first line corresponds to the term coming from the singularity in the $0$ direction, hence the $(p+\ell,p)$ contributions, while the second line corresponds to the $\pi$, and thus the $(p,p+\ell)$. See \cite{abs18} and section \ref{subsec:stok-bor} for further details.} \cite{bssv21}
\begin{align}
	\label{eq:asymptunn}
	u_{2g}^{(n,n)[0]}&\simeq \sum_{\ell=1}^{+\infty}\sum_{h=0}^{+\infty}\sum_{p=0}^{n}\sum_{k=0}^p-\frac{\mathsf{S}_{(n,n)\to(p+\ell,p)}}{2\pi\rmi}\frac{u_{2h}^{(p+\ell,p)[k]}}{(-2)^k}\widetilde{H}_k(g+n-h-\beta_{(p+\ell,p)}^{[k]},\ell A)\nonumber\\
	&\hspace{4mm}-\frac{\mathsf{S}_{(n,n)\to(p,p+\ell)}}{2\pi\rmi}\frac{u_{2h}^{(p,p+\ell)[k]}}{(-2)^k}(-1)^{h-g-n-\beta_{(p,p+\ell)}^{[k]}}H_k(g+n-h-\beta_{(p,p+\ell)}^{[k]},\ell A).
\end{align}
Notice that the term $\beta_{(p+\ell,p)}^{[k]}\sim (p-(p+\ell))\alpha=-\ell\alpha$, while the non-alpha dependent part is the same than for $\beta_{(p,p+\ell)}^{[k]}$. Now, let us assume without loss of generality that $\alpha>0$ from now on. This means that we will ignore the second line in \eqref{eq:asymptunn}. Notice that, in the case of $\alpha<0$, then we would ignore the first line. Following now the same steps as in subsection \ref{subsec:q-closed-form-asymp}, and recalling that $\beta_{(p+\ell,p)}^{[p]}=-\ell(\alpha-\frac 12)$, one arrives at
\begin{align}
	\label{eq:closed-form-asymp}
	N^{(\ell)}_1\delta_{n,0}= \sum_{p=0}^{n-\ell+1}\frac{(4s)^p}{p!}N^{(\ell)}_{\ell-n+p}B_p&\left\{\frac 1{\ell^0}\frac 1{2^1}\left(\psi^{(0)}\left(\alpha+\frac 12\right)+\psi^{(0)}(1)\right)-\log(8),\right.\nonumber\\
	&\hspace{3mm}\frac 1{\ell^1}\frac 1{2^2}\left(\psi^{(1)}\left(\alpha+\frac 12\right)+\psi^{(1)}(1)\right),\dots,\nonumber\\
	&\hspace{3mm}\left.\frac 1{\ell^{p-1}}\frac 1{2^{p}}\left(\psi^{(p-1)}\left(\alpha+\frac 12\right)+\psi^{(p-1)}(1)\right)\right\},
\end{align}
which, of course, coincides with the results in \cite{bssv21} in the limit $\alpha=0$. Also, notice that from the point of view of the matrix model, one always finds factors of $\alpha+\frac12$, which makes the above result very natural. The extra factors of digamma functions evaluated at $1$ were discussed in section \ref{subsec:q-closed-form-asymp}. It was conjectured that all the transcendental numbers yielding Stokes data in the large-$g$ limit were only appearing through the digamma functions and only the zeta function content. Therefore, we only need to determine the coefficients $N^{(\ell)}_1$. For that, we have a guess at the end of section \ref{subsec:numcheck} that comes from our numerical observations together with the previous results of \cite{bssv21}. These two results together, namely equations \eqref{eq:closed-form-asymp} and \eqref{eq:Ns1}, reproduce all our numerical data, which we present in the next section. 
%%%%%%%%%%%%%%%%%%%%%%%%%%%%%%%%%%%%%%%%%%%%%%%%%%%%%%%%%%%%%%%%%
%%%%%%%%%%%%%%%%%%%%%%%%%%%%%%%%%%%%%%%%%%%%%%%%%%%%%%%%%%%%%%%%%
\subsubsection{Numerical Checks}\label{subsec:numcheck}
%%%%%%%%%%%%%%%%%%%%%%%%%%%%%%%%%%%%%%%%%%%%%%%%%%%%%%%%%%%%%%%%%
%%%%%%%%%%%%%%%%%%%%%%%%%%%%%%%%%%%%%%%%%%%%%%%%%%%%%%%%%%%%%%%%%
Even though we have a closed-form conjecture for the Stokes coefficients, we are still missing two fundamental points: an analytic form for the $N^{(\ell)}_1$ Stokes coefficients, and extensive numerical checks on this conjecture. We will use both the method of Borel plane residues---see subsection \ref{subsec:qnumcheck}---as well as {\it classical} asymptotic checks, see \cite{asv11,sv13}. The main issue that one encounters is: how do we deal with the $\alpha$ dependence? We have taken two main approaches to this question, each of them motivated by different reasons. The first one is computing the Borel residues/Stokes coefficients by fixing $\alpha$ to some value. This approach has the advantage that computations are much faster and can be computed to much higher precision. The counterpart is that guessing a function from a list of values is neither a definitive construction nor an easy task, especially if the function is cumbersome. The second approach is computing them as power expansions in $\alpha$. This method is somehow the counterpart of the previous one. In particular, calculations are computationally heavy and precision is hard to obtain. On the other hand, a series expansion represents a function better than just a bunch of values. But, why is it more computationally heavy? For the asymptotic method, the only impediment is that, since Richardson transforms are based on constructing sets of values by taking combinations of coefficients\footnote{With the new values approximating the asymptotic limit much faster, {\it i.e.}, accelerating convergence.}, one cannot perform a lot of them if one keeps $\alpha$ arbitrary. This happens because computational times increase significantly due to the complicated final expressions (ratios of ratios of polynomials in $\alpha$). However, for fixed values of $\alpha$, this is not an issue since one can always simplify fractions without them getting very cumbersome. As an illustration, while for fixed $\alpha$ one can easily perform $60$ Richardson transforms; for arbitrary values, computational times increase significantly around the $6$-th one. Having solved this issue with the Richardson transform, the question repeats itself for the Pad\'e approximants. The moment either asymptotics or Borel plane residue's method need Pad\'e approximants, the story is again a bit complicated. Since Pad\'e approximants are also based on generating coefficients out of previous ones, it is computationally very heavy to compute them whenever they include symbolic factors. For that reason, one has to circumvent this problem by making a slight modification to the methods. For each $\alpha$ power, we have a polynomial in $s$ multiplying ($s$ being the Borel plane variable). We thus have Pad\'e approximated these polynomials in $s$ when the structure of the polynomial was reasonable\footnote{By reasonable we mean that in some cases the polynomials attached to the $\alpha$ powers were made out of a couple of monomials of large degree. In this case, the Pad\'e approximant is not very effective as an approximation.}. Therefore, we have separated all the $\alpha$ contributions and resummed them in the sense of \cite{bssv21}\footnote{This is, applying the Laplace transform to the (Pad\'e approximants of the) contributions.}. Hence, the result will be a set of values attached to their corresponding $\alpha$ power, {\it i.e.}, a numerical Taylor expansion in $\alpha$.

With these methods, one can guess the form of the first forward and backward Stokes coefficients. The fist one is given by
\begin{equation}
	\label{eq:S10}
	S^{(0)}_1(\alpha)=-\frac{2^{3\alpha-\frac 12}}{\Gamma\left(\alpha+\frac 12\right)}\rmi.
\end{equation}
The result for the backward Stokes coefficient $\widetilde{S}^{(0)}_{-1}$ is the expected one from the relation between the $(n,m)\leftrightarrow(m,n)$, which is
\begin{equation}
	\label{eq:tS10}
	\widetilde{S}^{(0)}_{-1}(\alpha)=-(-1)^{-\alpha}\frac{2^{-3 \alpha - \frac12}}{\Gamma\left(-\alpha + \frac12\right)}.
\end{equation}
Note that this agrees with the prediction from the matrix model \eqref{eq:tildeS10fromMM}, upon the usage of Euler reflection formula for the gamma function
\begin{equation}
	\Gamma(z)\Gamma(1-z)=\frac\pi{\sin(\pi z)},
\end{equation}
and acknowledging that the matrix model only makes sense for $\alpha\in\mathbb{Z}$.\\

\noindent\textbf{Asymptotics}: We have chosen the method of asymptotics\footnote{See \cite{abs18,asv11,bssv21,sv13} for details on this method. In particular, for understanding Richardson transforms.} to check the Stokes constants $S^{(0)}_{1}(\alpha)$ and $\widetilde{S}^{(0)}_{-1}(\alpha)$ since it has proven to be considerably fast and accurate---even when comparing with the Borel plane method. The reason being that we do not need to generate a lot of data for the $(1,0)$ and $(0,1)$ sectors\footnote{In fact, we only need one and it was fixed to be 1. So really, we only need to generate perturbative coefficients, which are considerably faster than the rest.}, and the fact that one can use Richardson transforms to improve convergence with astonishing results. For other vector coefficients, this method has proven to be considerably less efficient than the one in the next subsection \cite{bssv21}, but it is still worth for this case. Let us start from the asymptotic expression \eqref{eq:asymptunn} for the case $n=0$
\begin{align}
	\label{eq:asymptu00}
	u_{2g}^{(0,0)[0]}&\simeq \sum_{\ell=1}^{+\infty}\sum_{h=0}^{+\infty}\,-\,\frac{\mathsf{S}_{(0,0)\to(\ell,0)}}{2\pi\rmi}\,u_{2h}^{(\ell,0)[0]}\,\widetilde{H}_0\left(g-h-\beta_{(\ell,0)}^{[0]},\ell A\right)-\nonumber\\
	&\hspace{4mm}-\frac{\mathsf{S}_{(0,0)\to(0,\ell)}}{2\pi\rmi}\,u_{2h}^{(0,\ell)[0]}\,(-1)^{h-g+\beta_{(0,\ell)}^{[0]}}\,H_0\left(g-h-\beta_{(0,\ell)}^{[0]},\ell A\right).
\end{align}
We want to extract the leading contributions on the right-hand side. Due to the properties of the $\widetilde{H}$, $H$ functions \eqref{eq:HtildeHasymptoticgrowth}, these correspond to $\ell=1$, $h=0$. The expression then reads\footnote{Note that we have fixed the first coefficient of the $(1,0)$ and $(0,1)$ sectors to be 1; the fact that
	\begin{align}
		\beta_{(1,0)}^{[0]}=\frac12-\alpha,\qquad
		\beta_{(0,1)}^{[0]}=\frac12+\alpha;
	\end{align}
	and the map between Borel residues and Stokes coefficients \eqref{eq:qboreltostokes}
	\begin{align}
		\mathsf{S}_{(0,0)\to(1,0)}&=-N^{(1)}_1=-S_1^{(0)},\\
		\mathsf{S}_{(0,0)\to(0,1)}&=-N^{(-1)}_1=-\widetilde{S}_{-1}^{(0)}.
	\end{align}
}
\begin{align}
	\label{eq:asymptu00final}
	u_{2g}^{(0,0)[0]}&\simeq \frac{S^{(0)}_1}{2\pi\rmi}\widetilde{H}_0\left(g-\frac12+\alpha,A\right)-\frac{\widetilde{S}^{(0)}_{-1}}{2\pi\rmi}(-1)^{-g+\frac12+\alpha}H_0\left(g-\frac12-\alpha,A\right)+\cdots.
\end{align}
\noindent
Due to the $\alpha$ factor in the $\beta_{(1,0)}^{[0]}$ and $\beta_{(0,1)}^{[0]}$, it will be easy to isolate the Stokes coefficients. For $\alpha=0$ one has to take both sectors into account in the large $g$ limit. For $\alpha>0$, the dominant term will be proportional to $S_1^{(0)}(\alpha)$, so extracting the value from those cases would be easier and one would obtain more precision. The same is true for negative values of $\alpha$ and $\widetilde{S}_{-1}^{(0)}(\alpha)$. All of this can be seen from the asymptotic growths in \eqref{eq:HtildeHasymptoticgrowth}. However, in both cases it is possible to determine both constants. In tables \ref{tab:s10precasymp} and \ref{tab:sm10precasymp} we present extensive checks, for different values of $\alpha$, on $S_1^{(0)}(\alpha)$ and $\widetilde{S}_{-1}^{(0)}(\alpha)$, respectively\footnote{Note the the low precision of the value $\alpha=\pm0.2452$ is associated with the inaccuracy of the gamma functions in Mathematica for ``non-conventional'' values.}.\\

\begin{table}[hbt!]
	\begin{center}
		\scriptsize
		\begin{tabular}{||c||c|c||||c||c|c||||c||c|c||}
			\hline
			$\alpha$ & $S^{(0)}_1(\alpha)$ & Precision & $\alpha$ & $S^{(0)}_1(\alpha)$ & Precision & $\alpha$ & $S^{(0)}_1(\alpha)$ & Precision \\ [0.5ex] 
			\hline\hline
			& & & & & & & & \\
			0 & $-\displaystyle\frac{\rmi}{\sqrt{2\pi}}$ & $2.04\cdot10^{-54}$ & $\displaystyle \frac{3}{2}$ & $-16\rmi$ & $3.60\cdot10^{-50}$ & 3 & $-\displaystyle \frac{2048}{15}\rmi\displaystyle\sqrt\frac{2}{\pi}$ & $3.87\cdot10^{-45}$ \\
			& & & & & & & & \\
			$\displaystyle \frac{1}{2}$ & $-2\rmi$ & $1.40\cdot10^{-52}$ & 2 & $-\displaystyle\frac{128}{3}\rmi\displaystyle\sqrt\frac{2}{\pi}$ & $4.89\cdot 10^{-48}$ & $\displaystyle \frac 72$ & $-\displaystyle \frac{512}3 \rmi$ & $6.61\cdot 10^{-44}$ \\
			& & & & & & & & \\
			1 & $-8\rmi\displaystyle\sqrt\frac{2}{\pi}$ & $3.52\cdot10^{-51}$ & $\displaystyle\frac 52$ & $-64\rmi$ & $1.71\cdot 10^{-46}$ & 0.2452 & $-0.9557895\rmi$    & $5.11\cdot10^{-7}$ \\
			& & & & & & & & \\
			& & & & & & & & \\
			\hline\hline
		\end{tabular}
		\caption{Precision for the asymptotic checks of $S^{(0)}_1(\alpha)$ for different values of $\alpha$. The checks have been made with 400 data points and 60 Richardson transforms, with the exception of the point $\alpha=0.2452$, in which we have 3 Richardson transforms. \label{tab:s10precasymp}}
	\end{center}
\end{table}

\begin{table}[hbt!]
	\begin{center}
		\scriptsize
		\begin{tabular}{||c||c|c||||c||c|c||||c||c|c||} 
			\hline
			$\alpha$ & $\widetilde{S}^{(0)}_{-1}(\alpha)$ & Precision & $\alpha$ & $\widetilde{S}^{(0)}_{-1}(\alpha)$ & Precision & $\alpha$ & $\widetilde{S}^{(0)}_{-1}(\alpha)$ & Precision \\ [0.5ex] 
			\hline\hline
			& & & & & & & & \\
			0 & $-\displaystyle\frac{1}{\sqrt{2\pi}}$ & $2.04\cdot10^{-54}$ & $\displaystyle -\frac{3}{2}$ & $16\rmi$ & $4.88\cdot10^{-50}$ & $-3$ & $\displaystyle \frac{2048}{15}\displaystyle\sqrt\frac{2}{\pi}$ & $5.15\cdot10^{-45}$ \\
			& & & & & & & & \\
			$\displaystyle -\frac{1}{2}$ & $-2\rmi$ & $2.06\cdot10^{-53}$ & $-2$ & $-\displaystyle\frac{128}{3}\displaystyle\sqrt\frac{2}{\pi}$ & $6.54\cdot 10^{-48}$ & $-\displaystyle \frac 72$ & $\displaystyle \frac{512}3 \rmi$ & $8.79\cdot 10^{-44}$ \\
			& & & & & & & & \\
			$\displaystyle-1$ & $8\displaystyle\sqrt\frac{2}{\pi}$ & $1.78\cdot10^{-51}$ & $\displaystyle-\frac 52$ & $-64\rmi$ & $2.28\cdot 10^{-46}$ & -0.2452 & $0.9557895\rmi$    & $5.11\cdot10^{-7}$ \\
			& & & & & & & & \\
			& & & & & & & & \\
			\hline\hline
		\end{tabular}
		\caption{Precision for the asymptotic checks of $\widetilde S^{(0)}_{-1}(\alpha)$ for different values of $\alpha$. The checks have been made with 400 data points and 60 Richardson transforms, with the exception of the point $\alpha=0.2452$, in which we have 3 Richardson transforms.\label{tab:sm10precasymp}}
	\end{center}
\end{table}

\begin{figure}[hbt!]
	\centering
	%\begin{subfigure}[t]{0.5\textwidth}
	%\centering
	\includegraphics[height=2.5in]{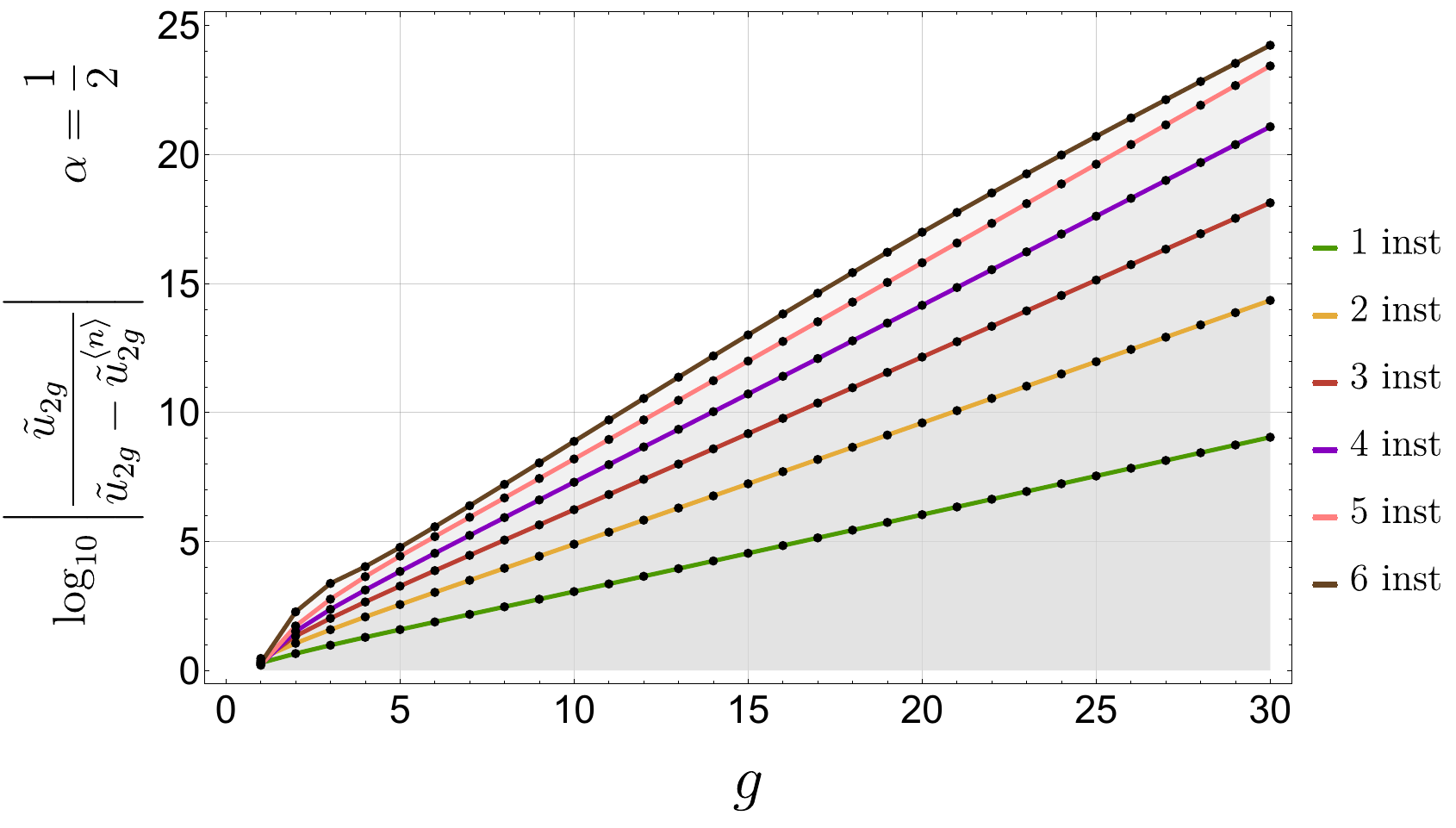}
	%\caption{$\alpha=\displaystyle\frac 12$}
	%\end{subfigure}%
	~ 
	%\begin{subfigure}[t]{0.5\textwidth}
	%\centering
	%\includegraphics[height=1.7in]{10.pdf}
	%\caption{$\alpha=1$}
	%\end{subfigure}\\
	%\begin{subfigure}[t]{0.5\textwidth}
	%\centering
	%\includegraphics[height=1.7in]{30.pdf}
	%\caption{\cite{asv11} computation }
	%\end{subfigure}
	\caption{Checks on the validity of large-order formula. Notice that even for relative small $g$ the formula agrees up to a decent precision. Here, we have gone up to $h=400,360,320,300,280,260$ for the $\ell=1,2,3,4,5,6$ instanton corrections.}
	\label{fig:asympcheck}
\end{figure}

\noindent\textbf{Resurgent Structure Checks via Asymptotics}: In continuation, one can assume the predicted analytic result of the Stokes coefficients to be true. Under this assumption, one can check the validity of the resurgent structure in the following way. Define the quantity 
\begin{equation}
	\label{eq:asympcheckexact}
	\tilde{u}_{2g}= \frac{2 \pi \rmi \,A^{g-\beta_{1,0}^{[0]}}}{\Gamma\left(g-\beta_{(1,0)}^{[0]}\right)}\,u^{(0,0)[0]}_{2g}.
\end{equation}
Its value is known {\it exactly} for any $g$, since the coefficients can be computed simply for the recursion relation. On the other hand, the asymptotic expansion \eqref{eq:asymptu00} yielding 
\begin{align}
	\label{eq:asympchecktransstruc}
	\tilde{u}_{2g}&\simeq \frac{A^{g-\frac12+\alpha}}{\Gamma\left(g-\frac12+\alpha\right)}\left\{ \sum_{\ell=1}^{+\infty}\sum_{h=0}^{+\infty}\left(S_1^{(0)}\right)^\ell u_{2h}^{(\ell,0)[0]}\frac{\Gamma\left(g-\beta_{(\ell,0)}^{[0]}-h+1\right)}{(\ell A)^{g-\beta_{(\ell,0)}^{[0]}-h+1}}\right.\nonumber\\
	&\hspace{4.5cm}\left.+\left(\widetilde{S}_{-1}^{(0)}\right)^\ell u_{2h}^{(0,\ell)[0]}\frac{\Gamma\left(g-\beta_{(0,\ell)}^{[0]}-h+1\right)}{(-\ell A)^{g-\beta_{(0,\ell)}^{[0]}-h+1}}\right\},
\end{align}
encodes a way of checking the transseries structure. This is, one can verify that, in the large-$g$ limit, the {\it exact} quantity \eqref{eq:asympcheckexact} agrees with the {\it asymptotic} expression \eqref{eq:asympchecktransstruc}. This describes the influence of the $(0,\ell)$, $(\ell,0)$ sectors on the asymptotics of the $(0,0)$. We have done this computation for the value $\alpha=\frac{1}{2}$ up to $g=30$. The results are presented in figure \ref{fig:asympcheck}. One sees that, as expected, higher instanton sectors give extra corrections to this quantity up to the moment in which the first instanton correction is not precise enough for it to be corrected by the next instanton contribution. \\

\noindent\textbf{Borel Plane Residues: First Diagonal}: This method has several advantages with respect to asymptotics, as depicted in \cite{bssv21}. For this reason, we shall rely on this method for the rest of the checks in the remaining parts of the paper. In particular, we will also use conformal transformations to improve our results \cite{bssv21}. The reason being that producing new coefficients for a parameter-dependent recursion relation is hard, so any tool that improves precision will shorten computational times. As an illustration, the Borel residue $\mathsf{S}_{(0,0)\to(0,1)}(\alpha=1)=-\widetilde{S}_{-1}^{(0)}(\alpha=1)$ has been computed with an agreement of $5.83\times 10^{-73}$ digits with the same number of coefficients as for the asymptotics method\footnote{Note that, for this specific case, this method requires double the number of coefficients since one needs both $(0,0)$ and $(0,1)$ sectors, whereas asymptotics only needs $(0,0)$ and the first coefficient of the $(0,1)$, which we chose to be one.}. The disadvantage of this method, however, is that the precision depends on the dampings\footnote{By dampings we refer to the point in the resurgent variable $x$ in which we impose continuity. The word ``damping'' refers to the fact that this $x$ value appears as the exponential weight in the Laplace transform of the sector and works as a parameter that zooms in on the singularity we are interested in---see \cite{bssv21} and subsection \ref{subsec:q-closed-form-asymp}.} one uses, and these also depend on the number of coefficients, so one has to explore a bit to find an almost ``optimal'' damping.

These numerical computations allow one to determine a significant amount of Borel residues for different values of $\alpha$ with a precision of $\sim$ 25-35 digits in a relatively small amount of time\footnote{As a benchmark, the data computed for guessing $N^{(1)}_0$ took around 1-1.5 weeks to produce.}. These are enough to get the Stokes vectors for the first diagonal. In addition to tables \ref{tab:NalphaOneHalf}, \ref{tab:NalphaOne}, \ref{tab:NalphaThreeHalves}, \ref{tab:NalphaTwo}, \ref{tab:NalphaFiveHalves}, \ref{tab:NalphaThree}, we have also computed $N^{(1)}_0$ for the corresponding negative values, as well as for the values $\alpha=\frac 72,\frac 34,\frac 14,\frac 15,\frac 18$. 

\begin{table}[hbt!]
	\begin{center}
		\scriptsize
		\begin{tabular}{||c||c|c||} 
			\hline
			$\alpha=\displaystyle \frac 12$ & Analytic Expression & Precision\\ [0.5ex] 
			\hline\hline
			& & \\
			$N^{(1)}_1$ & $-2\rmi$ & $5.46\cdot10^{-37}$\\
			& &  \\
			$N^{(1)}_0$ & $4 N^{(1)}_1 (\gamma_{\text{E}} + \log(8))$ & $8.68\cdot10^{-30}$\\
			& &  \\
			$N^{(1)}_{-1}$ & $\displaystyle{\frac 1{2!} \frac{N^{(1)}_0}{N^{(1)}_1} N^{(1)}_0 - \frac 12 4^2 N^{(1)}_1 \zeta(2) -\frac{4\pi^2}{3}\rmi} $ & $1.17\cdot10^{-26}$ \\
			& &  \\
			& &  \\
			\hline\hline
		\end{tabular}
		\caption{Analytic expressions and precision for the $N^{(\ell)}_x$ factors, for the case $\alpha=\frac 12$. The numerical checks correspond to taking 100 data points. This is, up to $g=200$.\label{tab:NalphaOneHalf}}
	\end{center}
\end{table}

\begin{table}[hbt!]
	\begin{center}
		\scriptsize
		\begin{tabular}{||c||c|c||} 
			\hline
			$\alpha=1$ & Analytic Expression & Precision  \\ [0.5ex] 
			\hline\hline
			& & \\
			$N^{(1)}_1$ & $-8\rmi\displaystyle\sqrt\frac{2}{\pi}$ & $1.76\cdot10^{-35}$  \\
			& &  \\
			$N^{(1)}_0$ & $4 N^{(1)}_1 (\gamma_{\text{E}} + \log(16) - 1)$ & $3.54\cdot10^{-28}$\\
			& & \\
			$N^{(1)}_{-1}$ & $\displaystyle{\frac 1{2!} \frac{N^{(1)}_0}{N^{(1)}_1} N^{(1)}_0 - \frac 12 4^2 N^{(1)}_1 \zeta(2) -64\rmi\sqrt{\frac 2\pi}} $ & $-1.41\cdot10^{-25}$  \\
			& & \\
			& & \\
			\hline\hline
		\end{tabular}
		\caption{Analytic expressions and precision for the $N^{(\ell)}_x$ factors, for the case $\alpha= 1$. The numerical checks correspond to taking 100 data points. This is, up to $g=200$.\label{tab:NalphaOne}}
	\end{center}
\end{table}

\begin{table}[hbt!]
	\begin{center}
		\scriptsize
		\begin{tabular}{||c||c|c||} 
			\hline
			$\alpha=\displaystyle \frac 32$ & Analytic Expression & Precision \\ [0.5ex] 
			\hline\hline
			& & \\
			$N^{(1)}_1$ & $-16\rmi$ & $1.36\cdot10^{-34}$ \\
			& &  \\
			$N^{(1)}_0$ & $\displaystyle 4 N^{(1)}_1 \left(\gamma_{\text{E}} + \log(8)-\frac 12\right)$ & $3.59\cdot10^{-27}$ \\
			$N^{(1)}_{-1}$ & $\displaystyle{\frac 1{2!} \frac{N^{(1)}_0}{N^{(1)}_1} N^{(1)}_0 - \frac 12 4^2 N^{(1)}_1 \zeta(2) - 32\rmi\left(1+\frac{\pi^2}{3}\right)} $ & $4.30\cdot10^{-23}$ \\
			& & \\
			& &  \\
			\hline\hline
		\end{tabular}
		\caption{Analytic expressions and precision for the $N^{(\ell)}_x$ factors, for the case $\alpha=\frac 32$. The numerical checks correspond to taking 100 data points. This is, up to $g=200$.\label{tab:NalphaThreeHalves}}
	\end{center}
\end{table}

\begin{table}[hbt!]
	\begin{center}
		\scriptsize
		\begin{tabular}{||c||c|c||} 
			\hline
			$\alpha=2$ & Analytic Expression & Precision \\ [0.5ex] 
			\hline\hline
			& & \\
			$N^{(1)}_1$ & $-\displaystyle\frac{128}{3}\rmi\displaystyle\sqrt\frac{2}{\pi}$ & $1.25\cdot10^{-30}$ \\
			& &  \\
			$N^{(1)}_0$ & $\displaystyle 4 N^{(1)}_1 \left(\gamma_{\text{E}} + \log(16) - \frac 43\right)$ & $2.73\cdot10^{-24}$\\
			& & \\
			$N^{(1)}_{-1}$ & $\displaystyle{\frac 1{2!} \frac{N^{(1)}_0}{N^{(1)}_1} N^{(1)}_0 - \frac 12 4^2 N^{(1)}_1 \zeta(2)-\frac{10240}{27}\rmi\sqrt{\frac{2}{\pi}}} $ & $-1.89\cdot10^{-24}$ \\
			& & \\
			& & \\
			\hline\hline
		\end{tabular}
		\caption{Analytic expressions and precision for the $N^{(\ell)}_x$ factors, for the case $\alpha= 2$. The numerical checks correspond to taking 100 data points. This is, up to $g=200$.\label{tab:NalphaTwo}}
	\end{center}
\end{table}

\begin{table}[hbt!]
	\begin{center}
		\scriptsize
		\begin{tabular}{||c||c|c||} 
			\hline
			$\alpha=\displaystyle \frac 52$ & Analytic Expression & Precision \\ [0.5ex] 
			\hline\hline
			& & \\
			$N^{(1)}_1$ & $-64\rmi$ & $4.66\cdot10^{-30}$ \\
			& &  \\
			$N^{(1)}_0$ & $\displaystyle 4 N^{(1)}_1 \left(\gamma_{\text{E}} + \log(8)-\frac 34\right)$ & $2.55\cdot10^{-25}$ \\
			$N^{(1)}_{-1}$ & $\displaystyle{\frac 1{2!} \frac{N^{(1)}_0}{N^{(1)}_1} N^{(1)}_0 - \frac 12 4^2 N^{(1)}_1 \zeta(2) -160 \rmi \left(1+\frac{4\pi^2}{15}\right)} $ & $2.52\cdot10^{-21}$ \\
			& & \\
			& &  \\
			\hline\hline
		\end{tabular}
		\caption{Analytic expressions and precision for the $N^{(\ell)}_x$ factors, for the case $\alpha=\frac 52$. The numerical checks correspond to taking 100 data points. This is, up to $g=200$.\label{tab:NalphaFiveHalves}}
	\end{center}
\end{table}

\begin{table}[hbt!]
	\begin{center}
		\scriptsize
		\begin{tabular}{||c||c|c||} 
			\hline
			$\alpha=3$ & Analytic Expression & Precision \\ [0.5ex] 
			\hline\hline
			& & \\
			$N^{(1)}_1$ & $-\displaystyle \frac{2048}{15}\rmi\displaystyle\sqrt\frac{2}{\pi}$ & $5.13\cdot10^{-28}$ \\
			& &  \\
			$N^{(1)}_0$ & $\displaystyle 4 N^{(1)}_1 \left(\gamma_{\text{E}} + \log(16) - \frac{23}{15}\right)$ & $2.22\cdot10^{-22}$\\
			& & \\
			$N^{(1)}_{-1}$ & $\displaystyle{\frac 1{2!} \frac{N^{(1)}_0}{N^{(1)}_1} N^{(1)}_0 - \frac 12 4^2 N^{(1)}_1 \zeta(2) + 8 N^{(1)}_1} $ & $-1.89\cdot10^{-24}$ \\
			& & \\
			& & \\
			\hline\hline
		\end{tabular}
		\caption{Analytic expressions and precision for the $N^{(\ell)}_x$ factors, for the case $\alpha=3$. The numerical checks correspond to taking 100 data points. This is, up to $g=200$.\label{tab:NalphaThree}}
	\end{center}
\end{table}

With all these results at hand and motivated by closed-form asymptotics, we were able to infer the expression
\begin{equation}
	N^{(1)}_0(\alpha)=4N^{(1)}_1(\alpha)\left(\frac 12 \psi^{(0)}\left(\alpha+\frac 12\right)+\frac 12 \psi^{(0)}(1	)-\log(8)\right),
\end{equation}
where $\psi^{(0)}(x)$ is the digamma function.\\

Given the results of this subsection, we are now left with two remaining goals. First, we need to guess a form for the general $N^{(\ell)}_1$. Second, we need to do extensive checks on the results given by the closed-form asymptotics method. \\

\noindent\textbf{Borel Plane Residues: Second Diagonal}: The second diagonal is much harder to produce due to the loss in precision and the amount of data needed to generate these results. The estimate\footnote{This rough estimate was obtained by doing a rough error propagation analysis about both the Pad\'e approximations and the numerical integrations. In addition, one should also take into account all the operations (multiplications and sums) needed to construct the Stokes coefficients from the Borel residues, but this exceeds the scope of this work---see \cite{bssv21} for further comments on the error analysis.} is that there is, approximately, a bit more than halving of precision when going to the second diagonal and then again to the third one and so on. Nevertheless, by using the data in table \ref{tab:N21prec} one can derive the following analytic form for $N^{(2)}_1$\footnote{Note that this way of writing this quantity is motivated by the generalization to any diagonal \eqref{eq:Ns1} discussed in the next paragraph.}
\begin{equation}
	N^{(2)}_1=\frac{(4 \pi)^{2-1}}{2^2} \,\rmi \left(\frac{-2^{3 \alpha - \frac 12}}{\Gamma\left(\alpha +\frac 12\right)}\right)^2.
\end{equation}

\begin{table}[hbt!]
	\begin{center}
		\scriptsize
		\begin{tabular}{||l|c|c||}
			\hline
			$\alpha$ & $\displaystyle N^{(2)}_1(\alpha)$ & Precision \\ [0.5ex] 
			\hline\hline
			& & \\
			0 \cite{bssv21}& $\displaystyle\frac{\rmi}{2}$ & $\sim 10^{-16}$  \\
			& & \\
			$\displaystyle \frac{1}{2}$ & $4\pi\rmi$ & $1.14\cdot10^{-12}$ \\
			& & \\
			1 & $128\rmi$ & $1.35\cdot10^{-7}$ \\
			& &  \\
			$\displaystyle \frac 32$ & $256\pi\rmi$ & $3.60\cdot10^{-10}$ \\
			& &  \\
			2 & $\displaystyle\frac{32768}9\rmi$ & $4.76\cdot10^{-4}$ \\
			& &  \\
			\hline\hline
		\end{tabular}
		\caption{Precision obtained for $N^{(2)}_1(\alpha)$  from the Borel residue method, for different values of $\alpha$. The checks have been made with 200 data points, this is, up to $g=400$. Notice that the decrease of precision is not that relevant since the quantities approximated are increasing, so in terms of digits of precision with respect to the original quantity the loss in precision is not that dramatic. Furthermore, as stated in subsection \ref{sec:origPII}, we expect the resurgent properties to be dominating later in $g$, the higher the value of $\alpha$ is---see figure \ref{fig:checksaction}.
			\label{tab:N21prec}}
	\end{center}
\end{table}
In the next paragraph, we will try to generalise the results in this and previous paragraphs by adding the conjecture in \cite{bssv21} as well. The aim will be to have a working conjecture on the structure of the $N^{(\ell)}_1$ coefficients. Upon having a proposal for this pattern, we will have a conjecture for {\it all} Stokes data associated with the double-cover solution of the inhomogeneous Painlev\'e II equation.\\

\textbf{Borel Plane Residues: Any Diagonal}: Due to the heavy computations that extra diagonal require\footnote{In particular, generating large-$g$ coefficients, which recall that for the perturbative sector alone are polynomials of order $g$ in $\alpha$.}---and given that we expect a bit less of a halving of the precision, so around $\sim$4-8 digits of precision---at this stage our conjecture is not without its caveats. It consists of a mere combination of previous guesses \cite{asv11,bssv21,sv13} alongside the structure found so far. From the results in the present work and trying to reproduce the structure in \cite{bssv21}, we conjecture a pattern on the first column given by 
\begin{equation}
	\label{eq:Ns1}
	N^{(\ell)}_1=\frac{(4 \pi)^{\ell - 1}}{\ell^2}\, \rmi \left(-\frac{2^{3 \alpha - \frac 12}}{\Gamma\left(\alpha +\frac 12\right)}\right)^\ell.
\end{equation}
Assume for now that these results are true. This means that we can produce {\it all Stokes coefficients} for the double-cover ansatz of the inhomogeneous Painlev\'e II equation. However, this is still not enough. The really interesting calculation would be to understand how these Stokes coefficients encode {\it transition functions} upon Stokes phenomena, and how to use them to construct global solutions. In particular, if we could find some type of monodromy in the transitions, this would be a much stronger check on our data. We therefore dedicate the next subsection to this exploration.

%%%%%%%%%%%%%%%%%%%%%%%%%%%%%%%%%%%%%%%%%%%%%%%%%%%%%%%%%%%%%%%%%
%%%%%%%%%%%%%%%%%%%%%%%%%%%%%%%%%%%%%%%%%%%%%%%%%%%%%%%%%%%%%%%%%
\subsubsection{Transition Functions and Stokes Phenomena}\label{subsec:transfunct}
%%%%%%%%%%%%%%%%%%%%%%%%%%%%%%%%%%%%%%%%%%%%%%%%%%%%%%%%%%%%%%%%%
%%%%%%%%%%%%%%%%%%%%%%%%%%%%%%%%%%%%%%%%%%%%%%%%%%%%%%%%%%%%%%%%%
The key to finding transition functions is obtaining the Stokes vector fields \cite{bssv21}
\begin{align}
	\underline{\boldsymbol{S}}_{0}&=\sum_{\ell=1}^{+\infty}\sum_{p=0}^{+\infty} \sigma_1^{p-1}\sigma_2^{p-1+\ell}N^{(\ell)}_{1-p}\left[\begin{array}{c}
		\sigma_1(p+\ell)\\
		\sigma_2(-p)
	\end{array}\right],\\
	\underline{\boldsymbol{S}}_{\pi}&=\sum_{\ell=1}^{+\infty}\sum_{p=0}^{+\infty} \sigma_1^{p-1+\ell}\sigma_2^{p-1}N^{(-\ell)}_{1-p}\left[\begin{array}{c}
		\sigma_1(-p)\\
		\sigma_2(p+\ell)
	\end{array}\right];
\end{align} 
and integrate them \cite{abs18,bssv21}. Therefore, we would first like to obtain a generating function for the $N^{(\ell)}_{1-p}$\footnote{Notice that the $N^{(-\ell)}_{1-p}$ can be obtained in terms of the $N^{(\ell)}_{1-p}$ via our backward-forward relation.}. For that, some massage of the expression \eqref{eq:closed-form-asymp} is necessary---see \cite{bssv21} for full details. By defining the ratio 
\begin{equation}
	R_p^{(\ell)}:=\frac{N^{(\ell)}_p}{N^{(\ell)}_1},
\end{equation}
one has 
\begin{align}
	\label{eq:closed-form-asymp-for-ratios}
	\delta_{p,0}= \sum_{q=0}^{p}R^{(\ell)}_{1-p+q}\frac{(4\ell)^q}{q!}B_q&\left\{\frac 1{\ell^0}\frac 1{2^1}\left(\psi^{(0)}\left(\alpha+\frac 12\right)+\psi^{(0)}(1)\right)-\log(8),\right.\nonumber\\
	&\hspace{3mm}\frac 1{\ell^1}\frac 1{2^2}\left(\psi^{(1)}\left(\alpha+\frac 12\right)+\psi^{(1)}(1)\right),\dots,\nonumber\\
	&\hspace{3mm}\left.\frac 1{\ell^{p-1}}\frac 1{2^{p}}\left(\psi^{(p-1)}\left(\alpha+\frac 12\right)+\psi^{(p-1)}(1)\right)\right\}.
\end{align}
This can be seen as a coefficient of a Cauchy product of two series
\begin{equation}
	R^{(\ell)}(\lambda)=\sum_{p=0}^{+\infty}R^{(\ell)}_{1-p}\lambda^p,\qquad\qquad B^{(\ell)}(\lambda)=\sum_{p=0}^{+\infty}\frac{(4\ell)^p}{p!}B_p\left\{\dots\right\} \lambda^p.
\end{equation}
Under this identification and upon taking the left-hand side to be 0\footnote{Note that the $p=0$ case has always been something that we needed to determine via other methods. More about this in the following.}, one has that the generating function of Stokes ratios $R^{(\ell)}(\lambda)$ can be determine by
\begin{equation}
	R^{(\ell)}(\lambda)=\frac1{B^{(\ell)}(\lambda)}.
\end{equation}
Henceforth, we need to understand the structure of the Bell polynomials with our particular entries. Using the properties of these polynomials one can write $B^{(\ell)}(\lambda)$ as the following exponential
\begin{equation}
	B^{(\ell)}(\lambda)=\exp\left\{-4\ell\lambda\log(8)+\ell\sum_{j=1}^{+\infty}\frac{(2\lambda)^j}{j!}\psi^{(j-1)}\left(\alpha+\frac12\right)+\ell\sum_{j=1}^{+\infty}\frac{(2\lambda)^j}{j!}\psi^{(j-1)}(1)\right\}.
\end{equation}
Since one can write the polygamma functions as derivatives of the logarithm of the gamma function $\psi^{(j-1)}(z)=\frac{\text{d}^j}{\text{d}z^j}\log \Gamma(z)$, we can rewrite the above formula as
\begin{equation}
	B^{(\ell)}(\lambda)=8^{-4\ell\lambda}\,\Gamma\left(1+2\lambda\right)^\ell \,\Gamma\left(\alpha+\frac12+2\lambda\right)^\ell,
\end{equation}
which yields the following expression for the generating function of Stokes ratios
\begin{equation}
	R^{(\ell)}(\lambda)=\frac{8^{4\ell\lambda}}{\Gamma\left(1+2\lambda\right)^\ell\, \Gamma\left(\alpha+\frac12+2\lambda\right)^\ell}.
\end{equation}
By using the conjecture \eqref{eq:Ns1}, one can now obtain the generating function of Stokes data\footnote{By generating function we mean 
	\begin{equation}
		N^{(\ell)}_{1-p}=\left.\frac1{p!}\,\frac{\partial^p}{\partial \lambda^p}N^{(\ell)}(\lambda)\right|_{\lambda=0}.
\end{equation}}
\begin{equation}
	\label{eq:gen-funct-first-form}
	N^{(\ell)}(\lambda)=\frac{(4\pi)^{\ell-1}}{\ell^2}\rmi\left(-2^{3\alpha-\frac12}\right)^\ell\left(\frac{8^{4\lambda}}{\Gamma\left(1+2\lambda\right) \Gamma\left(\alpha+\frac12+2\lambda\right)}\right)^\ell\equiv\sum_{p=0}^{+\infty}\lambda^pN^{(\ell)}_{1-p}.
\end{equation}
We can now apply the above result to the computation of the Stokes vector fields. Following \cite{bssv21}, we introduce $\upmu\equiv\sigma_1\sigma_2$ and write these as
\begin{align}
	\underline{\boldsymbol{S}}_{0}&=\sum_{\ell=1}^{+\infty} \left[\begin{array}{c}
		\sigma_2^{\ell-1}(\upmu\frac{\text{d}}{\text{d}\upmu}+\ell)N^{(\ell)}(\upmu)\\
		-\sigma_2^{\ell+1}\frac{\text{d}}{\text{d}\upmu}N^{(\ell)}(\upmu)
	\end{array}\right],\\
	\underline{\boldsymbol{S}}_{\pi}&=\sum_{\ell=1}^{+\infty} \left[\begin{array}{c}
		-\sigma_1^{\ell+1}\frac{\text{d}}{\text{d}\upmu}N^{(-\ell)}(\upmu)\\
		\sigma_1^{\ell-1}(\upmu\frac{\text{d}}{\text{d}\upmu}+\ell)N^{(-\ell)}(\upmu)
	\end{array}\right].
\end{align} 
Therefore, the quantity that we need to address is $\frac{\text{d}}{\text{d}\upmu}N^{(\ell)}(\upmu)$. In order to do that, we first write the generating function as\footnote{Notice that it is trivial to check that this expression agrees with \eqref{eq:gen-funct-first-form} upon substituting $N^{(1)}(\upmu)$.}
\begin{equation}
	N^{(\ell)}(\upmu)=\frac{(-4\pi\rmi)^{\ell-1}}{\ell^2}\left(N^{(1)}(\upmu)\right)^\ell.
\end{equation}
Hence, taking the derivative is trivial and one obtains  
\begin{equation}
	\frac{\text{d}}{\text{d}\upmu}N^{(\ell)}(\upmu)=\frac{(-4\pi\rmi)^{\ell-1}}\ell\left(N^{(1)}(\upmu)\right)^\ell\frac{\text{d}}{\text{d}\upmu}\log N^{(1)}(\upmu).
\end{equation}
Upon using the above results, one can perform the sums in $\ell$ to obtain the final form of our Stokes vector fields, as:
\begin{align}
	\underline{\boldsymbol{S}}_{0}&=\frac{\rmi}{4\pi}\log\left(1+4\pi\rmi\sigma_2N^{(1)}(\upmu)\right)\left[\begin{array}{c}
		\displaystyle{-\frac{\upmu\frac{\text{d}}{\text{d}\upmu}\log N^{(1)}(\upmu)+1}{\sigma_2}} \\
		\displaystyle{\sigma_2\frac{\text{d}}{\text{d}\upmu}\log N^{(1)}(\upmu)}
	\end{array}\right];\\
	\nonumber
	\\
	\underline{\boldsymbol{S}}_{\pi}&=\frac{\rmi}{4\pi}\log\left(1+4\pi\rmi\sigma_1N^{(-1)}(\upmu)\right)\left[\begin{array}{c}
		\displaystyle{\sigma_1\frac{\text{d}}{\text{d}\upmu}\log N^{(-1)}(\upmu)}\\
		\displaystyle{-\frac{\upmu\frac{\text{d}}{\text{d}\upmu}\log N^{(-1)}(\upmu)+1}{\sigma_1}}
	\end{array}\right].
\end{align} 

Let us now discuss the relation between $N^{(-1)}(\upmu)$ and $N^{(1)}(\upmu)$. This has to be given from the backward-forward relation \eqref{eq:back-forw}, which at the level of Stokes coefficients, for $\ell=1$, reads\footnote{We are now making the $\alpha$-dependence explicit, rather than $\upmu$.}
\begin{equation}
	N^{(-1)}_{1-p}(\alpha)=-\rmi\, (-1)^{p-\alpha}\sum_{q=0}^{p}\frac{(-4\pi\rmi)^q}{q!}N^{(1)}_{1-(p-q)}(-\alpha).
\end{equation}
Thus, by assuming that the ratio property $R^{(\ell)}(\lambda)=R^{(1)}(\lambda)^\ell$ is also fulfilled for the $-\pi$-Stokes-direction, one can obtain the other generating function\footnote{Recall that this generating function corresponds to the Stokes data associated with the $-\pi$ direction of the Borel plane.}
\begin{equation}
	N^{(-\ell)}(\upmu,\alpha)=-\rmi^{\ell}\,(-1)^{-\alpha\ell}\,\rme^{4\pi\rmi\upmu\ell}N^{(\ell)}(-\upmu,-\alpha),
\end{equation} 
where we have now slightly changed the notation to include the dependence on $\alpha$ to make the $\alpha\to-\alpha$ exchange explicitly. Having understood the generating function for Stokes data, we now want to obtain transition functions for our transseries parameters. In order to do this, and again following \cite{bssv21}, it is convenient to define the quantities\footnote{Note that for the $\pi$ transition, the $\upmu$ parameter is the same, whereas $\uprho_\pi=\sigma_1N^{(-1)}(\upmu)$ is a more suitable variable. One further realises that this map is invertible by means of \cite{bssv21}
	\begin{equation}
		\begin{array}{lll}
			\sigma_1=\displaystyle{\frac{\upmu \,N^{(1)}(\upmu)}{\uprho_0}}; &\qquad\qquad& \sigma_1=\displaystyle{\frac{\uprho_\pi}{N^{(-1)}(\upmu)}};\\
			\sigma_2=\displaystyle{\frac{\uprho_0}{N^{(1)}(\upmu)}};& & \sigma_2=\displaystyle{\frac{\upmu \,N^{(-1)}(\upmu)}{\uprho_\pi}}.
		\end{array}
\end{equation}}
\begin{align}
	\upmu,&\\
	\uprho_0&=\sigma_2\,N^{(1)}(\upmu).
\end{align}
From here, one we can compute the transition functions of transseries parameters from the Stokes vector fields. This is done via the system of equations \cite{bssv21}
\begin{align}
	\underline{\boldsymbol{S}}_{0}^{(1)}\frac{\partial}{\partial \sigma_1}\upmu+\underline{\boldsymbol{S}}_{0}^{(2)}\frac{\partial}{\partial \sigma_2}\upmu&=\upmu'(t),\\
	\underline{\boldsymbol{S}}_{0}^{(1)}\frac{\partial}{\partial \sigma_1}\uprho_0+\underline{\boldsymbol{S}}_{0}^{(2)}\frac{\partial}{\partial \sigma_2}\uprho_0&=\uprho_0'(t).
\end{align}
We can make the above expressions clear by computing the left-hand side of these equations explicitly. The result reads 
\begin{align}
	\frac1{4\pi\rmi}\log(1+4\pi\rmi\uprho_0)&=\upmu'(t),\\
	0&=\uprho_0'(t).
\end{align}
This implies that $\uprho_0$ is a constant of the transition, while $\upmu$ gets an additive term as:
\begin{align}
	\underline{\pmb{\mathbb{D}}}_0(\upmu,\uprho_0)&=\left(\upmu+\frac1{4\pi\rmi}\log(1+4\pi\rmi\uprho_0)\,\,,\,\,\uprho_0\right),\\
	\underline{\pmb{\mathbb{D}}}_\pi(\upmu,\uprho_\pi)&=\left(\upmu-\frac1{4\pi\rmi}\log(1-4\pi\rmi\uprho_\pi)\,\,,\,\,\uprho_\pi\right),
\end{align}
where the functions $\underline{\pmb{\mathbb{D}}}_{\theta}$ are defined via the Stokes automorphisms as
\begin{equation}
	\label{eq:stokes-transitions}
	\underline{\mathfrak{S}}_\theta \Phi(x;\,\upmu,\rho_\theta)= \Phi(x;\,\underline{\pmb{\mathbb{D}}}_{\theta}(\upmu,\rho_\theta)).
\end{equation}
As stated before, these variables uniquely determine the $\sigma_1$, $\sigma_2$, so we can easily undo the maps to obtain the transition functions in terms of our transseries parameters. The expressions are, however, very cumbersome, so we will not write them explicitly. 

This concludes the computation of the transition functions. A couple of comments have to be made. First, note that these transitions have been computed with a specific choice of logarithmic sheet---we will discuss this in the following. Also, notice that one has computed these transitions for very specific forms of the transseries sectors. In general, when discussing how to go around the complex plane, one still has to take into account the changes in the transseries sectors and transmonomials. In particular, we would like to understand if one can absorb specific changes into the transseries parameters.
Indeed, under the rotation $x\mapsto x\,\rme^{2\pi\rmi r}$, $r\in\mathbb{Z}$, we can reabsorb the phase of the resurgent $x$ variable in the $\sigma_1$, $\sigma_2$ parameters via \cite{bssv21}
\begin{equation}
	R_{r}(\sigma_1,\sigma_2)=\Big(\sigma_1 \rme^{\pi\rmi r-8\pi\rmi r\sigma_1\sigma_2-2\pi \rmi r\alpha}\,\,,\,\,\sigma_2\rme^{\pi\rmi r+8\pi\rmi r\sigma_1\sigma_2+2\pi\rmi r\alpha}\Big),
\end{equation} 	
while keeping the transseries evaluated at $x$---recall \eqref{eq:xtowrelation}. Thus, one can define the transition at any line as 
\begin{align}
	\underline{\pmb{\mathbb{D}}}_{2\pi r}(\sigma_1,\sigma_2)&=R_{r}(\underline{\pmb{\mathbb{D}}}_{0}(R_{- r}(\sigma_1, \sigma_2))),\\
	\underline{\pmb{\mathbb{D}}}_{(2r+1)\pi}(\sigma_1,\sigma_2)&=R_{r}(\underline{\pmb{\mathbb{D}}}_{\pi}(R_{-r}(\sigma_1, \sigma_2)));
\end{align} 	
With these formulae in hand, we can try to check if our transseries parameters $\sigma_1,\,\sigma_2$ fulfil some type of monodromy condition---see \cite{bssv21}. They do seem to do so, and we present an example of this computation in table \ref{tab:monodromy}. One sees that $\sigma_1,\,\sigma_2$ go back to themselves after 3 rotations in the $x=z^{-\frac32}$, plane, which corresponds to 2 rotations in the original variable. These monodromy properties of the transseries are further discussed in appendix \ref{app:iso}.
\begin{table}[hbt!]
	\begin{center}
			\tiny
			\begin{tabular}{||r||c|c||}
				\hline
				Quantity & $\sigma_1$ & $\sigma_2$  \\ [0.5ex] 
				\hline\hline
				$(\sigma_1,\sigma_2)$ & $-6.71606\dots$ & $-2.01245$ \\
				$R_{-1}\circ\underline{\pmb{\mathbb{D}}}_{-\pi}\circ\underline{\pmb{\mathbb{D}}}_0(\sigma_1,\sigma_2)$& $1.46\dots\, \times 10^{-16}-4.73\dots\, \times 10^{-17} \rmi$ & $7.99\dots\, \times 10^{16}+3.63\dots\, \times 10^{16} \rmi$ \\
				$R_{-2}\circ\underline{\pmb{\mathbb{D}}}_{-3\pi}\circ\underline{\pmb{\mathbb{D}}}_{-2\pi}\circ\underline{\pmb{\mathbb{D}}}_{-\pi}\circ\underline{\pmb{\mathbb{D}}}_0(\sigma_1,\sigma_2)$ & $4.98\dots\, \times10^{-17}-1.85\dots\, \times 10^{-16} \rmi$ & $1.79\dots\, \times10^{16}+6.68\dots\, \times 10^{16} \rmi$ \\
				$R_{-3}\circ\underline{\pmb{\mathbb{D}}}_{-5\pi}\circ\underline{\pmb{\mathbb{D}}}_{-4\pi}\circ\underline{\pmb{\mathbb{D}}}_{-3\pi}\circ\underline{\pmb{\mathbb{D}}}_{-2\pi}\circ\underline{\pmb{\mathbb{D}}}_{-\pi}\circ\underline{\pmb{\mathbb{D}}}_0(\sigma_1,\sigma_2)$ &$-6.71606\dots$ & $-2.01245\dots$ \\
				\hline\hline
			\end{tabular}
		\caption{Example on the monodromy properties of $\sigma_1$ and $\sigma_2$ after $2\pi\rmi$ rotations for a randomly generated value $\alpha=5.85356\dots$. Closure of direct monodromy is evident.
		\label{tab:monodromy}}
	\end{center}
\end{table}

%%%%%%%%%%%%%%%%%%%%%%%%%%%%%%%%%%%%%%%%%%%%%%%%%%%%%%%%%%%%%%%%%
%%%%%%%%%%%%%%%%%%%%%%%%%%%%%%%%%%%%%%%%%%%%%%%%%%%%%%%%%%%%%%%%%
\section{The Miura Transformation and Special Solutions}
\label{chap:Miura&SpecSol}
%%%%%%%%%%%%%%%%%%%%%%%%%%%%%%%%%%%%%%%%%%%%%%%%%%%%%%%%%%%%%%%%%
%%%%%%%%%%%%%%%%%%%%%%%%%%%%%%%%%%%%%%%%%%%%%%%%%%%%%%%%%%%%%%%%%

The $q$-deformed Painlev\'e II equation can be related to the inhomogeneous Painlev\'e II equation with parameter $\alpha\propto \frac12\pm q$\footnote{The fact that we write proportional is because this connection depends, of course, on the normalisation of both equations.} by means of the Miura transformation \cite{kms03}. In particular, we now expect their resurgent structures will be somehow related. Let us explore this a bit in this section.

It is well-known that the Miura map relates solutions of the Korteweg–de Vries (KdV) hierarchy to those of the modified KdV (mKdV) one \cite{djmw92}. Let us explain this in a bit of detail. The solutions $u(z,t_k)$ of the KdV hierarchy $\partial_{t_k}u(z,t_k)=\partial_z\mathcal R_{k+1}[u]$ must satisfy some physical constraints in order to represent a string theory. These translate into\footnote{For details, we highly recommend the reader to consult \cite{djmw92}.}
\begin{equation}
	\label{eq:phys_constr}
	u\mathcal{R}^2-\frac12\mathcal{R}\mathcal{R}''+\frac14(\mathcal{R}')^2=\Gamma^2,
\end{equation}
where $'\equiv \partial_z$ and $\mathcal R$ is a quantity that is defined from the Gel'fand--Dikii differential polynomials \cite{gd75} $\mathcal R_k[u]$ as
\begin{equation}
	\label{eq:Rexpansion}
	\mathcal R=\sum_{k=1}^{+\infty}\left(k+\frac12\right)t_k\mathcal R_k[u]-z,
\end{equation}
and $t_k$ are the KdV times. The case of 2d supergravity is obtained by taking $t_k=0$ for $k\geq2$. For simplicity, one normalises $\mathcal R\sim u$. This means, $t_1=-\frac83$ and $\mathcal R=u-z$. With this form, equation \eqref{eq:phys_constr} reads
\begin{equation}
	u(z)\Big(u(z)-z\Big)^2-\frac12\Big(u(z)-z\Big)\partial_z^2\Big(u(z)-z\Big)+\frac14\Big(\partial_z\big(u(z)-z\big)\Big)^2=\Gamma^2.
\end{equation}
By performing the change of variables $u(z)=f(z)^2+z$ \cite{kms03}, one obtains the $q$-deformed Painlev\'e II equation---notice that, in this case, one would say the ``$\Gamma$-deformed'' Painlev\'e II equation, as
\begin{equation}
	\label{eq:GammaPII}
	-\Gamma^2+zf(z)^4+f(z)^6-f(z)^3f''(z)=0.
\end{equation}

On the other hand, solutions $v(z,t_k)$ to the modified KdV hierarchy satisfy $\partial_{t_k}v(z,t_k)=\frac12\partial_z\mathcal S_{k}[v]$, where $\mathcal S_k[v]\equiv \mathcal R_k'[v^2+v']-v\, \mathcal R_k[v^2+v']$. The string equation in this case is given by
\begin{equation}
	\label{eq:string_eq_mKdV}
	\sum_{k=1}^{\infty}\left(k+\frac12\right)t_k\mathcal S_k[v]+z\,v(z)=\Lambda.
\end{equation}
Again, we are interested the case $t_k=-\frac83\delta_{k,1}$. Under these conditions, equation \eqref{eq:string_eq_mKdV} gives
\begin{equation}
	\label{eq:CPII}
	\frac{v''(z)}2-v(z)^3+zv(z)-\Lambda=0,
\end{equation}
which is the inhomogeneous Painlev\'e II equation with parameter $\Lambda$. The Miura map states that there is a {\it one-to-one} map between the solutions of \eqref{eq:phys_constr} and \eqref{eq:string_eq_mKdV} provided that $\Lambda=\frac12\pm\Gamma$. Naturally, it is given by $u(z)=v(z)^2+v'(z)$. Let us now show the explicit procedure to determine this correspondence, and then translate it into our case. Let $u(z)$ be a solution of \eqref{eq:phys_constr} and define
\begin{equation}
	X_\pm[u,v]=\frac12\mathcal R'[u]\mp\Gamma-v(z)\,\mathcal R[u].
\end{equation}
Setting $X_\pm[u,v]=0$, one can solve for $v(z)$. Furthermore, the quantity 
\begin{equation}
	X_\pm[u,v]^2\pm 2\Gamma X_\pm[u,v]-\mathcal R[u]X_\pm'[u,v]=0
\end{equation}
gets translated into
\begin{equation}
	-\Gamma^2+(v(z)^2+v'(z))\mathcal R[u]^2-\frac12\mathcal R[u]\mathcal R''[u]+\frac14(\mathcal R'[u])^2=0.
\end{equation}
Now, by using the identification $u(z)=v(z)^2+v'(z)$, the above equation recovers \eqref{eq:phys_constr}. On the other hand, if one includes the Miura map into the expression for $v(z)$ obtained from setting $X_\pm[u,v]=0$ one gets
\begin{equation}
	v(z)\mathcal R[v(z)^2+v'(z)]=\frac12\mathcal R'[v(z)^2+v(z)]\mp \Gamma.
\end{equation}
Including expression \eqref{eq:Rexpansion}, one can transform the above equation into
\begin{align}
	0&=\sum_{k=1}^{+\infty}\left(k+\frac12\right)t_k\left(\frac12\mathcal R'_k[v^2+v']-v\mathcal R_k[v^2+v']\right)-\frac12+z\,v(z)\mp\Gamma=\\
	&=\sum_{k=1}^{+\infty}\left(k+\frac12\right)t_k\mathcal S_k[v]+z\,v(z)-\left(\frac12\pm\Gamma\right),
\end{align}
which is \eqref{eq:string_eq_mKdV} provided that $\Lambda=\frac12\pm \Gamma$, as expected. Now that this is settled, we want to make the connection between the $\frac12\pm\Gamma$-Painlev\'e II \eqref{eq:CPII} equation and the $\Gamma$-deformed Painlev\'e II \eqref{eq:GammaPII} one. Note that we have two ways of writing the Miura map
\begin{align}
	\label{eq:miuramaprelation}
	f(z)+z&=v(z)^2+v'(z);\\
	v(z)&=\frac{f'(z)}{f(z)}\mp\frac{\Gamma}{f(z)^2},
\end{align}
where the second one is obtained when setting $X_\pm[f,v]=0$, and recall the relation between $u$ and $f$, $u(z)=f(z)^2+z$. By using these two equations alone, one can obtain both \eqref{eq:CPII} and \eqref{eq:GammaPII}, so one would expect to be able to relate their solutions. In particular, one even expects that their resurgent structures can be related. In the present paper, we have worked with the $q$-deformed Painlev\'e II solution that has a double-cover, this is, the one that behaves as $u(z)\sim\sqrt z$. Note that there is also another set of solutions that behave as $u(z)\sim z^{-1/4}$. These two determinations relate to each other in the following way: the four-cover (respectively, the double-cover) of the $q$-deformed Painlev\'e II equation is related to the double-cover (respectively, the single	-cover) of the inhomogeneous one. Here, we will show how this computation works and we will see how one can obtain the canonical Stokes coefficient $N^{(1)}_1$. Let us start by comparing the perturbative expansion on both sides of equation \eqref{eq:miuramaprelation}. We will set $\Gamma=q$ and $\Lambda=\frac12\pm\Gamma$, so that
\begin{align}
	f(z)+z&=\left(q^2-\frac14\right)z^{-2} +\left(\frac98- 5 q^2+2 q^4\right)z^{-5}+\cdots,\\
	v(z)^2+v'(z)&=\left(q^2-\frac14\right)z^{-2} +\left(\frac98- 5 q^2+2 q^4\right)z^{-5}+\cdots.
\end{align}
Taking the same quantities but now adding the transmonomial factors $\rme^{\frac{2\sqrt2}3\rmi z^{\frac32}}$, one has
\begin{align}
	f(z)+z&=\left(q^2-\frac14\right)z^{-2} +\dots+\sigma_1\,\rme^{\frac{2\sqrt2}3\rmi z^{\frac32}}\,2 (-2)^{\frac14}\,z^{\frac14}+\cdots,\\
	v(z)^2+v'(z)&=\left(q^2-\frac14\right)z^{-2}+\dots+\sigma_2\, \rme^{\frac{2\sqrt2}3\rmi z^{\frac32}}\,\rmi\sqrt2 \, z^{\frac14}+\cdots.
\end{align}
Comparing both sides of the equations one finds that 
\begin{equation}
	\sigma_1=2^{-\frac14}\left(\frac12+\frac\rmi2\right)\sigma_2.
\end{equation}
Thus, using that in the latter normalisation for $v(z)$, the Stokes coefficient reads $N^{(-1)}_{1}(\alpha)=-\frac{(-2)^{\frac14} \sin(\pi \Lambda)}{\sqrt\pi}$ and one has that
\begin{equation}
	N^{(1)}_1=\left(\frac12 + \frac\rmi2\right) 2^{-\frac14} \left(-\frac{(-2)^{\frac14} \sin(\pi \Lambda)}{\sqrt\pi}\right)=-\frac{\rmi}{\sqrt{2\pi}}\cos(\pi q),
\end{equation}
which agrees with our computation for the $q$-deformed Stokes coefficient. These results open a new door for future exploration, namely how one jumps from one transseries determination with its corresponding Stokes data to another determination, alongside its corresponding resurgent structure. Some preliminary works in this direction can be found in \cite{a99,t02} and references therein. Not only do we have the Miura map and all Stokes data for both expansions, but one could also explore known cases to see how this gets translated. By this we have in mind the Hastings--McLeod solution \cite{hm80}, the known special-function solutions or the rational solutions of the inhomogeneous Painlev\'e II equation, which we describe next.

%%%%%%%%%%%%%%%%%%%%%%%%%%%%%%%%%%%%%%%%%%%%%%%%%%%%%%%%%%%%%%%%%
%%%%%%%%%%%%%%%%%%%%%%%%%%%%%%%%%%%%%%%%%%%%%%%%%%%%%%%%%%%%%%%%%
\subsection{On the Analysis of Special Solutions}\label{sec:compspeccases}
%%%%%%%%%%%%%%%%%%%%%%%%%%%%%%%%%%%%%%%%%%%%%%%%%%%%%%%%%%%%%%%%%
%%%%%%%%%%%%%%%%%%%%%%%%%%%%%%%%%%%%%%%%%%%%%%%%%%%%%%%%%%%%%%%%%
This subsection is dedicated to the study of special solutions of the (in)homogeneous Painlev\'e II equation. In particular, we are interested in how our transseries encodes the information that these solutions are ``special''. More generally, what do these particular solutions correspond to in our transseries language? We will analyse three cases: the famous Hastings--McLeod solution \cite{hm80}, as well as special-function and rational solutions \cite{c06}.

\begin{figure}[hbt!]
	\centering
	\begin{subfigure}[t]{0.5\textwidth}
		\centering
		\includegraphics[height=2.5in]{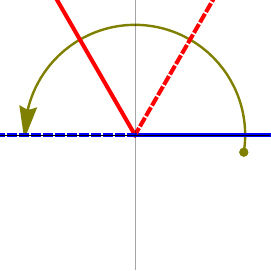}
		\caption{$z$-plane}
	\end{subfigure}%
	~ 
	\begin{subfigure}[t]{0.5\textwidth}
		\centering
		\includegraphics[height=2.5in]{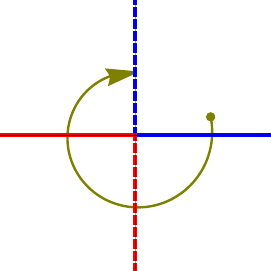}
		\caption{$x$-plane}
	\end{subfigure}
	\caption{Translation of the path we are following on the $z$-plane when constructing the Hastings--McLeod solution and the corresponding image in the $x$-plane. Recall that $z$ is the variable on the classical normalization of the inhomogeneous Painlev\'e II equation while $x$ is the resurgent variable---recall their relation $x=z^{-\frac32}$. We have chosen this path since the Stokes data we have computed are associated with walking on the $x$-plane in the anti-clockwise direction. Here, the solid (respectively, dashed) lines correspond to Stokes (respectively, anti-Stokes) lines. The blue colours represent the positive real or imaginary directions, whereas the red ones correspond to the negative ones. }
	\label{fig:HMpath}
\end{figure}

The Hastings--McLeod solution is a special solution of the homogeneous Painlev\'e II equation\footnote{Here, we will be working in the following normalisation
	\begin{equation}
		u''(z)-2u^3(z) + 2z\,u(z) = 0.
\end{equation}}
whose asymptotic expansions on the real line read \cite{hxz15,k04b,m08}
\begin{equation}
	\label{eq:HMexpansion}
	u(z)\sim\left\{\begin{array}{lll}
		\sqrt z+\dots+\displaystyle{\frac{\rmi}{2 \sqrt{2 \pi}} \rme^{-\frac43 z^{\frac32}}z^{-\frac14}}+\cdots& \qquad&z\to +\infty\\
		2^{\frac13}\text{Ai}\left(-2^{\frac13}z\right)+\dots\sim\displaystyle{\frac{\rme^{-\frac{2\sqrt2}3(-z)^{\frac32}}}{2^{\frac34}\sqrt\pi(-z)^{\frac14}}}+\cdots& &z\to -\infty
	\end{array}\right.,
\end{equation}
where $\text{Ai}(z)$ stands for the Airy function of the first kind. 
This solution has already been studied in the context of one-parameter transseries \cite{m08}. However, the analysis was done by expanding the equation twice, one time over the positive real line and one then over the negative real direction, and using a matrix model to compute these expansions. In our case, we would like to understand this solution by means of Stokes transitions, {\it i.e.}, rotating around the complex plane while applying the corresponding Stokes automorphisms. In figure \ref{fig:HMpath} we show the path both in the original $z$-plane as well as in the corresponding resurgent $x$-plane---see \cite{bssv21} for extra details. 

\begin{figure}[hbt!]
	\centering
	\begin{subfigure}[t]{\textwidth}
		\centering
		\includegraphics[height=1.3in]{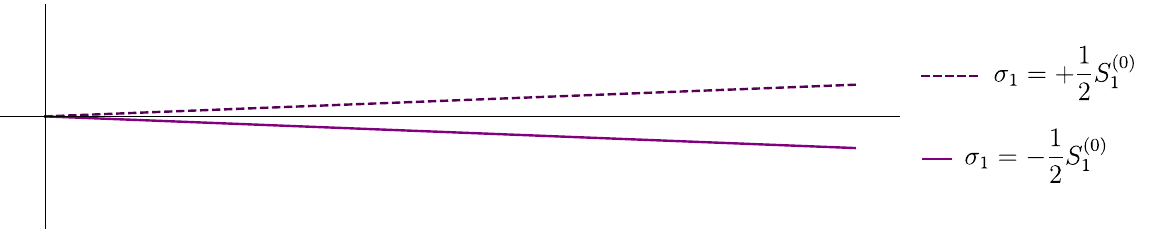}
		\caption{$z$-plane}
	\end{subfigure}%
	\\\vspace{0.5cm}
	\begin{subfigure}[t]{\textwidth}
		\centering
		\includegraphics[height=1.3in]{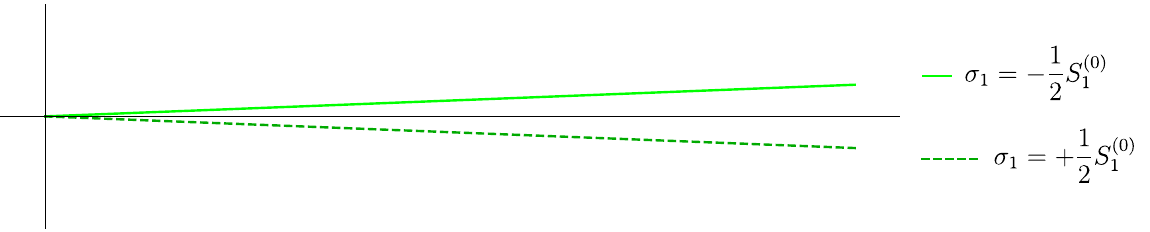}
		\caption{$x$-plane}
	\end{subfigure}
	\caption{Illustration of the value of $\sigma_1$ on both sides of the $0$-direction when doing median resummation \cite{as13} on the positive real line. Due to the change of variable \eqref{eq:change_of_var}, \eqref{eq:xtowrelation} the values for the original $z$-plane and the $x$-plane get inverted.}
	\label{fig:medresum}
\end{figure}

In the context of the two-parameter transseries, we have always worked on the $z\to +\infty$ limit. Thus, just as in \cite{m08}, the upper expansion in \eqref{eq:HMexpansion} is very easy to obtain. The result is, indeed, the one that comes out of applying median resummation \cite{as13} to construct a real solution on the real line. Even in the two-parameter case, the median resummation is obtained when applying the Stokes automorphism $\underline{\mathfrak{S}}_0^{-\frac 12}$ to the perturbative series\footnote{By perturbative series we mean the $(0,0)$ sector only.} and then resumming a bit above the real line on the $x$-plane\footnote{Alternatively, one can apply the Stokes automorphism $\underline{\mathfrak{S}}_0^{\frac 12}$ to the perturbative series and then resum a bit below the real line.}. This means that the contributions which enter this construction are only corrections associated with $\mathsf{S}_{(0,0)\to(n,0)}$ and $\Phi_{(n,0)}(z)$. Therefore, the results of median resumation are obtainable by merely using a one-parameter transseries. Looking back to the homogeneous case as studied in \cite{bssv21}, we see that this happens since we have a Stokes line on the real axis. We illustrate in figure \ref{fig:medresum} how median resummation looks like, both in the $z$-plane and the $x$-plane. 

Note that understanding which value $\sigma_1$ takes on both sides of the positive real line is crucial for walking around the complex plane. Once this first step is clear, we next want to address the expansion along the negative axis. We would like to obtain it by walking along the complex plane, crossing the corresponding Stokes lines and resuming the result into the expansion above---again, see figure \ref{fig:HMpath} to have an idea on the path we are following. Let us explain what happens when turning around the complex $x$-plane: if we start a bit below the real line, where we have $(\sigma_1,\sigma_2)=\left(\frac12S^{(0)}_1,0\right)$\footnote{Notice that starting from a bit above would mean that we should start with $(\sigma_1,\sigma_2)=\left(-\frac12S^{(0)}_1,0\right)$ and then cross the Stokes line in the $0$ direction, which gives $\underline{\pmb{\mathbb{D}}}_0\left(-\frac12S^{(0)}_1,0\right)=\left(\frac12S^{(0)}_1,0\right)$, as expected.}, then we need to cross the Stokes line in the $-\pi$ direction, which yields the transseries parameters---recall equation \eqref{eq:stokes-transitions}--
\begin{equation}
	\underline{\pmb{\mathbb{D}}}_{-\pi}\left(\frac12S^{(0)}_1,0\right)=\left(0,+\infty\right).
\end{equation}
This is not surprising to us. Indeed, note that we have always worked in the $u_0=1$ normalisation. However, the Hastings--McLeod solution at $-\infty$ corresponds to the choice $u_0=0$: we even find the instanton action described in footnote \ref{fnote:otherA}. Thus, it seems consistent that our automorphism breaks down, ruling out the validity of the transseries after the Stokes transition. This result seems to indicate that singularities of the transition functions correspond to jumping sheets in the Riemann surface of the transseries. This statement is, of course, a mere observational guess, but we will later see that for the special function solutions, the same happens. Summarising, this solution is a special case in which one crosses the branch-cut that separates the double-cover associated with $u_0=\pm1$ and the single-cover $u_0=0$---see \cite{a99,t02}.

Another case we would like to study is the case of special-function solutions. These correspond to solutions of---again, we include the equation for the reader to know the normalisation we are working on--- 
\begin{equation}
	\label{eq:PIIforSpecialFunctSol}
	u''(z)= zu(z) + 2u^3(z)+ n-\frac 12,\qquad \qquad n\in\mathbb{Z},\,n\geq1,
\end{equation}
{\it i.e.}, $\alpha=n-\frac 12$. In particular, we would like to reproduce the results in \cite{d18} in the context of resurgence. Taking our transseries back to the original variable we have the following expansions in terms of $n$
\begin{align}
	\Phi_{(0,0)}(z)&=\frac {\sqrt{-z}}{\sqrt 2} - \left(-\frac 14 + \frac n2\right)\frac 1{(-z)}-\left( \frac 5{16 \sqrt 2} - \frac {3 n }{4 \sqrt 2} + \frac{3 n^2}{4 \sqrt 2}\right)\frac 1{(-z)^{\frac 52}}+\cdots, \\
	\Phi_{(1,0)}(z)&=2^{-\frac n2} (-z)^{-\frac {3n}2-1}-\rmi2^{-\frac n2-1}\left(-\frac 73+5n-\frac{7n^2}{2}\right)(-z)^{-\frac {3n}2-\frac 52}+\cdots.
\end{align}
Let us focus on the Stokes phenomenon as stated in \cite{d18}. In this particular case, we would like to construct a solution that is real on the real axis. For that, we have already mentioned that we need to perform a median transseries resummation \cite{as13}. At leading order, this is given by 
\begin{equation}
	\underline{\mathfrak{S}}^{-\frac 12}_0\Phi_{(0,0)}(z)=\Phi_{(0,0)}(z)-\frac 12 S^{(0)}_1(n)\rme^{\frac{2\sqrt2}3 (-z)^{\frac 32}}\Phi_{(1,0)}(z)+\cdots,
\end{equation}
which, if we expand and stick to the first power in both sectors, reads
\begin{equation}
	\frac{(-z)^{\frac 12}}{\sqrt{2}}\left(1 + \frac{\rmi 2^{\frac 52 (n-1)}\, \rme^{\frac{2\sqrt2}3 (-z)^{\frac 32}}\, (-z)^{-\frac{3}2(n-1)}}{\Gamma(n)}+\cdots\right),
\end{equation}
which reproduces the findings in \cite{d18}. However, we would like to reproduce a more general result. This is, we would like to give a correspondence between the special-function solutions and our transseries parameters. For this, we will follow the description in \cite{c06}. The equation \eqref{eq:PIIforSpecialFunctSol} has a one-parameter family of solutions that can be written in terms of Airy functions\footnote{Hence the name of special-function solutions.} and their derivatives. Define the combination 
\begin{equation}
	\label{eq:Airycomb}
	\varphi(z)=C_1\, \text{Ai}\left(-2^{-\frac13}z\right)+C_2\,\text{Bi}\left(-2^{-\frac13}z\right),
\end{equation}
and set the quantity
\begin{equation}
	\label{eq:Phi}
	\Phi(z)\equiv \frac{\varphi'(z)}{\varphi(z)}.
\end{equation}
This one-parameter family of solutions that can be written in terms of $\Phi(z)$\footnote{Notice that, since these solutions can be written solely in terms of the ratio of \eqref{eq:Airycomb}, the only significant parameter is the ratio $\frac{C_1}{C2}$. Consequently, this family is, indeed, a one-parameter family of solutions.} can be obtained from different methods: $n\times n$ determinants, the Toda equation, among others \cite{c06}. We will be interested in obtaining a conjecture based on comparison against the first orders---in particular, we will infer a pattern from the $n=1,\dots,4$ examples. For this reason, we present these solutions for the first $n$'s in table \ref{tab:specialfunctionalsolutions}. The procedure now will be as follows: since both the transseries and transasymptotic\footnote{While transseries consist on a exponential contributions with an asymptotic series attached to it, transasymptotics is a method to rearrange the transseries that is based on summing up all exponential contributions at a fixed power of $z$, so heuristically is like switching the summation order of $n,m\leftrightarrow g$. See more details on section \ref{sec:concl} and appendix \ref{app:Hratio}. See \cite{cc01,bd15} for an in-depth analysis.} series are expanded around $z\to-\infty$\footnote{Note that in order to go from our resurgent variable, expanded around $+\infty$ to the normalisation in \eqref{eq:PIIforSpecialFunctSol}, we need to take $z\to-z$, so our expansions at $+\infty$ correspond to expansions of \eqref{eq:PIIforSpecialFunctSol} at $-\infty$.}, we will expand the special-function solutions at $z\to-\infty$. We will then compare against our transseries to see if we can get a map or a correspondence between $\{C_1,C_2\}\leftrightarrow\{\sigma_1,\sigma_2\}$.

\begin{table}[hbt!]
	\begin{center}
		\begin{tabular}{||l|l||}
			\hline
			$n$ & $u(z)$ \\ [0.5ex] 
			\hline\hline
			&\\
			1 & $-\Phi(z)$  \\
			&\\
			2 & $\displaystyle{\Phi(z)-\frac1{2\Phi(z)^2+z}}$\\
			&\\
			3 & $\displaystyle{\frac{2z\Phi(z)^2+\Phi(z)+z^2}{4\Phi(z)^3 + 2z\Phi(z) - 1 }+\frac1{2\Phi(z)^2+z}}$ \\
			&\\
			4& $\displaystyle{-\frac3z-\frac{2z\Phi(z)^2+\Phi(z)+z^2}{4\Phi(z)^3 + 2z\Phi(z) - 1 }+\frac{48\Phi(z)^3 + 8z^2\Phi(z)^2 + 28z\Phi(z) + 4z^3 - 9}{z(8z\Phi(z)^4 + 16\Phi(z)^3 + 8z^2\Phi(z)^2 + 8z\Phi(z) + 2z^3 - 3)}}$ \\
			& \\
			\hline\hline
		\end{tabular}
		\caption{Special-function solutions of the inhomogeneous Painlev\'e II equation \eqref{eq:PIIforSpecialFunctSol} for the first $n=1,\dots,4$ in terms of the quantity \eqref{eq:Phi}.
			\label{tab:specialfunctionalsolutions}}
	\end{center}
\end{table}

From the resurgence point-of-view, the speciality of these solutions is reflected by the fact that, for $\alpha=n-\frac12$, $n\in\mathbb{Z}^+$ and $\sigma_2=0$, the transitions at the $-\pi$ direction leave $\sigma_2=0$. This means, that the $(0,m)$ transseries sectors are ``non-dynamic''. More concretely, when turning around the complex plane, while $\sigma_1$ jumps at the Stokes lines, $\sigma_2$ does not get modified. Coming back to the analysis we want to perform, we would like to obtain the leading-order in each of the four cases and compare against our transseries ansatz\footnote{In fact, using the resummed version of the transseries---this is, summing the $\log$ contributions---we can determine both $\sigma_1$ and $\sigma_2$ just from the leading coefficient in the exponential contribution. This happens since we have two conditions: the prefactor of the $z$ monomial, $2^{-\frac n2-2 \sigma_1 \sigma_2}\, (-1)^{-1 + \frac{3 n}2+6 \sigma_1 \sigma_2}$, and its exponent $-1 + \frac{3 n}2+6 \sigma_1 \sigma_2$.}. Let us denote by $u_n(z)$ each of the solutions in table \ref{tab:specialfunctionalsolutions}. Thus, the exponential contribution of the asymptotic expansion reads
\begin{comment}
	\begin{align}
		u_1(z)&\sim\rmi\,\frac{C_1}{C_2}\,\frac{(-z)^{\frac12}}{\sqrt 2}\rme^{-\frac{2\sqrt2}3(-z)^{\frac32}}+\dots;\\
		u_2(z)&\sim8\rmi\frac{C_1}{C_2}\,\frac{(-z)^2}2\rme^{-\frac{2\sqrt2}3(-z)^{\frac32}}+\dots;\\
		u_3(z)&\sim32\rmi\,\frac{C_1}{C_2}\,\frac{(-z)^{\frac72}}{2\sqrt 2}\rme^{-\frac{2\sqrt2}3(-z)^{\frac32}}+\dots;\\
		u_4(z)&\sim\frac{256}3\rmi\,\frac{C_1}{C_2}\,\frac{(-z)^5}{4}\rme^{-\frac{2\sqrt2}3(-z)^{\frac32}}+\dots.
	\end{align}
\end{comment}
\begin{equation}
	\displaystyle{
		\begin{array}{*7{>{\displaystyle}c}}
			u_1(z)&\sim&-&\frac{C_1}{C_2}&\frac{(-z)^{\frac12}}{\sqrt 2}&\rme^{-\frac{2\sqrt2}3(-z)^{\frac32}}&+\cdots,\\
			\\
			u_2(z)&\sim&-8&\frac{C_1}{C_2}&\frac{(-z)^2}2&\rme^{-\frac{2\sqrt2}3(-z)^{\frac32}}&+\cdots,\\
			\\
			u_3(z)&\sim&-32&\frac{C_1}{C_2}&\frac{(-z)^{\frac72}}{2\sqrt 2}&\rme^{-\frac{2\sqrt2}3(-z)^{\frac32}}&+\cdots,\\
			\\
			u_4(z)&\sim&-\frac{256}3&\frac{C_1}{C_2}&\frac{(-z)^5}{4}&\rme^{-\frac{2\sqrt2}3(-z)^{\frac32}}&+\cdots.
	\end{array}}
\end{equation}
Comparing the latter expressions with the contributions from our $(1,0)$ sector, one can infer the following map between the coefficients $\{C_1,C_2\}$ and the transseries parameters $\{\sigma_1,\sigma_2\}$,
\begin{align}
	\sigma_1&=-\rmi\,\frac{S^{(0)}_1}2\,\frac{C_1}{C_2},\\
	\sigma_2&=0.
\end{align}
Thus, we found a map between special-function solutions and our transseries. Note that for the particular case $C_2=0$, the expansions of the special-function solutions corresponds to taking the negative branch of the double-cover for the perturbative sector and no exponentials---indeed, when $C_2=0$, then the $C_1$'s cancel and we have a 0-parameter transseries. This is consistent with the fact that exponentials for this branch are ill-behaved at $z\to+\infty$. This somehow resembles what happened in the case of the Hastings--McLeod solution: whenever $\sigma_1\sim\frac{C_1}{C_2}\sim\infty$ and $\sigma_2=0$, this was an indication that we were changing sheets in the Riemann surface of the transseries. Let us however keep in mind that this is just a perhaps coincidental observation, and much more exploration is needed along this direction. 

\begin{figure}[hbt!]
	\centering
	\includegraphics[height=3in]{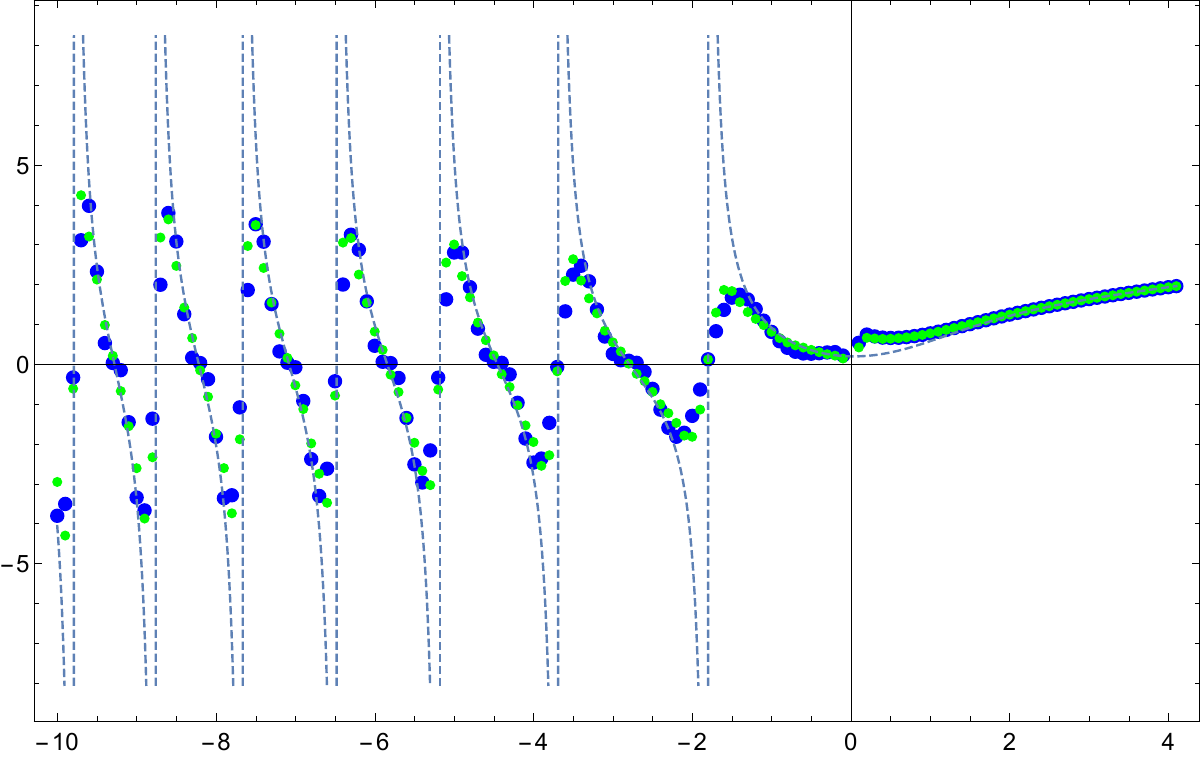}
	\caption{Agreement between the analytic special-function solution $u_1(z)$ (dashed line) and our transseries resummations (blue dots correspond to summing up to $(n,m)=(2,2)$ while green ones are associated with including up to the $(n,m)=(5,5)$ sector). Here, we have chosen $C_1=1=C_2$ and we have included 100 coefficients for the resummations in each of the sector. Note that the resummations do not quite capture the singularities of the solution, even when improving from $(n,m)=(2,2)$ to $(n,m)=(5,5)$, but it improves as $z$ goes to $-\infty$. It is expected that resummations will perform worse the closer we are to $z=0$, but theoretically if we included more coefficients and go to higher sectors, the approximation would start capturing these singularities.}
	\label{fig:specialfunctsolPIInequals1}
\end{figure}

Pay attention that this map is just a conjecture which we obtained at the level of the first exponential coefficient. Let us now explore this solution a bit more. We will start by noticing that, since $\sigma_2=0$, there are no other sectors than the $(k,0)$. This also means that we do not have any logarithm contributions and the whole problem {\it effectively becomes a one-parameter transseries}. In particular, not only we do not have $\sigma_2$ along the $z\to -\infty$ direction but furthermore, the transitions we undergo when rotating to the inverse direction do not turn on these sectors. In this sense, the special-function solutions are also special from the resurgence and transseries point of view. As an illustration, we present in figure \ref{fig:specialfunctsolPIInequals1} the resummations for the $u_1(z)$ solution.

The last special solution we would like to describe from the transseries point-of-view are the rational solutions of the inhomogeneous Painlev\'e II equation \cite{c06}
\begin{equation}
	\label{eq:PIIforRationalSol}
	u''(z)= zu(z) + 2u^3(z)+ n,\qquad \qquad n\in\mathbb{Z},\,n\geq0.
\end{equation}
We present in table \ref{tab:rationalsolutions} the first ones. 
\begin{table}[hbt!]
	\begin{center}
		\begin{tabular}{|l|l|}
			\hline
			$n$ & $u(z)$ \\ [0.5ex] 
			\hline\hline
			&\\
			0 & 0 \\
			&\\
			$1$ & $\displaystyle{-\frac1z}$  \\
			&\\
			$2$ & $\displaystyle{\frac1z-\frac{3z^2}{z^3+4}}$\\
			&\\
			$3$ & $\displaystyle{\frac{3z^2}{z^3+4}-\frac{6z^2(z^3+10)}{z^6+20z^3-80}}$ \\
			&\\
			\hline
		\end{tabular}
		\caption{Rational solutions of the inhomogeneous Painlev\'e II equation \eqref{eq:PIIforRationalSol} for the first $n=0,\dots,4$.		\label{tab:rationalsolutions}}
	\end{center}
\end{table}
It is easy to see that these solutions correspond to choosing $\sigma_1=0=\sigma_2$ on the single-cover determination. Indeed, we present in table \ref{tab:rationalsolutionsexpansions} the expansion of these solutions together with the first terms of the $(0,0)$ sector. It appears that these solutions need to keep the transseries parameters as they are when rotating around the complex plane. We do not have enough data to confirm this, but since the $(0,0)$ sector is an entire function for these values, we expect its Borel transform to be singularity free and therefore keep this condition at each Stokes line. It would be interesting to further study this problem in future work.
\begin{table}[hbt!]
	\begin{center}
		\begin{tabular}{|l|l|l|l|}
			\hline
			$n$ & $u(z)$ & Sol. Exp. & (0,0) Sector\\ [0.5ex] 
			\hline\hline
			0 & 0 & 0 & 0\\
			&&&\\
			$1$ & $\displaystyle{-\frac1z}$  & $\displaystyle{-\frac1z}$ & $\displaystyle{-\frac1z}$ \\
			&&&\\
			$2$ & $\displaystyle{\frac1z-\frac{3z^2}{z^3+4}}$ & $\displaystyle{\frac2z+\frac6{z^4}+\frac{12}{z^7}+\cdots}$ & $\displaystyle{\frac2z+\frac6{z^4}+\frac{12}{z^7}+\cdots}$ \\
			&&&\\
			$3$ & $\displaystyle{\frac{3z^2}{z^3+4}-\frac{6z^2(z^3+10)}{z^6+20z^3-80}}$ & $\displaystyle{\frac3z+\frac{24}{z^4}+\frac{408}{z^7}+\cdots}$& $\displaystyle{\frac3z+\frac{24}{z^4}+\frac{408}{z^7}+\cdots}$\\
			& & & \\
			\hline
		\end{tabular}
	\caption{Rational solutions of the inhomogeneous Painlev\'e II equation \eqref{eq:PIIforRationalSol} for the first $n=0,\dots,4$, their expansions around $z\to+\infty$, and the corresponding perturbative sector.		\label{tab:rationalsolutionsexpansions}}
	\end{center}
\end{table}
%%%%%%%%%%%%%%%%%%%%%%%%%%%%%%%%%%%%%%%%%%%%%%%%%%%%%%%%%%%%%%%%%
%%%%%%%%%%%%%%%%%%%%%%%%%%%%%%%%%%%%%%%%%%%%%%%%%%%%%%%%%%%%%%%%%
%\section{Inhomogeneous Painlev\'e II in Physical Systems}\label{sec:PIIinphys}
%%%%%%%%%%%%%%%%%%%%%%%%%%%%%%%%%%%%%%%%%%%%%%%%%%%%%%%%%%%%%%%%%
%%%%%%%%%%%%%%%%%%%%%%%%%%%%%%%%%%%%%%%%%%%%%%%%%%%%%%%%%%%%%%%%%

%%%%%%%%%%%%%%%%%%%%%%%%%%%%%%%%%%%%%%%%%%%%%%%%%%%%%%%%%%%%%%%%%
%%%%%%%%%%%%%%%%%%%%%%%%%%%%%%%%%%%%%%%%%%%%%%%%%%%%%%%%%%%%%%%%%
\section{Conclusions}\label{sec:concl}
%%%%%%%%%%%%%%%%%%%%%%%%%%%%%%%%%%%%%%%%%%%%%%%%%%%%%%%%%%%%%%%%%
%%%%%%%%%%%%%%%%%%%%%%%%%%%%%%%%%%%%%%%%%%%%%%%%%%%%%%%%%%%%%%%%%
In this paper we first studied the resurgent properties of the double-cover transseries solution of $q$-deformed Painlev\'e II equation, which appears in the description of $(2,4)$-super minimal string theory \cite{kms03}. In particular, we computed the transseries structure of its solution and the first Stokes data. The numerical computations are scarce, since the recurrence relation is computationally extremely heavy. In the second part of the paper, we studied the resurgent properties of the inhomogeneous Painlev\'e II equation. Starting from the description of the matrix model that double-scales to this equation, we ended up determining the first and simplest Stokes vector. After some preliminary analysis on the behaviour of the perturbative coefficients, we were motivated to change variables into what is commonly denoted by {\it resurgent variable}. There is no magic to this variable, but the fact that it makes the analysis cleaner. We have also {\it proposed} a structure for {\it all} Stokes vectors, based on the method of closed-form asymptotic and numerical observations \cite{bssv21}. This allowed us to determine transition functions of the transseries parameters at Stokes lines. Finally, we have dedicated a final section to a discussion of the Miura map and the study of the special solutions from the point-of-view of our transseries.

There are some interesting explorations to address in future work along the directions of the present paper. First of all, it is not clear how to relate (if possible) the transseries that one obtains upon choosing one out of the three\footnote{Note that, even though one would expect them to be 6 choices due to the power of 6 in the $q$-deformed Painlev\'e II equation, the reality is that we have 2 non-zero options (associated with the square-root determination we have used in this paper) and choosing the first coefficient to be 0, with multiplicity 4, that leads into a perturbative sector that starts with a fourth root.} possible choices for the first coefficient. In particular, we suspect that there should be some prescription telling us when we have to change our determination when going around the complex plane.

The real challenge, however, would be to resum the transseries into some known function or maybe a sum or product of them. A possible step in this direction is the aforementioned method of transasymptotics \cite{cc01,bd15}, which, in broad terms, consists of switching the sums in the transseries and finding the contributions at each $x^g$ factor \cite{bssv21}. The idea behind this procedure is that the equation is non-linear in the ``$g$-direction'', while being linear in the ``$n,m$-directions''. This means that, finding a close-form in the genus sum is much more complicated than trying to resum all instanton contributions at fixed genus. At leading orders, this is easy. For larger $g$, the amount of data that we need to resum increases considerably. Let us first discuss a couple of details on how to proceed. 

The first step is to resum the logarithmic sectors. Due to the property of the coefficients, the sectors $(n,m)[k]$ can be related to the $(n-k,m-k)[0]$. This has been already done in \cite{asv11,sv13,bssv21}. The result is
\begin{equation}
	\sum_{k=0}^{\infty} \sigma_1^k \sigma_2^k \log^k(x) \Phi_{(n+k,m+k)}^{[k]}(x)=x^{4(n-m)\sigma_1\sigma_2}\Phi_{(n,m)}^{[0]}(x).
\end{equation}
By denoting $\gamma=n-m$, it is then reasonable to expand our transseries as a sum of three contributions: $\gamma=0$, $\gamma\geq 1$ and $\gamma\leq -1$. Following \cite{bssv21}, we also define the quantities $\upmu=\sigma_1\sigma_2$ and $\xi_i=\sigma_i \, \rme^{(-1)^i\frac A x}\, x^{(-1)^i 4 \sigma_1 \sigma_2+\frac 12}$. With this in mind, one can follow the computations in the aforementioned papers to get
\begin{align}
	u(x,\sigma_1,\sigma_2)=&\sum_{g=0}^{\infty} x^g\left(\sum_{n=0}^g\upmu^n u^{(n, n)[0]}_{g-n}\right.
	+\nonumber\\
	&+\sum_{\gamma=1}^{\infty} \xi_1^\gamma \left(\sum_{m=0}^g \upmu^{2 m} u^{(\gamma + 2 m, 2 m)[0]}_{2 g - 2 m} +
	\sum_{m=0}^{g-1}\upmu^{2 m + 1} u^{(\gamma + 2 m + 1, 2 m + 1)[0]}_{2 g - 2 m - 2}\right)+\nonumber\\
	&\left.+\sum_{\gamma=1}^{\infty} \xi_2^\gamma \left(\sum_{m=0}^g\upmu^{2 m} u^{(2 m, 2 m + \gamma)[0]}_{2 g - 2 m} +
	\sum_{m=0}^{g-1}\upmu^{2 m + 1} u^{(2 m + 1, \gamma + 2 m + 1)[0]}_{2 g - 2 m - 2}\right)\right).
\end{align}
It is useful to introduce the notation
\begin{equation}
	\label{eq:transasymp}
	u(x,\sigma_1,\sigma_2)=\sum_{g=0}^{\infty} x^g\left(P^{(0)}_g+P^{(+)}_g+P^{(-)}_g\right).
\end{equation}
In this way, one can try to find a form for the three factors at each fixed genus. We will collect some results on appendix \ref{app:Hratio}, but at leading order we have 

\begin{align}
	u(x,\sigma_1,\sigma_2)&=1-\frac{2 \xi_1}{-2 + \xi_1}-\frac{2 \xi_2}{-2 + \xi_2}+\mathcal{O}(x)=\\
	&=1-\frac{2 \sqrt x \, \sigma_1 \, \rme^{-A/x}\, x^{-4 \sigma_1 \sigma_2}}{-2 + \sqrt x\, \sigma_1\, \rme^{-A/x}\, x^{-4 \sigma_1 \sigma_2}}-\frac{2 \sqrt x \,\sigma_2\, \rme^{A/x}\, x^{4 \sigma_1 \sigma_2}}{-2 + \sqrt x\, \sigma_2 \,\rme^{A/x}\, x^{4 \sigma_1 \sigma_2}}+\mathcal{O}(x).
\end{align}
Going again to the $z$ variable means taking $x=z^{-\frac 32}$ and adding a global $\sqrt z$. This is,
\begin{equation}
	u(z,\sigma_1,\sigma_2)=\sqrt z\left(1-\frac{2 \,\sigma_1\, \rme^{-Az^{\frac 32}}\, z^{-\frac 34+6 \sigma_1 \sigma_2}}{-2 + \sigma_1\, \rme^{-Az^{\frac 32}} \,z^{-\frac 34+6 \sigma_1 \sigma_2}}-\frac{2 \, \sigma_2\, \rme^{Az^{\frac 32}}\, z^{-\frac 34-6 \sigma_1 \sigma_2}}{-2 +  \sigma_2\, \rme^{Az^{\frac 32}}\, z^{-\frac 34-6 \sigma_1 \sigma_2}}+\mathcal{O}(x)\right).
\end{equation}

Another very interesting direction to explore is the connection with the isomonodromy problem. As computed in \cite{bssv21} for the case of the Painlev\'e I equation, all non-linear Stokes data can be {\it analytically computed} from the linear Stokes data associated with the isomonodromy system. The original computation for this case was done by Takei \cite{ta00} and translated to the two-parameter transseries by K. Iwaki in unpublished notes. Following these precedents, \cite{bssv21} contains an explicit derivation of the non-linear Stokes transition functions from the isomonodromy system. For the case of the inhomogeneous Painlev\'e II equation, one hopes that the computation is reproducible, albeit expecting some subtleties regarding the choice of $u_0$. In this sense, we point, {\it e.g.}, to \cite{i96,i03,ik03,jk92a,k04b,k94} and related works. These papers contain some discussion on the Riemann--Hilbert problem for the inhomogeneous Painlev\'e II equation and one also has a {\it plethora} of works by Kawai, Takei, Okamoto and co-authors, analysing the isomonodromic deformation of systems associated with the Painlev\'e transcendents ---in particular, we refer the reader to \cite{t02}, which we will quickly review and connect to our results in appendix \ref{app:iso}.

It would also be interesting to go one step further in the transasymptotics method and try to find the functions that generate the ratio of polynomials as given in appendix \ref{app:Hratio}. There have been important recent developments in the case of the Painlev\'e I equation \cite{sst}, while some other interesting preliminary results can be found in \cite{blmst17}. 

It is a well-established fact that the Painlev\'e equations appear in a {\it plethora} of Physical scenarios. In particular, we have mentioned the case of 2d quantum super-gravity \cite{dss90}. This is, $(2,4)$-super minimal string theory. There are two cases \cite{kms03}: the simplest one, in which the homogeneous Painlev\'e II equation describes the setup in which no RR-backgrounds nor charged D-branes are turned on; and the case of type 0A theory (alternatively, type 0B) with either $(N+q)$ D0-branes and $N$ anti D0-branes or insertions of even number of RR fields. The theory is described, in the large $N$ limit, by a $q$-deformed Painlev\'e II equation. The latter equation can be related to the inhomogeneous Painlev\'e II equation via the Miura map by taking $\alpha=\frac12\pm q$. Here, we want to point out the possible role of the special-function solutions. Recall that these are known to be given in terms of combinations of Airy functions and their derivatives. Furthermore, we showed that they correspond to specific choices of the transseries parameters. Via the Miura map, we know that these solutions can be translated into solutions of the $q$-deformed Painlev\'e II equation. In particular, these were the ones appearing whenever $\alpha\in \mathbb{Z}+\frac12$, which translates into $q\in \mathbb{Z}$. This is very interesting since \cite{kms03} gave several arguments to support that $q$ is somehow quantised. We first summarise the previous interpretations of the $q$ parameter. At the level of the matrix models, it appears in two ways. In that associated with the 0B theory, a unitary matrix model\footnote{Alternatively, a two-cut Hermitian matrix model.}, it plays the role of a mere integration constant. For the 0A theory, the matrix model is one of rectangular matrices of dimension $N\times(N+q)$ or vice-versa, so it appears as part of the dimension of the matrix. It is clear that for the case of the dimension, this quantity must be an integer. Surprisingly, at the level of the unitary matrix model, this quantity also exhibits conditions that it must be somehow ``quantised''.

At the level of string theory, the solution of \eqref{eq:PII_with_q2} was studied for large $z$ in both the positive and negative direction \cite{m08}. The $q$ parameter appears associated with specific powers of $z$ which were interpreted as worldsheet contributions with boundaries and handles. In particular, the $q$ dependence is solely determined by the number of boundaries at fixed number of handles. Now, making the connection with string theory, the interpretation is as follows: in one direction, where one has an expansion in $z^{-\frac32}$ one associates these powers with the number of boundaries in the spherical worldsheet, hence D-branes; in the other direction, with $z^{-3}$ expansion, one has spherical worldsheets with insertions of RR fields---each RR insertion has an associated weight of $z^{-\frac32}$, but they come in even numbers. Note that these expansions correspond to the different choices of $u_0^{(0,0)}$ that we have discussed throughout this paper. Therefore, we suspect that this type of open-closed string duality is associated with {\it jumping sheets on the Riemann surface of the transseries} rather than to Stokes transitions and the connection between both is rather mysterious, with the Hastings--McLeod solution being the closest example to these types of transitions. This analysis therefore highlights the importance of exploring when and how this ``jumping-sheets'' phenomenon happens. An interesting step in this direction would be to fully understand the correspondence between the physical theory and the transseries contributions. Some connections between the matrix model, D-branes and FZZT branes have already been made, albeit an understanding of the full non-perturbative picture is still missing. 

\newpage

\acknowledgments
The author would like to thank In\^es Aniceto, Salvatore Baldino, Alfredo Deaño, Ricardo Schiappa, Maximilian Schwick, for useful discussions, comments and/or correspondence. The author would further like to thank Paolo Gregori, Ricardo Schiappa, Maximilian Schwick, for comments on the first draft of this paper. The present paper represents the culmination of the author’s PhD thesis {\it Parametric Resurgences of the Second Painlev\'e Equation and the Case of 2 Dimensional Quantum Supergravity}, October 2022. The author would like to further thank the thesis advisor Ricardo Schiappa, and the members of the committee: Yoshitsugu Takei, Jos\'e Edelstein, João Pimentel Nunes, and Gabriel Lopes Cardoso. 

RV is supported by the LisMath Doctoral program and FCT-Portugal scholarship SFRH/ PD/BD/135514/2018. This research was supported in part by CAMGSD/IST-ID and via the FCT-Portugal grants UIDB/04459/2020, UIDP/04459/2020, PTDC/MAT-OUT/28784/2017. This paper is partly a result of the ERC-SyG project, Recursive and Exact New Quantum Theory (ReNewQuantum) funded by the European Research Council (ERC) under the European Union’s Horizon 2020 research and innovation programme, grant agreement 810573.

\newpage
%%%%%%%%%%%%%%%%%%%%%%%%%%%%%%%%%%%%%%%%%%%%%%%%%%%%%%%%%%%%%%%%%
%%%%%%%%%%%%%%%%%%%%%%%%%%%%%%%%%%%%%%%%%%%%%%%%%%%%%%%%%%%%%%%%%
\appendix
\section{Free Energy for $q$-Deformed and Inhomogeneous Painlev\'e II Equations}\label{app:freeEnergy}
%%%%%%%%%%%%%%%%%%%%%%%%%%%%%%%%%%%%%%%%%%%%%%%%%%%%%%%%%%%%%%%%%
%%%%%%%%%%%%%%%%%%%%%%%%%%%%%%%%%%%%%%%%%%%%%%%%%%%%%%%%%%%%%%%%%
In this appendix, we translate our findings into transseries for the free energy. Following the procedure in \cite{sv13}, one can perform the double-integration relating the free energy and the solution of the $q$-deformed Painlev\'e II equation. In terms of the normalisation \eqref{eq:PII_with_q2_final}, the relation reads\footnote{Notice, that in the normalisation \eqref{eq:original_PII_q2}, this equation would read 
	\begin{equation}
		F''(z)= \frac{u(z)^2}4,
	\end{equation}
	as in the original reference \cite{kms03}.}
\begin{equation}
	F''(z)= \frac{u(z)^2}2.
\end{equation}
By using standard notation for the transseries sectors---recall \eqref{eq:2ptransansatz}-\eqref{eq:asympsector}---the first couple of them associated with the free energy are given by
\begin{align}
	F^{(0,0)}(z)&=\frac{z^3}{12}+ \left(\frac{1}{16}-\frac{q^2}{4}\right) \log(z)+\left(-\frac{3}{256}+\frac{5q^2}{96}-\frac{q^4}{48}\right)z^{-3}+\cdots,\\
	F^{(1,0)}(z)&=\frac 14 z^{-\frac34} + \left(-\frac{65}{384} + \frac{q^2}{4}\right) z^{-\frac94}+ \left(\frac{19273}{73728} - \frac{221 q^2}{384} + \frac{q^4}{8}\right) z^{-\frac{15}4} +\cdots,\\
	F^{(1,1)}(z)&=-\frac{8}{3} z^{\frac32} + \left(-\frac{25}{48}+2q^2\right)z^{-\frac32} + \left(-\frac{6323}{12288} + \frac{437q^2}{192}-\frac{11}{12}\right) z^{-\frac92}+\cdots,\\
	F^{(2,0)}(z)&=\frac 1{16} z^{-\frac32} + \left(-\frac{59}{768} + \frac{q^2}{8} \right) z^{-3} + 
	\left(\frac{9745}{73728} - \frac{137q^2}{384}+ \frac{q^4}{8}\right)z^{-\frac92} +\cdots\\
	F^{(2,1)}(z)&=-\frac 38 z^{-\frac94}+ \left(
	\frac{51}{256} + \frac{23}8 \right)z^{-\frac{15}4}+\left(-\frac{4829}{49152}-\frac{8581 q^2}{768}+\frac{49q^4}{16}\right)z^{-\frac{21}4}+\cdots,\\
	F^{(2,2)}(z)&=6\log(z)+\left(-\frac{111}{32}+13q^2\right)z^{-3}+\left(-\frac{54507}{4096}+\frac{3755 q^2}{64}-\frac{94q^2}4\right)\alpha\,z^{-6}+\cdots.
\end{align}
Of course, Stokes data remains the same as long as one does not rescale either the sectors or the transseries parameters. For further discussion in this direction, we refer the reader to \cite{bssv21}.

Next, we would like to construct a double–scaled free energy from our inhomogeneous Painlev\'e II transseries. The connection is now given by
\begin{equation}
	F''(z)=-\frac14 u(z)^2.
\end{equation}
Following the procedure in \cite{sv13} to perform the double-integration of the squared transseries solution, we were able to obtain the following expressions for our free energy transseries sectors
\begin{align}
	F^{(0,0)}(z)&=-\frac{z^3}{24}+\frac \alpha 3 z^{\frac32}  - \left(\frac{1}{32}+\frac{\alpha}{8}\right) \log(z)+\left(\frac{\alpha}{24}(1+\alpha^2)\right)z^{-\frac32}+\nonumber\\
	&\hspace{7cm}+\left(\frac{3}{512}+\frac{19}{384}\alpha^2+\frac{\alpha^4}{48}\right)z^{-3}+\cdots,\\
	F^{(1,0)}(z)&=-\frac 18 z^{-\frac34+\frac{3\alpha}2} + \left(\frac{65}{768} - \frac{11}{64}\alpha + \frac{7}{64} \alpha^2\right) z^{-\frac94+\frac{3\alpha}2}\nonumber\\
	&+ \left(-\frac{19273}{147456} + \frac{1277}{3072}\alpha - \frac{2597}{6144} \alpha^2 + 
	\frac{59}{256} \alpha^3 - \frac{49}{1024} \alpha^4\right) z^{-\frac{15}4+\frac{3\alpha}2}+\cdots, \\
	F^{(1,1)}(z)&=\frac{4}{3} z^{\frac32} -\frac 32\alpha \log(z)+ \left(\frac{25}{96}+\frac 78\alpha\right)z^{-\frac32} +\alpha \left(\frac{179}{256} + \frac{41}{64}\alpha^2\right) z^{-3}+\cdots,\\
	F^{(2,0)}(z)&=-\frac 1{32} z^{-\frac32+3\alpha} + \left(\frac{59}{1536} - \frac 5{64} \alpha + \frac{7}{128}\alpha^2 \right) z^{-3+3\alpha} \nonumber\\&+ 
	\left(-\frac{9745}{147456} + \frac{2731}{12288}\alpha - \frac{1643}{6144} \alpha^2+\frac{181}{1024}\alpha^3-\frac{48}{1024}\alpha^4\right)z^{-\frac92+3\alpha}+\cdots, \\
	F^{(2,1)}(z)&=\left(\frac 3{16}+ \frac{17}{16}\alpha\right) z^{-\frac94+\frac{3\alpha}2}+ \left(
	-\frac{51}{512} - \frac{2641}{1536}\alpha + \frac{83}{32} \alpha^2-\frac{119}{128}\alpha^3 \right)z^{-\frac{15}4+\frac{3\alpha}2}+\cdots\\
	F^{(2,2)}(z)&=-3\log(z)+\frac{17}{4}\alpha z^{-\frac32}+\left(\frac{111}{64}+\frac{83}{16}\alpha^2\right)z^{-3}+\left(\frac{7805}{768}+\frac{835}{96}\alpha^2\right)\alpha\,z^{-\frac92}+\cdots,
\end{align}
which, of course, agree with the aforementioned work in the $\alpha\to0$ limit.

%%%%%%%%%%%%%%%%%%%%%%%%%%%%%%%%%%%%%%%%%%%%%%%%%%%%%%%%%%%%%%%%%
%%%%%%%%%%%%%%%%%%%%%%%%%%%%%%%%%%%%%%%%%%%%%%%%%%%%%%%%%%%%%%%%%
\section{More on the $\alpha$-Hermitian Matrix Model}\label{app:alphaMM}
%%%%%%%%%%%%%%%%%%%%%%%%%%%%%%%%%%%%%%%%%%%%%%%%%%%%%%%%%%%%%%%%%
%%%%%%%%%%%%%%%%%%%%%%%%%%%%%%%%%%%%%%%%%%%%%%%%%%%%%%%%%%%%%%%%%

\noindent\textbf{The $\boldsymbol{\alpha}$-Gaussian Matrix Model}: In order to properly normalize our computation of the one-instanton one-loop amplitude, we will have to find the corresponding corrections coming from the $\alpha$-Gaussian matrix model. The aim of this appendix is to obtain the first two orders of the free energy, $F^{\alpha\text{-G}}_0$ and $F^{\alpha\text{-G}}_1$, as well as the normalising term of the $\ell$-instanton contribution. These three quantities are the ones that will be necessary to complete the computation in subsection \ref{subsec:HMMSP}. The $\alpha$-Gaussian partition function for $N$ eigenvalues reads
\begin{equation}
	Z_{\alpha\text{-G}}(N)=\frac{1}{(2\pi)^N N!}\prod_{i=1}^N\int_{-\infty}^{+\infty}\text{d}\lambda_i\, \lambda_i^{2\alpha}\,\Delta^2(\lambda_i)\,\rme^{-\sum_{i=1}^{N}\frac{\lambda_i^2}{2g_{\text{s}}}}.
\end{equation}
Solving this model will give us both the term for normalizing the $\ell$-instanton contribution as well as the one for normalising the derivatives of the free energy. In order to do so, we use the method of orthogonal polynomials. The string equation and the initial condition read
\begin{equation}
	\label{eq:gaussianstringeq}
	h_n=g_{\text{s}}\left(n+\alpha\Big((-1)^{n+1}+1\Big)\right)h_{n-1}, \qquad  h_0=\frac{2^{\alpha-\frac 12}}\pi \, g_{\text{s}}^{\alpha+\frac 12}\; \Gamma\left(\frac 12 + \alpha\right).
\end{equation}
By plugging this recursion relation into Mathematica, one is able to obtain a general form for the $h_i$ coefficients. The partition function can be written in terms of these $h_i$'s as 
\begin{equation}
	Z_{\alpha\text{-G}}(N)=\prod_{i=0}^{N-1}h_i.
\end{equation}
Although result depends on the parity of $N$, the large-$N$ limit will not. For completeness, let us present both results 
\begin{align}
	&Z^{\mbox{\tiny even}}_{\alpha\text{-G}}(N)=\left\{\begin{array}{ll}
		\displaystyle{\frac{(2g_{\text{s}})^{2\alpha}}{(2\pi)^2}\Gamma\left(\alpha+\frac 12\right)\Gamma\left(\alpha+\frac 32\right)},	& \qquad N=2,\\
		& \\
		\displaystyle{\frac{(2g_{\text{s}})^{\frac 12 N(N+2\alpha)}}{(2\pi)^N}\frac{G_2\left(1+\frac N2\right)^2G_2\left(\frac{1+N}2+\alpha\right)G_2\left(\frac{3+N}2+\alpha\right)}{G_2\left(\alpha+\frac 12\right)^2\Gamma\left(\alpha+\frac 12\right)}},	& \qquad N\geq 4,
	\end{array}\right.\\
	&Z^{\mbox{\tiny odd}}_{\alpha\text{-G}}(N)=\left\{\begin{array}{ll}
		\displaystyle{\frac{(2g_{\text{s}})^{\alpha+\frac 12}}{2\pi}\; \Gamma\left(\frac 12 + \alpha\right)},	& \qquad N=1,\\ 
		& \\
		\displaystyle{\frac{(2g_{\text{s}})^{\frac 12 N(N+2\alpha)}}{(2\pi)^N}\frac{G_2\left(\frac {1+N}2\right)G_2\left(\frac{3+N}2\right)G_2\left(1+\frac{N}2+\alpha\right)^2}{G_2\left(\alpha+\frac 12\right)^2\Gamma\left(\alpha+\frac 12\right)}},	& \qquad N\geq 3,
	\end{array}\right.
\end{align}
where $G_2(z)$ stands for the Barnes $G$-function. 
In particular, we will be interested in four quantities: $Z_{\alpha\text{-G}}(1)$, $\partial_tF^{\alpha\text{-G}}_0(t)$, $\partial^2_tF^{\alpha\text{-G}}_0(t)$, and $\partial_tF^{\alpha\text{-G}}_1(t)$. The first one is easy to obtain
\begin{equation}
	\label{eq:ZalphaGauss(1)}
	Z_{\alpha\text{-G}}(1)=\frac{(2g_{\text{s}})^{\alpha+\frac 12}}{2\pi}\Gamma\left(\frac 12 + \alpha\right),
\end{equation}
while for the other two one has to take $F_{\alpha\text{-G}}=\log\left(Z_{\alpha\text{-G}}\right)$ and expand around $N\sim +\infty$\footnote{As stated before, in the large $N$ limit both even and odd expansions coincide, so we present the results together.}. Then, one takes the 't Hooft parameter $t=N\,g_{\text{s}}$ and obtains a power-series expansion in powers of $g_{\text{s}}$\footnote{Notice again that our expansion is {\it not} in even powers of $g_{\text s}$, since the $\alpha$-term of the matrix model is of order $g_{\text s}^{-1}$. One has to take this into account when comparing against other works in the literature.}
\begin{equation}
	F_{\alpha\text{-G}}\simeq\sum_{g=0}^{+\infty}F^{\alpha\text{-G}}_g(t)\, g_{\text{s}}^{g-2}.
\end{equation}
Hence, the first terms read
\begin{align}
	F^{\alpha\text{-G}}_0&=\frac{t^2}2 \left(-\frac{3}{2} +  \log(t)\right),\\
	F^{\alpha\text{-G}}_1&=t \alpha \left(-1 + \log(t)\right).
\end{align}
Note that, as a consistency check, these results agree with the ones of the classical Gaussian matrix model in the limit $\alpha\to 0$. Recall that in subsection \ref{subsec:HMMSP}, we have expanded our partition function in terms of {\it derivatives} of the free energy. Therefore, what we are really interested-in are the derivatives of these quantities with respect to the 't Hooft parameter. In particular, we have 
\begin{align}
	\partial^2_tF^{\alpha\text{-G}}_0(t)=\log(t);\qquad\qquad \partial_tF^{\alpha\text{-G}}_1(t)=\alpha \log(t).
\end{align}
This concludes the computation of the terms that will enter the normalisation of the $\alpha$-Hermitian matrix model in subsection \ref{subsec:HMMSP}.\\

\noindent\textbf{Loop Equations for the $\boldsymbol{\alpha}$-Hermitian Matrix Model}: In order to determine the large-$N$ expansion of the Hermitian matrix model, one of the key ingredients is finding the loop equations for the resolvent and the $s$-loop operators. The starting point is the partition function for this matrix model
\begin{equation}
	\int \text d M \det(M)^{2\alpha}\exp\left(-N \tr V(M)\right).
\end{equation}
The loop equations are obtained via inserting the resolvent operator 
\begin{equation}
	\left(\frac{1}{p-M}\right)_{ij}
\end{equation}
and taking derivatives with respect to the $k\ell$-matrix element $M_{k\ell}$. This is,
\begin{align}
	&\int \text dM \frac{\text d}{\text dM_{k\ell}}\left(\det(M)^{2\alpha}\left(\frac{1}{p-M}\right)_{ij}\rme^{-N\tr V(M)}\right)=\nonumber\\
	&\hspace{3.7cm}=\int  \text dM \left(2\alpha \,\frac{\frac{\text d}{\text dM_{k\ell}}\det(M)}{\det(M)}\left(\frac{1}{p-M}\right)_{ij}+\left(\frac{1}{p-M}\right)_{ik}\left(\frac{1}{p-M}\right)_{\ell j}\right.-\nonumber\\
	&\hspace{5.5cm}\left.-N\left(\frac{1}{p-M}\right)_{ij}(V'(M))_{\ell k} \right)\det(M)^{2\alpha}\,\rme^{-N\tr V(M)}.
\end{align}
The second and third term give the well-known result that one can find in \cite{a95,admn97}, so let us focus on the first term
\begin{align}
	\frac{\frac{\text d}{\text dM_{k\ell}}\det(M)}{\det(M)}&=\frac{\text d}{\text dM_{k\ell}}\log\det(M)=\frac{\text d}{\text dM_{k\ell}} \tr \log(M)=\\
	&=\frac{\text d}{\text dM_{k\ell}}\tr\sum_{n=1}^{\infty}\frac{(-1)^{n+1}}{n}(M-\mathbf{1})^n=\sum_{n=1}^{\infty}\frac{(-1)^{n+1}}{n}\tr\frac{\text d}{\text dM_{k\ell}}(M-\mathbf{1})^n\nonumber=
	\\&=\sum_{n=1}^{\infty}(-1)^{n+1}\tr\left((M-\mathbf{1})^{n-1}\frac{\text dM}{\text dM_{k\ell}}\right).
\end{align}
Now, the term $\frac{\text dM}{\text dM_{k\ell}}$ is a matrix with 1 at the $(k,l)$ position and 0 elsewhere. Any matrix multiplied by a matrix like this one will result into the following
\begin{equation}
	\left(\begin{array}{cccccc}
		a_{11}&\dots&\dots&a_{1N}\\
		\vdots&\ddots&\ddots&\vdots\\
		a_{N1}&\dots &\dots & a_{NN}
	\end{array}\right)
	\left(\begin{array}{cccccc}
		0&\dots&\dots&\dots&0\\
		\vdots&\ddots&1_{k\ell}&\ddots&\vdots\\
		0&\dots&\dots &\dots &0
	\end{array}\right)=\left(\begin{array}{ccccccc}
		0&\dots&0&a_{1k}&0&\dots&0\\
		0&\dots&0&a_{2k}&0&\dots&0\\
		\vdots&\ddots&\ddots&\vdots&\ddots&\ddots&\vdots\\
		0&\dots&0&a_{Nk}&0&\dots&0\\
	\end{array}\right),
\end{equation}
where there column lies at the $\ell$-th position. This means that, after taking the trace, one obtains
\begin{align}
	\sum_{n=1}^{\infty}(-1)^{n+1}\left[(M-\mathbf{1})^{n-1}\right]_{\ell k}&=\left[\sum_{n=0}^{\infty}(-1)^{n}(M-\mathbf{1})^{n}\right]_{\ell k}=\left[\frac{1}{\mathbf{1}+(M-\mathbf{1})}\right]_{\ell k}=\\
	&=\left[\frac{1}{M}\right]_{\ell k}.
\end{align}
Incorporating this into the previous equation, one obtains
\begin{align}
	&\int  \text dM \left(2\alpha \left(\frac{1}{M}\right)_{\ell k}\left(\frac{1}{p-M}\right)_{ij}+\left(\frac{1}{p-M}\right)_{ik}\left(\frac{1}{p-M}\right)_{\ell j}-N\left(\frac{1}{p-M}\right)_{ij}(V'(M))_{\ell k} \right)\times\nonumber\\
	&\hspace{2cm}\times\det(M)^{2\alpha}\,\rme^{-N\tr V(M)}.
\end{align}
Now, the loop equations are obtained by taking the quantity between brackets to vanish. Summing over $k=i$, $\ell=j$ and dividing by $N^2Z$ we get
\begin{equation}
	\frac{2\alpha}{N^2}\left\langle\tr \frac{1}{M(p-M)}\right\rangle+\left\langle\left(\frac{1}{N}\tr\frac{1}{p-M}\right)^2\right\rangle-\left\langle\frac{1}{N}\tr\frac{V'(M)}{p-M}\right\rangle=0.
\end{equation}
We introduce now the resolvent and the $s$-loop correlators by means of the application of the loop operators $\frac{\text d}{\text dV(p)}$, (see \cite{a95,admn97} for details) 
\begin{align}
	W(p)&=\left\langle \frac{1}{N}\tr \frac{1}{p-M}\right\rangle_{\mbox{\scriptsize conn}},\\
	W(p_1,\dots,p_s)&=\frac{\text d}{\text dV(p_s)}\dots\frac{\text d}{\text dV(p_{2})}W(p_1)=N^{s-2}\left\langle\tr \frac{1}{p_1-M}\dots\frac{1}{p_s-M}\right\rangle_{\mbox{\scriptsize conn}}.
\end{align}
With this, one has that 
\begin{align}
	\frac{1}{N}\left\langle\tr\frac{V'(M)}{p-M}\right\rangle&=\oint_\mathcal{C}\frac{\text d\omega}{2\pi \rmi}\,\frac{V'(\omega)}{p-\omega}\, W(\omega),\\
	\left\langle\left(\frac{1}{N}\tr\frac{1}{p-M}\right)^2\right\rangle&=\left\langle\left(\frac{1}{N}\tr\frac{1}{p-M}\right)^2\right\rangle_{\mbox{\scriptsize{conn}}}+\left\langle\left(\frac{1}{N}\tr\frac{1}{p-M}\right)\right\rangle_{\mbox{\scriptsize{conn}}}^2\nonumber\\
	&=\frac{\text d}{\text dV(p)}W(p)+(W(p))^2,\\
	\frac{2\alpha}{N^2}\left\langle\tr \frac{1}{M(p-M)}\right\rangle&=\frac{2\alpha}{N}\oint_\mathcal{C}\frac{\text d\omega}{2\pi \rmi}\,\frac{1}{\omega}\,\frac{1}{p-\omega}W(\omega),
\end{align}
where the $\mathcal{C}$ contour encloses all the cuts of the model. Note that one could have also taken
\begin{equation}
	\det(M)^{2\alpha}=\rme^{2\alpha\log(\det (M))}=\rme^{2\alpha\tr \log(M)}
\end{equation} 
and then consider this term as an added contribution to the potential of the form $-\frac{2\alpha}N\log(M)$, yielding the same result as the previous computation. Now, in order to solve the three-cut matrix model one can follow the same steps as in \cite{a95,admn97}. Let us only write down the loop equation for the resolvent factors. Power expanding $W(p)$ in powers of $N$\footnote{Notice that in our case we do not have an even power expansion, as expected from the computation of the inhomogeneous Painlev\'e II equation.}
\begin{align}
	W(p)\simeq\sum_{g=0}^{+\infty}\frac{1}{N^g}W_g(p),
\end{align}
we can insert this into our loop equation to obtain a recursion relation for the $W_g(p)$
\begin{align}
	\label{eq:MMrecrel}
	\left(\hat{\mathcal K}-2W_0(p)\right)W_g(p)=\sum_{g'=1}^{g-1}W_{g'}(p)\,W_{g-g'}(p)+\frac{\text d}{\text dV(p)}W_{g-2}(p)+2\alpha \oint_\mathcal{C}\frac{\text d\omega}{2\pi \rmi}\,\frac 1\omega\, \frac{1}{p-\omega}\,W_{g-1}(\omega),
\end{align}
where $\hat{\mathcal K}$ is the operator
\begin{align}
	\hat{\mathcal K}f(p)\equiv\oint_\mathcal{C}\frac{\text d\omega}{2\pi \rmi}\,\frac{V'(\omega)}{p-\omega}\,f(\omega).
\end{align}
\\

\noindent\textbf{Solving the Recursion Relation}: It is well established that the 0-th order of the resolvent is given by
\begin{align}
	W_0(p)=\frac 12 \oint_\mathcal{C}\frac{\text d\omega}{2\pi\rmi}\,\frac{V'(\omega)}{p-\omega}\,\frac{\phi^{(0)}(\omega)}{\phi^{(0)}(p)},
\end{align}
where 
\begin{equation}
	\phi^{(0)}(p)=\frac{1}{\sqrt{(p-x_1)\dots(p-x_{2s})}},\hspace{1cm}\mbox{for $s$ cuts,}
\end{equation}
and $x_{2i-1},x_{2i}$ are the endpoints of the $i$-th cut, $i=1,\dots,s$. In the case of our symmetric 3-cut case, the latter quantities take the form
\begin{align}
	W_0(p)=\frac 12\left(-\frac \lambda 6 \frac 1{\phi^{(0)}(p)}-p+\frac{\lambda}{6}p^3\right);\hspace{1cm}\phi^{(0)}(p)=\frac{1}{\sqrt{(p^2-x^2)(p^2-y^2)(p^2-z^2)}},
\end{align}
where the endpoints of the cuts $[-b,-a]\cup[-c,c]\cup[a,b]$ can be obtained from the endpoint equations
\begin{equation}
	\label{eq:endpointeq}
	\oint_{\mathcal C}\frac{\text d\omega}{2\pi\rmi}\,\omega^k\,V'(\omega)\,\phi^{(0)}(\omega)=2t\delta_{k,3}.
\end{equation}
The other quantity which is necessary to solve our problem is $W_1(p)$. For the case $g=1$, the recursion relation \eqref{eq:MMrecrel} reduces to
\begin{equation}
	\left(\hat{\mathcal K}-2W_0(p)\right)W_1(p)=2\alpha \oint_\mathcal{C}\frac{\text d\omega}{2\pi \rmi}\,\frac 1\omega\, \frac{1}{p-\omega}\,W_{0}(\omega).
\end{equation}
Plugging our expression for $W_0(p)$, we have that the right-hand side reads
\begin{align}
	2\alpha \oint_\mathcal{C}\frac{\text d\omega}{2\pi \rmi}\,\frac 1\omega\, \frac{1}{p-\omega}\, W_{0}(\omega)=\frac{\alpha\lambda}{6}\,\left(p^2-\frac{1}{p\phi^{(0)}}-\frac 12 (x^2+y^2+z^2)\right).
\end{align}
Therefore, we need to find terms that, upon being acted by the operator $\hat{\mathcal K}-2W_0(p)$, yield the desired result. The combination that does the trick is\footnote{In the general case, there is an extra term $\frac{\alpha}{2}\,(\sum_{i=1}^{6}x_i)\,p\,\phi^{(0)}(p)$ which disappears when the potential is symmetric.}
\begin{equation}
	W_1(p)=\alpha \left(p^2 \phi^{(0)}(p)  - \frac 1p\right),
\end{equation}
and one can just check this by deforming the contour to pick up the residues at infinity and $p$. In terms of the analysis in subsection \ref{subsec:HMMSP}, $W_1(p)$ determines the derivative of the free energy \cite{sv13}
\begin{equation}
	\partial_{s_i}F_1=(-1)^{i+1}\int_{x_{2i}}^{x_{2i+1}}\text dp\; W_1(p).
\end{equation}

These are all the required terms for the computation of the one-instanton one-loop amplitude of the matrix model performed in subsection \ref{subsec:HMMSP}.

\section{Some Technical Details}
\label{app:Hratio}
%%%%%%%%%%%%%%%%%%%%%%%%%%%%%%%%%%%%%%%%%%%%%%%%%%%%%%%%%%%%%%%%%
%%%%%%%%%%%%%%%%%%%%%%%%%%%%%%%%%%%%%%%%%%%%%%%%%%%%%%%%%%%%%%%%%

\textbf{Ratio of $\tilde{\boldsymbol{H}}$ Functions}: In order to obtain results from the closed-from asymptotics method, one of the most important steps is dealing with the ratio of $H$ and $\widetilde{H}$ functions \cite{bssv21}. These are the ones that encode most of the asymptotic growth information. Dealing with such expressions would blur the main text, so we dedicate this appendix to the computation of these ratios. We start by noticing that \cite{asv11}
\begin{align}
	\label{eq:Hexpansion}
	\widetilde{H}_p(g,s)&=\frac{\partial^p}{\partial g^p}\frac{\Gamma(g+1)}{s^{g+1}}=\nonumber\\
	&=\frac{\Gamma(g+1)}{s^{g+1}}\left(\delta_{p,0}+\Theta(p-1)\left(\widetilde{B}_s(g)+\partial_g\right)^{p-1}\widetilde{B}_s(g)\right),
\end{align}
with 
\begin{equation}
	\widetilde{B}_s(g)=\psi(g+1)-\log(s),
\end{equation}
where $\psi$ stands for the digamma function and $\Theta$ is the Heaviside step function. By choosing any primitive of $\widetilde{B}_s(g)$ one can rewrite the second row of equation \eqref{eq:Hexpansion} as
\begin{align}
	\left(\partial_g+\widetilde{B}_s(g)\right)^{p-1}\widetilde{B}_s(g)&=\exp\left(-\int\text dg\,\widetilde{B}_s(g)\right)\partial_g^{p-1}\left(\exp\left(\int \text dg\,\widetilde{B}_s(g)\right)\widetilde{B}_s(g)\right)=\\
	&=\exp\left(-\int\text dg\,\widetilde{B}_s(g)\right)\sum_{k=0}^{p-1}\left(\begin{array}{c}
		p-1\\
		k
	\end{array}\right)\partial_g^{k}\exp\left(\int \text dg\, \widetilde{B}_s(g)\right)\partial_g^{p-1-k}\widetilde{B}_s(g).
\end{align}
Now, we can make use of the Fa\`a di Bruno's formula to write the derivative in terms of Bell polynomials $B_{k,\ell}$ and complete Bell polynomials $B_k$
\begin{align}
	\partial^k_g\exp\left(\int\text dg\,\widetilde{B}_s(g)\right)&=\sum_{\ell=1}^{k}\exp^{(k)}\left(\int\text dg\,\widetilde{B}_s(g)\right)B_{k,\ell}\left(\widetilde{B}_s(g),\dots,\partial_g^{k-\ell}\widetilde{B}_s(g)\right)=\\
	&=\exp\left(\int \text dg\, \widetilde{B}_s(g)\right)B_k\left(\widetilde{B}_s(g),\dots,\partial_g^{k-1}\widetilde{B}_s(g)\right).
\end{align}
This finally gives us
\begin{align}
	\left(\partial_g+\widetilde{B}_s(g)\right)^{p-1}\widetilde{B}_s(g)&=\sum_{k=0}^{p-1}\left(\begin{array}{c}
		p-1\\
		k
	\end{array}\right)B_k\left(\widetilde{B}_s(g),\dots,\partial_g^{k-1}\widetilde{B}_s(g)\right)\partial_g^{p-1-k}\widetilde{B}_s(g)=\\
	&=B_p\left(\widetilde{B}_s(g),\dots,\partial_g^{p-1}\widetilde{B}_s(g)\right).
\end{align}
We can finally write the ratio of the $\widetilde{H}$ functions as\footnote{Notice that $B_0=1$.}
\begin{align}
	\label{eq:hratios}
	\frac{\widetilde{H}_p(g,s)}{\widetilde{H}_0(g,s)}&=B_p\left(\widetilde{B}_s(g),\dots,\partial_g^{p-1}\widetilde{B}_s(g)\right)\nonumber\\
	&=B_p\left(\psi(g+1)-\log(s),\psi^{(1)}(g+1),\dots,\psi^{(p-1)}(g+1)\right).
\end{align}
This is one of the ``key'' ingredients in the writing of closed-form asymptotics, so this compact expression will give us a way of easily generating analytical expressions for our Stokes data \cite{bssv21}.\\

\noindent\textbf{Some Results on Transasymptotics}: In this section, we present some results for the three quantities $P^{(0)}_g$, $P^{(+)}_g$, and $P^{(-)}_g$, in equation \eqref{eq:transasymp} for different values of $g$. Since these expressions get cumbersome very fast, we will only present the cases $g=0,1,2$. Note that the key feature of these coefficients is that each of them encodes {\it all} the exponential contributions at fixed $x^g$ power. Recall that we have used this approach because for the Hastings--McLeod solution there is no guarantee that there will be a hierarchy in the exponential weights. Hence, it is an educated guess to try to take them all into account at each order in $g$. Of course, one cannot ensure that and maybe the best approach is to prioritize going to higher genus instead of taking all exponentials, as we did for the positive real line. Or it could happen that we need both high genus and all exponentials in to reproduce it. In any case, it is an interesting exercise to perform this type of summations, so in the following we present some examples on how these quantities look like. For $g=0,1,2$, the coefficients in \eqref{eq:transasymp} read
\begin{align}
	P^{(0)}_0&=1,\\
	P^{(+)}_0&=-\frac{2 \xi_1}{-2 + \xi_1},\\
	P^{(-)}_0&=-\frac{2 \xi_2}{-2 + \xi_2},\\
	P^{(0)}_1&=-3\mu,\\
	P^{(+)}_1&=\frac{\xi_1}{48 (-2 + \xi_1)^2} (-34 + 72 \alpha - 168 \alpha^2 + 192 \mu - 
	1632 \alpha \mu - 3264 \mu^2 - 48 \xi_1 + 
	96 \alpha \xi_1 + 336 \mu \xi_1 + 3 \xi_1^2- \nonumber\\
	&\hspace{2.9cm}- 
	6 \alpha \xi_1^2 - 12 \mu \xi_1^2),\\
	P^{(-)}_1&=\frac{\xi_2}{48 (-2 + \xi_2)^2} (34 + 72 \alpha + 168 \alpha^2 + 192 \mu + 
	1632 \alpha \mu - 3264 \mu^2 + 48 \xi_2 + 
	96 \alpha \xi_2 + 336 \mu \xi_2 - 3 \xi_2^2 -\nonumber\\
	&\hspace{2.9cm}- 
	6 \alpha \xi_2^2 - 12 \mu \xi_2^2),\\
	P^{(0)}_2&=-\frac \alpha 2-\frac{51}{4}\mu^2,\\
	P^{(+)}_2&=\frac{\xi_1}{4608 (-2 + \xi_1)^3} (-3026+21600 \alpha - 25584 \alpha^2 + 35712 \alpha^3 - 
	14112 \alpha^4 + 88320 \mu-177600 \alpha \mu+\nonumber\\
	&\hspace{3.2cm} + 532224 \alpha^2 \mu - 
	274176 \alpha^3 \mu - 
	280320 \mu^2+2276352 \alpha \mu^2 - 1880064 \alpha^2 \mu^2+\nonumber\\
	&\hspace{3.2cm} + 2930688 \mu^3 -5326848 \alpha \mu^3- 5326848 \mu^4 - 
	17305 \xi_1  +46560 \alpha \xi_1 - 60024 \alpha^2 \xi_1 +\nonumber\\
	&\hspace{3.2cm}+ 
	26496 \alpha^3 \xi_1 - 7056 \alpha^4 \xi_1+ 145920 \mu \xi_1-500064 \alpha \mu \xi_1 + 309888 \alpha^2 \mu \xi_1-\nonumber\\
	&\hspace{3.2cm} - 
	137088 \alpha^3 \mu \xi_1- 938496 \mu^2 \xi_1 +1133568 \alpha \mu^2 \xi_1 - 940032 \alpha^2 \mu^2 \xi_1+\nonumber\\
	&\hspace{3.2cm} + 
	1354752 \mu^3 \xi_1-2663424 \alpha \mu^3 \xi_1 - 2663424 \mu^4 \xi_1 + 6714 \xi_1^2 -17244 \alpha \xi_1^2\nonumber+ \\
	&\hspace{3.2cm}+ 14904 \alpha^2 \xi_1^2 - 
	3024 \alpha^3 \xi_1^2 - 
	55224 \mu \xi_1^2 +120096 \alpha \mu \xi_1^2 - 35424 \alpha^2 \mu \xi_1^2\nonumber+\\
	&\hspace{3.2cm}+ 210816 \mu^2 \xi_1^2-117504 \alpha \mu^2 \xi_1^2- 
	117504 \mu^3 \xi_1^2 - 375 \xi_1^3+1290 \alpha \xi_1^3-\nonumber \\
	&\hspace{3.2cm}- 1332 \alpha^2 \xi_1^3 + 
	504 \alpha^3 \xi_1^3 + 4020 \mu \xi_1^3 -9648 \alpha \mu \xi_1^3 + 5904 \alpha^2 \mu \xi_1^3- 
	14976 \mu^2 \xi_1^3+\nonumber\\
	&\hspace{3.2cm}+19584 \alpha \mu^2 \xi_1^3 + 19584 \mu^3 \xi_1^3 - 9 \xi_1^4+36 \alpha \xi_1^4 - 36 \alpha^2 \xi_1^4 + 
	72 \mu \xi_1^4-144 \alpha \mu \xi_1^4-\nonumber \\
	&\hspace{3.2cm}- 144 \mu^2 \xi_1^4),\\
	P^{(-)}_2&=\frac{\xi_2}{4608 (-2 + \xi_2)^3} (-3026 - 88320 \mu - 
	280320 \mu^2 - 2930688 \mu^3 - 5326848 \mu^4 - 
	17305 \xi_2\nonumber- \\
	&\hspace{3.2cm} - 145920 \mu \xi_2- 938496 \mu^2 \xi_2 - 
	1354752 \mu^3 \xi_2 - 2663424 \mu^4 \xi_1 + 6714 \xi_2^2 \nonumber +\\
	&\hspace{3.2cm} + 
	55224 \mu \xi_2^2 + 210816 \mu^2 \xi_2^2+ 
	117504 \mu^3 \xi_2^2 - 375 \xi_2^3 - 4020 \mu \xi_2^3- \nonumber \\
	&\hspace{3.2cm}- 
	14976 \mu^2 \xi_2^3 - 19584 \mu^3 \xi_2^3 - 9 \xi_2^4 -
	72 \mu \xi_2^4 - 144 \mu^2 \xi_2^4),
\end{align}
where, in the last term, $P^{(-)}_2$, we have taken $\alpha=0$ for readability. In any case, the result for general $\alpha$ can be recovered from $P^{(+)}_2$ simply by taking $\alpha\to-\alpha$, $\mu\to -\mu$ and $\xi_1\to \xi_2$. As one can realise from these examples, there is a structure on these quantities. Namely, the $P^{(\pm)}_g$ are multiplied by a prefactor of $\frac{\xi_i}{(-2+\xi_i)^{g+1}}$ times a polynomial of order $2g$ in all $\mu$, $\xi$ and $\alpha$. Upon finding a generating function for those polynomials, one would have the {\it complete resummation} of the transseries, but this is an objective beyond the scope of this article.

%%%%%%%%%%%%%%%%%%%%%%%%%%%%%%%%%%%%%%%%%%%%%%%%%%%%%%%%%%%%%%%%%
%%%%%%%%%%%%%%%%%%%%%%%%%%%%%%%%%%%%%%%%%%%%%%%%%%%%%%%%%%%%%%%%%
\section{Connection to Isomonodromy}
\label{app:iso}
%%%%%%%%%%%%%%%%%%%%%%%%%%%%%%%%%%%%%%%%%%%%%%%%%%%%%%%%%%%%%%%%%
%%%%%%%%%%%%%%%%%%%%%%%%%%%%%%%%%%%%%%%%%%%%%%%%%%%%%%%%%%%%%%%%%
This final appendix is dedicated to understanding the relation between transition functions coming from the theory of isomonodromic deformations and our own, which were obtained by direct resurgence computations. In particular, we focus on the case of the single-cover determination of the transseries since the computations are much more straightforward. Unfortunately, these are also the ones for which we cannot construct the transition functions due to the scarce amount of data, but we still can check the first Stokes coefficient $N^{(1)}_1$. It would be very interesting to enlarge this analysis in future work, but for now, let us focus on this simpler example. In the context of the Painlev\'e I equation, the connection between the transitions functions for two-parameter transseries obtained from isomonodromy computations \cite{ta00} and the ones given by direct resurgent analysis was explored in \cite{bssv21,vv22}. 

For the case of the inhomogeneous Painlev\'e II equation, the isomonodromy method is based on the local reduction theorem---see \cite{kt98} and references therein. An explicit example on the application of this theorem can be found in \cite{t02} and we will reproduce this example here, but with some slight modifications. The idea behind this theorem is that one can reduce the transition functions of the Painlev\'e II equation to those of first one, at each Stokes line. The catch is that the normalisation of the two-parameter transseries in which these transition functions are valid is the one in which one considers one of the three turning points as the basis. In particular, we will be interested in the connection to our normalisation, in which $N^{(1)}_1=-\frac{(-2)^{\frac14}\sin(\pi\alpha)}{\sqrt\pi}$.

Following the notation in \cite{kt98,t02}, we define the Painlev\'e I and inhomogeneous Painlev\'e II equation as
\begin{equation}
	\lambda_{\text{I}}''(t)=\eta\,\,\Big(6\lambda_{\text{I}}^2(t)+t\Big),\qquad\qquad \lambda_{\text{II}}''(t)=\eta\,\,\Big(2\lambda_{\text{II}}^2(t)+t\,\lambda_{\text{II}}(t)+c\Big),
\end{equation}
with $\eta$ being a large parameter. The terms within the parenthesis are often defined as $F_{\text{J}}$, with $\text{J}=\text{I}, \text{II}$. The exponential terms for the two-parameter transseries ansatz have exponents $\pm\eta\phi_{\text{J}}$, where we have chosen the normalisation 
\begin{equation}
	\phi_{\text{I}}(t)=\int_{t_0=0}^t\sqrt{\frac{\partial F_{\text{I}}}{\partial\lambda_{\text{I}}}}\left(\lambda_{\text{I}}^{\text{\scriptsize pert}}(t),\,t\right),\qquad\qquad	\phi_{\text{II}}(t)=\int_{t_0=r}^t\sqrt{\frac{\partial F_{\text{I}}}{\partial\lambda_{\text{II}}}}\left(\lambda_{\text{II}}^{\text{\scriptsize pert}}(t),\,t\right),
\end{equation}
where $r$ is the turning point from which the Stokes line we are interested in emanates. Let us see how this theorem works: for the above normalisation of the second Painlev\'e II equation, connection formulae for the Painlev\'e I equation in the normalisation $t_0=0$ holds. If we would like to cross another Stokes line emanating from a different turning point, then we would need to change basis by means of 
\begin{equation}
	(\beta_1,\beta_2)=\left(-\tilde\beta_1\exp\left(\eta\int_{\text{\footnotesize initial t.p.}}^{\text{\footnotesize new t.p.}}\sqrt{\frac{\partial F_{\text{II}}}{\partial \lambda_{\text{II}}}}\text{d}t\right),-\tilde\beta_2\exp\left(-\eta\int_{\text{\footnotesize initial t.p.}}^{\text{\footnotesize new t.p.}}\sqrt{\frac{\partial F_{\text{II}}}{\partial \lambda_{\text{II}}}}\text{d}t\right)\right).
\end{equation} 
Let us explore the case for the single-cover: starting a bit above the negative real line (in the normalisation of \cite{t02}) the transition functions for the upper Stokes line are
\begin{align}
	\beta_1''\, \rme^{-\rmi \pi \frac{E''}4}\, \chi(E'') &= \beta_1' \, \rme^{-\rmi \pi \frac{E'}4}\,\chi(E');\\
	\rme^{\pi\rmi \frac{E''}2}- \rmi \beta_1'' \, \rme^{\rmi \pi \frac{E''}4}\, \chi(E'')&= \rme^{\rmi \pi \frac{E'}2},
\end{align}
where 
\begin{equation}
	E=-8\beta_1\beta_2,\qquad \qquad \chi(E)=\frac{\sqrt\pi 2^{1+\frac E4}}{\Gamma\left(1+\frac E4\right)}.
\end{equation}
The simplicity of the single-cover lies in the fact that the integration between turning points is very simple and yields factors of $\sim\pi\rmi\alpha$. Thus, the connection formula on the line below is obtained by changing
\begin{equation}
	(\beta_1,\beta_2)\to\left(-\beta_1 \rme^{2\pi\rmi\alpha},-\beta_2 \rme^{-2\pi\rmi\alpha}\right),
\end{equation}
yielding
\begin{align}
	\beta_1\, \rme^{-\rmi \pi \frac{E}4} \, \chi(E) &= \beta_1''\, \rme^{-\rmi \pi \frac{E''}4}\,\chi(E'');\\
	\rme^{\pi\rmi \frac{E}2}+ \rmi \beta_1\, \rme^{\rmi \pi \frac{E}4+2\pi\rmi\alpha}\, \chi(E)&= \rme^{\rmi \pi \frac{E}2}.
\end{align}	
Mapping to our normalisation implies the change of variables
\begin{equation}
	\beta_2 = -\frac{(1 - \rmi)}{2^{\frac34}}\,\rme^{-\rmi \pi \alpha_{\text{II}}}\, \sigma_1,\qquad \beta_1=-\frac{(1 - \rmi)}{2^{\frac34}}\, \rme^{\rmi \pi\alpha_{\text{II}}}\, \sigma_2.
\end{equation}
One can then easily solve the above systems of equations to find connection formulae between the upper and lower negative real axis for the single-cover solution of the inhomogeneous Painlev\'e II equation. It is interesting to series expand the result in order to compare against our first Stokes 
\begin{align}
	\sigma_1\sigma_2\mapsto\sigma_1\sigma_2 -\frac{(-2)^{\frac14}}{\sqrt\pi}\,\sin(\alpha_{\text{II}}\pi) \sigma_2+\cdots.
\end{align}
We see that, in fact, they coincide. This is a good hint although it would be interesting to understand the full connection. This is, either produce more resurgent data to obtain transition functions and fully compare against \cite{t02} or to perform the latter computations but for the case of the double-cover.

%%%%%%%%%%%%%%%%%%%%%%%%%%%%%%%%%%%%%%%%%%%%%%%%%%%%%%%%%%%%%%%%%
%%%%%%%%%%%%%%%%%%%%%%%%%%%%%%%%%%%%%%%%%%%%%%%%%%%%%%%%%%%%%%%%%
\newpage
%%%%%%%%%%%%%%%%%%%%%%%%%%%%%%%%%%%%%%%%%%%%%%%%%%%%%%%%%%%%%%%%%
%%%%%%%%%%%%%%%%%%%%%%%%%%%%%%%%%%%%%%%%%%%%%%%%%%%%%%%%%%%%%%%%%

%%%%% CLEAR DOUBLE PAGE!
\newpage{\pagestyle{empty}\cleardoublepage}

\addcontentsline{toc}{chapter}{\numberline{}\sf\bfseries{Bibliography}}

\begin{comment}
	
\bibliographystyle{plain}
%\bibliography{papers}

\begin{thebibliography}{100}

\bibitem{asv11} I.~Aniceto, R.~Schiappa, M.~Vonk,
\textit{The Resurgence of Instantons in String Theory},
Commun.\ Number\ Theor.\ Phys.\ \textbf{6} (2012) 339,
\texttt{arXiv:\arxivlink{1106.5922}[hep-th]}.
%%CITATION = 1106.5922;%%

\bibitem{bk90} E.~Br\'ezin, V.A.~Kazakov,
\textit{Exactly Solvable Field Theories of Closed Strings},
Phys.\ Lett.\ \textbf{B236} (1990) 144,
\texttt{DOI:\doilink{10.1016/0370-2693(90)90818-Q}}.
%%CITATION = PHLTA,B236,144;%%

\bibitem{ciny16} C--T. Chan, H. Irie, B. Niedner, C--H. Yeh, \textit{Wronskians, dualities and FZZT-Cardy branes}, J.\ Nucl.\ Phys.\ B \textbf{910} (2016) 55-177,
\texttt{DOI:\doilink{10.1016/j.nuclphysb.2016.06.014}}.

\bibitem{djmw92} S.~Dalley, C.~V.~Johnson, T.~Morris, A.~Watterstam,
\textit{Unitary Matrix Models and 2D Quantum Gravity (Virasoro constraints modified)},
Mod.\ Phys.\ Lett. \ \textbf{A7} (1992) 2753-2762, \texttt{arXiv:\arxivlink{hep-th/9206060}[hep-th]}.
%\texttt{DOI:\doilink{10.1142/S0217732392002226}}, 
%%CITATION = MPL,A7,2753;%%

\bibitem{djm92} S.~Dalley, C.~V.~Johnson, T.~Morris
\textit{Multicritical Complex Matrix Models and Nonperturbative 2-D Quantum Gravity},
Nucl.\ Phys.\ \textbf{B368} (1992) 625,
\texttt{DOI:\doilink{10.1016/0550-3213(92)90217-Y}}.
%%CITATION = NUPHA,B368,625;%%

\bibitem{d91} F.~David,
\textit{Phases of the Large $N$ Matrix Model and Nonperturbative Effects in 2D Gravity},
Nucl.\ Phys.\ \textbf{B348} (1991) 507,
\texttt{DOI:\doilink{10.1016/0550-3213(91)90202-9}}.
%%CITATION = NUPHA,B348,507;%%

\bibitem{d92} F.~David,
\textit{Nonperturbative Effects in Matrix Models and Vacua of Two-Dimensional Gravity},
Phys.\ Lett.\ \textbf{B302} (1993) 403,
\texttt{arXiv:\arxivlink{hep-th/9212106}}.
%%CITATION = HEP-TH/9212106;%%

\bibitem{d03} P.~Di~Francesco, P.H.~Ginsparg, J.~Zinn-Justin,
\textit{2D Gravity and Random Matrices},
Phys.\ Rept.\ \textbf{254} (1995) 1,
\texttt{arXiv:\arxivlink{hep-th/9306153}}.
%%CITATION = HEP-TH/9306153;%%

\bibitem{dss90} M.R.~Douglas, N.~Seiberg, S.H.~Shenker,
\textit{Flow and Instability in Quantum Gravity},
Phys.\ Lett.\ \textbf{B244} (1990) 381,
\texttt{DOI:\doilink{10.1016/0370-2693(90)90333-2}}.
%%CITATION = PHLTA,B244,381;%%

\bibitem{ds90} M.R.~Douglas, S.H.~Shenker,
\textit{Strings in Less Than One-Dimension},
Nucl.\ Phys.\ \textbf{B335} (1990) 635,
\texttt{DOI:\doilink{10.1016/0550-3213(90)90522-F}}.
%%CITATION = NUPHA,B335,635;%%

\bibitem{em08} B.~Eynard, M.~Mari\~no,
\textit{A Holomorphic and Background Independent Partition Function for Matrix Models and Topological Strings},
J.\ Geom.\ Phys.\ \textbf{61} (2011) 1181,
\texttt{arXiv:\arxivlink{0810.4273}[hep-th]}.
%%CITATION = ARXIV:0810.4273;%%

\bibitem{ez93} B.~Eynard, J.~Zinn-Justin,
\textit{Large Order Behavior of 2D Gravity Coupled to $d<1$ Matter},
Phys.\ Lett.\ \textbf{B302} (1993) 396,
\texttt{arXiv:\arxivlink{hep-th/9301004}}.
%%CITATION = PHLTA,B302,396;%%

\bibitem{gz91} P.H.~Ginsparg, J.~Zinn-Justin,
\textit{Large Order Behaviour of Nonperturbative Gravity},
Phys.\ Lett.\ \textbf{B255} (1991) 189,
\texttt{DOI:\doilink{10.1016/0370-2693(91)90234-H}}.
%%CITATION = PHLTA,B255,189;%%

\bibitem{gs21} P.~Gregori, R.~Schiappa,
\textit{From Minimal Strings towards Jackiw-Teitelboim Gravity: On their Resurgence, Resonance, and Black Holes}, \texttt{arXiv:\arxivlink{2108.11409}[hep-th]}.

\bibitem{gm90} D.J.~Gross, A.A.~Migdal,
\textit{Nonperturbative Two-Dimensional Quantum Gravity},
Phys.\ Rev.\ Lett.\ \textbf{64} (1990) 127,
\texttt{DOI:\doilink{10.1103/PhysRevLett.64.127}}.
%%CITATION = PRLTA,64,127;%%"

\bibitem{gm90b} D.J.~Gross, A.A.~Migdal,
\textit{A Nonperturbative Treatment of Two-Dimensional Quantum Gravity},
Nucl.\ Phys.\ \textbf{B340} (1990) 333,
\texttt{DOI:\doilink{10.1016/0550-3213(90)90450-R}}.
%%CITATION = NUPHA,B340,333;%%

\bibitem{kms03} I.~R.~Klebanov, J.~Maldacena, N.~Seiberg
\textit{Unitary and Complex Matrix Models as 1-d Type 0 Strings},
Commun. Math. Phys. \textbf{252} (2004) 275-323, \texttt{arXiv:\arxivlink{hep-th/0309168}}. %\texttt{DOI:\doilink{	10.1007/s00220-004-1183-7}},

\bibitem{lm93} R.~Lafrance, R.~C.~Myers,
\textit{Flows for rectangular matrix models},
Mod.\ Phys.\ Lett.\ \textbf{A9} (1994) 101-113, \texttt{arXiv:\arxivlink{9308113}[hep-th]}.
%\texttt{DOI:\doilink{10.1142/S0217732394000113}}, 
%%CITATION = MPL,A7,2753;%%

\bibitem{m06} M.~Mari\~no,
\textit{Open String Amplitudes and Large-Order Behavior in Topological String Theory},
JHEP\ \textbf{0803} (2008) 060,
\texttt{arXiv:\arxivlink{hep-th/0612127}}.
%%CITATION = HEP-TH/0612127;%%

\bibitem{m08} M.~Mari\~no,
\textit{Nonperturbative Effects and Nonperturbative Definitions in Matrix Models and Topological Strings},
JHEP\ \textbf{0812} (2008) 114,
\texttt{arXiv:\arxivlink{0805.3033}[hep-th]}.
%%CITATION = 0805.3033;%%

\bibitem{msw07} M.~Mari\~no, R.~Schiappa, M.~Weiss,
\textit{Nonperturbative Effects and the Large-Order Behavior of Matrix Models and Topological Strings},
Commun.\ Number\ Theor.\ Phys.\ \textbf{2} (2008) 349,
\texttt{arXiv:\arxivlink{0711.1954}[hep-th]}.
%%CITATION = ARXIV:0711.1954;%%

\bibitem{m90} T.~R.~Morris,
\textit{2-D Quantum Gravity, Multicritical Matter and Complex Matrices}, FERMILAB-PUB-90-136-T, \href{https://lss.fnal.gov/archive/1990/pub/Pub-90-136-T.pdf}{\texttt{https://lss.fnal.gov/archive/1990/pub/Pub-90-136-T.pdf}}.
%%CITATION = ARXIV:0809.2619;%%

\bibitem{ps09} S.~Pasquetti, R.~Schiappa,
\textit{Borel and Stokes Nonperturbative Phenomena in Topological String Theory and $c=\text{1}$ Matrix Models},
Ann.\ Henri\ Poincar\'e \textbf{11} (2010) 351,
\texttt{arXiv:\arxivlink{0907.4082}[hep-th]}.
%%CITATION = ARXIV:0907.4082;%%

\bibitem{sv13} R.~Schiappa, R.~Vaz,
\textit{The Resurgence of Instantons: Multi-Cut Stokes Phases and the Painlev\'e~II Equation},
Commun.\ Math.\ Phys.\ \textbf{330} (2014) 655,
\texttt{arXiv:\arxivlink{1302.5138}[hep-th]}.
%%CITATION = ARXIV:1302.5138;%%

\bibitem{zz01} A.~Zamolodchikov and Al.~Zamolodchikov, {\it Liouville field theory on a pseudosphere}, \texttt{arXiv:\arxivlink{hep-th/0101152}}.

\bibitem{fzz00} V.~Fateev, A.~B.~Zamolodchikov, A.~B.~Zamolodchikov, {\it Boundary Liouville Field Theory I. Boundary State and Boundary Two-point Function
}, \texttt{arXiv:\arxivlink{hep-th/0001012}}.

\bibitem{t00} J.~Teschner, {\it Remarks on Liouville theory with boundary}, PoS tmr2000 (2000) 041, \texttt{arXiv:\arxivlink{hep-th/0009138}}. %\texttt{DOI:\doilink{10.22323/1.006.0041}}

\bibitem{n04} Y. Nakayama, {\it Liouville Field Theory -- A decade after the revolution}, 	Int. J. Mod. Phys. A19, 2771-2930, (2004), \texttt{arXiv:\arxivlink{hep-th/0402009}}. %\texttt{DOI:\doilink{https://doi.org/10.1142/S0217751X04019500}}

\bibitem{t01} J. Teschner,  {\it Liouville theory revisited}
Class. Quantum Gravity, \textbf{18} (2001), p. R153, \texttt{arXiv:\arxivlink{hep-th/0104158}}. %\texttt{DOI:\doilink{10.1088/0264-9381/18/23/201}}

\bibitem{ss04} N. Seiberg, D. Shih,
\textit{Branes, Rings and Matrix Models in Minimal (Super)string Theory},
JHEP \textbf{0402} (2004) 021, \texttt{arXiv:\arxivlink{hep-th/0312170}[hep-th]}. %\texttt{DOI:\doilink{10.1088/1126-6708/2004/02/021}}.
%%CITATION = RU-90-47;%%

\bibitem{bssv21} S.~Baldino, R.~Schiappa, M.~Schwick, R.~Vega,
\textit{Resurgent Stokes Data for Painlev\'e Equations and 2D Quantum (Super) Gravity}, Commun.\ Number\ Theor.\ Phys.\ \textbf{17} (2023) 385, \texttt{arXiv:\arxivlink{2203.13726}}. %\texttt{DOI:\doilink{10.4310/CNTP.2023.v17.n2.a5}},

\bibitem{abs18} I.~Aniceto, G.~Ba\c sar, R.~Schiappa,
\textit{A Primer on Resurgent Transseries and Their Asymptotics}, Physics Reports
Volume \textbf{809}, 1-135, (2019) \texttt{arXiv:\arxivlink{1802.10441}[hep-th]}.  %\texttt{DOI:\doilink{10.1016/j.physrep.2019.02.003}},

%%CITATION = 1802.10441;%%

\bibitem{admn97} G.~Akemann, P.H.~Damgaard, U.~Magnea, S.M.~Nishigaki,
\textit{Multicritical Microscopic Spectral Correlators of Hermitian and Complex Matrices}, Nucl.Phys. \textbf{B519} (1998) 682-714, 
\texttt{arXiv:\arxivlink{hep-th/9712006}[hep-th]}. 
%%CITATION = 1802.10441;%%

\bibitem{a99} T. Aoki, 
\textit{Stokes geometry of Painleve equations with a large parameter}, RIMS Kokyuroku \textbf{1088} (1999) 39-54, \href{http://hdl.handle.net/2433/62844}{\texttt{http://hdl.handle.net/2433/62844}}.

\bibitem{blmst17} G.~Bonelli, O.~Lisovyy, K.~Maruyoshi, A.~Sciarappa, A.~Tanzini, {\it On Painlev\'e/gauge theory correspondence}. Lett. Math. Phys. \textbf{107}, 2359-2413 (2017), \texttt{arXiv:\arxivlink{1612.06235}}. %\texttt{DOI:\doilink{10.1007/s11005-017-0983-6}},

\bibitem{ck07} T.~Claeys, A.B.J.~Kuijlaars,
\textit{Universality in Unitary Random Matrix Ensables when the soft edge meets the hard edge},
Contemporary Mathematics \textbf{458} (2008), 265--280, \texttt{arXiv:\arxivlink{math-ph/0701003}[math-ph]}

\bibitem{ckv05} T.~Claeys, A.B.J.~Kuijlaars, M.~Vanlessen
\textit{Multi-critical unitary random matrix ensembles and the general Painlev\'e II equation},
Annals of Mathematics \textbf{168} (2008) 601--642, \texttt{arXiv:\arxivlink{math-ph/0508062}[math-ph]}

\bibitem{c03} P.A.~Clarkson
\textit{Painlev\'e Equations---Nonlinear Special Functions},
J.\ Comp.\ App.\ Math.\ \textbf{153} (2003) 127, \texttt{DOI:\doilink{10.1016/S0377-0427(02)00589-7}}.

\bibitem{c06} P.A.~Clarkson
\textit{Painlev\'e Equations---Nonlinear Special Functions},
in ``Orthogonal Polynomials and Special Functions'',
Lec.\ Notes\ Math.\ \textbf{1883} (2006) 331.

\bibitem{c19} P.A.~Clarkson
\textit{Open Problems for Painlev\'e Equations},
SIGMA\ \textbf{15} (2019) 006,
\texttt{arXiv:1901.10122[math.CA]}.
%%CITATION = 1901.10122;%%

\bibitem{d18} A.~Dea\~no,
\textit{Large $z$ Asymptotics for Special Function Solutions of Painlev\'e II in the Complex Plane}, SIGMA 14 (2018), 107, \texttt{arXiv:\arxivlink{1804.00563}[math.CA]}.
%\texttt{DOI:\doilink{10.3842/SIGMA.2018.107}},

\bibitem{fw14} B. Fornberg, J. A. C. Weideman, {\it A computational exploration of the second Painlev\'e equation}, Found. Comput. Math. \textbf{14} (2014), 985-1016, \texttt{DOI:\doilink{10.1007/s10208-013-9156-x}}.

\bibitem{fw15} B. Fornberg, J. A. C. Weideman, {\it A computational overview of the solution space of the imaginary Painlev\'e II equation}, Phys. D \textbf{309} (2015), 108-118, \texttt{DOI:\doilink{10.1016/j.physd.2015.07.008}}.

\bibitem{fw1b5} P. J. Forrester, N. S. Witte, {\it Painlev\'e II in random matrix theory and related fields}, Constr. Approx. \textbf{41} (2015), 589-613, \texttt{arXiv:\arxivlink{1210.3381}}.

\bibitem{gg18} A.~Grassi, J.~Gu,
\textit{Argyres--Douglas theories, Painlev\'e II and quantum mechanics},
J. High Energ. Phys. 2019, 60 (2019),
\texttt{DOI:\doilink{10.1007/JHEP02(2019)060}}, \texttt{arXiv:\arxivlink{1803.02320}[hep-th]}.

\bibitem{i96} A.R.~Its,
\textit{Connection Formulae for the Painlev\'e Transcendents},
in ``The Stokes Phenomenon and Hilbert's 16th Problem'' (1996) 139, \texttt{DOI:\doilink{10.1142/9789814531412}}.

\bibitem{i03} A.R.~Its,
\textit{The Riemann--Hilbert Problem and Integrable Systems},
Notices\ Amer.\ Math.\ Soc.\ \textbf{50} (2003) 1389, \href{https://www.ams.org/notices/200311/fea-its.pdf}{\texttt{https://www.ams.org/notices/200311/fea-its.pdf}}.

\bibitem{ik03} A. R.~Its, A. A.~Kapaev,
\textit{Quasi-Linear Stokes Phenomenon for the Second Painlev\'e Transcendent},
Nonlinearity\ \textbf{16} (2003) 363, \texttt{arXiv:\arxivlink{nlin/0108010} [nlin.SI]}. %\texttt{DOI:\doilink{10.1088/0951-7715/16/1/321}},

\bibitem{jk92a} N.~Joshi, M.D.~Kruskal,
\textit{The Painlev\'e Connection Problem: An Asymptotic Approach I},
Stud.\ App.\ Math.\ \textbf{86} (1992) 315, \texttt{DOI:\doilink{10.1002/sapm1992864315}}.

\bibitem{k04a} A. A.~Kapaev,
\textit{Quasi-Linear Stokes Phenomenon for the Painlev\'e First Equation},
J.\ Phys.\ \textbf{A37} (2004) 11149,
\texttt{arXiv:\arxivlink{nlin/0404026}[nlin.SI]}.
%%CITATION = NLIN/0404026;%%

\bibitem{k04b} A. A.~Kapaev,
\textit{Quasi-Linear Stokes Phenomenon for the Hastings--McLeod Solution of the Second Painlev\'e Equation},
\texttt{arXiv:\arxivlink{nlin/0411009}[nlin.SI]}.
%%CITATION = NLIN/0411009;%%

\bibitem{t02} Y.~Takei, {\it On an Exact WKB Approach to the Ablowitz-Segur's Connection Problem for the Second Painlev\'e Equation},
ANZIAM J. 44(2002), 111-119, \texttt{DOI:\doilink{10.1017/S1446181100007963}}

\bibitem{emm23} D. S. Eniceicu, R. Mahajan, C. Murdia, {\it Complex eigenvalue instantons and the Fredholm determinant expansion in the Gross-Witten-Wadia model}, \texttt{arXiv:\arxivlink{2308.06320}}.

\bibitem{msw08} M.~Mari\~no, R.~Schiappa, M.~Weiss,
\textit{Multi-Instantons and Multi-Cuts},
J.\ Math.\ Phys.\ \textbf{50} (2009) 052301,
\texttt{arXiv:\arxivlink{0809.2619}[hep-th]}.
%%CITATION = ARXIV:0809.2619;%%

\bibitem{a95} G.~Akemann,
\textit{Loop equations for multi-cut matrix models}, (1995)
\texttt{arXiv:\arxivlink{hep-th/9503185}[hep-th]}.

\bibitem{ackm93} J.~Ambj\o rn, L.~Chekhov, C.~F.~Kristjansen, Y.~Makeenko,
\textit{Matrix Model Calculations beyond the Spherical Limit},
Nucl.\ Phys.\ \textbf{B404} (1993) 127-172, \texttt{arXiv:\arxivlink{hep-th/9302014}[hep-th]}.%, \texttt{DOI:\doilink{10.1016/0550-3213(93)90476-6}}.
%%CITATION = NUPHA,B404,127;%%

\bibitem{hm80} S.P.~Hastings, J.B.~McLeod,
\textit{A boundary value problem associated with the second Painlev\'e transcendent and the Korteweg-de Vries equation},
Arch. Ration. Mech. Anal. \textbf{73}
(1980) 31,
\texttt{DOI:\doilink{https://doi.org/10.1007/BF00283254}}.

\bibitem{biz80} D.~Bessis, C.~Itzykson, J.~B.~Zuber, 
\textit{Quantum field theory techniques in graphical enumeration},
Advances in Applied Mathematics \textbf{1} (1980) 109-157,
\texttt{DOI:\doilink{10.1016/0196-8858(80)90008-1}}.

\bibitem{bo99} C.~M.~Bender, S.~A.~Orszag,	\textit{Advanced Mathematical Methods for Scientists and Engineers: Asymptotic Methods and Perturbation Theory}, \texttt{DOI:\doilink{10.1007/978-1-4757-3069-2}}.

\bibitem{s14} D.~Sauzin,
\textit{Introduction to 1-summability and resurgence},
\texttt{arXiv:\arxivlink{1405.0356}[math.DS]}.

\bibitem{gikm10} S.~Garoufalidis, A.~Its, A.~Kapaev, M.~Mari\~no,
\textit{Asymptotics of the Instantons of Painlev\'e~I},
Int.\ Math.\ Res.\ Notices\ \textbf{2012} (2012) 561,
\texttt{arXiv:\arxivlink{1002.3634}[math.CA]}.
%%CITATION = ARXIV:1002.3634;%%

\bibitem{mss22} M. Marino, R. Schiappa, M. Schwick, {\it New Instantons for Matrix Models}, \texttt{arXiv:\arxivlink{2210.13479}}.

\bibitem{sst23} R. Schiappa, M. Schwick, N. Tamarin, {\it All the D-Branes of Resurgence}, \texttt{arXiv:\arxivlink{2301.05214}}.

\bibitem{eggls23} B. Eynard, E. Garcia-Failde, P. Gregori, D. Lewanski, R. Schiappa, {\it Resurgent Asymptotics of Jackiw-Teitelboim Gravity and the Nonperturbative Topological Recursion}, \texttt{arXiv:\arxivlink{2305.16940}}.

\bibitem{hxz15} M.~Huang, S.-X.~Xu, L.~Zhang,
\textit{Location of Poles for the Hastings--McLeod Solution to the Second Painlev\'e Equation},
\texttt{arXiv:\arxivlink{1410.3338}[math.CA]}.

\bibitem{as13} I.~Aniceto, R.~Schiappa,
\textit{Nonperturbative Ambiguities and the Reality of Resurgent Transseries},
Commun.\ Math.\ Phys.\ \textbf{335} (2015) 183,
\texttt{arXiv:\arxivlink{1308.1115}[hep-th]}.
%%CITATION = 1308.1115;%%

\bibitem{cc01} O. Costin, R. D. Costin, {\it On the formation of singularities of solutions of nonlinear differential
systems in antistokes directions}, Invent. Math. \textbf{145}, 425-485 (2001), \texttt{DOI:\doilink{10.1007/s002220100153}}.

\bibitem{bd15} G. Ba\c{s}ar, G. Dunne, {\it Hydrodynamics, resurgence and trans-asymptotics}, (2015) \texttt{arXiv:\arxivlink{1509.05046}}.

\bibitem{ta00} Y.~Takei, {\it An Explicit Description of the Connection Formula for the First Painlev\'e Equation}, in ``Towards the Exact WKB Analysis of Differential Equations, Linear or Nonlinear'' (2000) Kyoto Univ. Press \textbf{204}, 271-296, ISBN: 9784876980895.

\bibitem{k94} A.V.~Kitaev,
\textit{Elliptic Asymptotics of the First and the Second Painlev\'e Transcendents},
Russ.\ Math.\ Surv.\ \textbf{49} (1994) 81, \texttt{DOI:\doilink{10.1070/RM1994v049n01ABEH002133}}.

\bibitem{sst} R. Schiappa, M. Schwick, N. Tamarin, {\it to appear}.

\bibitem{vv22} A. van Spaendonck, M. Vonk, {\it Painlev\'e I and exact WKB: Stokes phenomenon for two-parameter transseries}, \texttt{arXiv:\arxivlink{hep-th/2204.09062}}

\bibitem{kt98} T. Kawai, Y. Takei, {\it Algebraic Analysis of Singular Perturbation Theory}, Iwanami series in modern mathematics, \texttt{DOI:\doilink{10.1112/blms/bdn065}}.

\bibitem{gd75}
I. M. Gel'fand, L. A. Dikii, {\it Asymptotic behavior of the resolvent of Sturm-Liouville equations and the algebra of the Korteweg-De Vries equations}, Russ. Math. Surveys 30 (1975), \texttt{DOI:\doilink{10.1070/RM1975v030n05ABEH001522}}.


\end{thebibliography}

\begin{thebibliography}{10}
	\bibitem{a95}
	\label{a95}
	G.~Akemann,
	\textit{Loop equations for multi-cut matrix models},
	\texttt{arXiv:\arxivlink{hep-th/9503185}[hep-th]}.
	
	
	\bibitem{admn97}
	\label{admn97}
	G.~Akemann, P.H.~Damgaard, U.~Magnea, S.M.~Nishigaki,
	\textit{Multicritical Microscopic Spectral Correlators of Hermitian and Complex Matrices}, Nucl.Phys. \textbf{B519} (1998) 682-714, 
	\texttt{arXiv:\arxivlink{hep-th/9712006}[hep-th]}. 
	%%CITATION = 1802.10441;%%
	
	\bibitem{ackm93}
	J.~Ambj\o rn, L.~Chekhov, C.~F.~Kristjansen, Y.~Makeenko,
	\textit{Matrix Model Calculations beyond the Spherical Limit},
	Nucl.\ Phys.\ \textbf{B404} (1993) 127-172,
	\texttt{DOI:\doilink{10.1016/0550-3213(93)90476-6}}, \texttt{arXiv:\arxivlink{hep-th/9302014}[hep-th]}.
	%%CITATION = NUPHA,B404,127;%%
	
	\bibitem{abs18}
	I.~Aniceto, G.~Ba\c sar, R.~Schiappa,
	\textit{A Primer on Resurgent Transseries and Their Asymptotics}, Physics Reports
	Volume \textbf{809}, 1-135, (2019) \texttt{DOI:\doilink{10.1016/j.physrep.2019.02.003}},
	\texttt{arXiv:\arxivlink{1802.10441}[hep-th]}. 	
	%%CITATION = 1802.10441;%%
	
	\bibitem{as13}
	I.~Aniceto, R.~Schiappa,
	\textit{Nonperturbative Ambiguities and the Reality of Resurgent Transseries},
	Commun.\ Math.\ Phys.\ \textbf{335} (2015) 183,
	\texttt{arXiv:\arxivlink{1308.1115}[hep-th]}.
	%%CITATION = 1308.1115;%%
	
	\bibitem{asv11}
	\label{asv11}
	I.~Aniceto, R.~Schiappa, M.~Vonk,
	\textit{The Resurgence of Instantons in String Theory},
	Commun.\ Number\ Theor.\ Phys.\ \textbf{6} (2012) 339,
	\texttt{arXiv:\arxivlink{1106.5922}[hep-th]}.
	%%CITATION = 1106.5922;%%
	
	\bibitem{a99}	
	T. Aoki, 
	\textit{Stokes geometry of Painleve equations with a large parameter}, RIMS Kokyuroku \textbf{1088} (1999) 39-54, \href{http://hdl.handle.net/2433/62844}{\texttt{http://hdl.handle.net/2433/62844}}.
	
	
	\bibitem{bssv21}
	S.~Baldino, R.~Schiappa, M.~Schwick, R.~Vega,
	\textit{Resurgent Stokes Data for Painlev\'e Equations and 2D Quantum (Super) Gravity}, Commun.\ Number\ Theor.\ Phys.\ \textbf{17} (2023) 385, \texttt{DOI:\doilink{10.4310/CNTP.2023.v17.n2.a5}},  \texttt{arXiv:\arxivlink{2203.13726}}.
	
	
	\bibitem{bo99}
	C.~M.~Bender, S.~A.~Orszag,	\textit{Advanced Mathematical Methods for Scientists and Engineers: Asymptotic Methods and Perturbation Theory}, \texttt{DOI:\doilink{10.1007/978-1-4757-3069-2}}.
	
	
	
	\bibitem{mvb17}
	M.V.~Berry,
	\textit{Dingle's Self-Resurgence Formula},
	Nonlinearity \textbf{30} (2017) R25,
	\texttt{DOI:\doilink{10.1088/1361-6544/aa6c78}}.
	
	\bibitem{biz80}
	D.~Bessis, C.~Itzykson, J.~B.~Zuber, 
	\textit{Quantum field theory techniques in graphical enumeration},
	Advances in Applied Mathematics \textbf{1} (1980) 109-157,
	\texttt{DOI:\doilink{10.1016/0196-8858(80)90008-1}}.
	
	
	\bibitem{blmst17}
	G.~Bonelli, O.~Lisovyy, K.~Maruyoshi, A.~Sciarappa, A.~Tanzini, {\it On Painlev\'e/gauge theory correspondence}. Lett. Math. Phys. \textbf{107}, 2359–2413 (2017). \texttt{DOI:\doilink{10.1007/s11005-017-0983-6}}, \texttt{arXiv:\arxivlink{1612.06235}}.
	
	\bibitem{bde00}
	G.~Bonnet, F.~David, B.~Eynard,
	\textit{Breakdown of Universality in Multicut Matrix Models},
	J.\ Phys.\ \textbf{A33} (2000) 6739,
	\texttt{arXiv:\arxivlink{cond-mat/0003324}}.
	%%CITATION = COND-MAT/0003324;%%
	
	
	\bibitem{bk90}
	E.~Br\'ezin, V.A.~Kazakov,
	\textit{Exactly Solvable Field Theories of Closed Strings},
	Phys.\ Lett.\ \textbf{B236} (1990) 144,
	\texttt{DOI:\doilink{10.1016/0370-2693(90)90818-Q}}.
	%%CITATION = PHLTA,B236,144;%%
	
	\bibitem{bw73}
	C.M.~Bender, T.T.~Wu,
	\textit{Anharmonic oscillator. II. A study of perturbation theory in large order},
	Phys.\ Rev.\ D\ \textbf{7.6} (1973) 1620.
	
	\bibitem{ciny16}
	C--T. Chan, H. Irie, B. Niedner, C--H. Yeh, \textit{Wronskians, dualities and FZZT-Cardy branes}, J.\ Nucl.\ Phys.\ B \textbf{910} (2016) 55-177,
	\texttt{DOI:\doilink{10.1016/j.nuclphysb.2016.06.014}}.
	
	\bibitem{ck07}
	T.~Claeys, A.B.J.~Kuijlaars,
	\textit{Universality in Unitary Random Matrix Ensables when the soft edge meets the hard edge},
	Contemporary Mathematics \textbf{458} (2008), 265--280, \texttt{arXiv:\arxivlink{math-ph/0701003}[math-ph]}
	
	\bibitem{ckv05}
	T.~Claeys, A.B.J.~Kuijlaars, M.~Vanlessen
	\textit{Multi-critical unitary random matrix ensembles and the general Painlev\'e II equation},
	Annals of Mathematics \textbf{168} (2008) 601--642, \texttt{arXiv:\arxivlink{math-ph/0508062}[math-ph]}
	
	
	\bibitem{c03}
	P.A.~Clarkson
	\textit{Painlev\'e Equations---Nonlinear Special Functions},
	J.\ Comp.\ App.\ Math.\ \textbf{153} (2003) 127, \texttt{DOI:\doilink{10.1016/S0377-0427(02)00589-7}}.
	
	\bibitem{c06}
	P.A.~Clarkson
	\textit{Painlev\'e Equations---Nonlinear Special Functions},
	in ``Orthogonal Polynomials and Special Functions'',
	Lec.\ Notes\ Math.\ \textbf{1883} (2006) 331.
	
	\bibitem{c19}
	P.A.~Clarkson
	\textit{Open Problems for Painlev\'e Equations},
	SIGMA\ \textbf{15} (2019) 006,
	\texttt{arXiv:1901.10122[math.CA]}.
	%%CITATION = 1901.10122;%%
	
	\bibitem{cd20}
	N.~J.~Cleri, G.~V.~Dune,
	\textit{Resurgent Trans-Series for Generalized Hastings--McLeod Solutions},
	J. Phys.\ A \textbf{53} (2020) 35, \texttt{DOI:\doilink{10.1088/1751-8121/ab9fb8}}, \texttt{arXiv:\arxivlink{2002.06270}[math-ph]}.
	
	\bibitem{cch13}
	O.~Costin, R.D.~Costin, M.~Huang,
	\textit{Tronqu\'ee Solutions of the Painlev\'e Equation PI},
	\texttt{arXiv:\arxivlink{1310.5330}[math.CA]}.
	%%CITATION = 1310.5330;%%
	
	\bibitem{cd19}
	O.~Costin,G.V.~Dunne,
	\textit{Resurgent Extrapolation: Rebuilding a Function from Asymptotic Data. Painlev\'e I},
	J.\ Phys.\ A:\ Math.\ Theor.\ \textbf{52}\ (2019)\ 445205
	\texttt{arXiv:\arxivlink{1904.11593}[hep-th]}
	
	\bibitem{cesv13}
	R.~Couso-Santamar\'\i a, J.D.~Edelstein, R.~Schiappa, M.~Vonk,
	\textit{Resurgent Transseries and the Holomorphic Anomaly},
	Ann.\ Henri\ Poincar\'e\ \textbf{17} (2016) 331,
	\texttt{arXiv:\arxivlink{1308.1695}[hep-th]}.
	%%CITATION = 1308.1695;%%
	
	\bibitem{cesv14}
	R.~Couso-Santamar\'\i a, J.D.~Edelstein, R.~Schiappa, M.~Vonk,
	\textit{Resurgent Transseries and the Holomorphic Anomaly: Nonperturbative Closed Strings in Local $\BC\BP^2$},
	Commun.\ Math.\ Phys.\ \textbf{338} (2015) 285,
	\texttt{arXiv:\arxivlink{1407.4821}[hep-th]}.
	%%CITATION = 1407.4821;%%
	
	\bibitem{cms16}
	R.~Couso-Santamar\'\i a, M.~Mari\~no, R.~Schiappa,
	\textit{Resurgence Matches Quantization},
	J.\ Phys.\ \textbf{A50} (2017) 145402,
	\texttt{arXiv:\arxivlink{1610.06782}[hep-th]}.
	%%CITATION = 1610.06782;%%
	
	\bibitem{cs78}
	J.C.~Collins, D.E.~Soper,
	\textit{Large Order Expansion in Perturbation Theory},
	Annals\ Phys. \textbf{112} (1978) 209. 
	
	\bibitem{csv15}
	R.~Couso-Santamar\'\i a, R.~Schiappa, R.~Vaz,
	\textit{Finite N from Resurgent Large N},
	Annals\ Phys.\ \textbf{356} (2015) 1,
	\texttt{arXiv:\arxivlink{1501.01007}[hep-th]}.
	%%CITATION = 1501.01007;%%
	
	\bibitem{csv16}
	R.~Couso-Santamar\'\i a, R.~Schiappa, R.~Vaz,
	\textit{On Asymptotics and Resurgent Structures of Enumerative Gromov--Witten Invariants},
	Commun.\ Number\ Theor.\ Phys.\ \textbf{11} (2017) 707,
	\texttt{arXiv:\arxivlink{1605.07473}[math.AG]}.
	%%CITATION = 1605.07473;%%
	
	\bibitem{djmw92}
	S.~Dalley, C.~V.~Johnson, T.~Morris, A.~Watterstam,
	\textit{Unitary Matrix Models and 2D Quantum Gravity (Virasoro constraints modified)},
	Mod.\ Phys.\ Lett. \ \textbf{A7} (1992) 2753-2762,
	\texttt{DOI:\doilink{10.1142/S0217732392002226}}, \texttt{arXiv:\arxivlink{hep-th/9206060}[hep-th]}.
	%%CITATION = MPL,A7,2753;%%
	
	\bibitem{djm92}
	S.~Dalley, C.~V.~Johnson, T.~Morris
	\textit{Multicritical Complex Matrix Models and Nonperturbative 2-D Quantum Gravity},
	Nucl.\ Phys.\ \textbf{B368} (1992) 625,
	\texttt{DOI:\doilink{10.1016/0550-3213(92)90217-Y}}.
	%%CITATION = NUPHA,B368,625;%%
	
	\bibitem{d91}
	F.~David,
	\textit{Phases of the Large $N$ Matrix Model and Nonperturbative Effects in 2D Gravity},
	Nucl.\ Phys.\ \textbf{B348} (1991) 507,
	\texttt{DOI:\doilink{10.1016/0550-3213(91)90202-9}}.
	%%CITATION = NUPHA,B348,507;%%
	
	\bibitem{d92}
	F.~David,
	\textit{Nonperturbative Effects in Matrix Models and Vacua of Two-Dimensional Gravity},
	Phys.\ Lett.\ \textbf{B302} (1993) 403,
	\texttt{arXiv:\arxivlink{hep-th/9212106}}.
	%%CITATION = HEP-TH/9212106;%%
	
	\bibitem{d18}
	A.~Dea\~no,
	\textit{Large $z$ Asymptotics for Special Function Solutions of Painlev\'e II in the Complex Plane},
	\texttt{DOI:\doilink{10.3842/SIGMA.2018.107}}, \texttt{arXiv:\arxivlink{1804.00563}[math.CA]}.
	
	
	\bibitem{d14}
	E.~Delabaere,
	\textit{Resurgent Methods and the First Painlev\'e Equation},
	CIMPA Lecture Notes (2014).
	
	\bibitem{d16}
	E.~Delabaere,
	\textit{Divergent Series, Summability and Resurgence III: Resurgent Methods and the First Painlev\'e Equation},
	Lec.\ Notes\ Math.\ \textbf{2155} (2016).
	
	\bibitem{dgz93}
	P.~Di~Francesco,
	\textit{Rectangular Matrix Models and Combinatorics of Colored Graphs},
	Nucl.\ Phys.\ \textbf{B648} (2003) 461-496,
	\texttt{DOI:\doilink{10.1016/S0550-3213(02)00900-8}}, \texttt{arXiv:\arxivlink{cond-mat/0208037}}.
	%%CITATION = NUPHA,B648,461;%%
	
	\bibitem{d03}
	P.~Di~Francesco, P.H.~Ginsparg, J.~Zinn-Justin,
	\textit{2D Gravity and Random Matrices},
	Phys.\ Rept.\ \textbf{254} (1995) 1,
	\texttt{arXiv:\arxivlink{hep-th/9306153}}.
	%%CITATION = HEP-TH/9306153;%%
	
	\bibitem{dss90}
	M.R.~Douglas, N.~Seiberg, S.H.~Shenker,
	\textit{Flow and Instability in Quantum Gravity},
	Phys.\ Lett.\ \textbf{B244} (1990) 381,
	\texttt{DOI:\doilink{10.1016/0370-2693(90)90333-2}}.
	%%CITATION = PHLTA,B244,381;%%
	
	\bibitem{ds90}
	M.R.~Douglas, S.H.~Shenker,
	\textit{Strings in Less Than One-Dimension},
	Nucl.\ Phys.\ \textbf{B335} (1990) 635,
	\texttt{DOI:\doilink{10.1016/0550-3213(90)90522-F}}.
	%%CITATION = NUPHA,B335,635;%%
	
	\bibitem{e81}
	J.~\'Ecalle,
	\textit{Les Fonctions R\'esurgentes},
	Pr\'epub.\ Math.\ Universit\'e\ Paris-Sud\ \textbf{81-05} (1981), \textbf{81-06} (1981), \textbf{85-05} (1985).
	
	\bibitem{e08}
	B.~Eynard,
	\textit{Large $N$ Expansion of Convergent Matrix Integrals, Holomorphic Anomalies, and Background Independence},
	JHEP\ \textbf{0903} (2009) 003,
	\texttt{arXiv:\arxivlink{0802.1788}[math-ph]}.
	%%CITATION = ARXIV:0802.1788;%%
	
	\bibitem{em08}
	B.~Eynard, M.~Mari\~no,
	\textit{A Holomorphic and Background Independent Partition Function for Matrix Models and Topological Strings},
	J.\ Geom.\ Phys.\ \textbf{61} (2011) 1181,
	\texttt{arXiv:\arxivlink{0810.4273}[hep-th]}.
	%%CITATION = ARXIV:0810.4273;%%
	
	\bibitem{ez93}
	B.~Eynard, J.~Zinn-Justin,
	\textit{Large Order Behavior of 2D Gravity Coupled to $d<1$ Matter},
	Phys.\ Lett.\ \textbf{B302} (1993) 396,
	\texttt{arXiv:\arxivlink{hep-th/9301004}}.
	%%CITATION = PHLTA,B302,396;%%  
	
	\bibitem{fzz00}
	V.~Fateev, A.~B.~Zamolodchikov, A.~B.~Zamolodchikov, {\it Boundary Liouville Field Theory I. Boundary State and Boundary Two-point Function
	}, \texttt{arXiv:\arxivlink{hep-th/0001012}}.
	
	\bibitem{fw14}
	B. Fornberg, J. A. C. Weideman, {\it A computational exploration of the second Painlev´e equation}, Found. Comput. Math. \textbf{14} (2014), 985–1016.
	
	\bibitem{fw15}
	B. Fornberg, J. A. C. Weideman, {\it A computational overview of the solution space of the imaginary Painlev\'e II equation}, Phys. D \textbf{309} (2015), 108–118.
	
	\bibitem{fw1b5}
	P. J. Forrester, N. S. Witte, {\it Painlev\'e II in random matrix theory and related fields}, Constr. Approx. \textbf{41}	(2015), 589–613, \texttt{arXiv:\arxivlink{1210.3381}}.
	
	\bibitem{gikm10}
	S.~Garoufalidis, A.~Its, A.~Kapaev, M.~Mari\~no,
	\textit{Asymptotics of the Instantons of Painlev\'e~I},
	Int.\ Math.\ Res.\ Notices\ \textbf{2012} (2012) 561,
	\texttt{arXiv:\arxivlink{1002.3634}[math.CA]}.
	%%CITATION = ARXIV:1002.3634;%%
	
	\bibitem{gz91}
	P.H.~Ginsparg, J.~Zinn-Justin,
	\textit{Large Order Behaviour of Nonperturbative Gravity},
	Phys.\ Lett.\ \textbf{B255} (1991) 189,
	\texttt{DOI:\doilink{10.1016/0370-2693(91)90234-H}}.
	%%CITATION = PHLTA,B255,189;%%
	
	\bibitem{gg18}
	A.~Grassi, J.~Gu,
	\textit{Argyres--Douglas theories, Painlev\'e II and quantum mechanics},
	J. High Energ. Phys. 2019, 60 (2019),
	\texttt{DOI:\doilink{10.1007/JHEP02(2019)060}}, \texttt{arXiv:\arxivlink{1803.02320}[hep-th]}.
	
	\bibitem{gs21}
	P.~Gregori, R.~Schiappa,
	\textit{From Minimal Strings towards Jackiw-Teitelboim Gravity: On their Resurgence, Resonance, and Black Holes}, \texttt{arXiv:\arxivlink{2108.11409}[hep-th]}.
	
	\bibitem{gm90}
	D.J.~Gross, A.A.~Migdal,
	\textit{Nonperturbative Two-Dimensional Quantum Gravity},
	Phys.\ Rev.\ Lett.\ \textbf{64} (1990) 127,
	\texttt{DOI:\doilink{10.1103/PhysRevLett.64.127}}.
	%%CITATION = PRLTA,64,127;%%"
	
	\bibitem{gm90b}
	D.J.~Gross, A.A.~Migdal,
	\textit{A Nonperturbative Treatment of Two-Dimensional Quantum Gravity},
	Nucl.\ Phys.\ \textbf{B340} (1990) 333,
	\texttt{DOI:\doilink{10.1016/0550-3213(90)90450-R}}.
	%%CITATION = NUPHA,B340,333;%%
	
	\bibitem{hm80}
	S.P.~Hastings, J.B.~McLeod,
	\textit{A boundary value problem associated with the second Painlev\'e transcendent and the Korteweg-de Vries equation},
	Arch. Ration. Mech. Anal. \textbf{73}
	(1980) 31,
	\texttt{DOI:\doilink{https://doi.org/10.1007/BF00283254}}.	
	
	
	\bibitem{hxz15}
	M.~Huang, S.-X.~Xu, L.~Zhang,
	\textit{Location of Poles for the Hastings--McLeod Solution to the Second Painlev\'e Equation},
	\texttt{arXiv:\arxivlink{1410.3338}[math.CA]}.
	
	\bibitem{i96}
	A.R.~Its,
	\textit{Connection Formulae for the Painlev\'e Transcendents},
	in ``The Stokes Phenomenon and Hilbert's 16th Problem'' (1996) 139.
	
	\bibitem{i03}
	A.R.~Its,
	\textit{The Riemann--Hilbert Problem and Integrable Systems},
	Notices\ Amer.\ Math.\ Soc.\ \textbf{50} (2003) 1389.
	
	\bibitem{ik03}
	A.R.~Its, A.A.~Kapaev,
	\textit{Quasi-Linear Stokes Phenomenon for the Second Painlev\'e Transcendent},
	Nonlinearity\ \textbf{16} (2003) 363, \texttt{DOI:\doilink{10.1088/0951-7715/16/1/321}}, \texttt{arXiv:\arxivlink{nlin/0108010v1} [nlin.SI]}.
	
	\bibitem{jk01}
	N.~Joshi, A.V.~Kitaev,
	\textit{On Boutroux's Tritronqu\'ee Solutions of the First Painlev\'e Equation},
	Stud.\ Appl.\ Math.\ \textbf{107} (2001) 253,
	\texttt{DOI:\doilink{10.1111/1467-9590.00187}}.
	
	\bibitem{jk92a}
	N.~Joshi, M.D.~Kruskal,
	\textit{The Painlev\'e Connection Problem: An Asymptotic Approach I},
	Stud.\ App.\ Math.\ \textbf{86} (1992) 315.
	
	\bibitem{kny15}
	K.~Kajiwara, M.~Noumi, Y.~Yamada,
	\textit{Geometric Aspects of Painlev\'e Equations},
	J.\ Phys.\ \textbf{A50} (2017) 073001,
	\texttt{arXiv:\arxivlink{1509.08186}[nlin.SI]}.
	%%CITATION = 1509.08186;%%
	
	\bibitem{k88}
	A.A.~Kapaev,
	\textit{Asymptotic Behavior of the Solutions of the Painlev\'e Equation of the First Kind},
	Differ.\ Uravn.\ \textbf{24} (1988) 1684; Differ.\ Equ.\ \textbf{24} (1988) 1107.
	
	\bibitem{k04a}
	A.A.~Kapaev,
	\textit{Quasi-Linear Stokes Phenomenon for the Painlev\'e First Equation},
	J.\ Phys.\ \textbf{A37} (2004) 11149,
	\texttt{arXiv:\arxivlink{nlin/0404026}[nlin.SI]}.
	%%CITATION = NLIN/0404026;%%
	
	\bibitem{k04b}
	A.A.~Kapaev,
	\textit{Quasi-Linear Stokes Phenomenon for the Hastings--McLeod Solution of the Second Painlev\'e Equation},
	\texttt{arXiv:\arxivlink{nlin/0411009}[nlin.SI]}.
	%%CITATION = NLIN/0411009;%%
	
	\bibitem{kk93}
	A.A.~Kapaev, A.V.~Kitaev,
	\textit{Connection Formulae for the First Painlev\'e Transcendent in the Complex Domain},
	Lett.\ Math.\ Phys.\ \textbf{27} (1993) 243.
	
	\bibitem{kt98}
	T. Kawai, Y. Takei, {\it Algebraic Analysis of Singular Perturbation Theory}, Iwanami series in modern mathematics, ISBN: 0-8218-3547-5
	
	
	\bibitem{k94}
	A.V.~Kitaev,
	\textit{Elliptic Asymptotics of the First and the Second Painlev\'e Transcendents},
	Russ.\ Math.\ Surv.\ \textbf{49} (1994) 81.
	
	\bibitem{kms03}
	I.~R.~Klebanov, J.~Maldacena, N.~Seiberg
	\textit{Unitary and Complex Matrix Models as 1-d Type 0 Strings},
	Commun. Math. Phys. \textbf{252} (2004) 275-323, \texttt{DOI:\doilink{	10.1007/s00220-004-1183-7}}, 
	\texttt{arXiv:\arxivlink{hep-th/0309168}}.
	
	\bibitem{lm93}
	R.~Lafrance, R.~C.~Myers,
	\textit{Flows for rectangular matrix models},
	Mod.\ Phys.\ Lett.\ \textbf{A9} (1994) 101-113,
	\texttt{DOI:\doilink{10.1142/S0217732394000113}}, \texttt{arXiv:\arxivlink{9308113}[hep-th]}.
	%%CITATION = MPL,A7,2753;%%
	
	\bibitem{m06}
	M.~Mari\~no,
	\textit{Open String Amplitudes and Large-Order Behavior in Topological String Theory},
	JHEP\ \textbf{0803} (2008) 060,
	\texttt{arXiv:\arxivlink{hep-th/0612127}}.
	%%CITATION = HEP-TH/0612127;%%
	
	\bibitem{m08}
	M.~Mari\~no,
	\textit{Nonperturbative Effects and Nonperturbative Definitions in Matrix Models and Topological Strings},
	JHEP\ \textbf{0812} (2008) 114,
	\texttt{arXiv:\arxivlink{0805.3033}[hep-th]}.
	%%CITATION = 0805.3033;%%
	
	\bibitem{msw07}
	M.~Mari\~no, R.~Schiappa, M.~Weiss,
	\textit{Nonperturbative Effects and the Large-Order Behavior of Matrix Models and Topological Strings},
	Commun.\ Number\ Theor.\ Phys.\ \textbf{2} (2008) 349,
	\texttt{arXiv:\arxivlink{0711.1954}[hep-th]}.
	%%CITATION = ARXIV:0711.1954;%%
	
	\bibitem{msw08}
	M.~Mari\~no, R.~Schiappa, M.~Weiss,
	\textit{Multi-Instantons and Multi-Cuts},
	J.\ Math.\ Phys.\ \textbf{50} (2009) 052301,
	\texttt{arXiv:\arxivlink{0809.2619}[hep-th]}.
	%%CITATION = ARXIV:0809.2619;%%
	
	\bibitem{m90}
	T.~R.~Morris,
	\textit{2-D Quantum Gravity, Multicritical Matter and Complex Matrices}, FERMILAB-PUB-90-136-T.
	%%CITATION = ARXIV:0809.2619;%%
	
	\bibitem{n04}
	Y. Nakayama, {\it Liouville Field Theory -- A decade after the revolution}, 	Int. J. Mod. Phys. A19, 2771-2930, (2004), \texttt{DOI:\doilink{https://doi.org/10.1142/S0217751X04019500}} \texttt{arXiv:\arxivlink{hep-th/0402009}}.
	
	\bibitem{olbc10}
	F.W.J.~Olver, D.W.~Lozier, R.F.~Boisvert, C.W.~Clark,
	\textit{NIST Handbook of Mathematical Functions},
	Cambridge University Press (2010).
	
	\bibitem{ps09}
	S.~Pasquetti, R.~Schiappa,
	\textit{Borel and Stokes Nonperturbative Phenomena in Topological String Theory and $c=\text{1}$ Matrix Models},
	Ann.\ Henri\ Poincar\'e \textbf{11} (2010) 351,
	\texttt{arXiv:\arxivlink{0907.4082}[hep-th]}.
	%%CITATION = ARXIV:0907.4082;%%
	
	
	\bibitem{s14}
	D.~Sauzin,
	\textit{Introduction to 1-summability and resurgence},
	\texttt{arXiv:\arxivlink{1405.0356}[math.DS]}.
	
	
	\bibitem{sv13}
	R.~Schiappa, R.~Vaz,
	\textit{The Resurgence of Instantons: Multi-Cut Stokes Phases and the Painlev\'e~II Equation},
	Commun.\ Math.\ Phys.\ \textbf{330} (2014) 655,
	\texttt{arXiv:\arxivlink{1302.5138}[hep-th]}.
	%%CITATION = ARXIV:1302.5138;%%
	
	\bibitem{ss04}
	N. Seiberg, D. Shih,
	\textit{Branes, Rings and Matrix Models in Minimal (Super)string Theory},
	JHEP \textbf{0402} (2004) 021. \texttt{DOI:\doilink{10.1088/1126-6708/2004/02/021}}.
	%%CITATION = RU-90-47;%%
	
	\bibitem{s90}
	S.H.~Shenker,
	\textit{The Strength of Nonperturbative Effects in String Theory},
	in ``The Large $N$ Expansion in Quantum Field Theory and Statistical Physics'' (1990) 809.
	%%CITATION = RU-90-47;%%
	
	
	\bibitem{ta00}	
	Y.~Takei, {\it An Explicit Description of the Connection Formula for the First Painlev\'e Equation}, in
	“Towards the Exact WKB Analysis of Differential Equations, Linear or Nonlinear” (2000) Kyoto Univ. Press \textbf{204}, 271-296, ISBN: 9784876980895.
	
	\bibitem{t02}	
	Y.~Takei, {\it On an Exact WKB Approach to the Ablowitz-Segur's Connection Problem for the Second Painlev\'e Equation},
	ANZIAM J. 44(2002), 111-119
	
	\bibitem{t00}
	J.~Teschner, {\it Remarks on Liouville theory with boundary}, PoS tmr2000 (2000) 041, \texttt{DOI:\doilink{10.22323/1.006.0041}} \texttt{arXiv:\arxivlink{hep-th/0009138}}.
	
	\bibitem{t01}
	J. Teschner,  {\it Liouville theory revisited}
	Class. Quantum Gravity, \textbf{18} (2001), p. R153, \texttt{DOI:\doilink{10.1088/0264-9381/18/23/201}} \texttt{arXiv:\arxivlink{hep-th/0104158}}.
	
	\bibitem{vv22}
	A. van Spaendonck, M. Vonk, {\it Painlev\'e I and exact WKB: Stokes phenomenon for two-parameter transseries}, \texttt{arXiv:\arxivlink{hep-th/2204.09062}}
	
	\bibitem{zz01}
	A.~Zamolodchikov and Al.~Zamolodchikov, {\it Liouville field theory on a pseudosphere}, \texttt{arXiv:\arxivlink{hep-th/0101152}}.
\end{thebibliography}

\end{comment}
\end{document}